\documentclass[3p,sort&compress]{elsarticle}

\usepackage[T1]{fontenc}
\usepackage{lineno}
\usepackage[hidelinks]{hyperref}
\modulolinenumbers[5]
\usepackage{subcaption}
\usepackage{array, multirow, booktabs, bm, color, todonotes, amsmath, hhline}

\numberwithin{equation}{section}
\newcommand\numberthis{\addtocounter{equation}{1}\tag{\theequation}}

\makeatletter
\renewcommand\p@subfigure{\thefigure\,}

\DeclareCaptionLabelFormat{mysublabelfmt}{(\alph{sub\@captype})}
\makeatother

\journal{Physics Reports}

\begin{document}

\begin{frontmatter}

\title{Electrical Detection of Magnetization Dynamics via Spin Rectification Effects}

\author{Michael Harder\corref{memail}}
\ead{michael.harder@umanitoba.ca}

\author{Yongsheng Gui\corref{memail}}
\ead{ysgui@physics.umanitoba.ca}

\author{Can-Ming Hu\corref{mycorrespondingauthor}}
\ead{hu@physics.umanitoba.ca}
\cortext[mycorrespondingauthor]{Corresponding author at: Dynamic Spintronics Group, University of Manitoba, http://www.physics.umanitoba.ca/$\sim$hu/}
\address{Department of Physics and Astronomy, University of Manitoba, Winnipeg, Canada R3T 2N2}

\begin{abstract}
The purpose of this article is to review the current status of a frontier in dynamic spintronics and contemporary magnetism, in which much progress has been made in the past decade, based on the creation of a variety of micro- and nano-structured devices that enable electrical detection of magnetization dynamics.  The primary focus is on the physics of spin rectification effects, which are well suited for studying magnetization dynamics and spin transport in a  variety of magnetic materials and spintronic devices.  Intended to be intelligible to a broad audience, the paper begins with a pedagogical introduction, comparing the methods of electrical detection of charge and spin dynamics in semiconductors and magnetic materials respectively.  After that it provides a comprehensive account of the theoretical study of both the angular dependence and line shape of electrically detected ferromagnetic resonance (FMR), which is summarized in a handbook formate easy to be used for analyzing experimental data.  We then review and examine the similarity and differences of various spin rectification effects found in ferromagnetic films, magnetic bilayers and magnetic tunnel junctions, including a discussion of how to properly distinguish spin rectification from the spin pumping/inverse spin Hall effect generated voltage.  After this we review the broad applications of rectification effects for studying spin waves, nonlinear dynamics, domain wall dynamics, spin current, and microwave imaging.  We also discuss spin rectification in ferromagnetic semiconductors.  The paper concludes with both historical and future perspectives, by summarizing and comparing three generations of FMR spectroscopy which have been developed for studying magnetization dynamics.
\end{abstract}


\end{frontmatter}

\tableofcontents

%%%%%%%%%%%%%%%%%%%%%%%%%%%%%%%%%%%%%%%%%%%%%%%%%%%%%%%%%%%%%%%%%%%%%%%%%%%%%%%%%%%%%%%%%%%%%%%%%%%%%%%%%%%%%%%%%%%%%%%%%%%%%%%%%%%%%%%%%%%%%%%%%%%%%%%%%%%%%%%%%%%%%%%%%%%%%%%%%%%%%%%%%%%%%%%%%%%%%%%%%%%%%%%%%%%%%%%%%%%

\section{Introduction}

Magnetization dynamics is a venerable subject in solid state physics, originating with the discovery of ferromagnetic resonance (FMR) absorption \cite{Arkadyev1912, Dorfman1923, Griffiths1946, Kittel1947, Kittel1948, VonsovskiiBook} in the first half of the 20$^\text{th}$ century.  Yet still today, well into the 21$^\text{st}$ century, the study of magnetization dynamics continues to be an active and fast paced research field \cite{HillebrandsBookVolI, HillebrandsBookVolII, HillebrandsBookVolIII}.  Early understanding of magnetization dynamics was based solely on the phenomenological Landau-Lifshitz-Gilbert (LLG) equation that does not consider magneto-transport effects, with experimental studies employing microwave absorption measurements.  However this has changed recently.  The application of modern material growth and nano-lithographic techniques, which was extended from the semiconductor to magnetism community, has revealed a wealth of new and rich physics demonstrating that charge transport and spin dynamics are strongly correlated in magnetic materials.  On one hand charge transport can be controlled through magnetization dynamics (or magnetization configurations) through, for example, giant magnetoresistance (GMR) in magnetic multilayer devices \cite{Baibich1988, Binasch1989}.   On the other hand, the magnetization in nano-structured magnetic multilayers patterned with a sub-micrometer lateral size can be reversed by a dc current \cite{Katine2000, Huai2004}.  Further studies have even found that dc currents and dc voltages can be generated by the rectification effect of high-frequency magnetization dynamics \cite{Tsoi2000, Kiselev2003, Gui2005a, Tulapurkar2005, Costache2006a, Costache2006, Saitoh2006, Sankey2006, Kubota2008, Gui2007, Mecking2007, Yamaguchi2007, Gui2007a, Bedau2007, Sankey2007, Bai2008, Hui2008, Wirthmann2008, Gui2009, Wirthmann2010, Boone2009, Bonetti2010, Atsarkin2010, Mosendz2010a, Saraiva2010, Kajiwara2010, Sandweg2010, Liu2011, Azevedo2011, Harder2011a, Feng2012, Wang2013a, Deorani2014, Soh2014, Wang2014c}.  This so-called spin rectification effect provides a novel technique for the study of magnetization dynamics -- electrical detection.  

Electrical detection techniques are not new to the study of dynamic material properties, having long been used in semiconductors and photovoltaics.  For example, it is well known that the dynamic interaction between photons and electron charge can generate non-thermal electron diffusion, thus converting some of the photon energy into electric energy when an electromagnetic wave is incident upon a material surface. This so-called photovoltaic effect was first observed by French physicist Edmond Becquerel in 1839 \cite{Becquerel1839} using a pair of platinum electrodes dipped in an electrolyte solution \cite{Williams1960}.  Nowadays, exploitation of the photovoltaic effect in solar cells provides the third most important renewable energy source after hydro and wind power.  The key element of a solar cell is a semiconducting device in which electrons in the valence band can absorb photon energy, jump into the conduction band, and then generated electron-hole pairs which are separated by a p-n junction and collected by different electrodes \cite{Gratzel2001}.  In addition to the photovoltage produced in this manner, the generation of electron-hole pairs also results in photoconductivity \cite{Petritz1956} -- the enhanced electrical conductivity of semiconducting materials.  Hence this effect is widely used in photodetectors over a broad frequency range, from infrared up to gamma radiation \cite{Rogalski2002, Konstantatos2006, Soci2007, KnollBook}.

The high conversion efficiency from photon energy to an electric signal in semiconductors also makes photovoltage and photoconductivity a highly sensitive detection method to study charge dynamics (especially in two-dimensional electron systems).  These techniques have provided insight into, for example, inter-band transitions \cite{Maan1982}, inter-subband transitions \cite{Schneider1991}, magneto-plasmonics \cite{Holland2002} and spin-flip excitations \cite{Hu2003}.  Such success naturally begs the question, could electrical detection provide insight into spin dynamics in magnetically ordered materials?  On the surface it seems like the transition to spin dynamics in ferromagnetic materials would be difficult. First, the excitation of spin dynamics occurs in d-shell electrons, which have less influence on the electric response of a material. Early work on spin-induced current and voltage in semiconductors \cite{Ganichev2002} and ferromagnetic metal films \cite{Egan1963} indicated that the conversion efficiency between photon energy and electricity through spin excitations was very poor, often requiring high power (up to kW) GHz and THz sources.  Second, the conductivity of ferromagnetic metals is at least four orders of magnitude larger than that of semiconductors.  Hence the impact of microwave photons on the conductivity of ferromagnetic metals was expected to be negligible. 

However, starting from the middle 2000s, triggered by the rapid development of micro/nano-structuring techniques applied to ferromagnetic devices, there has been a surge of research interest in the study of spin dynamics in ferromagnetic materials by means of electrical detection.  A vast number of experimental and theoretical investigations have been carried out, not only on all aspects of traditional spin waves in ferromagnetic monolayers, but also on novel spin phenomena such as spin pumping (SP) and spin-transfer torque (STT) in ferromagnetic multilayers.  Through this surge of research, which was sometimes accompanied by intense scientific debate, a comprehensive understanding of one of the key mechanisms responsible for dc voltage generation in a variety of magnetic devices, the spin rectification effect (SRE), has been developed.  

The simplest way to understand the SRE is to highlight the close analogy between spin rectification and the well known optical rectification which occurs in nonlinear media with large second-order susceptibility \cite{Bass1962, ShenBook}.  In optical rectification the nonlinear optical response of the time-dependent electric field $e_0\cos\left(\omega t\right)$ is governed by the trigonometric relation $\cos^2\left(\omega t\right)=\left[1+\cos\left(2\omega t\right)\right]/2$, and hence results in optical rectification and second harmonic generation. Similarly, spin rectification is the generation of a dc voltage/current due to the nonlinear coupling between an oscillating current and an oscillating resistance in magnetic structures.  Despite the similarities, including that both optical and spin rectification are linearly dependent on the electromagnetic power, spin rectification has several unique features.  First, the dc signal is linearly proportionally to the resistance of the sample, which makes it a powerful tool to investigate spin dynamics in monolayer nano-structured samples, which are notoriously difficult to study using traditional transmission/reflection measurements since such a signal is proportional to sample volume. Second, the oscillating resistance caused by the dynamic magnetization can be driven by a time dependent magnetic field, $h_0 \cos\left(\omega t + \phi\right)$, where $\phi$ is the relative phase between microwave $e$ and $h$ fields and plays an important role in the curious line shape of electrically detected ferromagnetic resonance. Due to its high sensitivity and feasibility of implementation, spin rectification has significant impact on the study of spin dynamics. Over the past decade, and with the great effort of many research groups, the SRE has been widely employed to study ferromagnetic resonance (FMR), spin wave resonance, and domain wall resonance in various ferromagnetic materials including ferromagnetic metals, ferromagnetic semiconductors, ferrimagnetic insulators as well as multilayer structures.  Furthermore, properly analyzing the SRE has been found essential for studying novel spin dynamics including phase-resolved FMR, spin-transfer torque, spin pumping, and the spin Hall effect. Parallel to such a path of basic research on spintronics, the SRE has also found its way into novel microwave applications, such as spintronic microwave imaging and wireless microwave energy harvesting.

This article provides a review of the rectification effects in magnetic structures and their applications. We will examine the physical basis of electrical detection techniques as they apply to magnetization dynamics, and examine a variety of applications in which such techniques have been employed.  In Sec. \ref{sec:ingredients} we begin by examining the ``ingredients" of spin rectification, namely magnetoresistance, magnetization dynamics and their nonlinear coupling through the generalized Ohm's law.  In Sec. \ref{structures} we turn to various device structures, starting with early pulsed microwave studies followed by an examination of rectification in ferromagnetic monolayers, bilayers and magnetic tunnel junctions.  Along the way we discuss spin pumping and spin Hall effects and the important role of line shape and angular analyses in distinguishing the many competing voltage producing effects in magnetic devices.  Having explained the basic techniques of electrical detection of magnetization dynamics, in Sec. \ref{sec:applications} we turn to several important applications.  On the side of fundamental physics we look at the detection of spin waves and nonlinear magnetization dynamics, dc electrical detection of ac spin current, magnetization dynamics in ferromagnetic semiconductors and studies of domain wall dynamics.  On the applied side, we examine the role of spin rectification in novel microwave sensing, imaging, and wireless energy harvesting technologies.  The article is concluded in Sec. \ref{sec:conclusions} including both historical and future perspectives.  By reviewing the historical evolution of three generations of FMR spectroscopy, including cavity transmission, flip chip absorption, and electrical detection techniques, we also glimpse into the future of FMR spectroscopy, where the venerable science of magnetism is merging with the modern physics of cavity QED, leading us to a new field of cavity spintronics \cite{Hu2015}.   

\section{The Ingredients of Spin Rectification} \label{sec:ingredients}

Spin rectification (SR) is the production of a dc voltage $V_\text{dc}$ in a ferromagnetic structure due to the nonlinear coupling between a dynamic resistance $R\left[{\bf H } \left(t\right)\right]$ and a dynamic current ${\bf I} \left(t\right)$.  To understand the origin of this effect, consider a ferromagnetic structure excited by a microwave field at angular frequency $\omega$.  The electric and magnetic fields inside the structure therefore take the form ${\bf e}\left(t\right)  = {\bf e}_0 e^{-i\omega t}$ and ${\bf h}\left(t\right) = {\bf h}_0 e^{-i\left(\omega t - \Phi\right)}$, where $\Phi$ is a phase shift associated with losses in the system \cite{JacksonBook}.  The ${\bf e}\left(t\right)$ field will drive a current, ${\bf I}\left(t\right) = \textbf{I}_0 e^{-i \omega t} = \sigma {\bf e}\left(t\right)$ and the ${\bf h}\left(t\right)$ field will drive magnetization precession, ${\bf m}\left(t\right) = \chi {\bf h}\left(t\right)$, where $\sigma$ and $\chi$ are the high frequency material response functions, the conductivity and susceptibility respectively.  If the structure is a simple monolayer, ${\bf h}\left(t\right)$ alone will drive the magnetization (Recently it has actually been found that spin-orbit torques may also be present in single layer devices \cite{Kurebayashi2014}.  We will discuss spin-orbit torques briefly in Sec. \ref{standshe}, however for the purpose of this review we will focus on field driven magnetization dynamics in single layer devices).   In bilayer devices or magnetic tunnel junctions where a spin current ${\bf j}_s\left(t\right)$ is present, an additional spin torque will contribute to the magnetization precession so that ${\bf m}\left(t\right) = \chi {\bf h}\left(t\right) + \chi_s {\bf j}_s\left(t\right)$, where $\chi_s$ is the frequency dependent effective susceptibility due to the spin current.  Due to the magnetoresistance (MR) of the heterostructure, the effect of the magnetization precession is to produce a dynamic resistance, $R\left(t\right) = R\left[{\bf H}\left(t\right)\right]$, dependent on the magnetic field ${\bf H}\left(t\right) = {\bf H}_0 + {\bf h}\left(t\right)$.  As a consequence, a nonzero time averaged voltage can be measured along the current direction,     
\begin{equation}
V = \langle \text{Re}\left\{I\left(t\right)\right\} \text{Re}\left\{R\left[\textbf{H}\left(t\right)\right]\right\}\rangle 
=\frac{1}{2} I_0 \textbf{h}_0 \cdot \nabla R \left(\textbf{H}_{0}\right) \cos \Phi. \label{genrect}
\end{equation}
\begin{figure}[!ht]
\centering
\includegraphics[width=13cm]{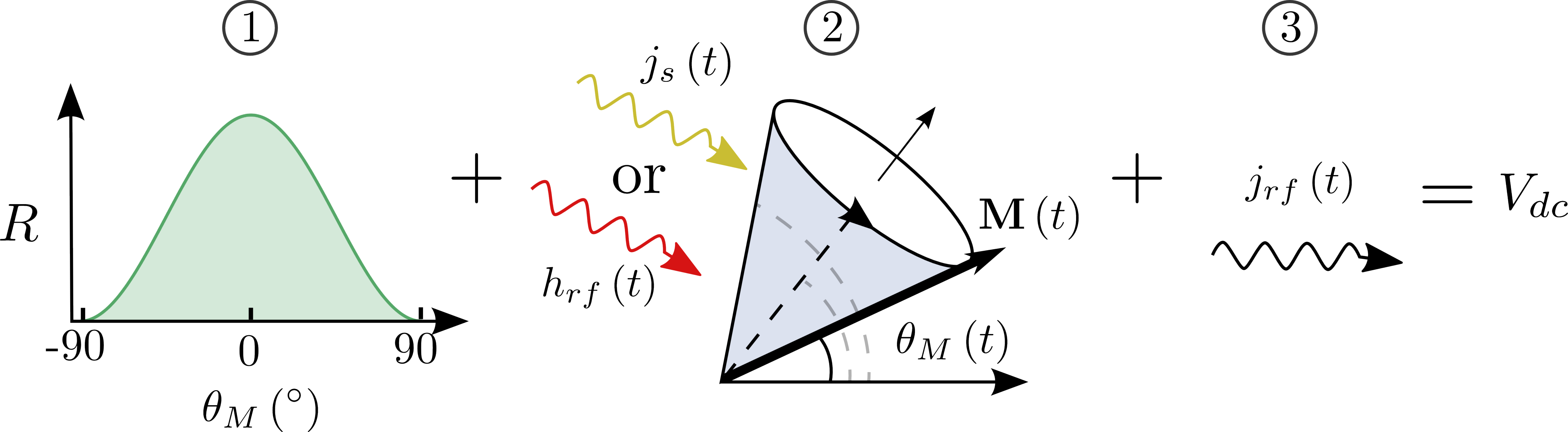}
\caption{\footnotesize{The ingredients of spin rectification.  Magnetoresistance effects result in a magnetization orientation dependence in the resistance of the ferromagnetic heterostructure.  The origin of the magnetoresistance will depend on the device structure and various forms of magnetoresistance are discussed in Sec. \ref{magnetotransport}.  By driving the magnetization with microwave frequency fields the resulting precession will produce an rf resistance.  The magnetization may be driven by a field torque produced by an rf magnetic field, $h_\text{rf}$, or from a spin torque produced by a spin polarized current $j_s$.  The rf resistance can then couple nonlinearly to an rf current and produce a rectified voltage.}}
\label{ingred}
\end{figure}
Here we have expanded the resistance in a Taylor series around the dc field ${\bf H}_0$ to first order in the rf field ${\bf h}\left(t\right)$, $R\left({\bf H}\left(t\right)\right) = R\left({\bf H}_0\right) + {\bf h}\left(t\right) \cdot \nabla R\left({\bf H}_0\right)$ ($\nabla R\left({\bf H}_0\right) = \partial R/\partial {\bf H} |_{{\bf H}_0}$ is the field gradient of the resistance evaluated at ${\bf H}_0$), and taken the time average over one period, denoted by $\langle\rangle$.  Based on this discussion we see that spin rectification requires three ingredients: 1) Some form of magnetoresistance, whereby the resistance of the heterostructure depends on the orientation of the magnetization. 2) A torque which drives the rf magnetization, resulting in a time varying magnetoresistance. 3) An rf current which can couple to the rf resistance and produce a dc voltage.  These three requirements are summarized in Fig. \ref{ingred}.      

In this section we provide a general review of the ingredients of spin rectification: Sec. \ref{magnetotransport} contains a discussion of the relevant magnetoresistance effects while in Sec. \ref{magnetizationdynamics} we review magnetization dynamics and the Landau-Lifshitz-Gilbert equation.  For simplicity we only concern ourselves with field torques in Sec. \ref{magnetizationdynamics}, leaving the treatment of spin torques to Sec. \ref{standshe} and Sec. \ref{srinmtjs} where we focus on bilayer devices and magnetic tunnel junctions (MTJs).  Finally in Sec. \ref{spinrectification} we present the simplest example of spin rectification, the voltage produced in a ferromagnetic monolayer, which will serve as an illustrative example which sets the stage for the detailed description of SR in more complex structures presented in Sec. \ref{structures}.

%%%%%%%%%%%%%%%%%%%%%%%%%%%%%%%%%%%%%%%%%%%%%%%%%%%%%%%%%%%%%%%%%%%%%%%%%%%%%%%%%%%%%%%%%%%%%%%%%%%%%%%%%%%%%%%%%%%%%%%%%%%%%%%%%%%%%%%%%%%%%%%%%%%%%%%%%%%%%%%%%%%%%%%%%%%%%%%%%%%%%%%%%%%%%%%%%%%%%%%%%%%%%%%%%%%%%%%%%%%

\subsection{Magnetotransport} \label{magnetotransport}

Magnetotransport -- the modification of transport properties due to magnetic fields -- can depend on both the charge and spin of the carrier and as such is responsible for a myriad of phenomena in metals and semiconductors \cite{StohrBook}.  However the magnetotransport effects which result in spin rectification are those which cause magnetoresistance, where the electrical resistance of the material depends on an applied magnetic field, $R = R(\textbf{H})$.  Although even non-magnetic metals display so called ordinary magnetoresistance (OMR) due to charged-induced magnetotransport (the Lorentz force), the maximum magnetoresistance ratio of such effects, MR = $\left[R\left({\bf H}\right) - R\left(0\right)\right]/R\left(0\right)$, is too small to be of technological interest. 
\begin{table}[ht]
\def\arraystretch{1.5}
 \caption{\footnotesize{Summary of static magnetoresistance effects discussed in detail in Sec. \ref{magnetotransport}.  The magnetoresistance effects for ferromagnetic monolayers are considered for an in-plane $\textbf{H}$ field and in-plane magnetization ${\bf M}$.  $\theta_m$ denotes the angle between the magnetization and the current flow, $\theta$ is the angle between the magnetization of the two ferromagnetic layers, and $\theta_s$ is the angle between the magnetization and the spin polarization.}}
 \centering
\begin{tabular}{>{\centering\arraybackslash}m{0.9cm}m{2.5cm}m{1.9 cm}m{7cm}c}
  \toprule                       
MR Effect & \multicolumn{1}{c}{Origin} & \multicolumn{1}{>{\centering\arraybackslash}m{1.9cm}}{Device Structure} & \multicolumn{1}{c}{Angular Dependence} & MR ratio\\ \bottomrule
  AMR & SO Coupling & FM monolayer & $R_\text{AMR} \left(\theta_m\right) = R\left(0\right) - \Delta R \sin^2 \theta_m$ & $\sim$ a few \%\\ 
  PHE & SO coupling & FM monolayer & $R_\text{PHE} \left(\theta_m\right) = \frac{1}{2}\Delta R \sin2\theta_m$ & $\sim$ a few \% \\
  AHE & SO coupling & FM monolayer & $R_\text{AHE} \left(\theta_m\right) = 0$& $\sim$ a few \%  \\
  GMR & Spin accumulation &FM/NM multilayers & 
  $G_\text{GMR}\left(\theta\right) = G_\text{P} + \Delta G \sin^2\left(\theta/2\right)$
  or, if GMR ratio is small,
  $R_\text{GMR}\left(\theta\right) = R_\text{P} + \Delta R \sin^2\left(\theta/2\right)$& $\sim$ 100 \% \\
  & Nonaligned FM inclusions & Granular magnetic solids & 
   & $\sim$ 10 \% \\
  TMR & Spin dependent tunnelling & FM/Insulator multilayer (MTJ)& Same as GMR& $\sim$ 600 \% \\
  SMR & SHE and spin relaxation through STT & FMI/NM bilayer & $R_\text{SMR}\left(\theta_s\right) = R\left(0\right) + \Delta R \sin^2\theta_s$ & $\sim$ 0.01 \% \\ \bottomrule
\end{tabular}
 \label{mrsummary}
\end{table}
On the other hand in magnetic heterostructures large magnetoresistance due to spin-dependent transport dominates and as a result MR effects in magnetic materials have a rich history of applications.  The anisotropic magnetoresistance (AMR), first discovered by William Thomson in 1856 \cite{Thomson1856}, was the first MR effect used by IBM to build hard disk drive read heads in 1991.  The chief limitation of AMR for memory applications was the low MR ratio of $\sim$ a few \%.  For this reason the celebrated discovery of giant magnetoresistance (GMR), independently by Fert \cite{Baibich1988} and Gr\"unberg \cite{Binasch1989}, with room temperature MR ratios up to $\sim$ 100\%, revolutionized the magnetic recording industry.  By 1997 IBM had used spin valve sensors based on GMR to replace the AMR sensors in magnetoresistive hard disk drive read heads, providing an increase in areal density of up to 100\% per year.  While GMR in multilayer structures has become important for memory applications, it is important to note that from a physics perspective GMR is not limited to multilayer structures but is more generally due to inhomogeneities and can be realized e.g. in granular magnetic solids \cite{Xiao1992}.

Even further increases in the MR ratio became possible with advances in the fabrication of tunnel barriers made of crystalline magnesium oxide (MgO) which made tunnel magnetoresistance (TMR) an attractive possibility for next generation memory devices.  TMR, originally discovered in 1975 by Julli\`ere in Fe/Ge-O/Co junctions at 4.2 K \cite{Julliere1975}, has recently been used to achieve MR ratios of 600\% at room temperature and over 1100\% at 4.2 K in CoFeB/MgO/CoFeB magnetic tunnel junctions (MTJs)\cite{Ikeda2008}.  The large MR ratio of MTJs allows the creation of non volatile, high endurance memories with fast random access making magnetic random access memory (MRAM)  a good candidate for a `universal memory' \cite{Parkin2003}.  

All of these memory applications we have described rely on the $static$ properties of MR, which are summarized in Table \ref{mrsummary} and which we will discuss further in this section as the first ingredient of spin rectification.  However, when combined with the second ingredient, magnetization dynamics, spin rectification opens the broader possibility of novel applications utilizing the $dynamic$ properties of magnetoresistance. 

%%%%%%%%%%%%%%%%%%%%%%%%%%%%%%%%%%%%%%%%%%%%%%%%%%%%%%%%%%%%%%%%%%%%%%%%%%%%%%%%%%%%%%%%%%%%%%%%%%%%%%%%%%%%%%%%%%%%%%%%%%%%%%%%%%%%%%%%%%%%%%%%%%%%%%%%%%%%%%%%%%%%%%%%%%%%%%%%%%%%%%%%%%%%%%%%%%%%%%%%%%%%%%%%%%%%%%%%%%%

\subsubsection{Two-Current Model}

The two-current model, which attributes the resistivity in ferromagnetic materials to two independent spin channels, provides the basis for understanding OMR, AMR, GMR and TMR \cite{Tsymbal2001}. This model is based on three key properties of 3d ferromagnets.  First, as noted by Mott \cite{Mott1936}, the low effective mass of the valence $sp$ band electrons, compared to the high effective mass of the $d$ band electrons, means that $sp$ electrons carry most of the electric current.  Second, because spin-flip scattering in ferromagnetic materials is negligible \cite{Fert2012} the current can be modelled as two independent spin channels -- spin-up (majority) and spin-down (minority) electrons.  And finally, due to exchange splitting, the density of states (DOS) of $d$ band electrons is greater for spin-up than spin-down electrons.  Since the scattering rate is proportional to the DOS, this final observation means that the scattering of up spins will be greater, and therefore the up spin channel will provide the greatest resistance contribution (hence the term majority electrons).  Although originally described for collinear systems, spin rectification often occurs under non-collinear conditions and the two-current model has also been extended to such situations \cite{Kovalev2002, Barnas2005}.

In the case of OMR the electron current corresponds to the $sp$ band electrons while the hole current corresponds to the $d$ band holes and it is sufficient to only consider the effect of the Lorentz force on these two currents, ignoring the spin degree of freedom.  However for AMR, GMR and TMR the spin dependence of the two currents is necessary.  Insight into the origin of this spin dependence can be gained by considering the Drude conductivity per spin for free electrons \cite{Tsymbal2001}
\begin{equation}
\sigma = \frac{e^2}{h} \frac{k_F^2}{3\pi} \lambda. \label{drude}
\end{equation}
Here $k_F$ is the Fermi momentum and $\lambda$ is the mean free path which depends on the Fermi velocity and the relaxation time $\tau$, $\lambda = v_F \tau$.  The relaxation time can be estimated by Fermi's golden rule
\begin{equation}
\tau^{-1} = \frac{2\pi}{\hbar} \langle V_\text{scat}^2\rangle N\left(\epsilon_F\right) \label{relaxation}
\end{equation}
where $ \langle V_\text{scat}^2\rangle$ is the average value of the scattering potential and $N\left(\epsilon_F\right)$ is the DOS at the Fermi energy.  
From Eqs. \ref{drude} and \ref{relaxation} we see that the conductivity has an intrinsic spin dependence caused by the spin dependence of $k_F, v_F$ and $N(\epsilon_F)$ which is due to the band structure of the material, but there is also a contribution from the scattering potential which is not an intrinsic property of the ferromagnetic material and can be due to e.g. defects, impurities or lattice vibrations.  In the case of AMR the scattering potential, due to the spin-orbit interaction, is fundamental, however in GMR and TMR the effect of the scattering potential may be averaged out by interface effects in the multilayer structure, leaving the band structure as the key contribution to the spin dependent conductivity \cite{Tsymbal2001}.
%%%%%%%%%%%%%%%%%%%%%%%%%%%%%%%%%%%%%%%%%%%%%%%%%%%%%%%%%%%%%%%%%%%%%%%%%%%%%%%%%%%%%%%%%%%%%%%%%%%%%%%%%%%%%%%%%%%%%%%%%%%%%%%%%%%%%%%%%%%%%%%%%%%%%%%%%%%%%%%%%%%%%%%%%%%%%%%%%%%%%%%%%%%%%%%%%%%%%%%%%%%%%%%%%%%%%%%%%%%

\subsubsection{Anisotropic Magnetoresistance}

The earliest discovered magnetoresistance effect in ferromagnetic metals was AMR \cite{Thomson1856}.  The key characteristic of AMR is the angular dependence, 
\begin{equation}
R\left(H\right) =  R(0) - \Delta R \sin^2 \left(\theta_M\right) \label{amrresistance}
\end{equation}
where $\theta_M$ is the angle between the current and magnetization direction and $\Delta R = R_\parallel - R_\perp$ is the difference in resistance between the current and magnetization aligned ($R_\parallel$) and orthogonal ($R_\perp$).  
\begin{figure}[!b]
\centering
\includegraphics[width=10cm]{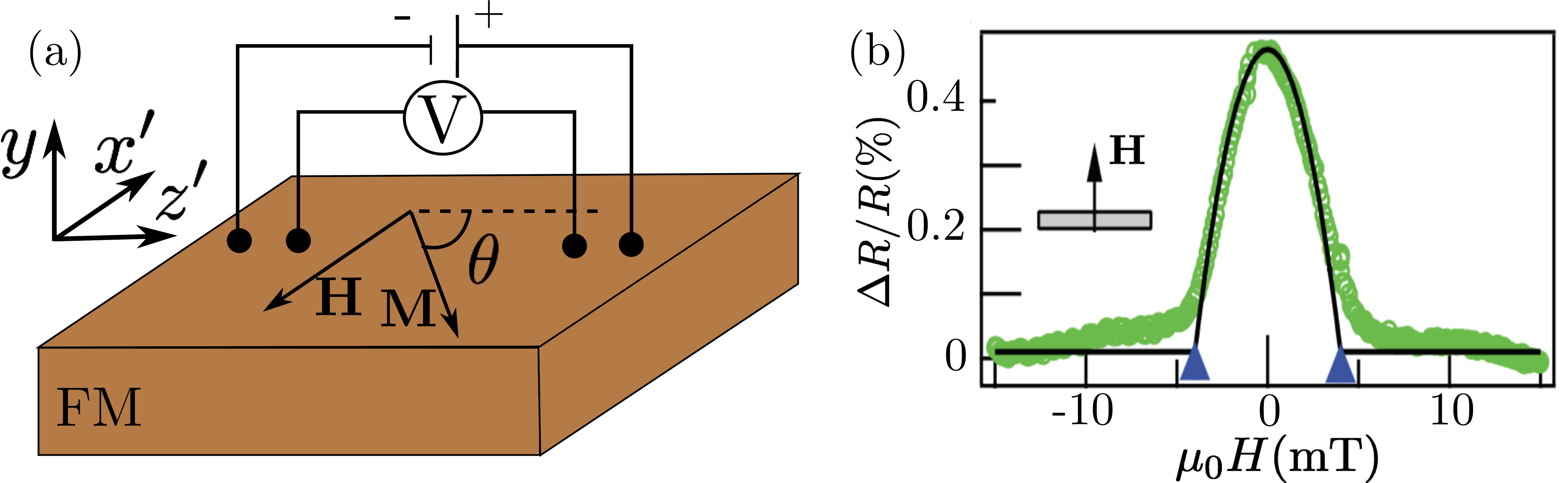}
\caption{\footnotesize{(a) AMR measurement configuration.  A dc current is sent through the ferromagnetic sample and the voltage is measured as a function of the angle $\theta$ between the magnetization and current direction.  Typically $\theta$ is controlled by changing the field strength with $\theta_H = 90^\circ$ fixed.  With zero applied field the magnetization will lie along the easy axis of the sample, in the current direction, and the resistance will be large.  As the magnetic field strength is increased toward saturation, the magnetization rotates to align with the field and the resistance decreases.  (b) AMR of $\sim 0.4\%$ in a permalloy (Py) microstrip at $\theta_H = 90^\circ$.  The arrows denote the anisotropic field $\mu_0 H_A = \mu_0 N_{x^\prime} M_0 = 4.0$ mT.  The open circles are experimental data and the solid curve is a fit using $R(0) = 112.66~\Omega, \Delta R = 0.47~\Omega$ and $H_A = 4.0$ mT.  $Source:$ Adapted from Ref. \cite{Harder2011a}.}}
\label{coordandamr}
\end{figure}
It should be noted that Eq. \ref{amrresistance} may not be appropriate for single crystalline films due to the band structure influence on AMR \cite{McGuire1975}.  In such samples AMR can exhibit additional four-fold symmetry \cite{Ramos2008, Hu2012, Ding2013, Xiao2015}, asymmetric behaviour \cite{Chen2015a}, and strong dependence on the angle between the current and crystalline axis \cite{Ramos2008, Hu2012, Ding2013, Xiao2015, Chen2015a, Xiao2015a}.  In both poly and single crystalline films typically $\Delta R >0$ \cite{Potter1974}, although there are exceptions \cite{Jaoul1977, McGuire1975}, and while the effect of AMR is small compared to GMR or TMR in multilayer structures, in monolayers AMR is the dominate effect and is typically as large as a few percent, although this depends on, amongst other factors, the material, sample geometry and temperature \cite{McGuire1975}, for example as the film thickness decreases, AMR decreases \cite{Mayadas1974}.  

Experimentally $\theta_M$ is controlled by an applied magnetic field $\textbf{H}$ and depends on the anisotropy fields in the sample \cite{Harder2011a}.  Therefore the magnetic field range over which AMR takes place is determined by the field needed to change the direction of the magnetization, typically on the order of 10 mT.  The exact relationship between $\theta_M$ and $\textbf{H}$ can be determined by minimizing the free energy, $F \propto \textbf{H} \cdot \textbf{M}$ with respect to $\theta_M$, taking into account the anisotropy fields.  For an in-plane magnetization with only shape anisotropy $\theta_M$ satisfies 
\begin{equation*}
H \sin\left(\theta_H - \theta_M\right) - M \cos \theta_M \sin \theta_M \left(N_{x^\prime} - N_{z^\prime}\right) = 0.
\end{equation*}
Here $\theta_H$ is the angle of $\textbf{H}$ with respect to the current direction, and $N_{x^\prime}$ and $N_{z^\prime}$ are the demagnetization factors in the coordinate system defined in Fig. \ref{coordandamr} (a), where ($\widehat{x}^\prime, \widehat{y}, \widehat{z}^\prime$) are fixed with the sample length along $\widehat{z}^\prime$ and the sample width along $\widehat{x}^\prime$ and the measurement coordinate system ($\widehat{x}, \widehat{y}, \widehat{z}$) rotates with the $\textbf{H}$ direction along $\widehat{z}$.  Typically the dimension of the sample in the $\widehat{x}^\prime$ direction is much less than in the $\widehat{z}^\prime$ direction, so the $N_{z^\prime}$ factor can be ignored.  Fig. \ref{coordandamr} (b) shows the AMR measured in a Py microstrip of dimension 300 $\mu$m $\times$ 20 $\mu$m $\times$ 50 nm at an angle $\theta_H = 90^\circ$ in which case $\sin\theta_M = H/N_{x^\prime} M$.  The measured data (symbols) has been fit using $R(0) = 112.66~ \Omega, \Delta R = 0.47 ~\Omega, \mu_0 H_A = 4.0$ mT and $N_{x^\prime} = 0.004$ and the AMR ratio is found to be $\Delta R/R(0) \sim 0.4 \%$.

The physical origin of AMR is spin-dependent scattering in ferromagnetic metals due to their band structure and the spin-orbit interaction \cite{McGuire1975, Smit1951, Potter1974}.  This means that to understand the magnitude and sign of the effect a detailed knowledge of the band structure is required.  However the angular dependence can be determined either by symmetry arguments \cite{McGuire1975} or through the generalized Ohm's law \cite{Juretschke1960, Jan1956},
\begin{equation}
\textbf{E} = \rho_\perp \textbf{J} + \frac{\Delta \rho_O}{\textbf{H}^2} \left(\textbf{J}\cdot \textbf{H}\right) \textbf{H} + \frac{\Delta \rho}{\textbf{M}^2} \left(\textbf{J}\cdot \textbf{M}\right) \textbf{M}- \frac{\rho_H}{|\textbf{H}|}\textbf{J} \times \textbf{H} -\frac{\rho_{AHE}}{|\textbf{M}|} \textbf{J}\times \textbf{M} \label{genohms}
\end{equation}
which is a phenomenological modification of the usually linear Ohm's law to include the nonlinear effects introduced by the presence of magnetic fields and a nonzero magnetization.  The second and third terms describe OMR and AMR respectively while the third and fourth terms describe the ordinary Hall effect (OHE) and the anomalous Hall effect (AHE).  OMR and the OHE are included here for completeness.  Although the angular dependence of OMR is the same as AMR, as we will see, the spin rectified voltage is proportional to $\Delta R$ which in ferromagnetic monolayers is much smaller for OMR than it is for AMR.  Also, as illustrated in Fig. \ref{coordandamr} (b) AMR is measured by sweeping the magnetic field and therefore in such measurements OMR would just contribute a constant background since $\theta_H$ is fixed.  Likewise the AHE dominates over the OHE.  We note that recently additional galvanometric effects have been predicted \cite{Zhang2016a}. 

The resistivity along a general direction $\widehat{n}$ is defined as
\begin{equation}
\rho_n = \frac{\widehat{n} \cdot \textbf{E}}{|\textbf{J}|}. \label{resdef}
\end{equation}
Using this definition the generalized Ohm's law immediately leads to the angular dependence of AMR described by Eq. \ref{amrresistance}.  The in-plane resistance measured transverse to the current direction can be determined by taking $\widehat{n}$ along the width of the microstrip so that $\widehat{n} \cdot \textbf{J} = 0$.  This transverse AMR is know as the planar Hall effect (PHE) and has an angular dependence
\begin{equation*}
R_{PHE}\left(\theta_M\right) = \frac{\Delta R}{2} \sin\left(2\theta_M\right).
\end{equation*} 
The AHE will not contribute when the voltage is measured along the strip due to the cross product.  However, although there is no static transverse contribution either, the AHE will contribute to the rectified transverse voltage. 

%%%%%%%%%%%%%%%%%%%%%%%%%%%%%%%%%%%%%%%%%%%%%%%%%%%%%%%%%%%%%%%%%%%%%%%%%%%%%%%%%%%%%%%%%%%%%%%%%%%%%%%%%%%%%%%%%%%%%%%%%%%%%%%%%%%%%%%%%%%%%%%%%%%%%%%%%%%%%%%%%%%%%%%%%%%%%%%%%%%%%%%%%%%%%%%%%%%%%%%%%%%%%%%%%%%%%%%%%%%

\subsubsection{Giant Magnetoresistance}

GMR is a spin accumulation effect present in multilayer structures consisting of alternating ferromagnetic (FM) and normal metal (NM) layers and as the name suggests results in much larger magnetoresistance ratios than AMR, up to nearly 100\%.  The simplest description of GMR is the resistor model \cite{Edwards1991}, 
\begin{figure}[!ht]
\centering
\includegraphics[width=7.1cm]{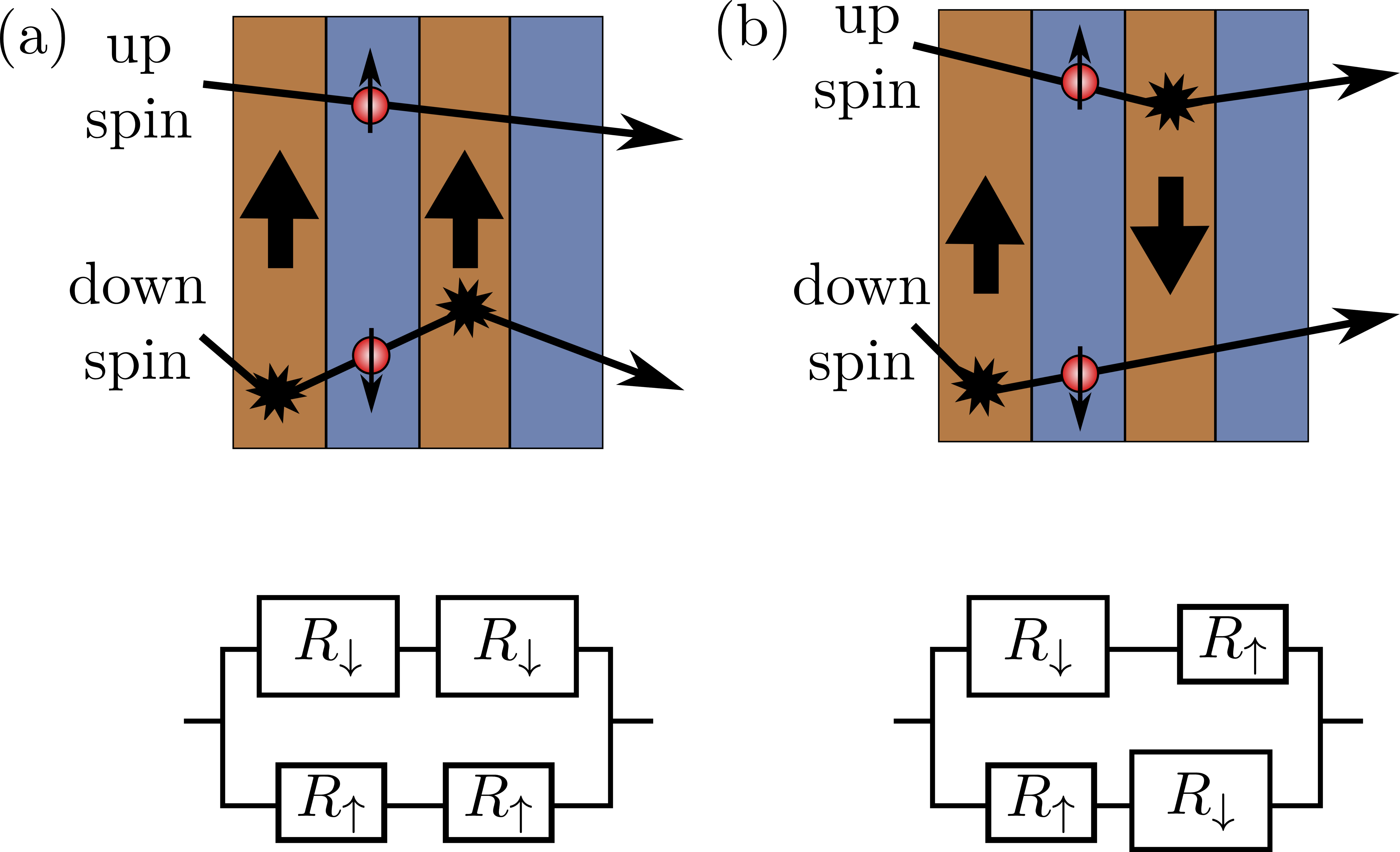}
\caption{\footnotesize{Resistor model of GMR.  Due to the small probability of spin flip scattering the majority (up) and minority (down) electrons can be treated as two independent conduction channels.  (a) When the magnetization of the FM layers is aligned, the majority electrons with spin along the magnetization direction can move with little scattering and therefore have an overall low resistance, while the minority electrons are strongly scattered in each FM layer and have a resulting high resistance.  Adding the resistance from the two channels in parallel leads to a low resistance state.  (b) When the magnetization of the FM layers is antiparallel, the minority and majority electrons are scattered equally after passing through two FM layers.  Adding the resistance for the two channels in parallel leads to a high resistance state.  $Source:$ Adapted from Ref. \cite{Tsymbal2001}.}}
\label{gmrmodel}
\end{figure}
based on the idea of the two band model discussed above, that the current is carried by two non-interacting spin channels.  If the magnetizations of the alternating FM layers are aligned, the majority electrons will experience a resistance $R_\uparrow$ in each layer, whereas the minority electrons which are more strongly scattered will have resistance $R_\downarrow > R_\uparrow$.  Adding these resistances in parallel, the P state has a resistance
\begin{equation}
R_P = \frac{2R_\uparrow R_\downarrow}{R_\uparrow + R_\downarrow}.
\end{equation}
On the other hand, if the magnetizations are anti parallel, the minority and majority electrons will experience the same resistance after travelling through the two layers, so that the AP state has resistance
\begin{equation}
R_{AP} = \frac{R_\uparrow + R_\downarrow}{2}
\end{equation}
and therefore the GMR ratio is
\begin{equation}
\text{GMR} = \frac{\Delta R}{R} = \frac{R_{AP}- R_P}{R_P} = \frac{\left(R_\downarrow - R^\uparrow\right)^2}{4 R_\downarrow R_\uparrow} = \frac{\left(\alpha-1\right)^2}{4\alpha}
\end{equation}
where $\alpha = R_\downarrow/R_\uparrow = \rho_\downarrow/\rho_\uparrow$ is the spin-asymmetry parameter.  Therefore in order to achieve GMR the multilayer structure must be capable of achieving two resistance states -- the parallel, P and antiparallel, AP, states where the magnetization of the alternating FM layers is aligned or anti aligned respectively.  The P state can always be achieved by applying sufficiently large magnetic fields, however the AP state must be achieved by sample design.  Generally $R_\text{P} > R_\text{AP}$, although there are exceptions for multilayers comprised of different FM layers \cite{George1994}.  One limitation of this resistor model is that surface scattering at the FM/NM interface has been ignored \cite{Bass2012}.  We should note that here we have adopted the most optimistic definition of the GMR ratio.  An alternative definition, $\text{GMR}  = \left(R_\text{AP}-R_\text{P}\right)/\left(R_\text{AP} + R_\text{P}\right)$ is also used by many authors.

\begin{figure}[!ht]
\centering
\includegraphics[width=12cm]{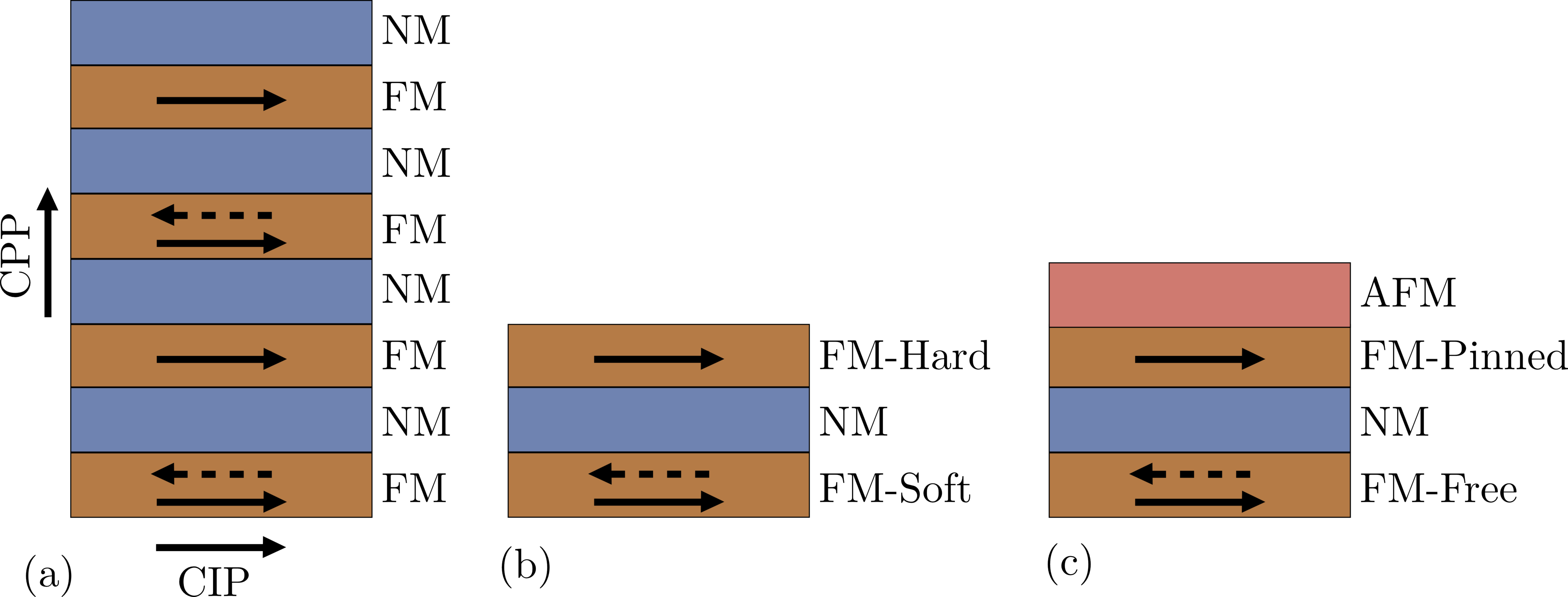}
\caption{\footnotesize{Sample structures which display GMR.  The resistance state in all structures is controlled with an in-plane magnetic field.  (a) A multilayer structure composed of alternating ferromagnetic (FM) and nonmagnetic (NM) layers.  When the magnetization of all FM layers are aligned (all solid arrows) the structure will be in the low resistance P-state, while the high resistance AP-state will occur when alternating FM layers point in opposite directions (dashed arrows).  The FM layers may or may not be exchange locked.  Also shown are the CIP and CPP current directions.  In both cases the resistance is measured in the current direction.  (b) A pseudo spin valve composed of two FM layers of high (hard) and low (soft) coercivities separated by a NM layer.  (c) A spin valve where a FM layer is pinned by exchange coupling to an antiferromagnetic (AF) layer.  The thickness of the NM layer is adjusted so that the magnetization of the free FM layer can still be adjusted.  $Source:$ Adapted from Ref. \cite{Tsymbal2001}.}}
\label{gmrstructures}
\end{figure}

Initial studies of GMR achieved the AP state using antiferromagnetic (AFM) coupling between the FM layers \cite{Grunberg1986} at $H=0$.  Adjusting the thickness of the NM layer produces oscillations in the exchange coupling between layers \cite{Parkin1990, Parkin1991b, Parkin1991c, Mosca1991} and therefore it is possible to choose a thickness such that the layers are AFM coupled.  The disadvantage of the AFM coupled multilayer structures is that it takes a large $H$ to switch between the AP and P states.  Fortunately the AP states can be achieved in other ways.  Fig. \ref{gmrstructures} 
summarizes the different structures that give GMR.  One option is to have a [FM/NM]$_n$ multilayer repeating $n$ identical FM and NM layers.  The AP state can then be realized by AFM exchange coupling between layers or by dipolar coupling, which can be made the dominant coupling effect by increasing the thickness of the NM layer to reduce the exchange interactions. Alternatively a FM/NM/FM multilayer can be used.  This so called pseudo spin valve contains both a soft FM layer with small coercivity field, $H_c$, and a hard FM layer \cite{Barnas1990, Shinjo1990, Dupas1990}.  An applied field will switch the soft layer first, providing the AP state.  A true spin valve structure, AF/FM/NM/FM contains a free FM layer that can be switched in a small field and a FM layer which is pinned by an antiferromagnetic layer and therefore only reverses under a large field \cite{Dieny1991}.

\begin{figure}[!b]
\centering
\includegraphics[width=15cm]{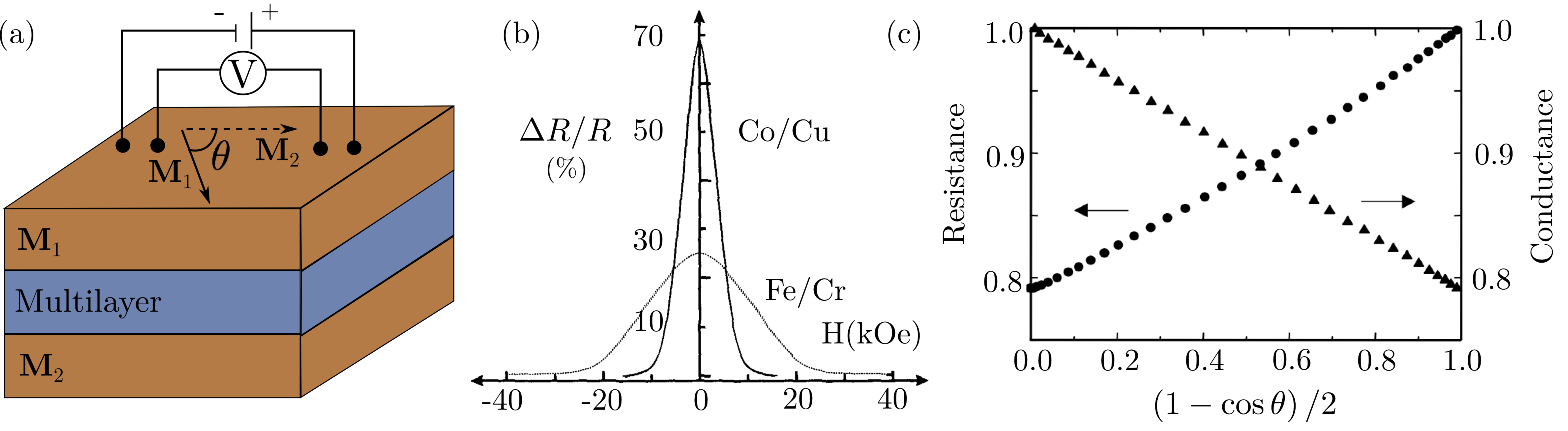}
\caption{\footnotesize{(a) GMR measurement setup, which is similar to that used for AMR measurements.  The key difference is that for a multilayer sample the resistance could be measured using a current applied in the sample plane (shown) or the current could be applied perpendicularly.  Typically the magnetoresistance ratio for the CPP geometry will be greater than for the CIP geometry, however the absolute resistance measured will be less.  (b) GMR curve for Fe/Cr and Co/Cu multilayer structures \cite{Parkin1995}.  The MR ratio is up to 2 orders of magnitude greater than AMR.  (c) The angular dependence of GMR.  While the conductance is linear in $\left(1-\cos\theta\right)$ the resistance has slight deviations from linearity, however these variations are higher order in the GMR ratio \cite{Steren1995a}.}}
\label{gmrdata}
\end{figure}

As shown in Fig. \ref{gmrstructures} (a), GMR experiments are normally performed with the current in the layer plane (CIP) or current perpendicular to the layer plane (CPP), with the resistance measured in the current direction.  However experiments have been performed with the current at an angle to the layer plane (CAP) \cite{Ono1995, Gijs1995}.  For CIP GMR the relevant length scale is the mean free path, $\lambda$ of the NM and FM, however for CPP GMR due to spin dependent accumulation effects the longer (compared to $\lambda$) spin diffusion length is important \cite{Valet1993}.  Although CPP GMR $>$ CIP GMR for a sample with fixed lateral dimension and thickness, the CPP resistance will be as much as 8 orders of magnitude less than the CIP resistance \cite{Pratt1991}.  Therefore initially GMR experiments were performed using the CIP geometry due to the 0.01 - 1 $\Omega$ resistances which can easily be achieved.  However improvements in fabrication techniques have lead to CPP becoming standard, with a best MR ratio at room temperature of 80\% as reported by Jung et al. \cite{Jung2016}, which is even higher than TMR in Al-O based MTJs.  For a good review of experimental techniques see Ref. \cite{Bass2012}.

The angular dependence of GMR has been studied in both CIP \cite{Dieny1991, Chaiken1990} and CPP geometries \cite{Dauguet1996} and was found to follow
\begin{equation}
R(\theta) = R_P + \frac{1}{2}\left(R_{AP} - R_P\right)\left(1-\cos\theta\right) = R_\text{P} + \Delta R \sin^2\left(\theta/2\right)\label{gmrtheta}
\end{equation}
where $\theta$ is the angle between the magnetization of the two layers and $\Delta R = R_\text{AP} - R_\text{P}$.  Different approaches to a theoretical analysis of the angular dependence predict that the conductance, not the resistance, varies as $\cos\theta$ \cite{Vedyayev1994, Vedyayev1997,Blaas1999}.  Although the deviations of the resistance from the $\cos\theta$ dependence is second order in the GMR ratio they are noticeable in the data shown in Fig. \ref{gmrdata} (c).  This deviation is more pronounced in the CPP geometry.  

%%%%%%%%%%%%%%%%%%%%%%%%%%%%%%%%%%%%%%%%%%%%%%%%%%%%%%%%%%%%%%%%%%%%%%%%%%%%%%%%%%%%%%%%%%%%%%%%%%%%%%%%%%%%%%%%%%%%%%%%%%%%%%%%%%%%%%%%%%%%%%%%%%%%%%%%%%%%%%%%%%%%%%%%%%%%%%%%%%%%%%%%%%%%%%%%%%%%%%%%%%%%%%%%%%%%%%%%%%%

\subsubsection{Tunnel Magnetoresistance}

\begin{figure}[!b]
\centering
\includegraphics[width=10cm]{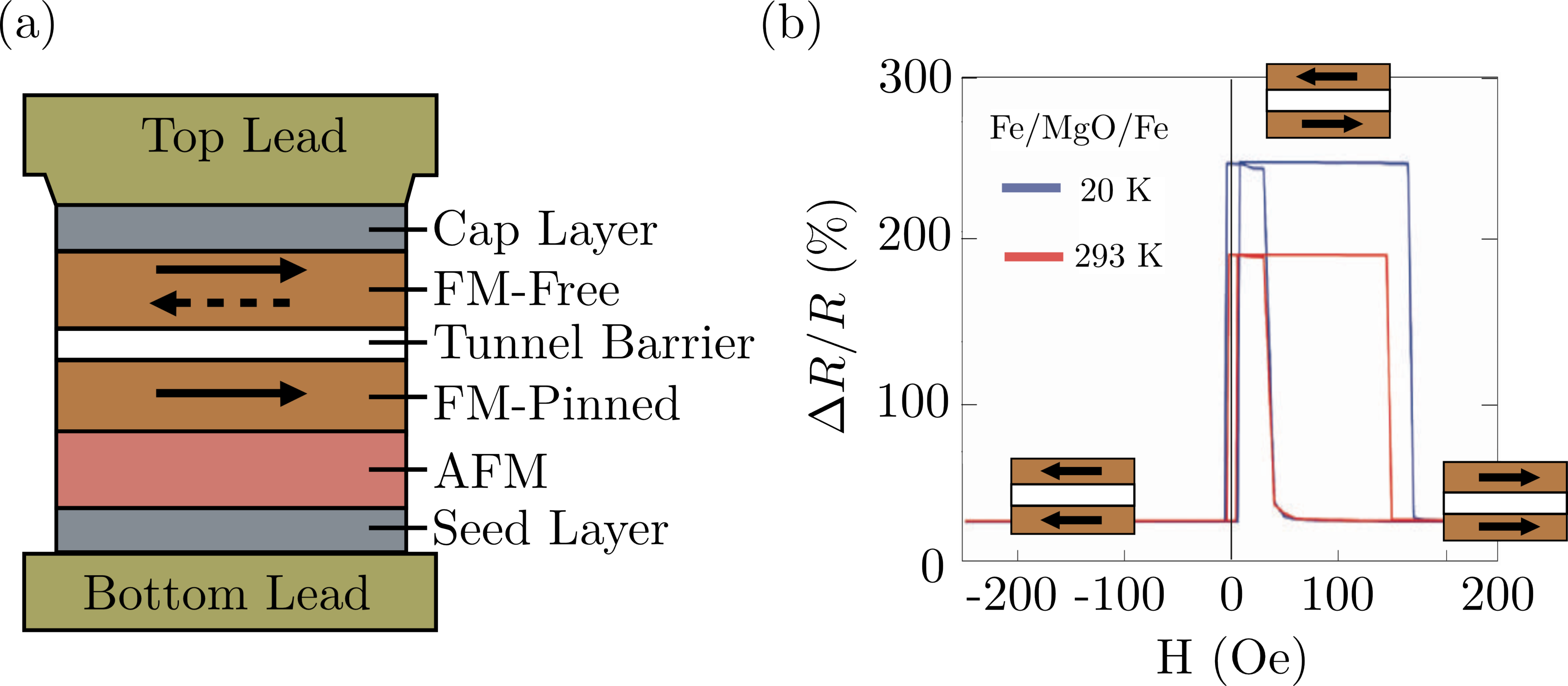}
\caption{\footnotesize{(a) Schematic cross section of a typical spin valve MTJ.  Exchange coupling with the AF layer pins the magnetization of the bottom FM electrode and a tunnel barrier separates the free and pinned layers. (b) Typical TMR data taken from a Fe(001)/MgO(001)/Fe(001) MTJ \cite{Yuasa2004}.  The resistance is largest in the antiparallel state and a MR ratio of nearly 200\% is found at room temperature.}}
\label{tmrstructure}
\end{figure}

Similar to GMR, TMR is due to a resistance difference between P and AP configurations in a multilayer structure.  However in TMR the FM layers are separated by an insulator and therefore TMR is a result of spin polarized tunnelling, rather than spin accumulation effects.  A typical schematic structure of a spin valve magnetic tunnel junction (MTJ) in which TMR is observed is shown in Fig. \ref{tmrstructure}.   
Although a pseudo spin valve may also be used, the spin valve is preferred because the resistance change occurs near zero magnetic field and the spin valve has greater magnetic stability \cite{Gider1998}.  The ability to control the tunnelling probability of the majority and minority electrons through high quality interfaces and a choice of electrode and barrier materials is what allows the high MR ratios observed in MTJs.

Early observations of small TMR ratios ($\sim 14\%$ at 4.2 K) were made as early as 1975 \cite{Julliere1975, Maekawa1982}, however as these were not reproducible due to the difficulty of the fabrication process, little research was done on the effect until the discovery of GMR provided a renewed interest in the possibility of large MR ratios.  The first reproducible experiments used a pseudo spin valve MTJ with an amorphous aluminum oxide tunnel barrier and achieved MR of $20\%$ at room temperature \cite{Moodera1995, Miyazaki1995}.  Further optimization of fabrication techniques led to TMR ratios of $\sim 70\%$ in Al-O MTJs, however this was not high enough for modern technological applications.  To increase TMR the amorphous tunnel barrier had to be replaced with a crystalline barrier such as MgO which is capable of symmetry selection.  Improvements in the structure quality of MgO based MTJs quickly led to the giant TMR results of up to 200\% in 2004 \cite{Parkin2004, Yuasa2004} and modern devices are capable of achieving MR ratios of over 500\% at RT \cite{Lee2007, Ikeda2008}.  

The increased TMR due to an MgO tunnel barrier can be understood, at least qualitatively, by considering the phenomenological model of Julli\`ere \cite{Julliere1975} where the TMR is due to spin-dependent electron tunnelling.  In Julli\`ere's model spin is assumed to be conserved during tunnelling so that the two-current model can be applied and the tunnelling probability will be proportional to the spin dependent DOS of the two electrodes.  Also the tunnelling probability is assumed to be spin independent, which means that spin dependent tunnelling rates will only be a result of the spin dependent DOS.  Since the conductance $G$ is proportional to the DOS, $N_j^i(E_F)$, where $j = \text{L}, \text{R}$ indicates the left and right electrodes and $i = \uparrow, \downarrow$ indicates the majority and minority electrons in the electrode, we have
\begin{equation*}
G_\text{P} \propto N_\text{L}^\uparrow N_\text{R}^\uparrow + N_\text{L}^\downarrow N_\text{R}^\downarrow, ~~ G_\text{AP} \propto N_\text{L}^\uparrow N_\text{R}^\downarrow + N_\text{L}^\downarrow N_\text{R}^\uparrow
\end{equation*}  
and so the TMR ratio is
\begin{equation}
\text{TMR} = \frac{G_\text{P} - G_\text{AP}}{G_\text{AP}} = \frac{R_\text{AP} - R_\text{P}}{R_\text{P}} = \frac{2 P_\text{L} P_\text{R}}{1-P_\text{L} P_\text{R}} \label{jullieremodel}
\end{equation}
where
\begin{equation*}
P_j = \frac{N_j^\uparrow - N_j^\downarrow}{N_j^\uparrow + N_j^\uparrow}, ~~ j = \text{L}, \text{R}.
\end{equation*}

TMR ratios estimated from measured spin polarizations using Julli\`ere's model agree with the measured TMR ratios fairly well.  Therefore using the experimental results for polarization in 3d ferromagnetic metals and their alloys, which typically range from $0<P<0.6$ below 4.2 K \cite{Parkin2003, Meservey1994}, Eq. \ref{jullieremodel} can be used to estimate a maximum TMR of $\sim 100 \%$.  This value is further reduced at room temperature due to thermal spin fluctuations. 

One way to increase the polarization of the electrode would be to use half-metallic ferromagnetic materials which act as metals for one spin direction and insulators for the other, and therefore theoretically have polarizations of 100\% \cite{Bowen2003, Ishikawa2006}.  Half metal electrodes have been made using e.g. CrO$_2$ or certain Heusler alloys, however large TMR ratios have only been achieved at low temperatures \cite{Yuasa2012}.  Another way to increase the polarization of the electrodes would be to exploit another limitation of Julli\`ere's model -- the assumption that the tunnelling is completely incoherent.  This assumption is incorrect even for amorphous insulating layers and is clearly incorrect for crystalline tunnel barriers such as MgO where the tunnelling states have specific symmetries.  The different symmetry states experience different decay rates determined by their symmetry matching with the lattice, while all states tunnel at the same rate through the amorphous Al-O.  This characteristic can be exploited to allow only highly spin polarized states to tunnel through, dramatically increasing the TMR ratio.  

The magnitude of TMR depends on the electrode and barrier materials, the fabrication technique and quality of interfaces.  These details will not be discussed in detail here, however for more information see e.g. \cite{LeClair2012, Yuasa2012, Belashchenko2012}.  It is useful to mention though that for device applications the Fe/MgO/Fe MTJs we have discussed cannot be used.  This is because the pinned layer used in spin valve MTJs uses a synthetic ferrimagnetic tri-layer that has an fcc(111) orientation on which the bcc(001) Fe/MgO/Fe structure cannot be grown.  The reliability of the standard fcc(111) pinned layer is essential and it is therefore easier to use a new MTJ structure.  The standard alternative to the Fe/MgO/Fe MTJ is a CoFeB/MgO/CoFeB structure \cite{Djayaprawira2005}.  The amorphous CoFeB electrode layer can be grown on the synthetic pinning layer and also allows the growth of a MgO(001) tunnel barrier.  

The angular dependence for small TMR ratios is the same as Eq. \ref{gmrtheta} which was already discussed for GMR \cite{Moodera1996}.  However for larger TMR ratios, the deviations from $\cos\theta$ dependence may become pronounced \cite{Jaffres2001} and the correct angular dependence is most easily expressed in terms of the conductance, $G$ as \cite{Slonczewski1989, Suzuki2008}
\begin{equation}
G\left(\theta\right) = R\left(\theta\right)^{-1} = \frac{G_\text{P} + G_\text{AP}}{2} + \frac{G_\text{P} - G_\text{AP}}{2} \cos \theta = G_\text{P} + \Delta G \sin^2\left(\theta/2\right) \label{tmrtheta}.
\end{equation}
Here $G_\text{P} = G\left(\theta = 0\right) = R^{-1}_\text{P}$, $G_\text{AP} = G\left(\theta = \pi\right) = R^{-1}_\text{AP}$ and $\Delta G = G_\text{AP} - G_\text{P}$.  Eq. \ref{tmrtheta} and Eq. \ref{gmrtheta} agree to $\mathcal{O}\left(\text{TMR}^2\right)$.

%%%%%%%%%%%%%%%%%%%%%%%%%%%%%%%%%%%%%%%%%%%%%%%%%%%%%%%%%%%%%%%%%%%%%%%%%%%%%%%%%%%%%%%%%%%%%%%%%%%%%%%%%%%%%%%%%%%%%%%%%%%%%%%%%%%%%%%%%%%%%%%%%%%%%%%%%%%%%%%%%%%%%%%%%%%%%%%%%%%%%%%%%%%%%%%%%%%%%%%%%%%%%%%%%%%%%%%%%%%

\subsubsection{Spin Hall Magnetoresistance} \label{sec:shmr}

The MR effects discussed so far all require an electrical current to pass through the FM material in which the resistance change is observed, meaning these effects cannot be used to study the magnetic properties of ferromagnetic insulators (FMIs).  However the discovery of spin Hall magnetoresistance (SMR) in FMI/NM hybrid structures \cite{Nakayama2013, Lu2013, Chen2013, Hahn2013, Althammer2013, Vlietstra2013, Weiler2013, Vlietstra2013a, Miao2014, Marmion2014, Isasa2014, Grigoryan2014, Vlietstra2014, Lin2014} can be used to study FMI since this effect produces a magnetoresistance in the NM layer through the influence of the magnetization in the FMI.  This is dramatically different than the MR effects already discussed where the magnetoresistance occurs in the FM layer.  The key distinguishing features of SMR, compared to AMR, is the greater perpendicular resistance, $R_T \sim R_\parallel  > R_\perp$ for SMR while $R_\parallel > R_\perp \sim R_T$ for AMR, where $R_T$ is the perpendicular MR with $\textbf{M} \perp \textbf{I}$ and $\textbf{M}$ perpendicular to the sample plane \cite{Nakayama2013}.  Most studies have focussed on Yttrium-Iron-Garnet (YIG)/Pt structures and the removal of Pt or the use of a non magnetic insulator is found to remove the effect.

\begin{figure}[!ht]
\centering
\includegraphics[width=15cm]{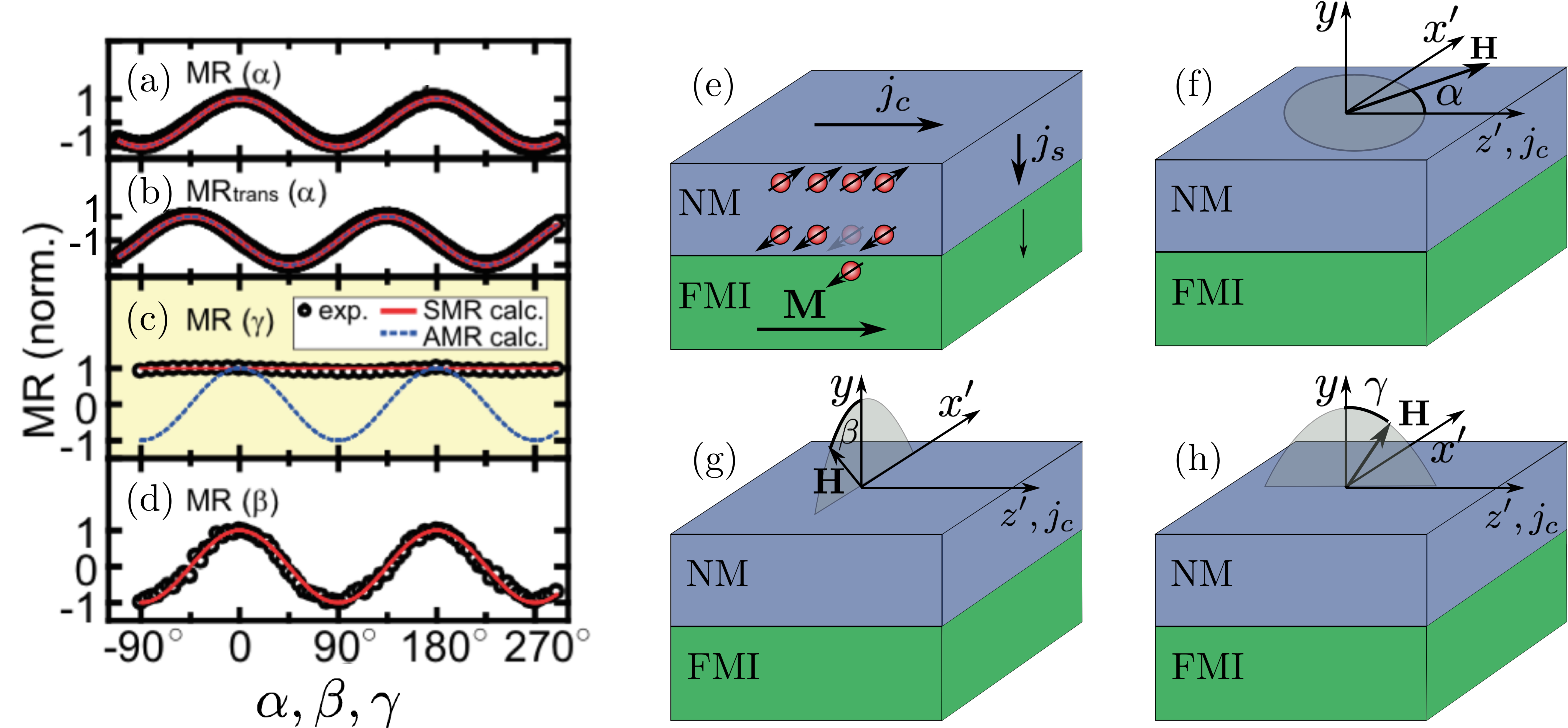}
\caption{\footnotesize{(a) - (d) Systematic measurements of the SMR in a Pt$|$YIG bilayer showing the $\alpha, \beta$ and $\gamma$ (defined in (f)-(h) respectively) dependence of the magnetoresistance \cite{Nakayama2013}.  Black circles are experimental data while red and blue curves are the expected curves due to SMR and AMR respectively.  (b) Schematic illustration of spin Hall magnetoresistance in a FMI(green)/NM(blue) hybrid structure.  The charge current in the NM generates a spin current through the SHE.  If the magnetization in the FMI is perpendicular to the spin polarization at the FMI/NM interface, a spin-transfer torque will act on the magnetization, reducing the current in the NM and increasing the resistance.  When the spin polarization and magnetization are aligned, the spins are reflected and the resistance decreases \cite{Brataas2013}.  (a) shows the $\alpha$ dependence of the longitudinal MR, (b) shows the $\alpha$ dependence of the transverse MR, (c) shows the $\beta$ dependence of the longitudinal MR.  For these cases the angular dependence of SMR and AMR is the same.  (d) shows the $\gamma$ dependence of the longitudinal MR which shows the key difference between SMR and AMR. $Source$: Data from Ref. \cite{Nakayama2013}.  Schematic adapted from Ref. \cite{Brataas2013}.}}
\label{smrfig}
\end{figure}

SMR is due to a combination of the spin Hall effect (SHE) \cite{Dyakonov1971, Hirsch1999, Hoffmann2013, Sinova2014} and spin dependent scattering at the FMI/NM interface.  A charge current in the NM will generate a spin current through the SHE.  This results in an accumulation of spin at the NM/FMI interface.  If the spin polarization is parallel to the magnetization, then the spin currents do not enter the FMI and are reflected back at the interface.  The inverse spin Hall effect (ISHE) then generates an electric current which is parallel to the original current, increasing the current in the NM, resulting in a decreased resistance.  On the other hand when the polarization of the spin current is perpendicular to the magnetization, spin diffusion into the FMI can occur via spin-transfer torque (discussed more in Secs. \ref{standshe} and \ref{srinmtjs}).  When spin current does cross the interface, spin accumulation is reduced and the charge current in the NM is reduced, increasing the resistance.  Therefore the net result of the spin Hall effect in the metal, and spin dependent scattering at the metal-insulator interface is that the resistance will be high (low) when the spin current polarization is perpendicular (parallel) to the magnetization of the FMI and the resistance change may be written as
\begin{equation*}
R = R_0 + \Delta R_{\text{max}} \sin^2 \theta_s
\end{equation*}
where $\theta_s$ is the angle between the magnetization and the spin polarization \cite{Hahn2013}.  $\Delta R_{\text{max}}$ will depend on the direction of magnetization rotation as shown in Fig. \ref{smrfig} (a) - (d).   Since SMR is due in part to spin diffusion at the FMI/NM interface, the sample size must be comparable to the spin diffusion length.

SMR is much smaller than AMR, $\sim 0.01\%$ in YIG/Pt \cite{Nakayama2013}, though this value depends on both temperature \cite{Marmion2014} and the Pt thickness \cite{Vlietstra2013}.  SMR has been used to measure the spin mixing conductance in YIG/Pt \cite{Vlietstra2013a} and comparative studies of spin pumping, the spin Seebeck effect and SHR have been performed \cite{Weiler2013, Vlietstra2014}.  Since the initial work on YIG/Pt samples, different FMI/NM hybrid structures have also been studied using for example Fe$_3$O$_4$, NiFe$_2$O$_4$ \cite{Althammer2013} and CoFe$_2$O$_4$ \cite{Isasa2014} as FMI and Au, Cu \cite{Althammer2013} and Ta \cite{Hahn2013} as NM. It has been found that SMR is enhanced in trilayer (FMI/NM/FMI) structures and it exhibits a different angular dependence from AMR  \cite{Chen2013}.  However, it should be noted that for thin ferromagnetic polycrystalline films the angular dependent magnetoresistance is much more complex than that generally accepted for isotropic bulk samples \cite{Campbell1982, McGuire1975, Rijks1997, Gil2005, Kobs2011}. In particular, single layer polycrystalline Fe thin films sandwiched between insulating nonmagnetic layers \cite{Zou2016} have recently been found to reproduce all AMR correlations previously reported in polycrystalline films, including the ones that resembles SMR. These findings suggest that caution should be taken when using angular dependent MR to ascertain the prescence of SMR.

To rule out the possibility that the MR was caused by known proximity effects, which cause NM layers to acquire FM properties when deposited on a FM layer \cite{Lu2013a, Huang2012}, studies have been performed with Cu inserted between the NM and FMI.  Since the proximity effect is not present in Cu and exchange forces will not be effective over the thick Cu layer used, the observation of MR in Pt/Cu/YIG confirms the explanation due to SMR, since the long diffusion length of copper would still enable SMR.  Nevertheless proximity effects may still be important \cite{Lu2013} and in fact both effects may be responsible \cite{Miao2014}.

%%%%%%%%%%%%%%%%%%%%%%%%%%%%%%%%%%%%%%%%%%%%%%%%%%%%%%%%%%%%%%%%%%%%%%%%%%%%%%%%%%%%%%%%%%%%%%%%%%%%%%%%%%%%%%%%%%%%%%%%%%%%%%%%%%%%%%%%%%%%%%%%%%%%%%%%%%%%%%%%%%%%%%%%%%%%%%%%%%%%%%%%%%%%%%%%%%%%%%%%%%%%%%%%%%%%%%%%%%%

\subsection{Magnetization Dynamics} \label{magnetizationdynamics}

The second requirement of spin rectification is some form of magnetization dynamics which can be controlled to produce a dynamic magnetoresistance.  Magnetization dynamics may be produced by either a \textit{field torque}, where an applied rf magnetic field interacts with the magnetization through a Zeeman type interaction, or by a \textit{spin torque}, where a spin polarized current influences the magnetization motion through an exchange interaction.  Depending on device structure either or both of these torques may be present and it is necessary to know the response function of the magnetization motion to the field or spin torque.  Determining such a response function in the case of field torques will be the focus of this section, with spin torques discussed in Secs. \ref{standshe} and \ref{srinmtjs}.

%%%%%%%%%%%%%%%%%%%%%%%%%%%%%%%%%%%%%%%%%%%%%%%%%%%%%%%%%%%%%%%%%%%%%%%%%%%%%%%%%%%%%%%%%%%%%%%%%%%%%%%%%%%%%%%%%%%%%%%%%%%%%%%%%%%%%%%%%%%%%%%%%%%%%%%%%%%%%%%%%%%%%%%%%%%%%%%%%%%%%%%%%%%%%%%%%%%%%%%%%%%%%%%%%%%%%%%%%%%

\subsubsection{Field Torque Induced Magnetization Dynamics}

One way to excite a dynamic magnetization is by applying a microwave field which will cause the magnetization to precess.  This precessional motion is described by the phenomological Landau-Lifshitz-Gilbert (LLG) equation.  The Landau-Lifshitz (LL) equation \cite{Landau1935} without damping follows from the Heisenberg equation of motion for the spin operator $\textbf{S}$ in the presence of a magnetic field $\textbf{H}_i$ with a Zeeman type interaction \cite{HillebrandsBookVolI},
\begin{equation}
\frac{d\textbf{M}}{dt} = - \gamma \textbf{M} \times \textbf{H}_i \label{llgnodamp}
\end{equation}
where $\gamma = \mu_0 g |q|/2 m >0$ is the gyromagnetic ratio of the material undergoing FMR.  $g$ is the Land\'e $g$-factor which is approximately 2 for a free electron and in general depends on the ferromagnetic material and the frequency of the applied microwave field, typically ranging from 1.7 to 2.3.  $\textbf{H}_i$ is the internal magnetic field which contains contributions from the external applied field, $\textbf{H}$, the anisotropy fields and the demagnetization fields.  Together these magnetic fields produce a torque which causes the magnetization to precess around $\textbf{H}_i$.  
\begin{figure}[!ht]
\centering
\includegraphics[width=8.5cm]{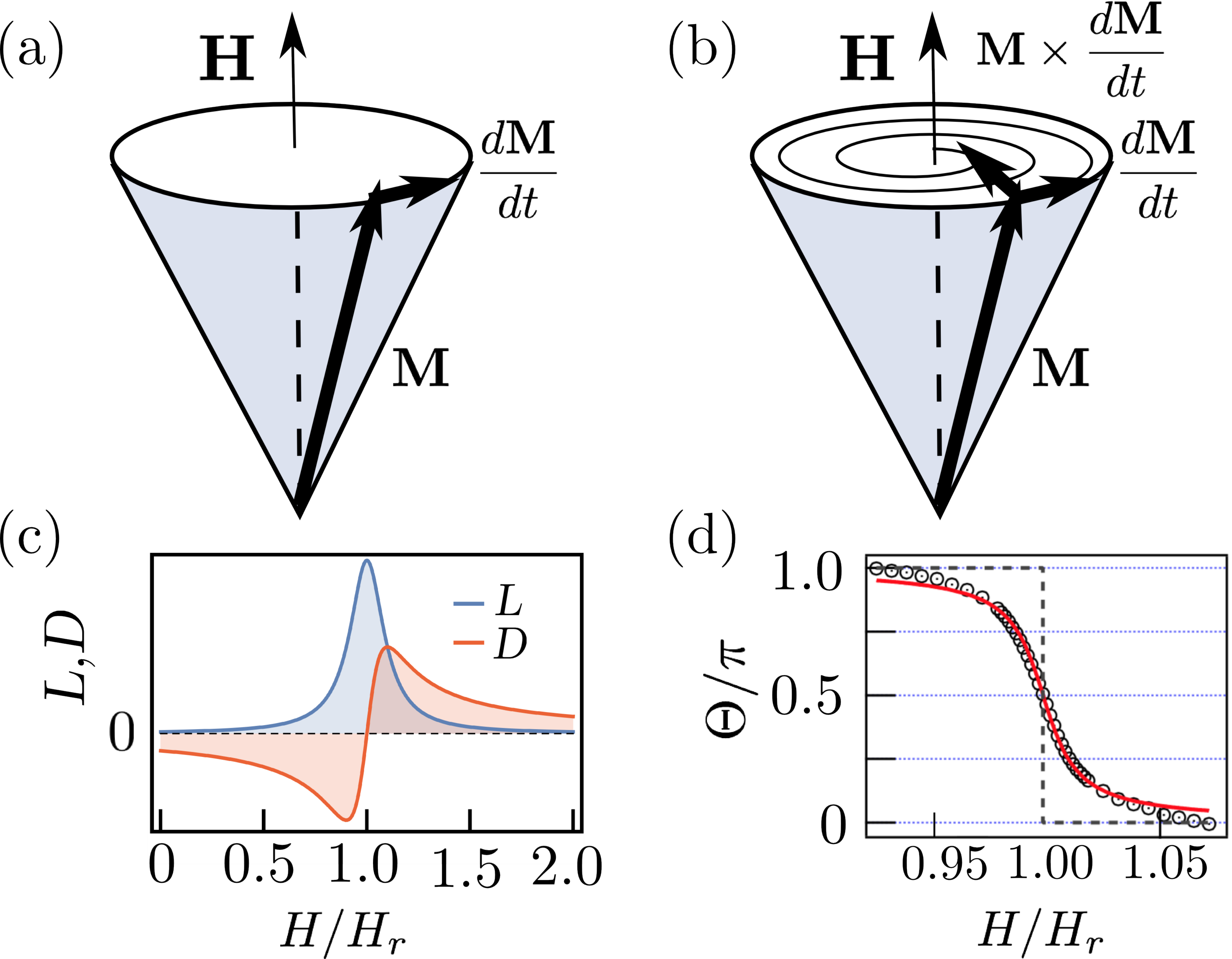}
\caption{\footnotesize{(a) Without damping the magnetization experiences a torque which causes it to precess at a constant cone angle around the $\textbf{H}$ field direction.  (b) Damping produces an inward torque which reduces the cone angle and causes the magnetization to align with $\textbf{H}$.  (c) The Lorentz and dispersive line shape components which both contribute to the susceptibility determined by solving the LLG equation.  $L$ is symmetric around the resonance field, while $D$ is antisymmetric.  (d) The spin resonance phase, $\Theta$, which is another key characteristic (along with the line shape components) of the magnetization dynamics.  Near resonance the spin resonance phase changes from $\pi$ to 0 within a range on the order of the line width $\Delta H$.}}
\label{mdynamics}
\end{figure}

Without damping the precession described by the LL equation will continue regardless of the magnitude of \textbf{H}.  Of course this behaviour is not observed experimentally and instead hysteresis curves show that for large enough \textbf{H} the magnetization should saturate and align with the magnetic field.  This behaviour can be introduced by the addition of a damping term that will produce a torque towards the magnetic field.  Such damping can be introduced in various ways.  When it is necessary to distinguish the spin-lattice relaxation from the spin-spin relaxation Bloch damping may be used \cite{Bloch1946}.  However the simplest way to introduce damping is via the Gilbert damping parameter $\alpha$ \cite{Gilbert2004},
\begin{equation}
\frac{d\textbf{M}}{dt} = - \gamma \textbf{M} \times \textbf{H}_i + \frac{\alpha}{M} \left(\textbf{M} \times \frac{ d\textbf{M}}{dt} \right). \label{llg}
\end{equation}
The Gilbert damping term produces a torque which causes the magnetization to move inward and align with the field and is applicable over a wide power range.  Since this torque is perpendicular to the magnetization direction the LLG equation of Eq. \ref{llg} describes precession and damping that keep the magnitude of $\textbf{M}$ constant. Note that, although the Gilbert damping is introduced phenomenologically, it has a microscopic origin in the spin-orbit interaction \cite{Hickey2009}.

The internal magnetic field $\textbf{H}_i$ has both a dc and rf contribution, $\textbf{H}_i = \textbf{H}_{0i} + \textbf{h}_i e^{-i\omega t}$.  As mentioned the internal field consists of the externally applied field and any anisotropy fields which are present.  If the dominant effect is the shape anisotropy, then the internal field can be related to the applied field and the magnetization through the demagnetization factors, $H_{0ik} =  H_{k} - N_k M_{0k}$ and $h_{ik} = h_k - N_k m_k$.  Here the $k$ subscript denotes the $k^{\text{th}}$ component of the respective fields, $\textbf{H}_{0i}$ is the internal dc field, $\textbf{H}$ is the applied dc field, $\textbf{h}_{i}$ is the internal rf field, $\textbf{h}$ is the applied rf field, $N_k$ is the demagnetization factor in the $k^{\text{th}}$ direction, and $\textbf{M} = \textbf{M}_0 + \textbf{m}(t)$ where $\textbf{M}_{0}$ and $\textbf{m}(t)$ are the equilibrium and non-equilibrium magnetizations respectively.  The demagnetization factors, which relate the internal magnetic fields to the externally applied magnetic fields, are geometry dependent and obey the sum rule $N_x + N_y + N_z = 1$ \cite{SodhaBook}.  We define our coordinate system so that the external magnetic field is always along the $z$ axis.  By doing so both the dc field in and perpendicular to the plane can be described by the same solution to the LLG equation by choosing the correct demagnetization factors.  $\textbf{M}_0$ will also be along $\widehat{z}$ and since $M$ is constant $\textbf{m}(t)$ will only have dynamical components along $\widehat{x}$ and $\widehat{y}$ (actually for elliptical precession ${\bf m}$ will have a $2\omega$ dynamic component along $\widehat{z}$, however we will not consider this higher order term here).  Using the dc and rf components in the LLG equation and keeping only linear terms in $\textbf{m}$ and $\textbf{h}$ (motivated by the low microwave power used by typical experiments, but see Sec. \ref{nonlinear}) the Polder tensor $\chi$, which relates the rf magnetization to the rf magnetic field, can be determined,
\begin{equation}
\textbf{m} = \chi \textbf{h} = \left( 
\begin{array}{ccc}
\chi_{xx} & i \chi_{xy}&0 \\
-i \chi_{xy} & \chi_{yy} &0 \\
0&0&0 
\end{array} \right) 
\textbf{h} = \left( \begin{array}{ccc}
|\chi_{xx}| & |\chi_{xy}|e^{i\pi/2}&0 \\
|\chi_{xy}| e^{-i\pi/2} & |\chi_{yy}| &0 \\
0&0&0 
\end{array} \right) 
\textbf{h} e^{i\Theta} \label{polder}
\end{equation}
where $\left(\chi_{xx}, \chi_{xy}, \chi_{yy}\right) = \left(D+i L\right) \left(A_{xx}, A_{xy}, A_{yy}\right)$ with \cite{Mecking2007}
\begin{align*}
A_{xx} &= \frac{\gamma M_0\left[M_0 N_y + \left(H- N_z M_0\right)\right]}{\alpha \omega \left[2 \left(H-N_z M_0\right) + M_0\left(N_z + N_y\right)\right]}\\
A_{xy} &= - \frac{M_0}{\alpha \left[2\left(H-N_z M_0\right) + M_0 \left(N_z + N_y\right)\right]} \\
A_{yy} &= \frac{\gamma M_0 \left[M_0 N_x + \left(H-N_z M_0\right)\right]}{\alpha \omega \left[2\left(H- N_z M_0\right) + M_0 \left(N_z + N_y\right)\right]}
\end{align*}
and
\begin{align*}
L = \frac{\Delta H^2}{\left(H-H_r\right)^2 + \Delta H^2}, ~~ D = \frac{\Delta H \left(H-H_r\right)}{\left(H-H_r\right)^2 + \Delta H^2}.
\end{align*}
Here $H = |\bf{H}|$ and $M_0 = |\bf{M}_0|$.  Also note that the rf magnetic field $\textbf{h}$ contains a possible phase shift with respect to the electric field which will drive the rf current in the sample. Here this phase shift is kept implicit for simplicity, but we will have to include it explicitly in the next section when we analyze spin rectification.  $H_r$ is the resonance field which depends on $\omega$ and the sample configuration according to the Kittel formula
\begin{equation}
\omega^2 = \gamma^2 \left[H_r + M_0 \left(N_y - N_z\right)\right]\left[H_r + M_0 \left(N_x - N_z\right)\right], \label{eq:FMRRes}
\end{equation}
and $\Delta H$ is a measure of the resonance line width (for the symmetric Lorentz function $L$, $\Delta H$ is the half width half maximum) which in general depends on $\omega, H$ and the sample configuration,
\begin{equation*}
\Delta H = \frac{2\left(H-N_z M_0\right) + M_0 \left(N_x + N_y\right)}{H+ H_r + M_0 \left(N_x + N_y - 2 N_z\right)} \frac{\alpha \omega}{\gamma}
\end{equation*}
but reduces to $\Delta H \approx \alpha \omega/\gamma$ near resonance.  The form of the line width which follows directly from the LLG equation does not take into account the effect of magnetization inhomogeneities and for comparison to experimental results the expression $\Delta H = \Delta H_\text{in} + \alpha \omega/\gamma$ is typically used, where $\Delta H_\text{in}$ is the inhomogeneous line width broadening.  $\Theta$ is the spin resonance phase which describes the phase shift between the rf driving force and the dynamic magnetization response.  It is defined by $\tan\Theta = \Delta H/\left(H-H_r\right) = L/D$ so that $\cos\Theta = D/\sqrt{L^2 + D^2}$ and $\sin\Theta = L/\sqrt{L^2 + D^2}$.  Near FMR $\Theta$ changes sign over a field range of order $\Delta H$ from $\pi$ for $H<H_r$ to 0 for $H>H_r$ as shown in Fig. \ref{mdynamics} (d).  

An important feature of $\chi$ is that each element contains both a Lorentz and dispersive component which have opposite symmetries around $H_r$ with $L\left(H_r + \delta H\right) = L\left(H_r - \delta H\right)$ and $D\left(H_r + \delta H\right) = - D \left(H_r - \delta H\right)$ where $\delta H$ is the field detuning.  This symmetry which is illustrated in Fig. \ref{mdynamics} (c) is strongly dependent on the spin resonance phase, as we can see from the fact that for any element of $\chi$, $|\chi_{ij}|^2 \propto L^2 + D^2 = L$.  In addition, both the Lorentz and dispersive components are symmetric under a field rotation of $\pi$, $\textbf{H} \to - \textbf{H}$ (note that under this rotation $H_r \to -H_r$).

%%%%%%%%%%%%%%%%%%%%%%%%%%%%%%%%%%%%%%%%%%%%%%%%%%%%%%%%%%%%%%%%%%%%%%%%%%%%%%%%%%%%%%%%%%%%%%%%%%%%%%%%%%%%%%%%%%%%%%%%%%%%%%%%%%%%%%%%%%%%%%%%%%%%%%%%%%%%%%%%%%%%%%%%%%%%%%%%%%%%%%%%%%%%%%%%%%%%%%%%%%%%%%%%%%%%%%%%%%%

\subsection{The Method of Analyzing Spin Rectification: Basic Examples} \label{spinrectification}

Having discussed magnetoresistance and magnetization dynamics we are now in a position to consider analyzing spin rectification \cite{Mecking2007,Harder2011a, Gui2012a}. The dc voltage produced by SR results from the nonlinear coupling between a time dependent resistance caused by magnetoresistance and a time dependent current driven by an rf $\textbf{e}$ field.  SR is nonlinear in the sense that only contributions of $\mathcal{O}\left(\textbf{h}^2, \textbf{j}^2, \textbf{h} \textbf{j}\right)$ will contribute to the time average which is taken over one period.  In this section we present the \textit{simplest example} of spin rectification, which occurs in ferromagnetic monolayers where the magnetoresistance is caused by magnetization precession driven by a torque from an rf $h$ field.  Therefore we will name this effect \textit{field torque rectification} to distinguish it from SR due to spin torque which will be discussed in Secs. \ref{srinbilayers} and \ref{srinmtjs}.  This simplest case of SR illustrates the key ideas that can be used to investigate SR in more complex device structures and may be analyzed using Eq. \ref{genrect}.  However if the voltage is not measured parallel to the rf current in general the planar, ordinary and extraordinary Hall effects will also contribute to the measured voltage.  It is therefore easier to determine the voltage line shape directly from the generalized Ohm's law of Eq. \ref{genohms}.  This method also provides another way to see the nonlinearity of the SR effect since the rf current generates corrections to the electric field due to AMR and the Hall effects.  We will consider only the AMR and AHE contributions in Eq. \ref{genohms} which are dominant in ferromagnetic metals (the OMR and OHE could be treated analogously). 

An important feature of the voltage line shape is its angular dependence, which allows the different physical contributions and the different driving fields to be separated.  In FM monolayers both the static behaviour, summarized in Table \ref{mrsummary}, and the rf behaviour, characterized by a coordinate change, will contribute to the angular dependence.  Even before going through a detailed derivation of the line shape it is possible to note the key angular dependencies due to the static behaviour from the magnetoresistance summarized in Table \ref{mrsummary}.  For AMR, $R_{AMR} \left(\theta_m\right) = R\left(0\right) - \Delta R \sin^2 \theta_m$ so if $\textbf{M}_0$ and $\textbf{H}$ are collinear, $V_{AMR} \propto \partial R/\partial \theta_H = \sin 2\theta_H$.  Proceeding analogously we expect $V_{PHE} \propto \cos2\theta_H$ and that $V_{AHE}$ will have no angular dependence from the static behaviour.  Each of these effects will also have an angular contribution which is unique to each component of the driving rf field, which must be determined by an analysis of the generalized Ohm's law.

%%%%%%%%%%%%%%%%%%%%%%%%%%%%%%%%%%%%%%%%%%%%%%%%%%%%%%%%%%%%%%%%%%%%%%%%%%%%%%%%%%%%%%%%%%%%%%%%%%%%%%%%%%%%%%%%%%%%%%%%%%%%%%%%%%%%%%%%%%%%%%%%%%%%%%%%%%%%%%%%%%%%%%%%%%%%%%%%%%%%%%%%%%%%%%%%%%%%%%%%%%%%%%%%%%%%%%%%%%%

\subsubsection{In-Plane Angular Dependence}

\begin{figure}[!ht]
\centering
\includegraphics[width=14.2cm]{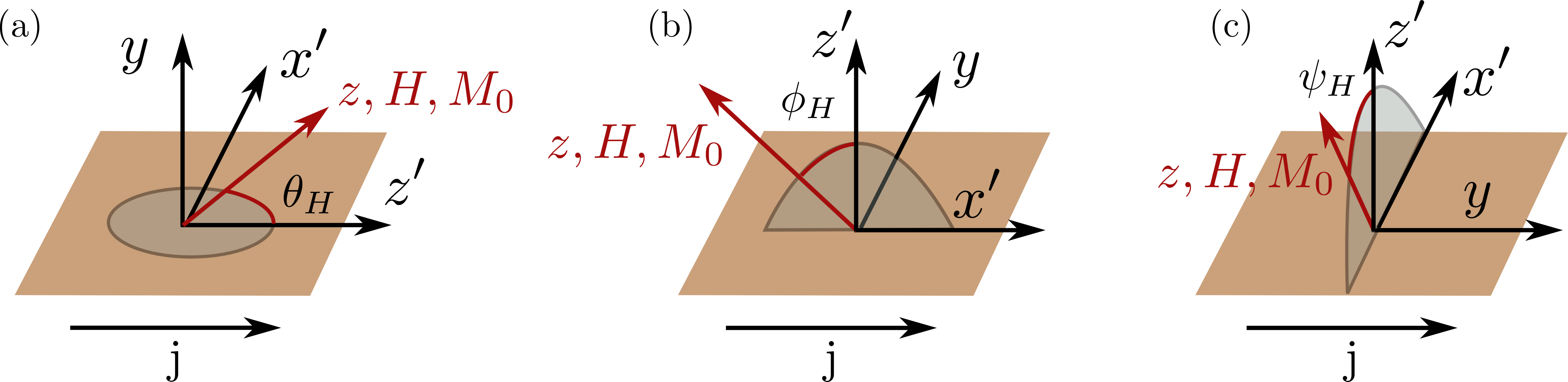}
\caption{\footnotesize{(a) Coordinate systems for an in-plane static magnetic field \textbf{H}.  The $z^\prime$ and $x^\prime$ axes are fixed along the sample's length and width respectively with the $y$ axis normal to the sample, and current is assumed to flow along $z^\prime$.  The $z$ axis is along the static field $\textbf{H}$ and static magnetization $\textbf{M}_0$ which is rotated by an angle $\theta_H$ with respect to the $z^\prime$ axis.  (b) Coordinate systems for the first out-of-plane rotation where the field is rotated along the sample length.  The $x^\prime$ axis is now along the length of the sample, the $y$ axis is along the width of the sample and the $z^\prime$ axis is normal to the sample, with current along $x^\prime$.  Again the static field and magnetization are along the $z$ axis which is rotated by an angle $\phi_H$ with respect to $z^\prime$.  (c) Coordinate systems for the second out-of-plane rotation where the field is rotated along the sample width.  The $y$ axis is now along the length of the sample, the $x^\prime$ axis is along the width of the sample and the $z^\prime$ axis is normal to the sample and the current is along $y$.  Again the static field and magnetization are along the $z$ axis which is rotated by an angle $\psi_H$ with respect to $z^\prime$.}}\label{srgeo}
\end{figure}

We first consider the case where $\textbf{H}$ is applied in the sample plane.  The coordinate system used is shown in Fig. \ref{srgeo} (a).  $(\widehat{x}^\prime, \widehat{y}, \widehat{z}^\prime)$ are the sample coordinates, with $\widehat{z}^\prime$ fixed along the long axis of the sample, $\widehat{x}^\prime$ along the width and $\widehat{y}$ normal to the sample plane.  $(\widehat{x}, \widehat{y}, \widehat{z})$ rotate with $\textbf{H}$ so that $\textbf{H} = H \widehat{z}$ makes an angle $\theta_H$ with $\widehat{z}^\prime$.  Note that since we always consider field rotations in the $x^\prime-z^\prime$ plane, the $\widehat{y}$ direction will remain fixed in all coordinate systems and there is no need to define both sample and field $\widehat{y}$ directions.  The in-plane coordinate systems are related by $\left(\widehat{x}, \widehat{y}, \widehat{z}\right) = \left(\cos\theta_H \widehat{x}^\prime - \sin\theta_H \widehat{z}^\prime, \widehat{y}, \sin \theta_H \widehat{x}^\prime + \cos\theta_H \widehat{z}^\prime\right)$, which can be conveniently written as $\vec{\textbf{x}}=U \vec{\textbf{x}}^\prime$ with $U$ being a rotation about $\widehat{y}$ by $\theta_H$.  If the sample length (along $\widehat{z}^\prime$) is much greater than its width (along $\widehat{x}^\prime$), the current will flow along $\widehat{z}^\prime$.  If in addition there is no applied dc current then $\textbf{J} = j^t \widehat{z}^\prime = \text{Re}(j_{z^\prime}e^{-i\omega t})\widehat{z}^\prime$ (a dc current will lead to a photoresistance \cite{Mecking2007} which we do not discuss here).  Splitting the magnetization into its dc and rf components, $\textbf{M} = M_0 \widehat{z}+ \textbf{m}^t = M_0\widehat{z} + \text{Re}\left(\textbf{m} e^{-i\omega t}\right)$, the time average of the electric field is
\begin{equation}
\textbf{E}_{MW} = -\frac{\Delta \rho}{M_0^2} 
\langle  \textbf{j}^t \cdot \textbf{M}_0 \textbf{m}^t + \textbf{j}^t \cdot \textbf{m}^t \textbf{M}_0 \rangle + R_{AHE} \langle \textbf{j}^t \times \textbf{m}^t\rangle.
\label{emprelim}
\end{equation}
Here we have used the fact that to lowest order $|\textbf{M}| = M_0$ is constant and therefore $\textbf{m} \cdot \textbf{M} = 0$.  From Eq. \ref{emprelim} we can see that, due to the cross product, only a magnetization component perpendicular to the plane will produce a signal due to the AHE.  This is why, when $\textbf{H}$ is in-plane, there is no static angular dependence associated with the AHE, and also why the AHE will not contribute when the voltage is measured longitudinally i.e. along the length of the sample.  We also see that only $m_x$ will contribute to longitudinal measurements.  These expectations are realized when we calculate the longitudinal voltage by integrating $\textbf{E}_{MW}$ along the length of the strip,
\begin{equation}
\left(V_\text{SR}\right)_\theta^l = \frac{\Delta R}{M_0} \sin \left(2\theta_H\right) \langle I_{z^\prime}^t m_{x}^t\rangle = \frac{\Delta R}{2 M_0} I_{z^\prime} \sin \left(2\theta_H\right) \text{Re}(m_{x}). \label{inplanev1}
\end{equation}
Here $\Delta R = \rho l/A$ and $I_{z^\prime} = j_{z^\prime}A$ where $A$ and $l$ are the cross sectional area and length of the strip respectively.  As expected the $\sin\left(2\theta_H\right)$ dependence is the key characteristic of the AMR induced SR.  

Even though only $m_x$ contributes to the SR voltage, since $m_x$ can be related to the rf $\textbf{h}$ field using the susceptibility in Eq. \ref{polder} both $h_x$ and $h_y$ can drive $m_x$.  Furthermore $\textbf{h}$ must be written in the primed coordinates, $\textbf{m} = \chi \textbf{h} = \chi U \textbf{h}^\prime$ which will introduce an additional $\theta_H$ dependence which is different for each component of $\textbf{h}^\prime$ and will also allow $h_{z^\prime}$ to contribute.  Using the susceptibility the photovoltage is 
\begin{equation}
\left(V_\text{SR}\right)_\theta^l = \frac{\Delta R}{2 M_0} I_{z^\prime} \left[A_L^{\theta x} L + A_D^{\theta x} D\right]  \label{inplanev2}
\end{equation}
where
\begin{align*}
A_L^{\theta x} &= \sin\left(2\theta_H\right) \left[-A_{xx} h_{x^\prime} \cos\theta_H \sin \Phi_{x^\prime} - A_{xy} h_y \cos \Phi_y + A_{xx} h_{z^\prime} \sin \theta_H \sin \Phi_{z^\prime}\right], \\
A_D^{\theta x} &= \sin\left(2\theta_H\right) \left[A_{xx} h_{x^\prime} \cos\theta_H \cos\Phi_{x^\prime} - A_{xy} h_y \sin\Phi_y - A_{xx} h_{z^\prime} \sin\theta_H \cos\Phi_{z^\prime}\right].
\end{align*}
Here we have introduced the notation $A_L^{ij}$ and $A_D^{ij}$ for the amplitudes of the Lorentz and dispersive contributions respectively where $i = \theta, \phi, \psi$ corresponding to the $\theta_H$, $\phi_H$ and $\psi_H$ rotations as defined in Fig. \ref{srgeo} and $j = x, y$ indicating from which component of the dynamic magnetization, $m_x$ or $m_y$ this amplitude arises.  We will use this same notation when discussing out-of-plane rotations in the next section.  We also denote the voltage as $\left(V_\text{SR}\right)_i^j$ again with $i = \theta, \phi, \psi$ and $j = l, t$ indicating that it is a voltage measured along the length or width of the sample respectively.  Similar notation will be used when we discuss spin pumping.

The relative phase $\Phi$ between the rf electric and magnetic fields, which may be different in the $\widehat{x}^\prime, \widehat{y}$ and $\widehat{z}^\prime$ directions, is now explicit.  The amplitudes $A_L^{\theta x}$ and $A_D^{\theta x}$ of the Lorentz and dispersive line shapes show the key characteristics of the AMR SR: $V \propto \sin\left(2\theta_H\right) \cos\theta_H$ for precession driven by $h_{x^\prime}$,  $V \propto \sin\left(2\theta_H\right)$ for precession driven by $h_{y}$ and $V \propto \sin\left(2\theta_H\right) \sin\theta_H$ for precession driven by $h_{z^\prime}$.  The different angular dependence allows the contribution from each $\textbf{h}^\prime$ component to be separated, and the ratio of Lorentz and dispersive amplitudes for each component, $A_L^{\theta x}/A_D^{\theta x}$, allows the relative phase to be determined for each $\textbf{h}^\prime$ component.  

For transverse measurements, across the width $w$ in the $\widehat{x}^\prime$  direction, both the AMR and AHE will contribute,
\begin{equation}
\left(V_\text{SR}\right)_\theta^t = -\frac{\Delta R}{2M_0} I_{z^\prime} \cos\left(2\theta_H\right) \text{Re}\left(m_x\right) - \frac{w R_{AHE}}{2} j_{z^\prime} \text{Re}\left(m_y\right). \label{outofplanev}
\end{equation}
Again only $m_x$ contributes to the SR due to AMR.  On the other hand only $m_y$ contributes to the AHE as expected.  Again the magnetization is related to $\textbf{h}$ through the susceptibility and an additional angular dependence is introduced by changing from $\textbf{h}$ to $\textbf{h}^\prime$,
\begin{equation*}
\left(V_\text{SR}\right)_\theta^t= -\frac{\Delta R}{2M_0} I_{z^\prime} \frac{\cos\left(2\theta_H\right)}{\sin\left(2\theta_H\right)} \left[A_L^{tx} L + A_D^{tx} D\right] - \frac{wR_{AHE}}{2} j_{z^\prime} \left[A_L^{ty} L + A_D^{ty} D\right]
\end{equation*}
where
\begin{align*}
A_L^{\theta y}&= A_{xy} h_{x^\prime}\cos\theta_H  \cos\Phi_{x^\prime} - A_{yy} h_{y} \sin\Phi_{y} - A_{xy} h_{z^\prime}\sin\theta_H  \cos\Phi_{z^\prime}, \\
A_D^{\theta y}&= A_{xy} h_{x^\prime}\cos\theta_H  \sin \Phi_{x^\prime} + A_{yy} h_{y} \cos\Phi_{y} - A_{xy} h_{z^\prime}\sin\theta_H  \sin\Phi_{z^\prime}
\end{align*}
$\Delta R = \Delta \rho w/A$ and $I_{z^\prime} = j_{z^\prime}A$.  The characteristic of the SR due to the PHE is the $\cos \left(2\theta_H\right)$ dependence, whereas the SR due to AHE has no angular dependence, other than that due to the coordinate change of $\textbf{h}$.

%%%%%%%%%%%%%%%%%%%%%%%%%%%%%%%%%%%%%%%%%%%%%%%%%%%%%%%%%%%%%%%%%%%%%%%%%%%%%%%%%%%%%%%%%%%%%%%%%%%%%%%%%%%%%%%%%%%%%%%%%%%%%%%%%%%%%%%%%%%%%%%%%%%%%%%%%%%%%%%%%%%%%%%%%%%%%%%%%%%%%%%%%%%%%%%%%%%%%%%%%%%%%%%%%%%%%%%%%%%

\subsubsection{Out-of-Plane Angular Dependence}

For an out-of-plane magnetic field we now have two planes of rotation to consider, as shown in Fig. \ref{srgeo} (b) and (c), however the treatment is almost identical to the in-plane case.  For both out-of-plane rotations we choose a coordinate system where again $\widehat{y}$ is parallel to $\widehat{y}^\prime$, $\widehat{z}^\prime$ is normal to the sample plane, and $\widehat{z}$ is directed along the external magnetic field direction.  The latter condition allows us to use the same solution of the LLG equation as the in-plane case (though the demagnetization factors will change).  This choice of coordinates means that for a rotation along the long axis of the sample, as shown in Fig. \ref{srgeo} (b), $\widehat{x}^\prime$ is along the long axis of the sample, $\widehat{y}$ is along the width of the sample and we choose $\phi_H$ to denote the angle between $\widehat{z}$ and $\widehat{z}^\prime$.  For a rotation along the short axis of the sample, as shown in Fig. \ref{srgeo} (c) $\widehat{x}^\prime$ is now along the width of the sample, $\widehat{y}$ is along the length and we choose $\psi_H$ to denote the angle between $\widehat{z}$ and $\widehat{z}^\prime$.  In both cases the ${\bf x}$ and ${\bf x}^\prime$ coordinates are related by $\vec{\textbf{x}} = U^\dagger \vec{\textbf{x}}^\prime$ (with $\theta_H$ replaced by $\phi_H$ or $\psi_H$ as appropriate).  

Proceeding as previously for the in-plane case, the photo voltage for the out-of-plane rotation by $\phi_H$ measured along the strip in the $\widehat{x}^\prime$ direction is
\begin{equation}
\left(V_\text{SR}\right)_\phi^l = \frac{\Delta R}{2M_0} I_{x^\prime} \sin\left(2\phi_H\right) \text{Re}\left(m_x\right)= \frac{\Delta R}{2M_0} I_{x^\prime} \left[A_L^{\phi x} L + A_D^{\phi x} D\right] \label{eq:outofplaneV}
\end{equation}
where
\begin{align*}
A_L^{\phi x} &= \sin\left(2\phi_H\right)\left[-A_{xx} h_{x^\prime} \cos\phi_H \sin\Phi_{x^\prime} - A_{xy} h_y \cos\Phi_y - A_{xx} h_{z^\prime} \sin\phi_H \sin\Phi_{z^\prime}\right], \\
A_D^{\phi x} &= \sin\left(2\phi_H\right)\left[A_{xx} h_{x^\prime} \cos\phi_H \cos\Phi_{x^\prime} - A_{xy} h_y \sin\Phi_y + A_{xx} h_{z^\prime} \sin\phi_H \cos\Phi_{z^\prime}\right].
\end{align*}
Again for measurements made along the width of the sample, now in the $\widehat{y}$ direction, both the AMR and AHE will contribute.  The form of the photo voltage is
\begin{align*}
\left(V_\text{SR}\right)_\phi^t &= \frac{\Delta R}{2M_0} I_{x^\prime} \sin\phi_H\text{Re}\left(m_y\right) - \frac{wR_{AHE}}{2} j_{x^\prime} \sin\phi_H\text{Re}\left(m_x\right) \\
&=\frac{\Delta R}{2M_0} I_{x^\prime} \left[A_L^{\phi y} L + A_D^{\phi y} D\right] - \frac{wR_{AHE}}{4} j_{x^\prime} \sec\phi_H\left[A_L^{\phi x} L + A_D^{\phi x} D\right]
\end{align*}
where
\begin{align*}
A_L^{\phi y}&= \sin\phi_H\left[A_{xy} h_{x^\prime}\cos\phi_H  \cos\Phi_{x^\prime} - A_{yy} h_{y} \sin\Phi_{y} + A_{xy} h_{z^\prime}\sin\phi_H  \cos\Phi_{z^\prime}\right], \\
A_D^{\phi y}&= \sin\phi_H\left[A_{xy} h_{x^\prime}\cos\phi_H  \sin \Phi_{x^\prime} + A_{yy} h_{y} \cos\Phi_{y} + A_{xy} h_{z^\prime}\sin\phi_H  \sin\Phi_{z^\prime}\right].
\end{align*}

For the other out-of-plane rotation by angle $\psi_H$ along the width of the strip, the longitudinal voltage measured along the strip length in the $\widehat{y}$ will vanish, $\left(V_\text{SR}\right)_\psi^l = 0$, since the static magnetization has no component along the measurement direction and therefore the average dynamic magnetization in this direction will be zero.  This fact can be used to separate spin pumping and spin rectification as we will discuss in Sec. \ref{sec:spsrseparation}.  On the other hand, in the traverse direction along $\widehat{x}^\prime$ the voltage will be 
\begin{align*}
\left(V_\text{SR}\right)_\psi^t &= \frac{\Delta R}{2M_0} I_{y} \sin\left(\psi_H\right) \text{Re}\left(m_y\right) + \frac{w R_{AHE}}{2} j_y \sin\psi_H \text{Re}\left(m_x\right) \\
&= \frac{\Delta R}{2M_0} I_{y} \left[A_L^{\psi y} L + A_D^{\psi y} D\right] + \frac{w R_{AHE}}{2} j_y \sec\psi_H \left(A_L^{\psi x} L + A_D^{\psi x} D\right) \numberthis\label{eq:outofplanepsiV}
\end{align*}
where
\begin{align*}
A_L^{\psi x} &= \sin\left(2\psi_H\right)\left[-A_{xx} h_{x^\prime} \cos\psi_H \sin\Phi_{x^\prime} - A_{xy} h_y \cos\Phi_y - A_{xx} h_{z^\prime} \sin\psi_H \sin\Phi_{z^\prime}\right], \\
A_D^{\psi x} &= \sin\left(2\psi_H\right)\left[A_{xx} h_{x^\prime} \cos\psi_H \cos\Phi_{x^\prime} - A_{xy} h_y \sin\Phi_y + A_{xx} h_{z^\prime} \sin\psi_H \cos\Phi_{z^\prime}\right], \\
A_L^{\psi y}&= \sin\psi_H\left[A_{xy} h_{x^\prime}\cos\psi_H  \cos\Phi_{x^\prime} - A_{yy} h_{y} \sin\Phi_{y} + A_{xy} h_{z^\prime}\sin\psi_H  \cos\Phi_{z^\prime}\right], \\
A_D^{\psi y}&= \sin\psi_H\left[A_{xy} h_{x^\prime}\cos\psi_H  \sin \Phi_{x^\prime} + A_{yy} h_{y} \cos\Phi_{y} + A_{xy} h_{z^\prime} \sin\psi_H \sin\Phi_{z^\prime}\right].
\end{align*}

The angular dependence of the rectified voltage in all six measurement/rotation configurations is summarized in Table \ref{srangularip} for both AMR and AHE.  In general when $V_\text{SR}$ is measured as a function of angle, both Lorentz and dispersive amplitudes will contribute, but angular fitting will allow the $h_{x^\prime}, h_y$ and $h_{z^\prime}$ components as well as the AMR and AHE contributions to be separated.  The line shape also has distinct symmetries under the rotation $\textbf{H} \to - \textbf{H}$ which are summarized in Table \ref{lineshapetable} in Sec. \ref{srinbilayers} after spin pumping has been discussed.

In closing our detailed discussion of field torque spin rectification we should note that our analysis here assumes that the static magnetization is fully saturated and aligned with the magnetic field.  If the sample has large anisotropy fields, there will be a regime where this is not the case. This is particularly true in the out-of-plane configurations, although non collinear dynamics can also be observed for an in-plane static field \cite{Huo2015}.  Such non collinear behaviour can clearly be observed in $\omega - H$ dispersion curves as deviations from the normal Kittel-like behaviour.  Despite this assumption, any formulas determined above will be exact if we simply replace $\theta_H \to \theta_M$.  The resulting expression can then be related to experimentally accessible parameters by determining the relationship, $\theta_M\left(\theta_H\right)$.  For a detailed discussion of such an analysis see Ref. \cite{Huo2015}.  It should also be mentioned that the effect of eddy currents has recently been investigated in FM/NM bilayers with thin NM layers, and can influence the line shape, as well as induce screening effects in such systems \cite{Flovik2015, Flovik2016}.

%%%%%%%%%%%%%%%%%%%%%%%%%%%%%%%%%%%%%%%%%%%%%%%%%%%%%%%%%%%%%%%%%%%%%%%%%%%%%%%%%%%%%%%%%%%%%%%%%%%%%%%%%%%%%%%%%%%%%%%%%%%%%%%%%%%%%%%%%%%%%%%%%%%%%%%%%%%%%%%%%%%%%%%%%%%%%%%%%%%%%%%%%%%%%%%%%%%%%%%%%%%%%%%%%%%%%%%%%%%

\section{Spin Rectification Effects in Magnetic Structures}\label{structures}

In Section \ref{magnetotransport} we discussed the physical origins of various magnetoresistance effects which occur in ferromagnetic heterostructures.  Since SR depends on both the underlying MR effect and the cause of magnetization dynamics, the origin and characteristics of spin rectification will also vary between different device structures.  Spin rectification in ferromagnetic monolayers, bilayer devices and MTJs will be the focus of this section.  Initial studies of field torque induced SR in ferromagnetic monolayers began as early as 1960 using pulsed microwave sources and will be the first topic of discussion.  These initial studies only became noticed by the community when advances in device fabrication techniques allowed the direct integration of ferromagnetic heterostructures and coplanar wave guides (CPWs).  For monolayer devices this led to the so called spin dynamo which will be considered next.  Parallel to the rediscovery of SR via the spin dynamo, SR became known by the spin pumping (SP) community since both effects will contribute to voltage measurements made on FM/NM bilayer samples.  Our next topic of discussion will therefore be the generation of dc voltages in bilayer samples, due to SR, SP and spin-transfer torque (STT).  Finally STT in MTJs and the so called spin diode will be considered.  

%%%%%%%%%%%%%%%%%%%%%%%%%%%%%%%%%%%%%%%%%%%%%%%%%%%%%%%%%%%%%%%%%%%%%%%%%%%%%%%%%%%%%%%%%%%%%%%%%%%%%%%%%%%%%%%%%%%%%%%%%%%%%%%%%%%%%%%%%%%%%%%%%%%%%%%%%%%%%%%%%%%%%%%%%%%%%%%%%%%%%%%%%%%%%%%%%%%%%%%%%%%%%%%%%%%%%%%%%%%

\subsection{Ferromagnetic Films: Early Studies Using Pulsed Microwave Sources}

\begin{figure}[!ht]
\centering
\includegraphics[width=11cm]{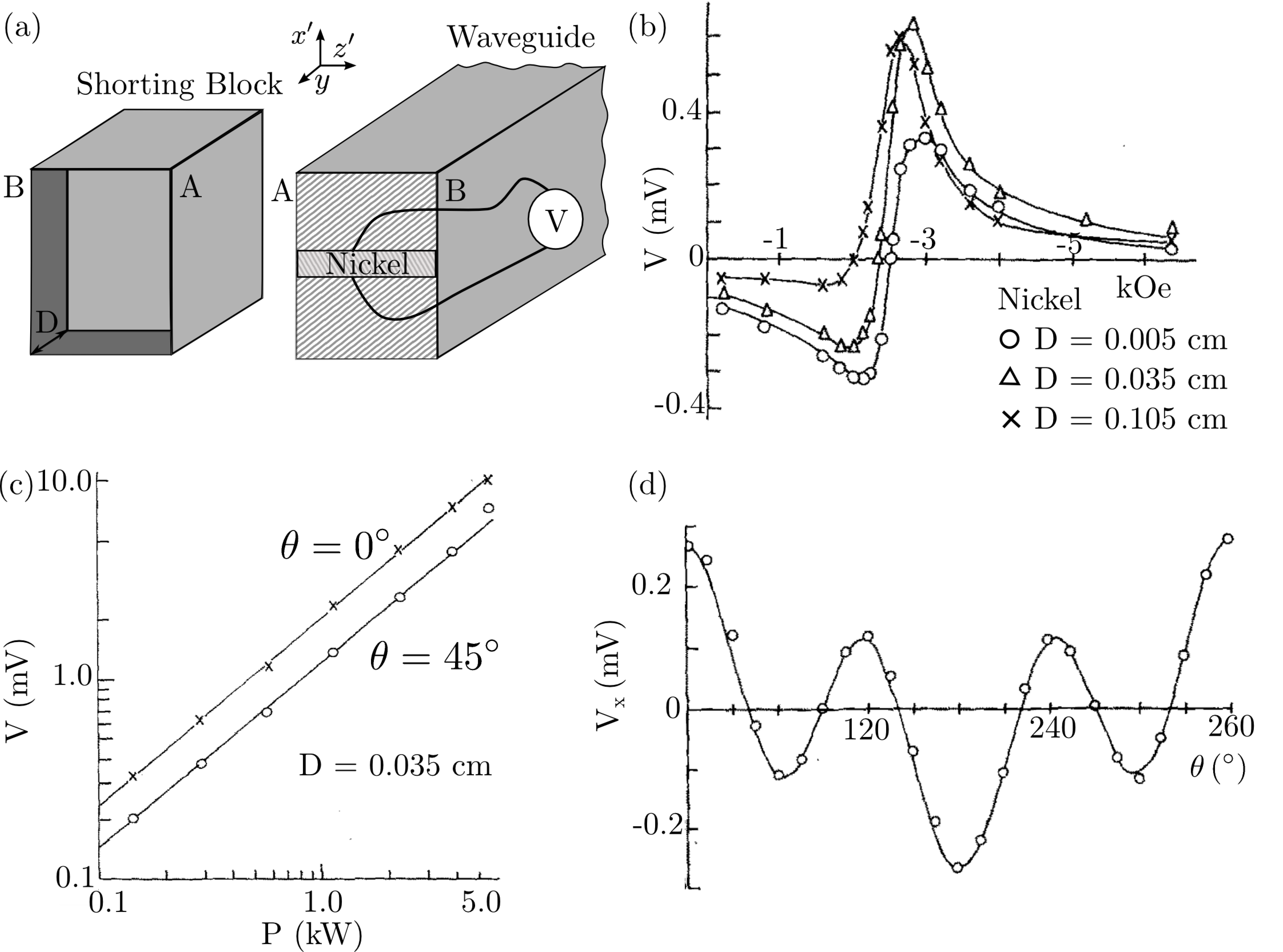}
\caption{\footnotesize{(a) The experimental setup used for early pulsed microwave studies.  A Ni thin film was placed at the end of a rectangular waveguide.  The sample holder was made of the same material as the film to simplify the field profile in the waveguide and a recessed shorting block allowed control over the distance $D$ between the short and the sample.  (b) Typical voltage curves measured for several values of $D$, showing both dispersive and Lorentz contributions to the line shape.  (c) Typical pulsed microwave power sensitivity of $\sim$ 1 nV/mW illustrating the high powers required to obtain a measurable voltage signal.  (d) The angular dependence of the transverse rectified voltage, corresponding to magnetization precession driven by a single $h$ field component ($h_{x^\prime}$) and containing both AMR and AHE contributions.  $Source:$ Adapted from Ref. \cite{Egan1963}.}}
\label{pulseddata}
\end{figure}

The pioneering work in the study of SR was performed by Juretschke in the early 1960's using pulsed microwave sources \cite{Egan1963, Juretschke1963} with a theoretical basis similar to the one discussed in Section \ref{spinrectification} \cite{Juretschke1960} (but using the Bloch equations).   A systematic study of the power, field strength and angular dependence of the rectified signal confirmed that the dc voltage accompanying FMR in thin FM films arose from the nonlinear coupling between the rf current and field torque induced magnetoresistance -- the dc signal was linearly dependent on the microwave power, see Fig. \ref{pulseddata} (c), and the angular dependence followed $V \left(\theta_H\right) = \left(V_1 \cos 2\theta_H + V_2\right) \cos \theta_H$, see Fig. \ref{pulseddata} (d).  This angular dependence is consistent with the discussion in Sec. \ref{spinrectification}
of a transverse measurement driven by a single field component, $h_{x^\prime}$ with $\Phi_{x^\prime} = 0$, where both AMR, $V_1$, and the AHE, $V_2$, contribute and the line shape consists of both Lorentz and dispersive contributions.  The typical measurement setup and data for pulsed microwave studies is shown in Fig. \ref{pulseddata}.  This early work established a novel way to study the GHz frequency properties of thin FM materials -- the initial samples studied varied in thickness from $\sim$ 10 - 100 nm and had lateral dimensions $\sim$ cm $\times$ cm. 

While the experimental design itself was sufficiently simple to facilitate adoption within the community, Juretschke's ideas did not become widespread until some 40 years later, due to the following experimental and theoretical limitations: 
\begin{enumerate}
\item The experimental configuration was not optimal.  In a transverse measurement, both AMR and the AHE contribute to the measured voltage, leading to a more complex signal.  Recent works have allowed for the independent study of AMR and the AHE using either a longitudinal configuration where only AMR contributes \cite{Mecking2007} or special samples where the AHE dominates and AMR is ignorable in the transverse direction \cite{Chen2013a}.  Also the precise control of the field configuration required to justify a theoretical description with a single $h$ field component was difficult to achieve experimentally.  This issue has been resolved through the development of a full theoretical description including all $h$ field components \cite{Mecking2007} and the development of novel on chip device structures with integrated CPWs, allowing for precise field orientation control.
\item Very high microwave powers were required.  Due to the low power sensitivities of $\sim$ 1 nV/mW shown in Fig. \ref{pulseddata} (c), large microwave powers of up to 5 kW were required to produce measurable voltage signals of $\sim$ mV.  This has been greatly improved in recent years due to sample design/fabrication, which allows the FM material to be directly integrated into a CPW chip, leading to typical power sensitivities of $\sim \mu$V/mW.  Sensitive detection techniques, such as lock-in amplification, capable of measuring signals down to the nV scale have also helped the situation.
\item The line shape was not fully understood.  Fig. \ref{pulseddata} (b) shows the complex line shape observed in these early experiments.  At the time the role of the relative phase in controlling such a line shape (discussed in Sec. \ref{spinrectification}) had not been discovered and therefore the precise meaning of the Lorentz and dispersive contributions to the voltage line shape was unknown.
\end{enumerate}

Despite these difficulties there was a small amount of early work performed using high power pulsed microwaves following either Juretschke's waveguide technique \cite{Seavey1960} or an analogous approach with microwave cavities \cite{Heinz1962} (which allowed mV voltages at slightly lower power).  This very early work is briefly summarized in Ref. \cite{Juretschke1963}.  Pulsed microwave techniques were also used to study spin wave resonances \cite{Moller1970} and applied to FM semiconductors, including in the study of the AHE \cite{Toda1970, Chazalviel1972, Chazalviel1975} and also used to study the AHE in paramagnetic metals \cite{Wald1971}.  

However, aside from the few works mentioned, Juretschke's work went largely unappreciated until the rectification effect was rediscovered four decades later \cite{Costache2006a, Gui2007}. Two very different paths led to these experiments \cite{Costache2006a, Gui2007}. One path came from the semiconductor community studying electrical detection of spin dynamics, which led to the extension of such powerful experimental methods from semiconductor \cite{Ganichev2002, Hu2003, Yang2006} to ferromagnetic materials \cite{Gui2005a, Gui2007}.  The other path emerged from theoretical progress in studying magnetism, motivated by the prediction of a dc voltage induced by spin pumping in magnetic bilayers (see section 3.3). The spin pumping community, having observed the voltage produced in bilayer samples at FMR, needed to ensure the source of such voltage was in fact spin pumping, which required careful analysis of the voltage produced by rectification effects \cite{Costache2006a}. In 2006, these two paths crossed during the development of dynamic spintronic devices which integrated micro-structured ferromagnetic monolayers with microwave CPWs.  

%%%%%%%%%%%%%%%%%%%%%%%%%%%%%%%%%%%%%%%%%%%%%%%%%%%%%%%%%%%%%%%%%%%%%%%%%%%%%%%%%%%%%%%%%%%%%%%%%%%%%%%%%%%%%%%%%%%%%%%%%%%%%%%%%%%%%%%%%%%%%%%%%%%%%%%%%%%%%%%%%%%%%%%%%%%%%%%%%%%%%%%%%%%%%%%%%%%%%%%%%%%%%%%%%%%%%%%%%%%

\subsection{Micro-Structured Monolayers: The Spin Dynamo} \label{spindynamo}

\begin{table}[!b]
\def\arraystretch{1}
\caption{\footnotesize{A summary of key experimental results using single ferromagnetic devices.  In Refs. \cite{Costache2006a, Gui2007, Wirthmann2008} the current in the FM material was due to capacitive and/or inductive coupling with the current in the CPW, while in Ref. \cite{Chen2013a} the rf magnetic field was due to the current in the Py.}}
\centering
\begin{tabular}{>{\centering\arraybackslash}m{4cm}>{\centering\arraybackslash}m{4cm}>{\centering\arraybackslash}m{6cm}}
\toprule 
\multicolumn{2}{l}{{\bf Spin Rectification due to Field Torque}} \\ \toprule               
Reference & Device Structure & Result \\ \midrule \\
Costache et al. \cite{Costache2006a} & \includegraphics[width=2.4cm]{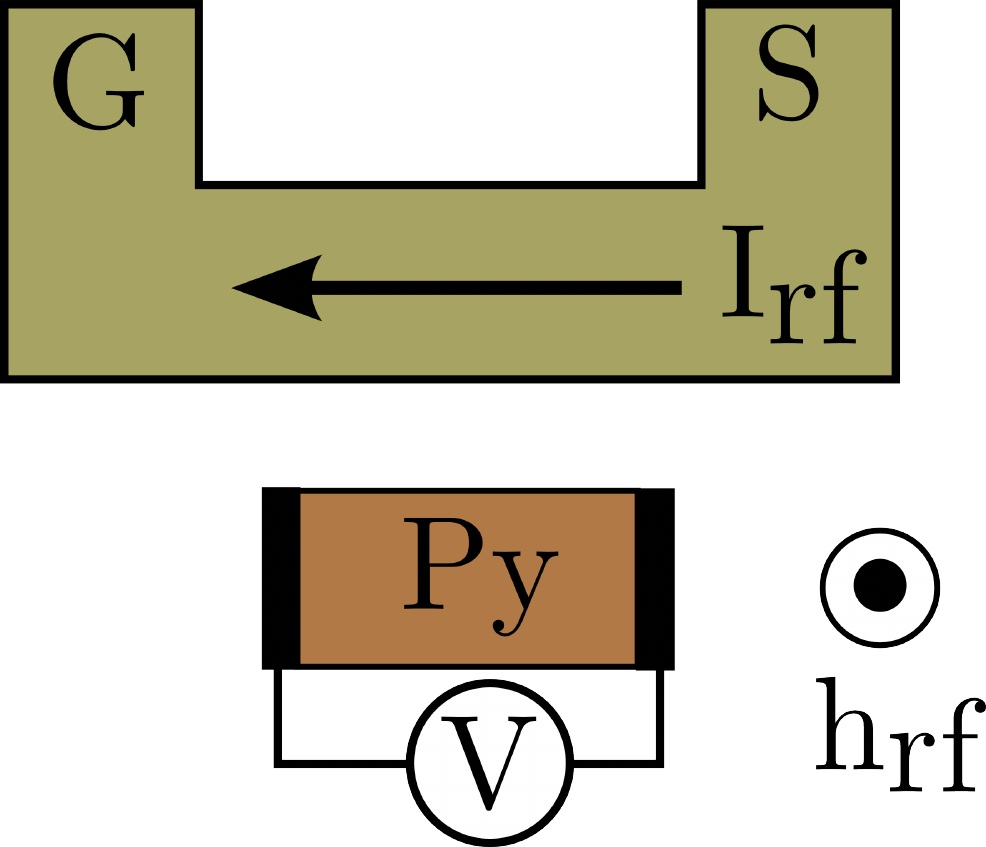} & Measurement of dc voltage due to SR in Py\\ \\
Gui et al. \cite{Gui2007} & \includegraphics[width=2.4cm]{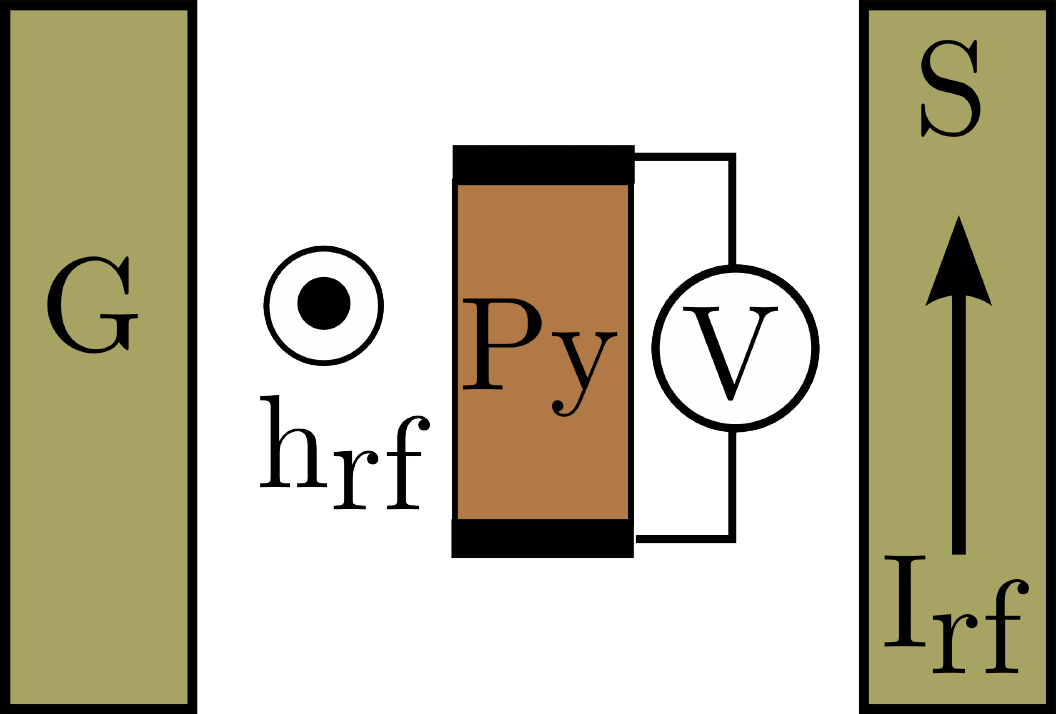} &  Measurement of bipolar dc current due to SR in Py\\ \\
Wirthmann et al. \cite{Wirthmann2008} & \includegraphics[width=2.8cm]{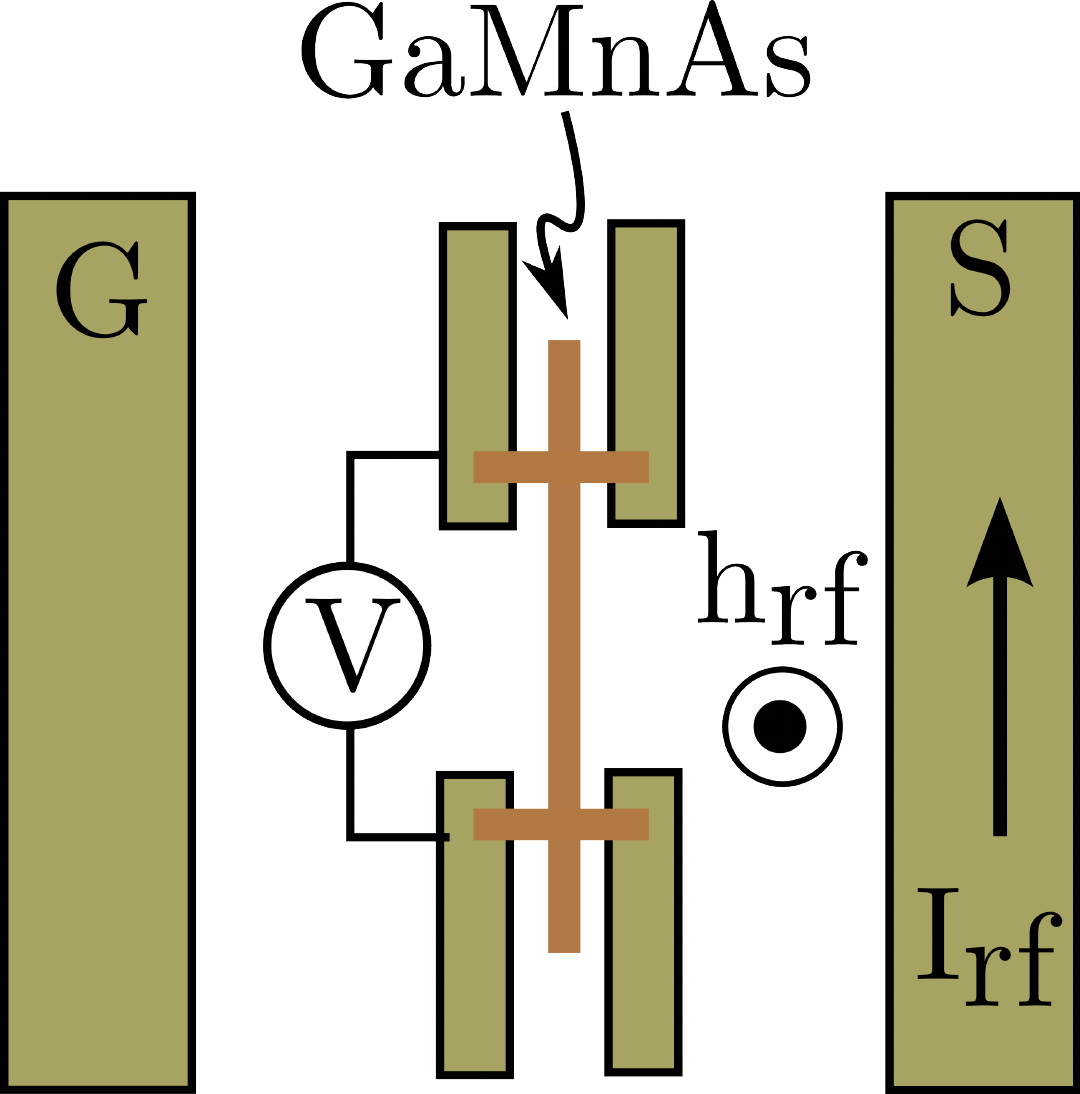} & Measurement of SR in ferromagnetic semiconductor GaMnAs\\ \\
Chen et al. \cite{Chen2013a} &\includegraphics[width=2.8cm]{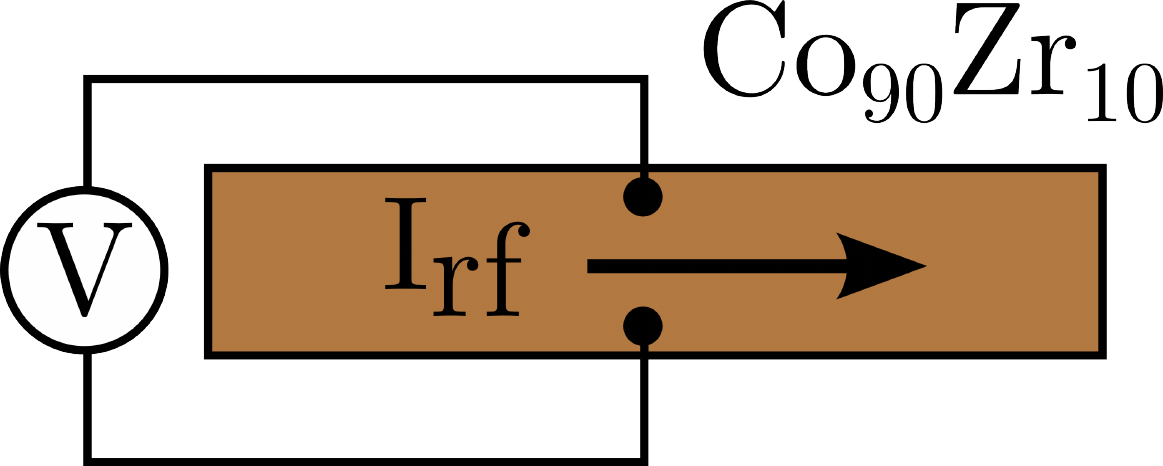} &  Measurement of SR due to AHE in Co$_{90}$Zr$_{10}$ which has suppressed AMR\\ \\ 
\toprule
\multicolumn{2}{l}{{\bf Spin Rectification due to Spin-Transfer Torque}} \\ \toprule                 
Reference & Device Structure & Result \\ \midrule \\
Yamaguchi et al. \cite{Yamaguchi2007} &\includegraphics[width=2.8cm]{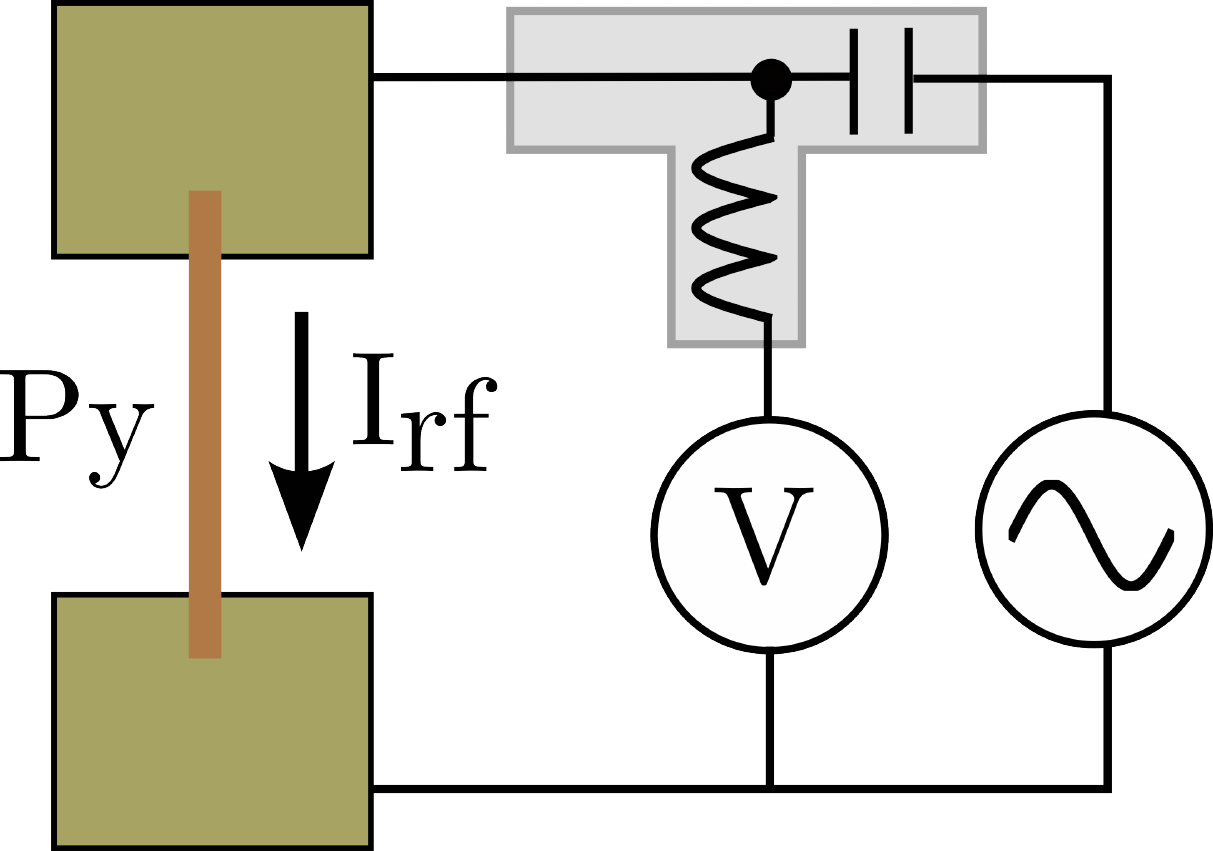} &  Measurement of dc voltage due to SR in Py nanowire\\ \\
Bedau et al.  \cite{Bedau2007} &\includegraphics[width=4cm]{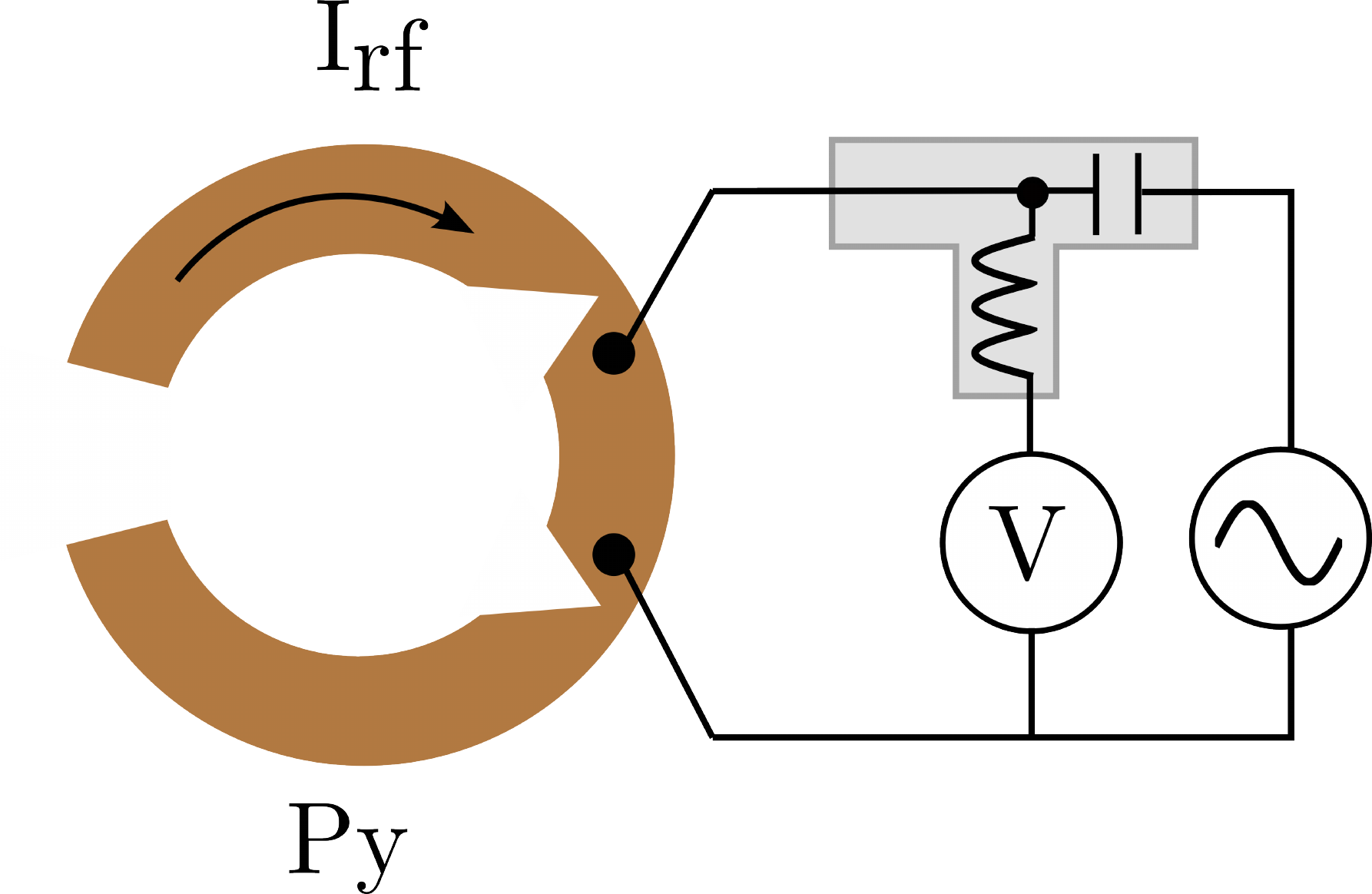} & Detection of domain wall resonance using SR 
\\ \bottomrule
\end{tabular}
\label{spindynamosummary}
\end{table}  
 
\begin{figure}[!b]
\centering
\includegraphics[width=7cm]{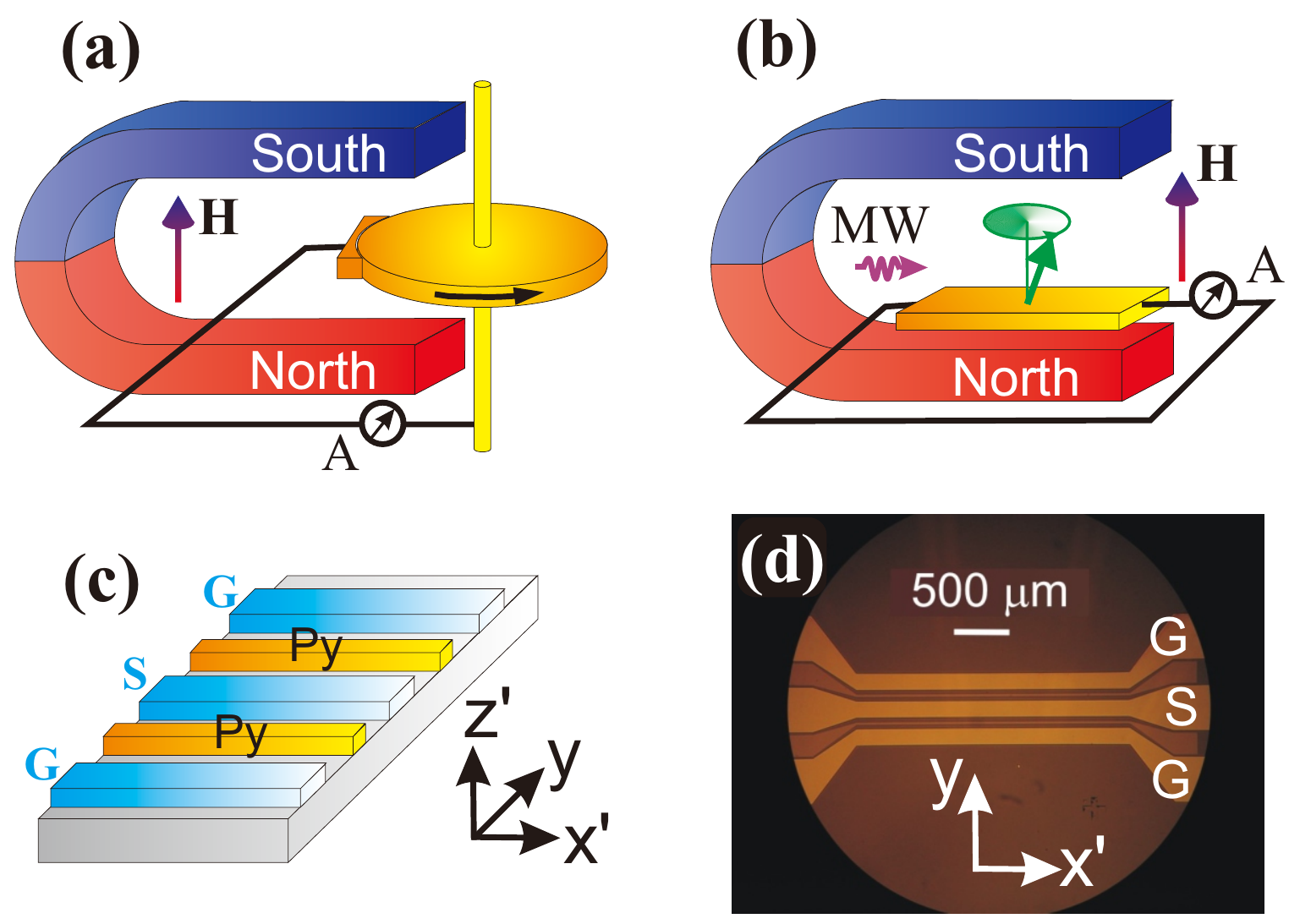}
\caption{\footnotesize{A striking analogy exists between Faraday's dynamo and the spin dynamo.  (a) In Faraday's dynamo a revolving copper disk converts energy from rotation to an electrical current. (b) Meanwhile a spin dynamo, with a FM strip, converts energy from spin precession to a bipolar current of electricity, which can be rectified into a dc voltage.  $Source:$ Adapted from Ref. \citenum{Gui2007}. (c) A schematic illustration and (d) micrograph of the CPW structure used to provide the driving rf magnetic field in the spin dynamo.}}
\label{sdynamo}
\end{figure} 

In this section we consider spin rectification in ferromagnetic monolayers.  In such devices the merger of advanced fabrication techniques and integration with coplanar wave guides allows for enhanced control of the driving magnetic field and increased power sensitivity compared to pulsed microwave studies.  The first observation of SR in such devices was made independently by Costache et al. \cite{Costache2006a} and Gui et al. \cite{Gui2007} using permalloy (Py) in a longitudinal configuration where the rectification is due to AMR.  AMR based SR has also been observed in ferromagnetic semiconductors where it allows for broadband material characterization \cite{Wirthmann2008}.  Although SR in single ferromagnetic devices is most commonly due to field torque driven magnetization dynamics and AMR, in certain single layer devices, such as FM nanowires and notched FM rings where large spatial variations of the magnetization exist \cite{Yamaguchi2007, Bedau2007}, or homogenous ferromagnets with certain crystalline symmetries \cite{Kurebayashi2014}, as it is also possible to observe spin torque induced rectification.  Also, in certain FM materials (such as Co$_{90}$Zr$_{10}$) AMR may be suppressed allowing studies of pure AHE SR \cite{Chen2013a}.  Some of the key experimental work related to SR in single ferromagnetic devices is summarized in Table \ref{spindynamosummary}.  Our discussion here will focus on ferromagnetic monolayers where SR is due to AMR and field torque driven magnetization dynamics, with spin torque discussed later in the context of bilayer devices and MTJs.

%%%%%%%%%%%%%%%%%%%%%%%%%%%%%%%%%%%%%%%%%%%%%%%%%%%%%%%%%%%%%%%%%%%%%%%%%%%%%%%%%%%%%%%%%%%%%%%%%%%%%%%%%%%%%%%%%%%%%%%%%%%%%%%%%%%%%%%%%%%%%%%%%%%%%%%%%%%%%%%%%%%%%%%%%%%%%%%%%%%%%%%%%%%%%%%%%%%%%%%%%%%%%%%%%%%%%%%%%%%

\subsubsection{Method and Device Structure}

\begin{figure}[!b]
\centering
\includegraphics[width=15cm]{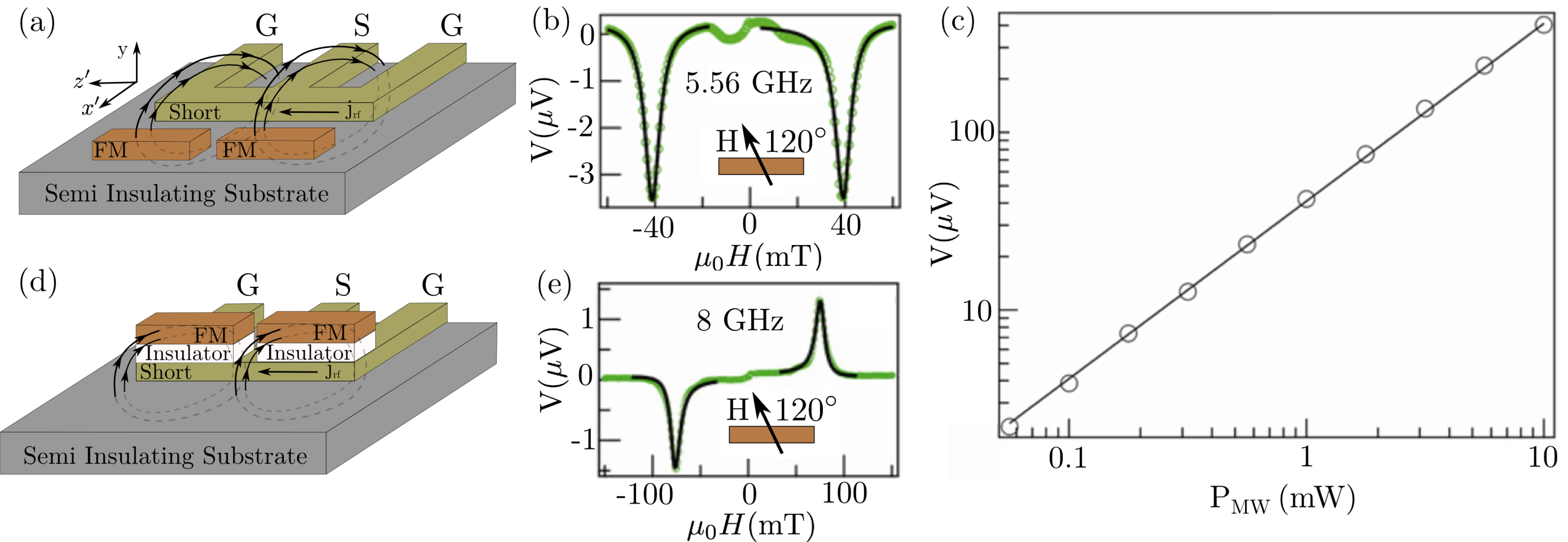}
\caption{\footnotesize{(a) and (d) Schematic diagrams of a first and second generation spin dynamo respectively.  Circulating arrows indicate the direction of the Oersted field.  In the first generation device the ferromagnetic microstrip is placed beside the short of the CPW, therefore the dominant rf magnetic field in the FM is in the $y$ direction.  In the second generation spin dynamo the FM is placed above (or below) the CPW (with an electrically insulating layer in-between) and therefore the dominant rf magnetic field in the FM is in the $x^\prime$ direction.  (b) and (e) The line shape measured at an in-plane angle of $\theta_H = 120^\circ$ at a frequency of $\omega/2\pi = 5.56$ GHz for the first generation and $\omega/2\pi = 8$ GHz for the second generation spin dynamo respectively \cite{Harder2011a}.  The first generation device used a Py microstrip of dimension 300 $\mu$m $\times$ 20 $\mu$m $\times$ 50 nm while the second generation device used a Py strip of dimension 300 $\mu$m $\times$ 7 $\mu$m $\times$ 100 nm with a 200 nm thick SiO$_2$ insulating layer.  In both cases the line shape is nearly perfect Lorentz, which means for the first generation dynamo that the relative phase is nearly zero, while for the second generation dynamo the relative phase is nearly 90$^\circ$.  The fits (black curves) are done using Eq. \ref{inplanev2} with (b) $\mu_0 \Delta H = 3.6$ mT and $\mu_0 H_r = 39.1$ mT and (e) $\mu_0 \Delta H = 6.0$ mT and $\mu_0 H_r = 76.5$ mT.  In both cases the line shapes follow the theoretical expectation summarized in Table \ref{lineshapetable}.  (c) Linear dependence of the photo voltage on microwave power in a first generation spin dynamo.  The power sensitivity for this device is 41 $\mu$V/mW, typical for microstructured spin dynamos.  $Source:$ Adapted from Refs. \cite{Harder2011a} and \cite{Gui2012a}.}}
\label{srdevice}
\end{figure}

A spin dynamo \cite{Gui2007, Mecking2007} is an on-chip device consisting of a FM microstrip, such as Py, placed in the microwave field of a ground-signal-ground (GSG) CPW, so named due to its analogy with a Faraday dynamo as shown in Fig. \ref{sdynamo}.  A first and second generation spin dynamo are shown in Fig. \ref{srdevice} (a) and (d).  In the first generation device Py microstrips are placed either beside the short of the CPW or between the GS lines.  
As shown in Fig. \ref{srdevice} (a) two FM strips may be located on one device.  The key difference between the first and second generation devices is: 1) a different component of the microwave field will drive precession ($h_y$ vs $h_{x^\prime}$ for a first and second generation device respectively) and 2) the field intensity in the FM layer will be greater in the second generation device since it can be located closer to the CPW -- a second generation spin dynamo has $\sim$ 200 nm spacing between the CPW and microstrip, compared to $\sim$ 40 $\mu$m in the first generation structure, which results in an increase of the in-plane $h$ field by two orders of magnitude.

Fig. \ref{srdevice} (b) shows the field dependent voltage measured using a first generation spin dynamo consisting of a Cu/Cr CPW fabricated beside a Py microstrip of dimensions 300 $\mu$m $\times$ 20 $\mu$m $\times$ 50 nm on a SiO$_2$/Si substrate.  The line shape is measured at an in-plane angle of $\theta_H = 120^\circ$ at a frequency of $\omega/2\pi = 5.56$ GHz.  The line shape follows Eq. \ref{inplanev2} and since the relative phase is nearly zero it has a dominant Lorentz contribution, following the symmetry expected from Table \ref{lineshapetable}.  In comparison, Fig. \ref{srdevice} (e) shows the voltage measured in a second generation device where the Py is above the CPW short, being electrically isolated by a 200 nm thick SiO$_2$ layer.  Again the in-plane line shape in Fig. \ref{srdevice} (e) is nearly Lorentz, but as can be seen from Eq. \ref{inplanev2}, for an $h_{x^\prime}$ driving field a Lorentz line shape means that the relative phase is nearly 90$^\circ$.

Fig. \ref{srdevice} (c) shows a typical linear power sensitivity curve for a first generation device.  In this device the power sensitivity is 41 $\mu$V/mW -- a three order of magnitude improvement over the power sensitivity of pulsed microwave studies.  The magnitude of this improvement is typical of spin dynamos, whose power sensitivity typically ranges from 10 - 100 $\mu$V/mW \cite{Gui2007}.  This large power sensitivity has enabled studies across a large range of powers, allowing electrical detection of very weak modes due to spin waves, as well as the study of nonlinear effects \cite{Gui2009, Gui2009a} (see Sec. \ref{spinwaves} and \ref{nonlinear} respectively).

%%%%%%%%%%%%%%%%%%%%%%%%%%%%%%%%%%%%%%%%%%%%%%%%%%%%%%%%%%%%%%%%%%%%%%%%%%%%%%%%%%%%%%%%%%%%%%%%%%%%%%%%%%%%%%%%%%%%%%%%%%%%%%%%%%%%%%%%%%%%%%%%%%%%%%%%%%%%%%%%%%%%%%%%%%%%%%%%%%%%%%%%%%%%%%%%%%%%%%%%%%%%%%%%%%%%%%%%%%%

\subsubsection{Lineshape Analysis} \label{lineshapeanalysis} 

An analysis of spin rectification experiments requires understanding both the line shape and the angular dependence of the rectified voltage.  Experimentally the line shape is studied by controlling the magnitude of the applied static magnetic field, while the angular dependence of course involves changing the direction of the field.  The importance of a careful line shape analysis is emphasized by considering the subtlety of the relative phase dependence of the line shape derived in Eq. \ref{inplanev2} and shown experimentally in Fig. \ref{srdevice} (b) and (e), where it is clear that even though the line shape is nearly Lorentz, there is still a small dispersive component.  The fact that the spin rectification may have both Lorentz and dispersive contributions is especially important when other voltage producing effects, such as spin pumping, need to be distinguished.  In early studies it was assumed that the rectification line shape was dispersive -- a fact that was used to distinguish the effect from spin pumping \cite{Mosendz2010, Mosendz2010a}.  This assumption is correct when the rf currents are dominated by capacitative coupling, e.g. when the sample is in close proximity to the waveguide and FMR is driven by the $h_{x^\prime}$ field \cite{Hoffmann2013}.  Due to the importance of this special case, the line shape symmetry for $\Phi_{x^\prime} = \Phi_{y} = \Phi_{z^\prime} = 0$ is summarized in Table \ref{lineshapetable}.  The SP line shape is also summarized in this table and will be discussed more in Section \ref{srinbilayers}.  The key features summarized in Table \ref{lineshapetable} are the Lorentzian and dispersive nature of the voltage line shape as well as the symmetry of the line shape under the change $\textbf{H} \to - \textbf{H}$.  

\begin{figure}[!b]
\centering
\includegraphics[width=7cm]{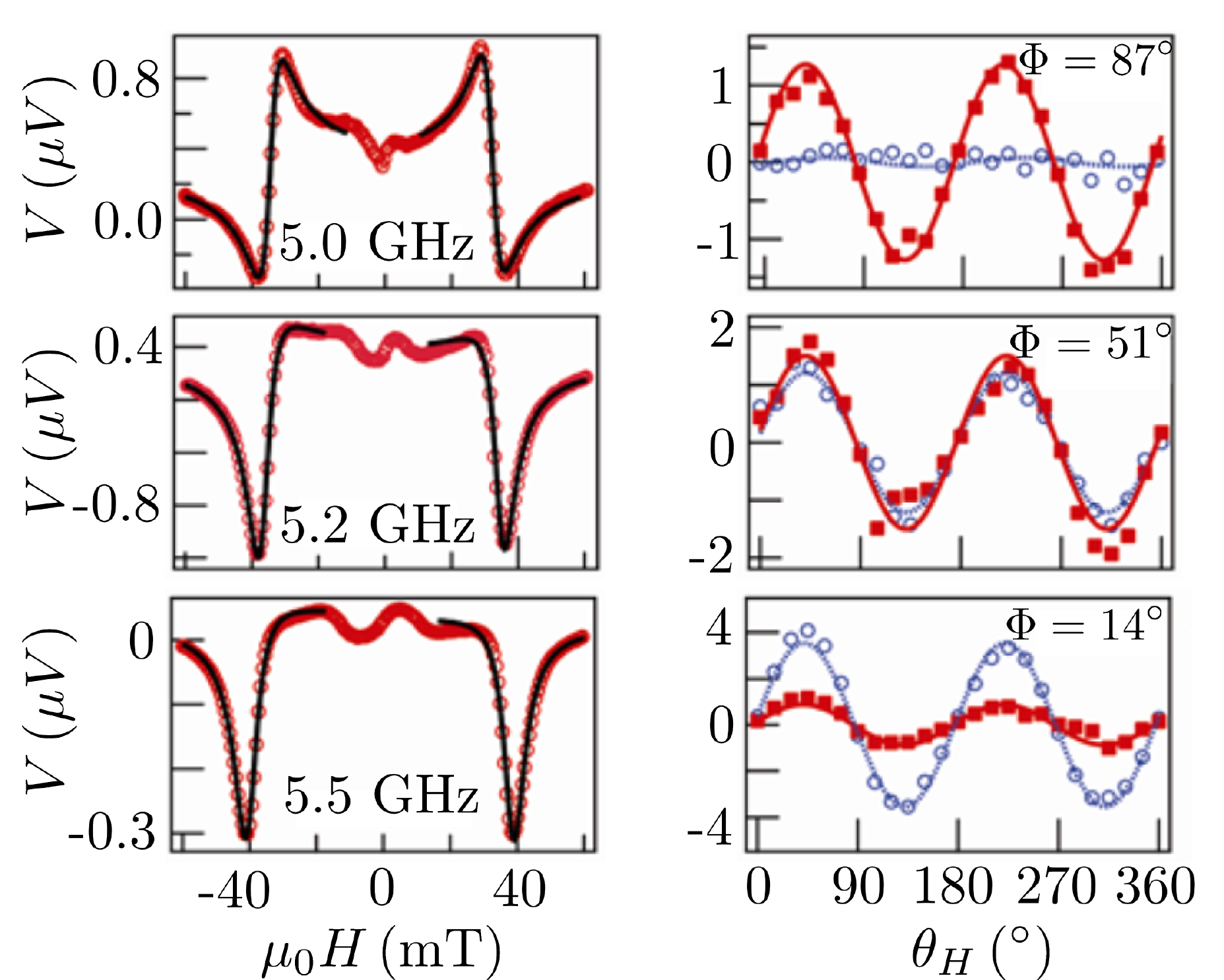}
\caption{\footnotesize{A demonstration of the frequency dependence of the relative phase.  The data was collected using the first generation spin dynamo shown in Fig. \ref{srdevice} (a).  The left panel shows FMR spectra at $\theta_H = 120^\circ$ for several frequencies ranging from 5.0 to 5.5 GHz.  The line shape fits (black lines) are performed using Eq. \ref{inplanev2}.  The right panel shows the Lorentz (squares) and dispersive (circles) amplitudes of the voltage as a function of $\theta_H$.  The solid curves are a fit to $\sin2\theta_H$.  Both panels clearly show the line shape changing between dominantly Lorentz at 5.5 GHz and dispersive at 5.0 GHz.  $Source:$ Adapted from Ref. \cite{Harder2011a}.}}
\label{relativephase}
\end{figure}
This latter symmetry is determined both from the fact that $L(H, H_r) = L(-H, -H_r)$ and $D(H, H_r) = -D(-H, - H_r)$ and from the angular dependence of the line shape and its symmetry under the change $\theta_H \to \theta_H + \pi$.  Despite the importance of this simplified case, in general both dispersive and Lorentz contributions will be present except in certain carefully designed devices and both should be considered when performing a line shape analysis \cite{Harder2011a, Azevedo2011}.  This means that the relative phase, which in general depends in a complicated way on e.g. the waveguide, coaxial cables, bonding wires and sample  holder, must be calibrated.  This can be done by first separating the field components by fitting the angular dependence (see Sec. \ref{angulardependence}), and then using these results to determine the $L$ and $D$ fitting.   

\begin{table}[!b]
\def\arraystretch{1}
\caption{\footnotesize{Voltage line shapes of SR and SP.  Theoretical voltage curves for SR are calculated under the simplification $\Phi_{x^\prime} = \Phi_y = \Phi_{z^\prime} = 0$ and illustrate whether the line shape is Lorentz or dispersive, as well as the symmetry under a $180^\circ$ rotation $\textbf{H} \to - \textbf{H}$.  $\textbf{H}$ is directed at any angle in the indicated rotation plane, other than integer multiples of $90^\circ$; at these high symmetry points the voltage may vanish.  For SP the line shape in all rotation planes is identical: Lorentz and symmetric in $\textbf{H}$, however the signal may vanish in certain configurations as summarized in Table \ref{srangularip}.}}
\centering
\begin{tabular}{>{\centering\arraybackslash}m{2cm}>{\centering\arraybackslash}m{1.94cm}>{\centering\arraybackslash}m{1.94cm}>{\centering\arraybackslash}m{1.94cm}>{\centering\arraybackslash}m{1.94cm}>{\centering\arraybackslash}m{1.94cm}>{\centering\arraybackslash}m{1.94cm}}
\cmidrule[0.75pt](l{0.3em}r{0.3em}){1-7}                       
& \includegraphics[width=1.7cm]{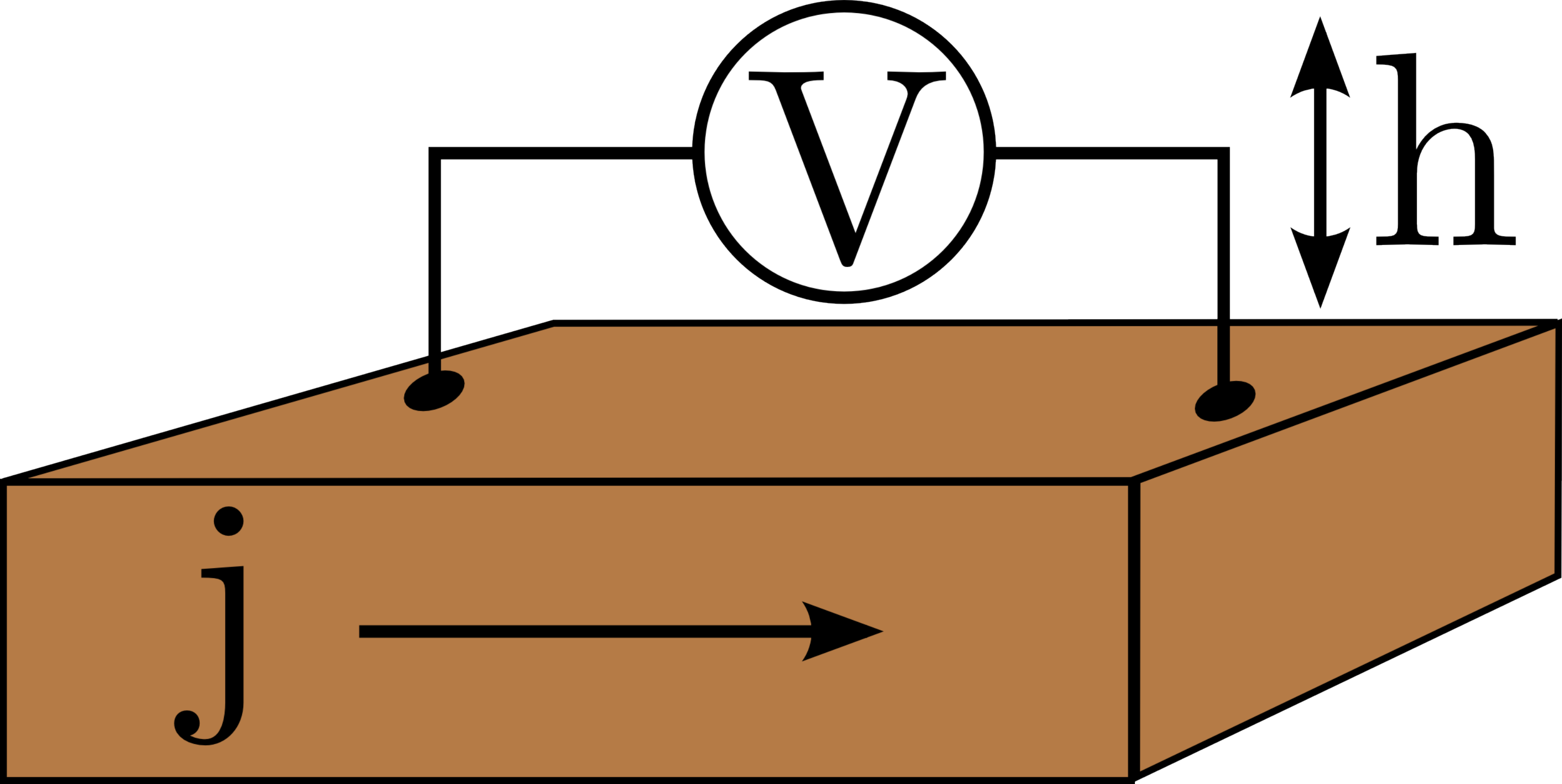} & \includegraphics[width=1.7cm]{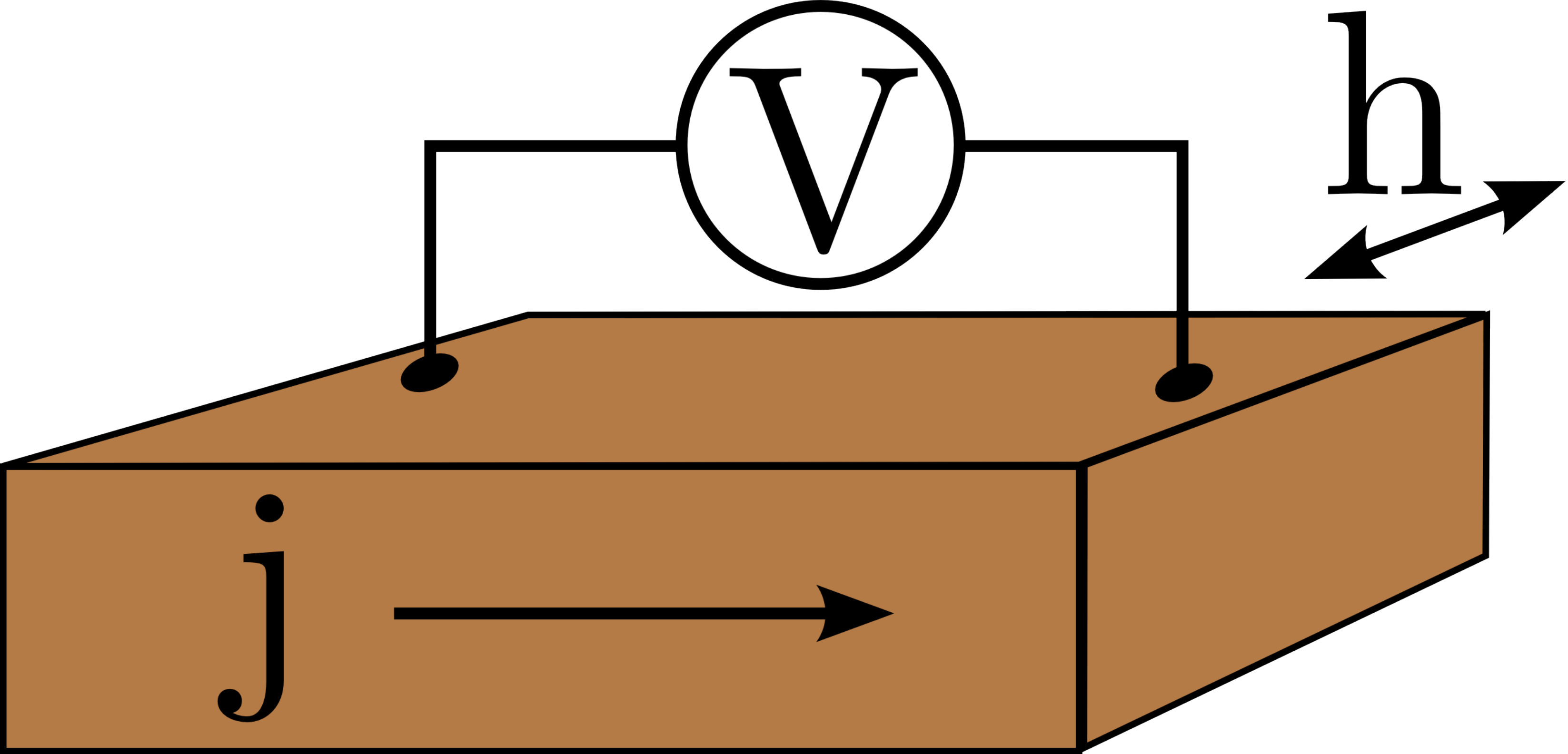} &\includegraphics[width=1.7cm]{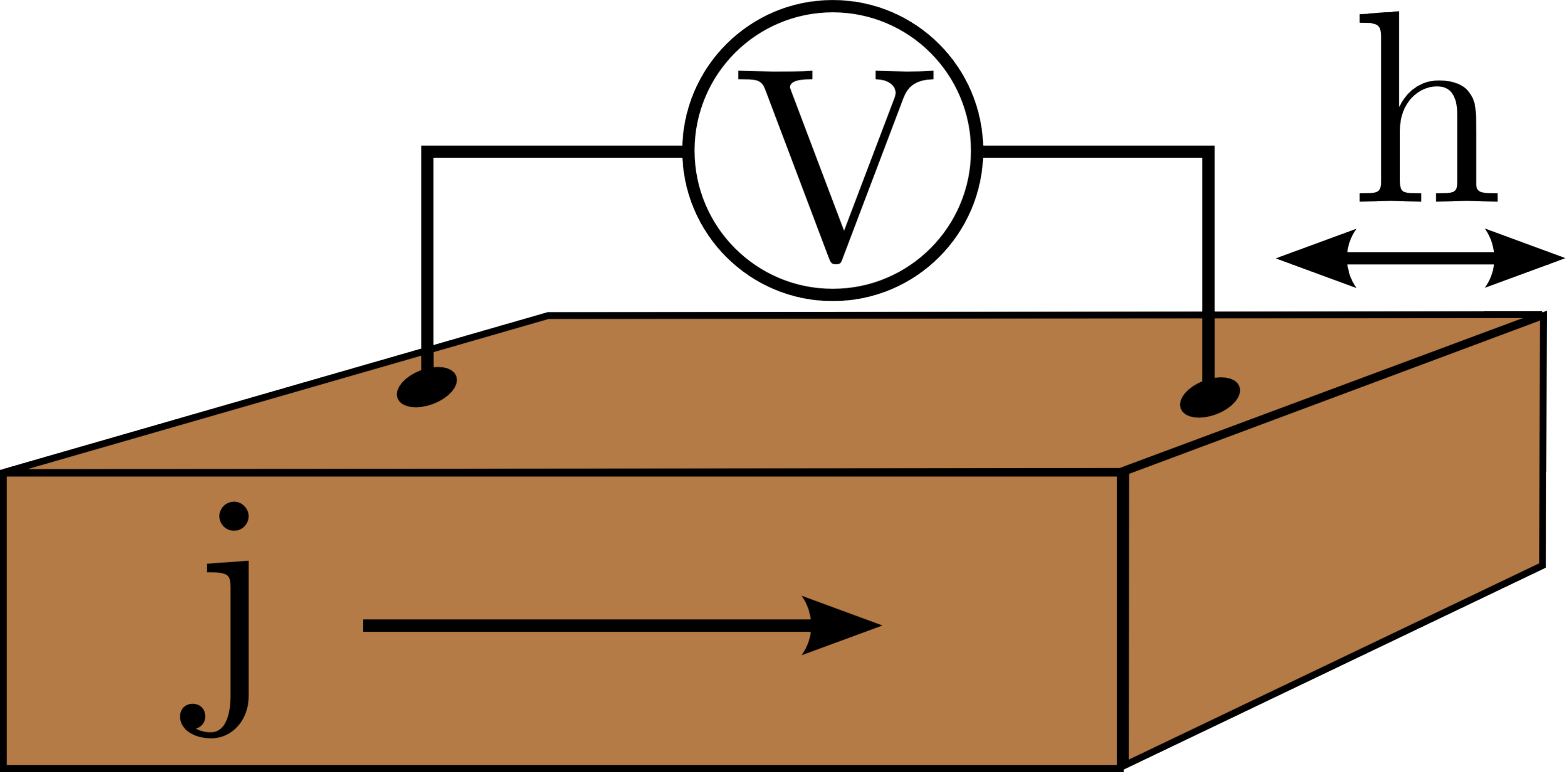}&\includegraphics[width=1.7cm]{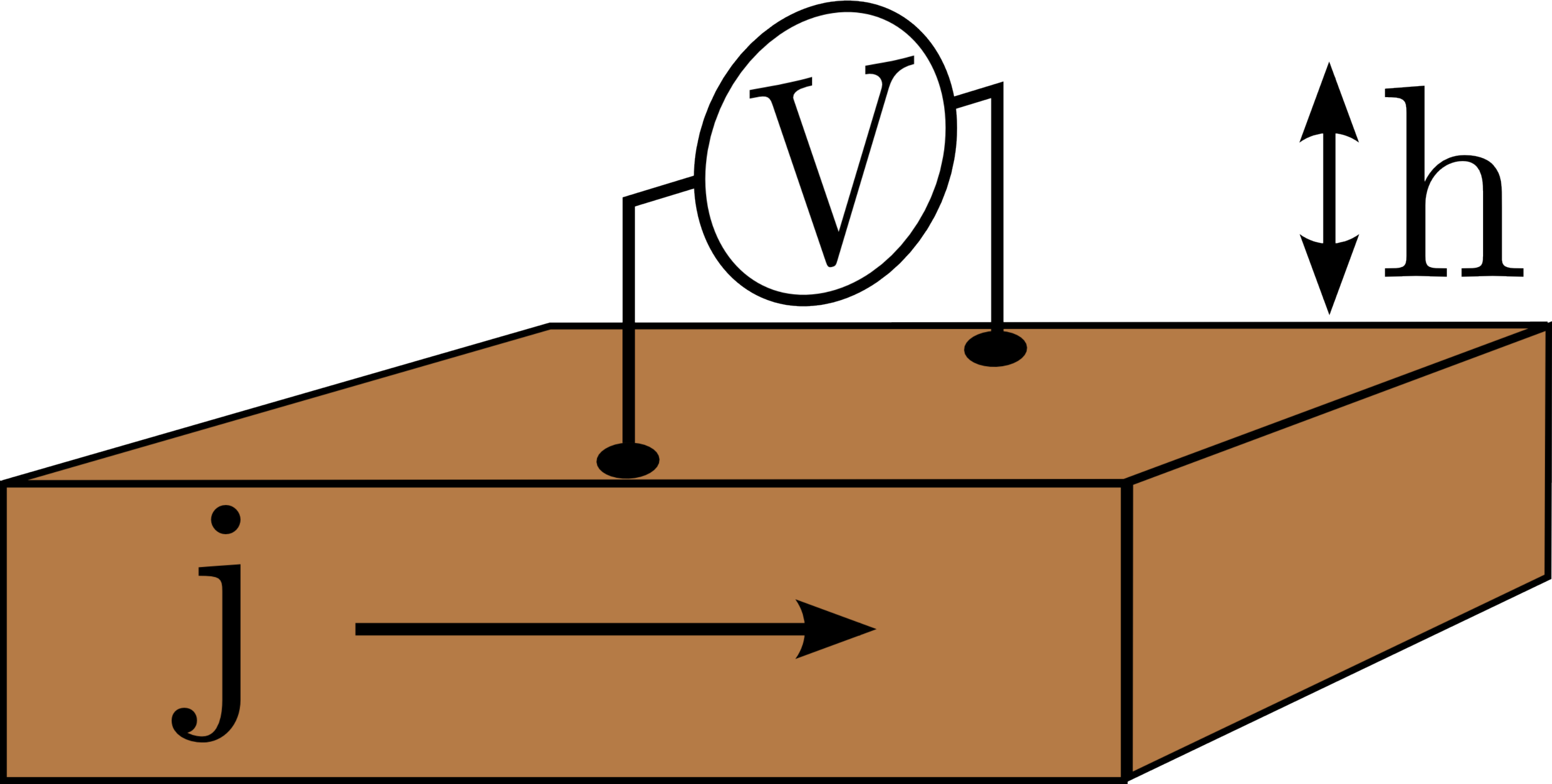}&\includegraphics[width=1.7cm]{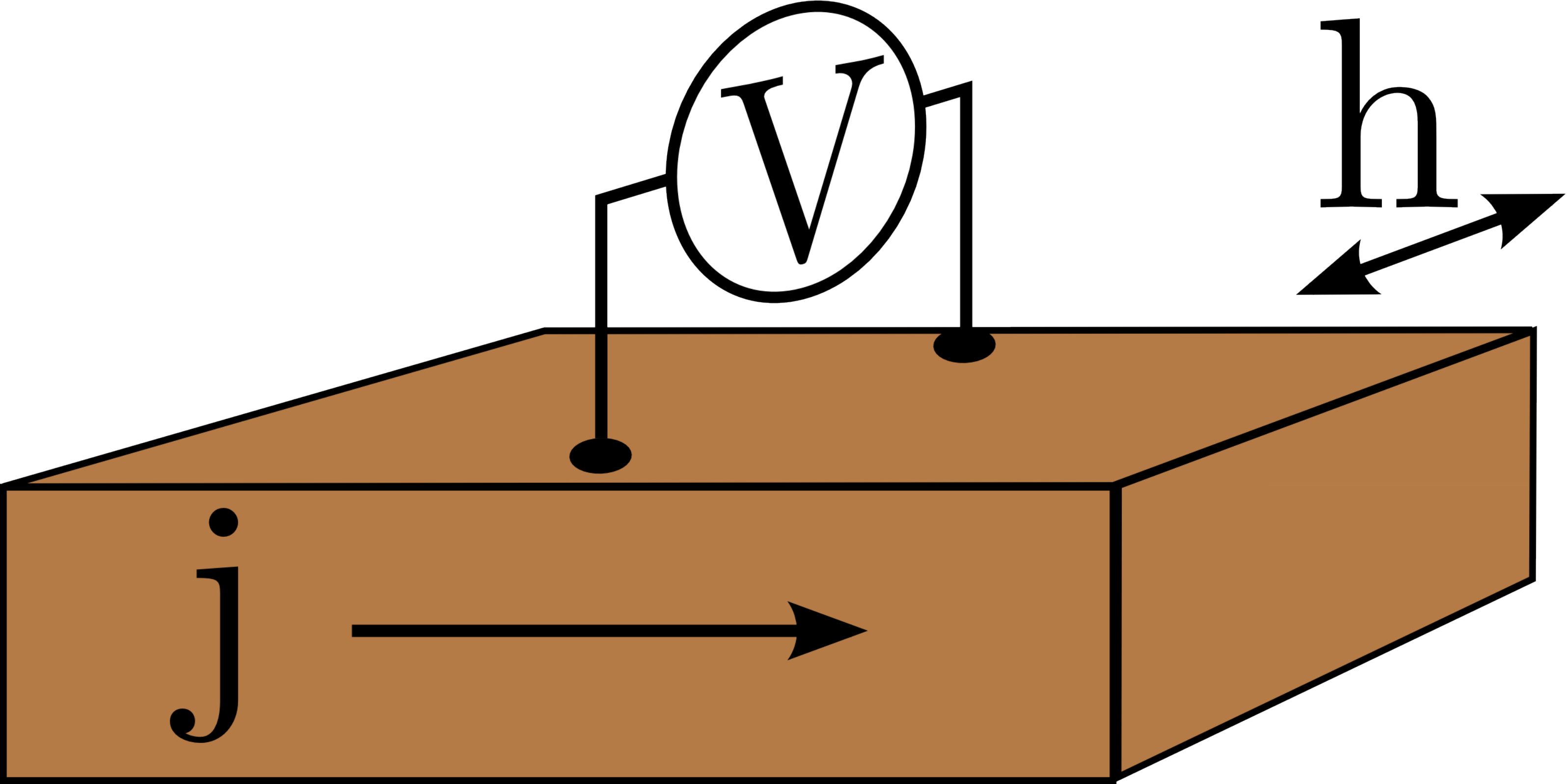}&\includegraphics[width=1.7cm]{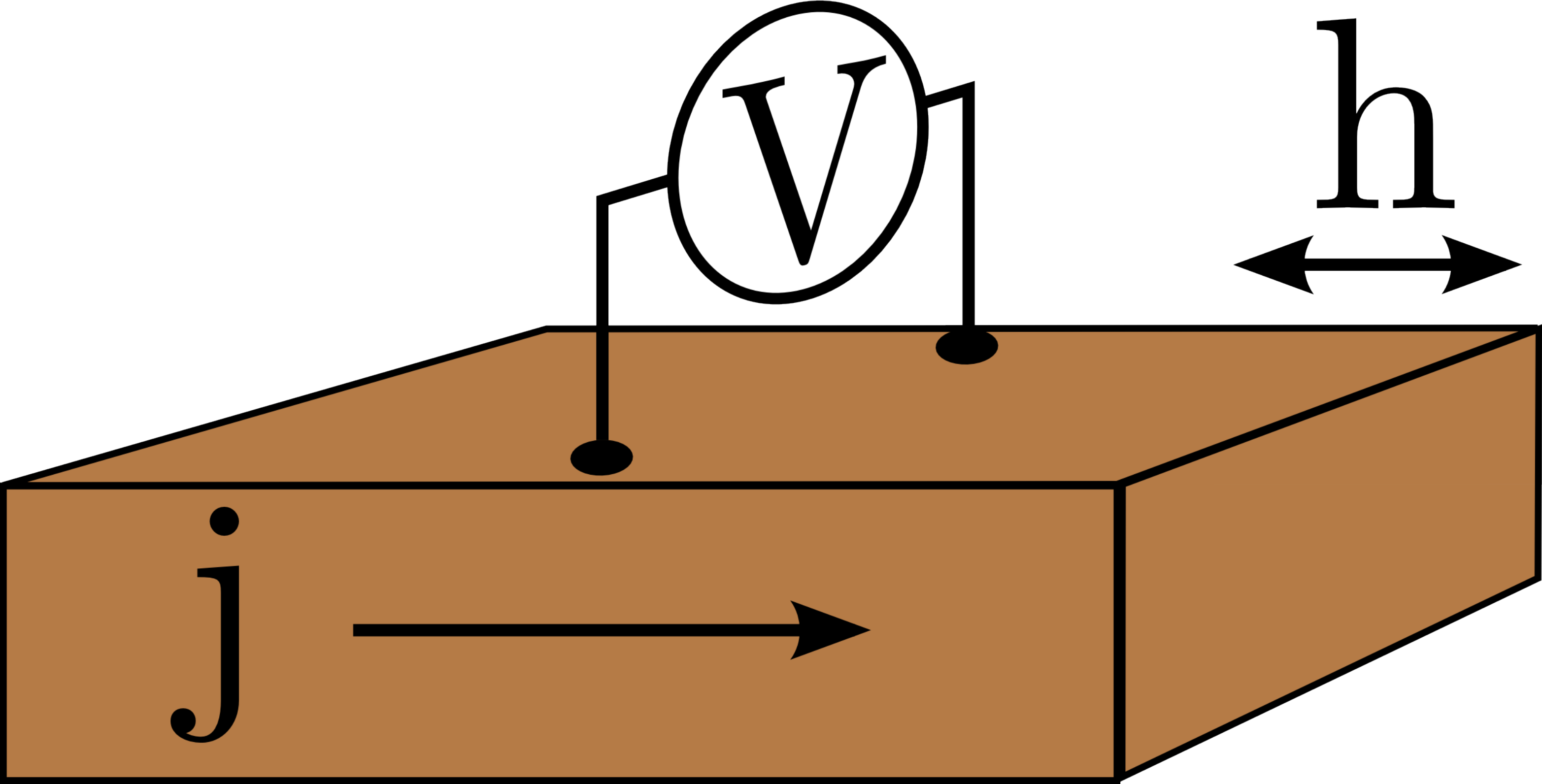}\\ \cmidrule(l{0.3em}r{0.3em}){1-7}
\includegraphics[width=2cm]{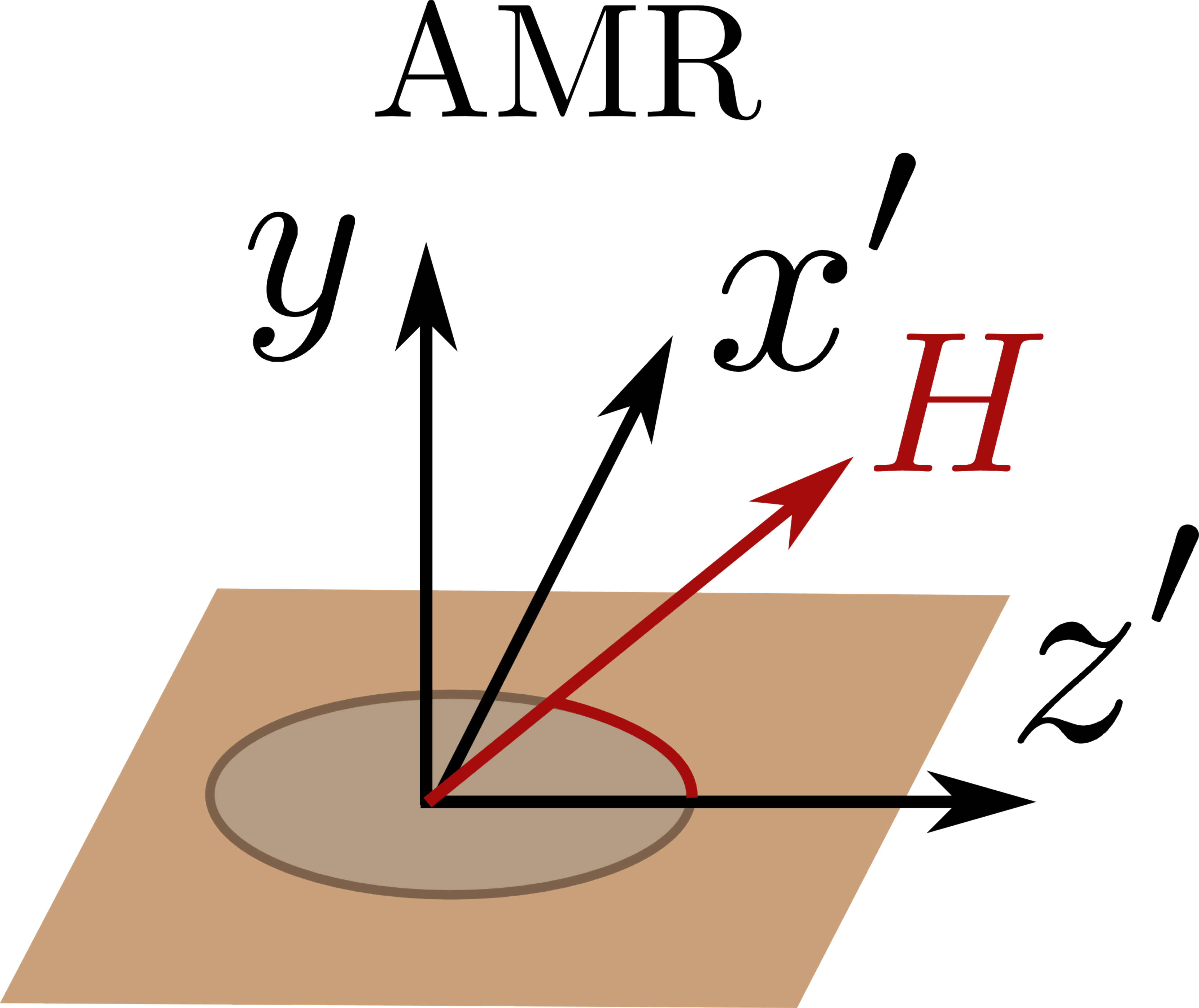}&\includegraphics[width=1.94cm]{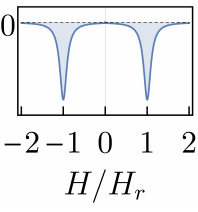}&\includegraphics[width=1.94cm]{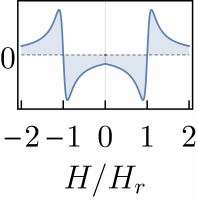}&\includegraphics[width=1.94cm]{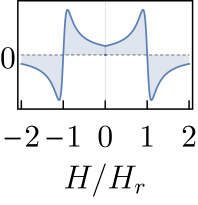} &\includegraphics[width=1.94cm]{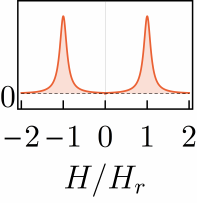}&\includegraphics[width=1.94cm]{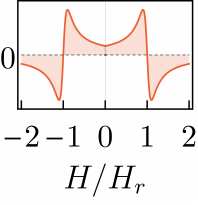}&\includegraphics[width=1.94cm]{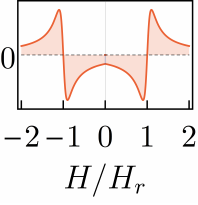}\\ \cmidrule{2-7}
\includegraphics[width=2cm]{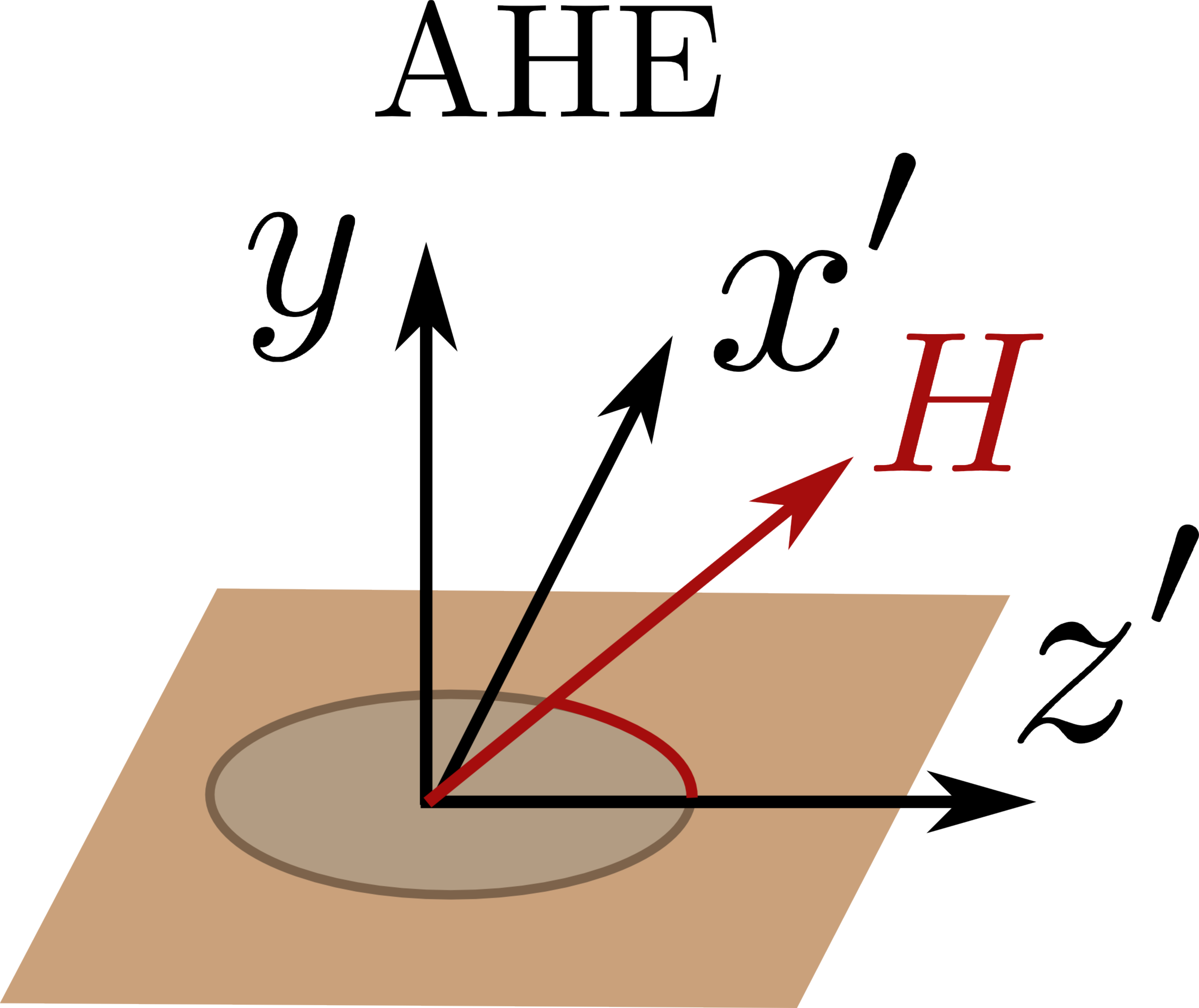}&\includegraphics[width=1.94cm]{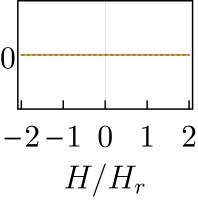}&\includegraphics[width=1.94cm]{Table3RTheta7.pdf}&\includegraphics[width=1.94cm]{Table3RTheta7.pdf} &\includegraphics[width=1.94cm]{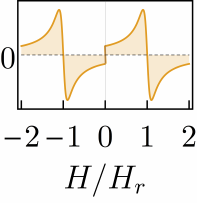}&\includegraphics[width=1.94cm]{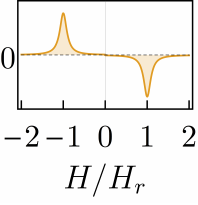}&\includegraphics[width=1.94cm]{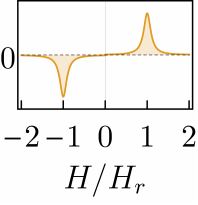} \\ \cmidrule{1-7}
\includegraphics[width=2cm]{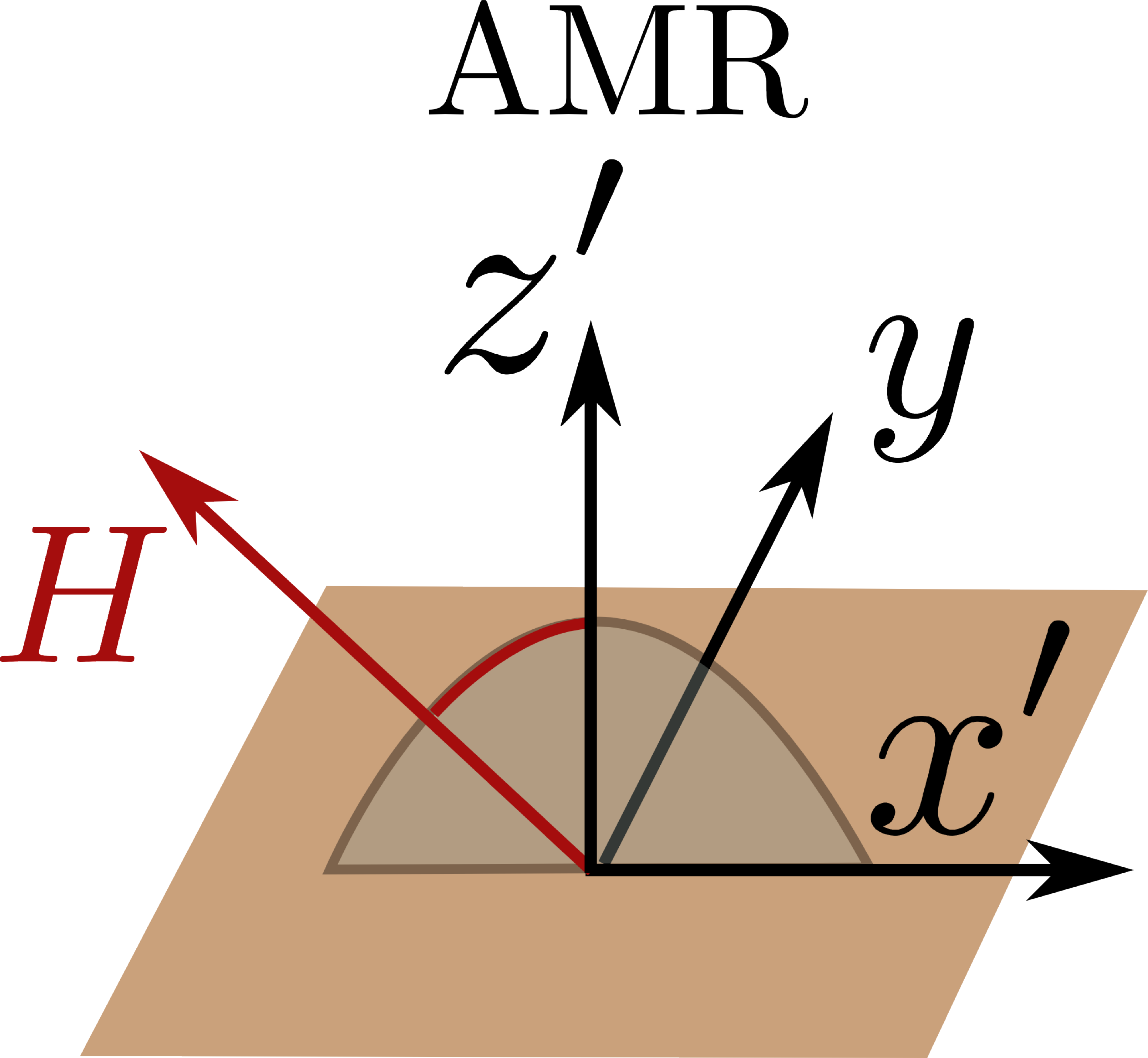}&\includegraphics[width=1.94cm]{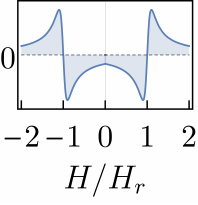}&\includegraphics[width=1.94cm]{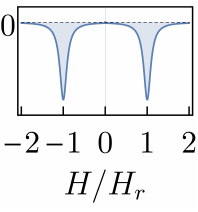}&\includegraphics[width=1.94cm]{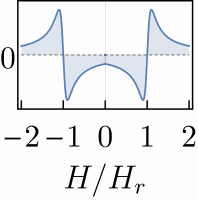} &\includegraphics[width=1.94cm]{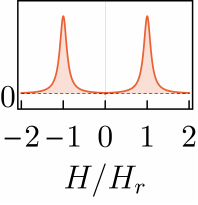}&\includegraphics[width=1.94cm]{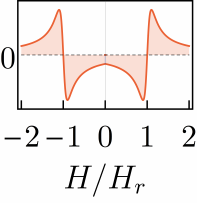}&\includegraphics[width=1.94cm]{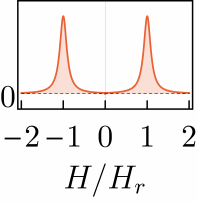}\\ \cmidrule{2-7}
\includegraphics[width=2cm]{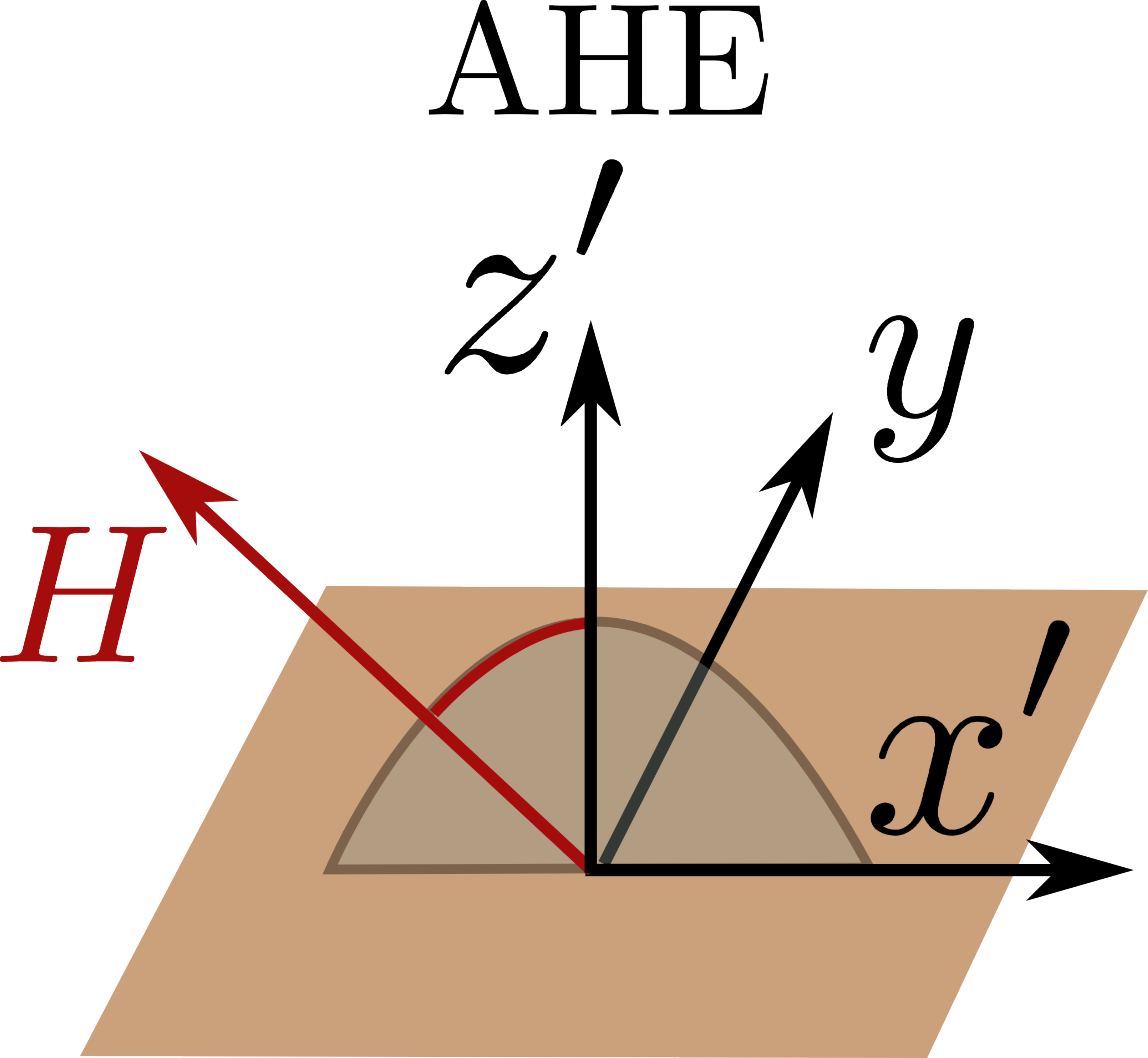}&\includegraphics[width=1.94cm]{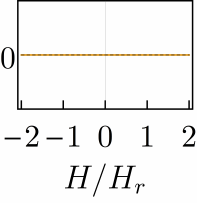}&\includegraphics[width=1.94cm]{Table3RPhi7.pdf}&\includegraphics[width=1.94cm]{Table3RPhi7.pdf} &\includegraphics[width=1.94cm]{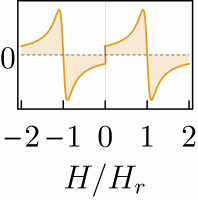}&\includegraphics[width=1.94cm]{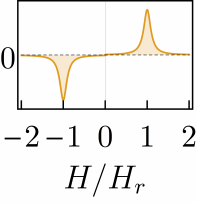}&\includegraphics[width=1.94cm]{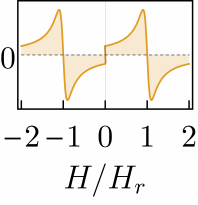} 
\\ \cmidrule{1-7}
\includegraphics[width=2cm]{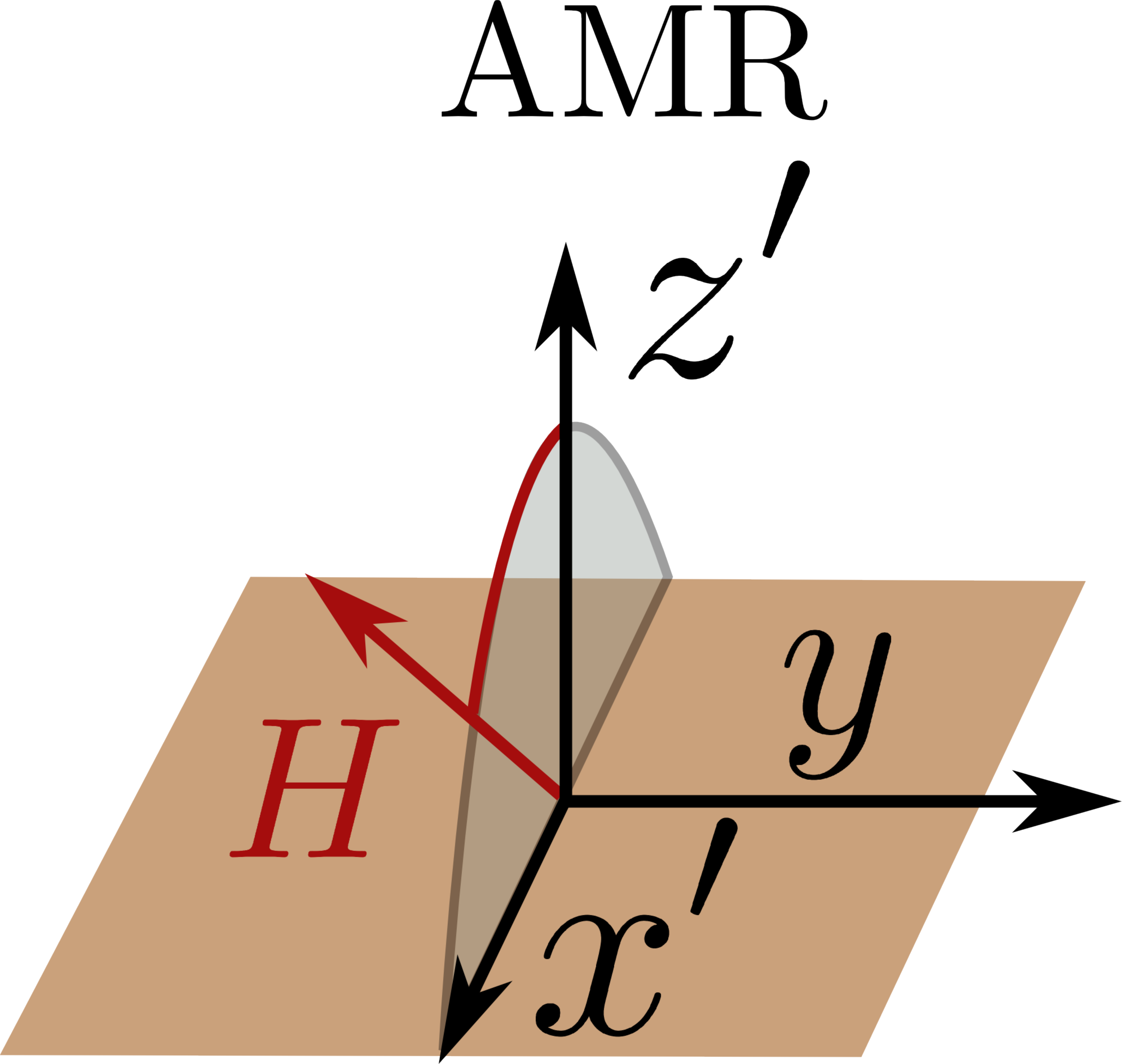}&\includegraphics[width=1.94cm]{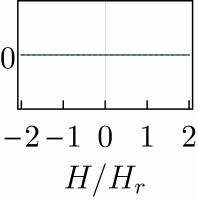}&\includegraphics[width=1.94cm]{Table3RPsi1.pdf}&\includegraphics[width=1.94cm]{Table3RPsi1.pdf} &\includegraphics[width=1.94cm]{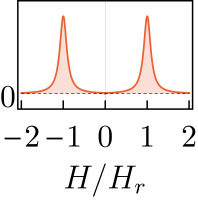}&\includegraphics[width=1.94cm]{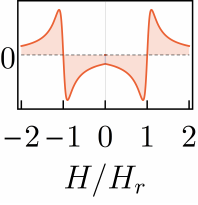}&\includegraphics[width=1.94cm]{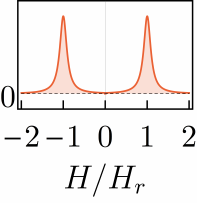}\\ \cmidrule{2-7}
\includegraphics[width=2cm]{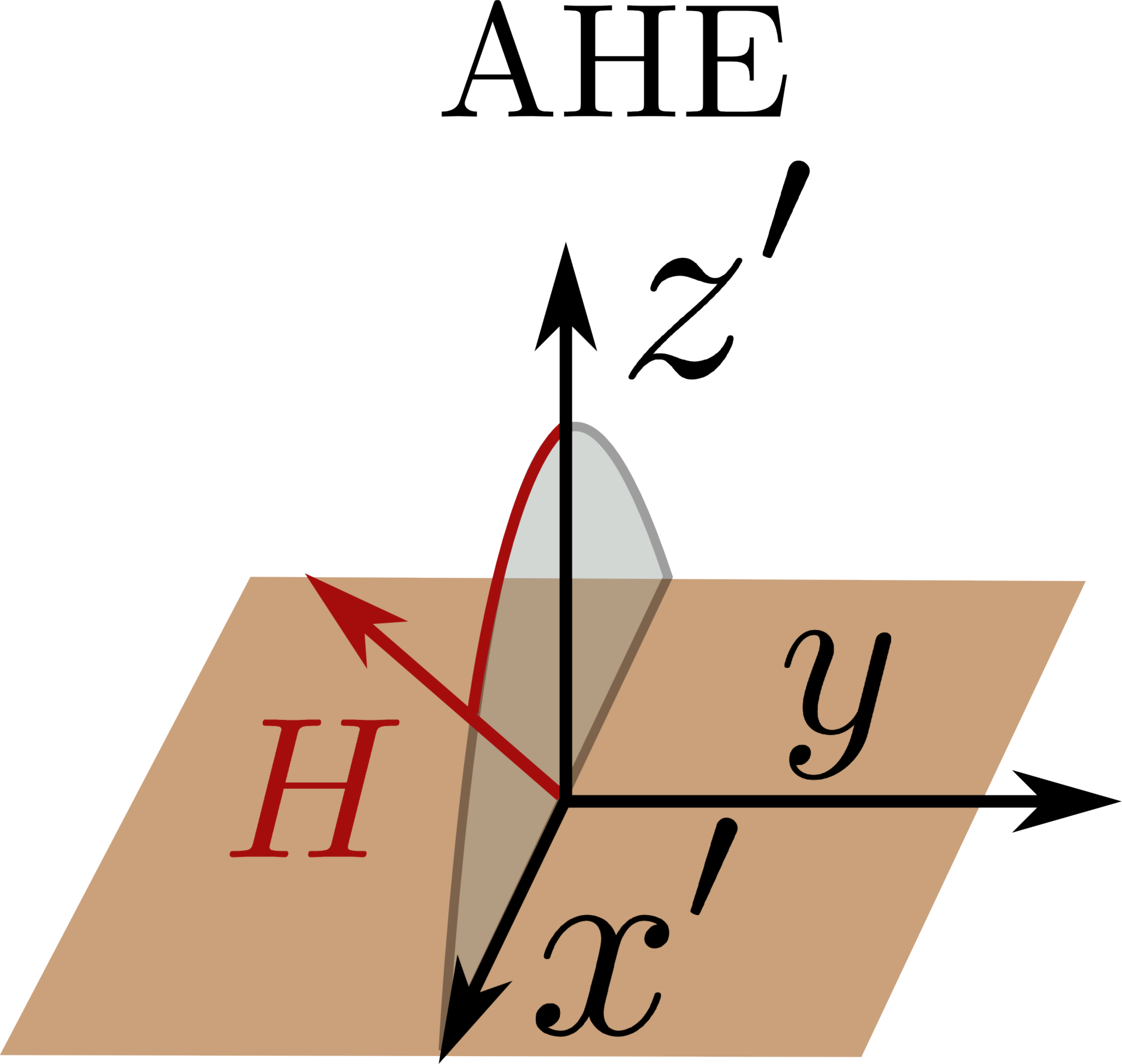}&\includegraphics[width=1.94cm]{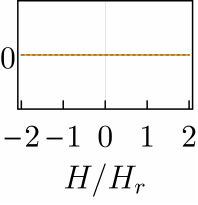}&\includegraphics[width=1.94cm]{Table3RPsi7.pdf}&\includegraphics[width=1.94cm]{Table3RPsi7.pdf} &\includegraphics[width=1.94cm]{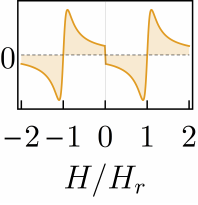}&\includegraphics[width=1.94cm]{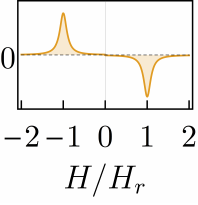}&\includegraphics[width=1.94cm]{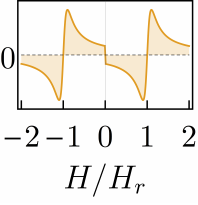} \\
\cmidrule[0.75pt](l{0.3em}r{0.3em}){1-7} \\[-0.7cm]
\cmidrule[0.75pt](l{0.3em}r{0.3em}){1-7} \\
SP&\includegraphics[width=1.94cm]{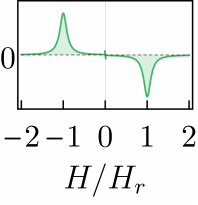}&\includegraphics[width=1.94cm]{Table3RSP1.pdf}&\includegraphics[width=1.94cm]{Table3RSP1.pdf} &\includegraphics[width=1.94cm]{Table3RSP1.pdf}&\includegraphics[width=1.94cm]{Table3RSP1.pdf}&\includegraphics[width=1.94cm]{Table3RSP1.pdf} \\
\cmidrule[0.75pt](l{0.3em}r{0.3em}){1-7}
\end{tabular}
\label{lineshapetable}
\end{table}

\begin{table}[!t]
\def\arraystretch{1}
\caption{\footnotesize{The angular dependence of SR for in-plane and both out-of-plane configurations.  The only contribution to longitudinal measurements comes from AMR, however both AMR and the AHE contribute to the transverse voltage.}}
\centering
\begin{tabular}{>{\centering\arraybackslash}m{2cm}>{\centering\arraybackslash}m{1.84cm}>{\centering\arraybackslash}m{1.84cm}>{\centering\arraybackslash}m{1.84cm}>{\centering\arraybackslash}m{1.84cm}>{\centering\arraybackslash}m{1.84cm}>{\centering\arraybackslash}m{1.84cm}>{\centering\arraybackslash}m{0pt}@{}}
\cmidrule[0.75pt](l{0.3em}r{0.3em}){1-8}                       
& \includegraphics[width=1.7cm]{Table4M1.pdf} & \includegraphics[width=1.7cm]{Table4M2.pdf} &\includegraphics[width=1.7cm]{Table4M3.pdf}&\includegraphics[width=1.7cm]{Table4M4.pdf}&\includegraphics[width=1.7cm]{Table4M5.pdf}&\includegraphics[width=1.7cm]{Table4M6.pdf}\\ \cmidrule(l{0.3em}r{0.3em}){1-8} 
 AMR & $\sin2\theta_H$ & $\sin2\theta_H\cos\theta_H$ & $\sin2\theta_H\sin\theta_H$ & $\cos2\theta_H$ & $\cos2\theta_H\cos\theta_H$ & $\cos2\theta_H\sin\theta_H$ &\\[0.5cm]
\includegraphics[width=2cm]{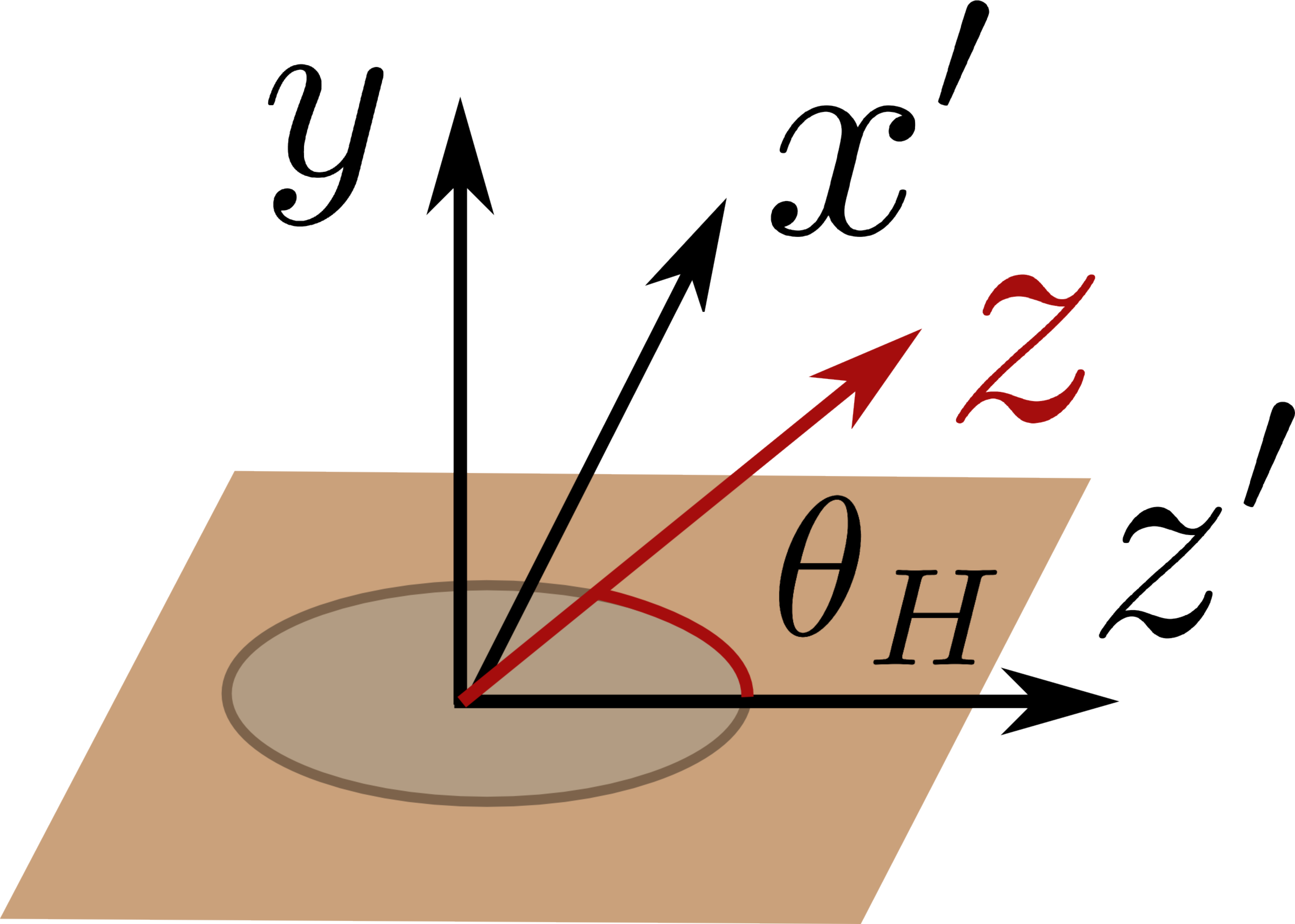}&\includegraphics[width=1.84cm]{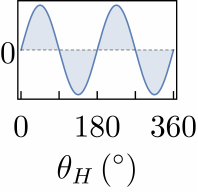}&\includegraphics[width=1.84cm]{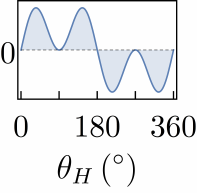}&\includegraphics[width=1.84cm]{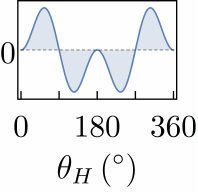} &\includegraphics[width=1.84cm]{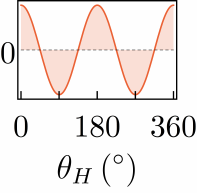}&\includegraphics[width=1.84cm]{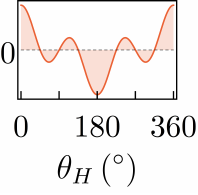}&\includegraphics[width=1.84cm]{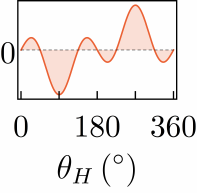}\\ \cmidrule{2-8}
AHE&$0$ & $0$ & $0$ & constant & $\cos\theta_H$ & $\sin\theta_H$ &\\[0.5cm]
\includegraphics[width=2cm]{Table45C1.pdf}&\includegraphics[width=1.84cm]{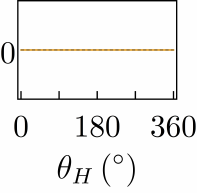}&\includegraphics[width=1.84cm]{Table4RTheta7.pdf}&\includegraphics[width=1.84cm]{Table4RTheta7.pdf} &\includegraphics[width=1.84cm]{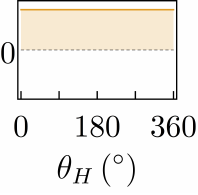}&\includegraphics[width=1.84cm]{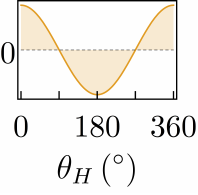}&\includegraphics[width=1.84cm]{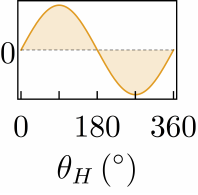} \\ \cmidrule{1-8}
AMR& $\sin2\phi_H\sin\phi_H$ & $\sin2\phi_H$ & $\sin2\phi_H\cos\phi_H$ & $\sin^2\phi_H$ & $\sin\phi_H$ & $\sin\phi_H\cos\phi_H$ &\\[0.5cm]
\includegraphics[width=2cm]{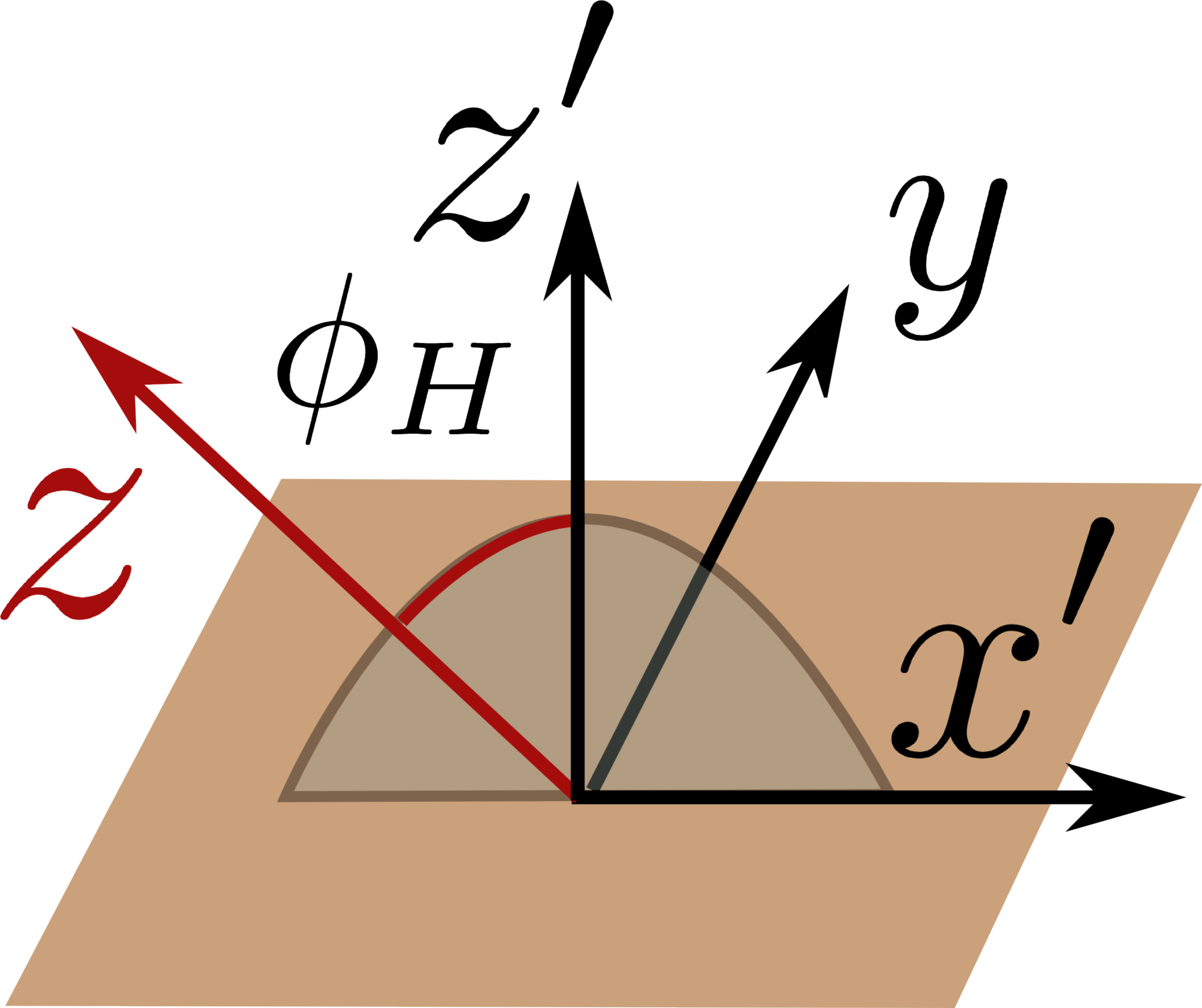}&\includegraphics[width=1.84cm]{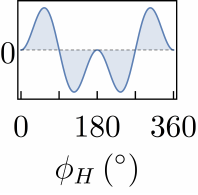}&\includegraphics[width=1.84cm]{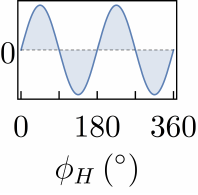}&\includegraphics[width=1.84cm]{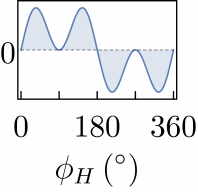} &\includegraphics[width=1.84cm]{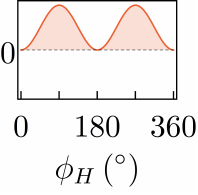}&\includegraphics[width=1.84cm]{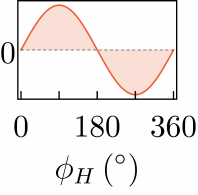}&\includegraphics[width=1.84cm]{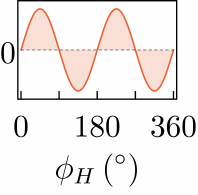}\\ \cmidrule{2-8}
AHE&$0$ & $0$ & $0$ & $\sin^2\phi_H$ & $\sin\phi_H$ & $\sin\phi_H\cos\phi_H$ &\\[0.5cm]
\includegraphics[width=2cm]{Table45C2.pdf}&\includegraphics[width=1.84cm]{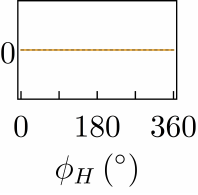}&\includegraphics[width=1.84cm]{Table4RPhi7.pdf}&\includegraphics[width=1.84cm]{Table4RPhi7.pdf} &\includegraphics[width=1.84cm]{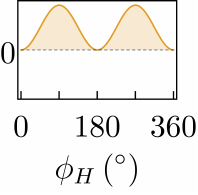}&\includegraphics[width=1.84cm]{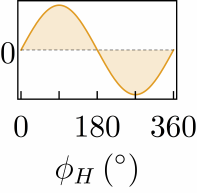}&\includegraphics[width=1.84cm]{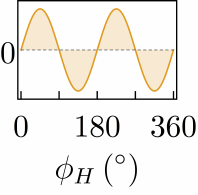} 
\\ \cmidrule{1-8}
AMR & $0$ & $0$ & $0$ & $\sin^2\psi_H$ & $\sin\psi_H\cos\psi_H$ & $\sin\psi_H$ &\\[0.5cm]
\includegraphics[width=2cm]{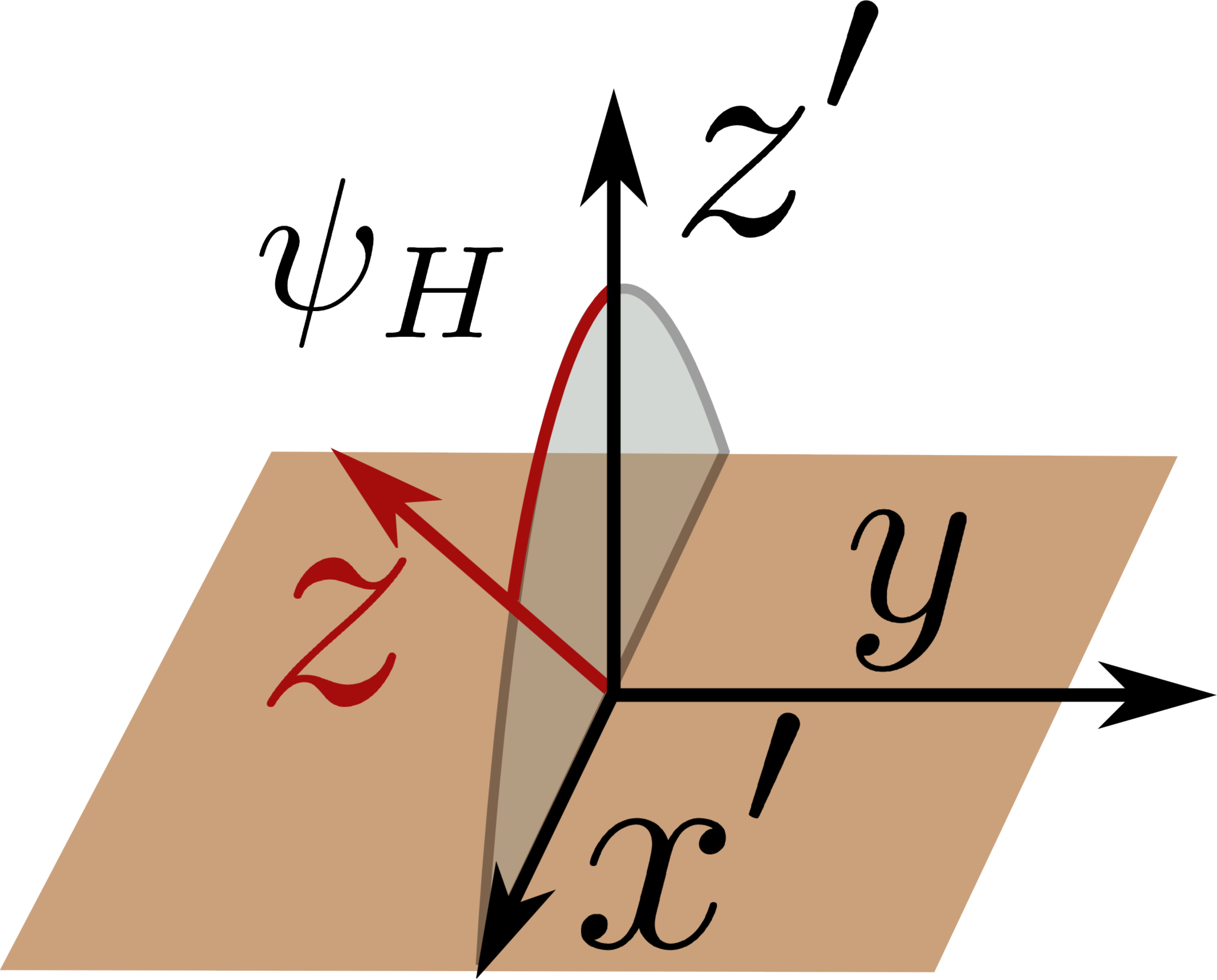}&\includegraphics[width=1.84cm]{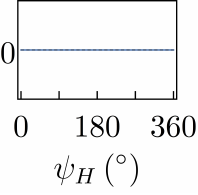}&\includegraphics[width=1.84cm]{Table4RPsi1.pdf}&\includegraphics[width=1.84cm]{Table4RPsi1.pdf} &\includegraphics[width=1.84cm]{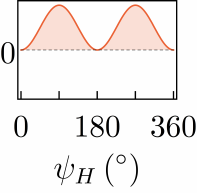}&\includegraphics[width=1.84cm]{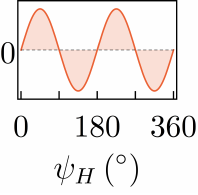}&\includegraphics[width=1.84cm]{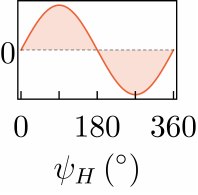}\\ \cmidrule{2-8}
AHE&$0$ & $0$ & $0$ & $\sin^2\psi_H$ & $\sin\psi_H\cos\psi_H$ & $\sin\psi_H$&\\[0.5cm]
\includegraphics[width=2cm]{Table45C3.pdf}&\includegraphics[width=1.84cm]{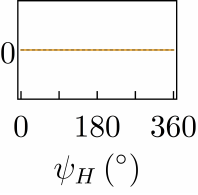}&\includegraphics[width=1.84cm]{Table4RPsi7.pdf}&\includegraphics[width=1.84cm]{Table4RPsi7.pdf} &\includegraphics[width=1.84cm]{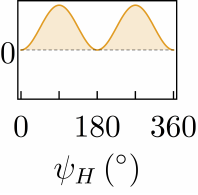}&\includegraphics[width=1.84cm]{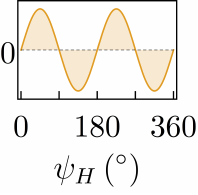}&\includegraphics[width=1.84cm]{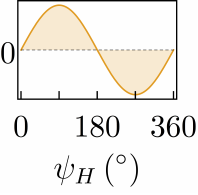}\\
\cmidrule[0.75pt](l{0.3em}r{0.3em}){1-8} 
\end{tabular}
\label{srangularip}
\end{table}

The importance of both Lorentz and dispersive contributions to the spin rectification voltage has been demonstrated by systematically controlling the relative phase between the rf current and magnetic field using a phase shifter inserted in one path through the technique of spintronic Michelson interferometry \cite{Wirthmann2010, Harder2011a}.  Fig. \ref{relativephase} shows representative data from this study, which demonstrates the frequency dependence of the relative phase in the first generation spin dynamo shown in Fig. \ref{srdevice} (a).  The line shape clearly changes from almost purely Lorentz, at $\omega/2\pi = 5.5$ GHz where $\Phi_{x^\prime} = 14^\circ$ to almost purely dispersive at $\omega/2\pi = 5.0$ GHz where $\Phi_{x^\prime} = 87^\circ$.  This indicates the device dependent nature of the SR line shape, as the relative phase will generally be a complicated function of device material, cabling setup, sample holder geometry etc.  The fact that the line shape depends on the relative phase has been exploited to perform microwave imaging by studying the phase shift produced when an object to be imaged is placed between the microwave source and the spin dynamo \cite{Zhu2010a, Cao2012a}.  This approach has also been used to perform dielectric measurements \cite{Zhu2011b} (see Section \ref{imagingsec}).

%%%%%%%%%%%%%%%%%%%%%%%%%%%%%%%%%%%%%%%%%%%%%%%%%%%%%%%%%%%%%%%%%%%%%%%%%%%%%%%%%%%%%%%%%%%%%%%%%%%%%%%%%%%%%%%%%%%%%%%%%%%%%%%%%%%%%%%%%%%%%%%%%%%%%%%%%%%%%%%%%%%%%%%%%%%%%%%%%%%%%%%%%%%%%%%%%%%%%%%%%%%%%%%%%%%%%%%%%%% 
\subsubsection{Angular Dependence} \label{angulardependence} 

In Sec. \ref{lineshapeanalysis} it was pointed out that in order to characterize the relative phase, which may be different for each $\textbf{h}$ component, the voltage must be separated by driving field.  This requirement is most clearly emphasized by the experimental data shown in Fig. \ref{srdevice} (b) and (e) which were taken for a first and second generation spin dynamo respectively.  Notice that while the line shape is nearly Lorentz in both cases, the relative phase is very different -- for these devices $\Phi_y = -9^\circ$ and $\Phi_{x^\prime} = -102^\circ$ have been measured respectively \cite{Harder2011a}.  This is in agreement with the line shape analysis -- from the summary provided in Table \ref{lineshapetable} with zero relative phase the $h_y$ driven FMR will be Lorentz, however the $h_{x^\prime}$ driven FMR will be dispersive.  This means that the $h_{x^\prime}$ driven FMR must have a relative phase near $90^\circ$ if the line shape will be Lorentzian.  This highlights the fact that for $\Phi$ to be correctly determined, the $\textbf{h}$ component driving FMR must first be identified, which can be done by performing an analysis of the angular dependence.

The key angular dependencies, based on the analysis of Sec. \ref{spinrectification}, are summarized in Table \ref{srangularip}.  The AMR and  AHE will have different angular dependencies for each $\textbf{h}$ component in each measurement configuration, and therefore by measuring the rectified voltage as a function of the static field orientation, the driving field can be determined, which then allows one to proceed with a line shape analysis.  Therefore to characterize the spin rectification both line shape and angular experiments are necessary, which are carried out by varying $|\textbf{H}|$ and the field direction respectively.  Table \ref{srangularip} summarizes the angular dependence for all measurement/angular configurations discussed in Sec. \ref{spinrectification} however it should be noted that in single crystalline films, where the angular dependence of AMR may differ from Eq. \ref{amrresistance}, the angular dependence of SR may differ from Table \ref{srangularip} \cite{McGuire1975, Ramos2008, Hu2012, Ding2013, Xiao2015, Chen2015a, Xiao2015a}.

%%%%%%%%%%%%%%%%%%%%%%%%%%%%%%%%%%%%%%%%%%%%%%%%%%%%%%%%%%%%%%%%%%%%%%%%%%%%%%%%%%%%%%%%%%%%%%%%%%%%%%%%%%%%%%%%%%%%%%%%%%%%%%%%%%%%%%%%%%%%%%%%%%%%%%%%%%%%%%%%%%%%%%%%%%%%%%%%%%%%%%%%%%%%%%%%%%%%%%%%%%%%%%%%%%%%%%%%%%%

\subsection{Magnetic Bilayers: The Spin Battery} \label{srinbilayers}

In ferromagnetic/normal metal (FM/NM) bilayers a dc voltage may still be generated by field torque rectification, however an additional signal due to spin-transfer torque rectification and also spin pumping may be present.   SP and STT require a conversion between charge and spin currents within the device and therefore rely on the spin Hall effects \cite{Sinova2014} (although the recently studied spin torques arising from Rashba/Dresselhaus spin-orbit interactions will not require the spin Hall effect, and will be discussed in Sec. \ref{standshe}, for simplicity here we generally exclude spin-orbit torques from our definition of spin torque).  As a result, the magnitude of the ST and SP voltage will depend on the strength of the spin-orbit coupling in the normal metal layer, which is the physical origin of the spin Hall effect.  Therefore to observe large spin-transfer torque rectification strong spin-orbit coupling is required, which makes platinum (Pt) a popular choice for the NM layer in such devices.  In Sec. \ref{standshe} we will discuss recent studies which have shown that current driven spin-torque may arise from the Rashba/Dresselhaus spin-orbit interactions.
\begin{figure}[!ht]
\centering
\includegraphics[width=16cm]{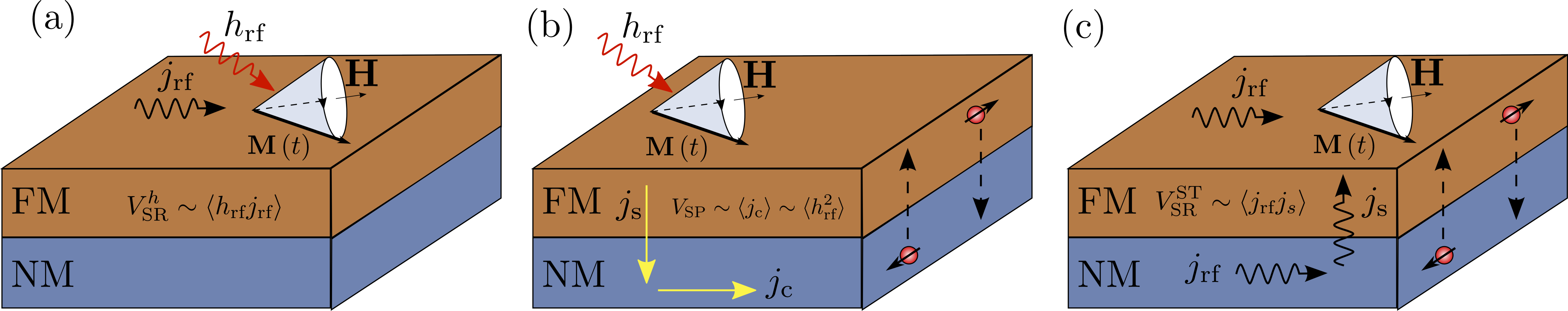}
\caption{\footnotesize{In FM/NM bilayer structures the dc voltage may be due to (a) field induced spin rectification, (b) spin pumping or (c) spin-transfer torque induced rectification.  (a) In field induced rectification a microwave $h$ field excites FMR in the ferromagnetic layer.  The magnetization precession generates an rf magnetoresistance which couples nonlinearly to the rf current flowing in the FM layer, producing a dc voltage which can have both Lorentz and dispersive components and depends on both the relative and spin resonance phase.  (b) In spin pumping the microwave induced magnetization precession generates a non-equilibrium spin distribution at the FM/NM interface which equilibrates by sending a spin current into the NM.  Via the inverse spin Hall effect this spin current is converted into a charge current, and the dc contribution then generates a dc voltage which is purely Lorentzian.  (c) In spin-transfer torque rectification an rf charge current in the NM is converted into an ac spin current through the spin Hall effect and enters the FM layer.  The misalignment between the FM magnetization and the spin current polarization produces a torque on the magnetization, resulting in an rf magnetoresistance which can produce a dc voltage by coupling to the rf charge current already present in the FM.}}
\label{vinbilayers}
\end{figure}

The three voltage producing effects in bilayer devices are shown in Fig. \ref{vinbilayers}.  In rectification (both field and spin-transfer torque induced) the role of the magnetization dynamics is to generate a dynamic magnetoresistance which can nonlinearly couple to an rf current and produce a dc voltage.  This process is shown in Fig. \ref{vinbilayers} (a) for field induced magnetization dynamics.  Note that since the rf $h$ field directly excites FMR in the  FM layer the NM layer is not active in this process.   However for both STT and SP the NM is necessary since it is where the spin/charge current conversion takes place.  In spin pumping, schematically illustrated in Fig. \ref{vinbilayers} (b), the role of magnetization precession its very different from its purpose in rectification.  In SP the role of the microwave excited FMR is to set up a non-equilibrium spin distribution at the FM/NM interface.  Via a diffusive process this spin distribution can equilibrate by injecting a spin current from the FM into the NM.  The inverse spin Hall effect will then convert this spin current into a charge current and the dc component will then produce a dc voltage.  Finally STT rectification is shown in Fig. \ref{vinbilayers} (c).  An rf current flowing in the NM layer is converted into an ac spin current via the spin Hall effect.  The ac spin current then flows into the FM layer and produces a torque on the magnetization due to the exchange interaction.  This torque drives magnetization precession, which will generate an rf magnetoresistance which can couple to the rf current and produce a dc voltage.  Since the line shape of SP is determined by the form of the spin current generation, $j_s \sim h^2$, there is no relative phase dependence and the SP line shape will be purely Lorentzian.  Similarly, for STT rectification since it is $j_s$ which drives FMR, and not $h$, there is no relative phase dependence.  The presence of these three different physical effects in bilayer devices makes their study intriguing from both a practical and fundamental physics viewpoint and as a result there has been a large number of studies performed on bilayers in the past decade, which we will discuss in detail in the next five sections.  We start our discussion with a brief review of spin pumping and spin Hall effects, followed by a summary of experimental observations of spin pumping in both the  "vertical" and "transverse" configurations.  We then discuss the separation of spin rectification and spin pumping and look at spin-transfer torque rectification.  To provide some guidance to the relevant literature a summary of several key studies is provided in Table \ref{spexperiments}.   
\noindent

%%%%%%%%%%%%%%%%%%%%%%%%%%%%%%%%%%%%%%%%%%%%%%%%%%%%%%%%%%%%%%%%%%%%%%%%%%%%%%%%%%%%%%%%%%%%%%%%%%%%%%%%%%%%%%%%%%%%%%%%%%%%%%%%%%%%%%%%%%%%%%%%%%%%%%%%%%%%%%%%%%%%%%%%%%%%%%%%%%%%%%%%%%%%%%%%%%%%%%%%%%%%%%%%%%%%%%%%%%%

\subsubsection{The Basic Physics of Spin Pumping and Spin Hall Effects}

The kernel of spintronics is the generation and manipulation of spin currents, making experimental techniques which reliably generate spin currents of the utmost importance.  One of the widely used methods for spin current generation is the transport of non-equilibrium magnetization pumped by FMR, known as spin pumping (see e.g. Ref. \cite{Tserkovnyak2005, Brataas2011} for comprehensive reviews).  
\begin{figure}[!b]
\centering
\includegraphics[width=6cm]{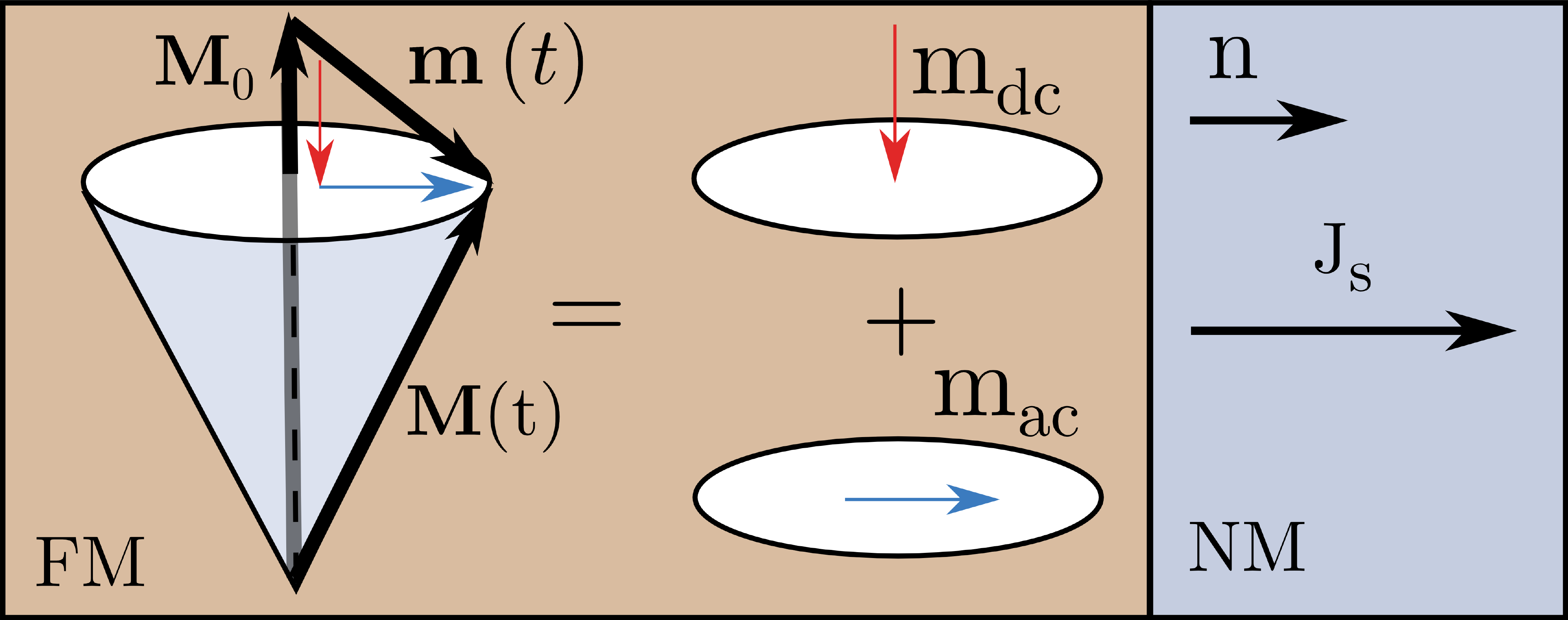}
\caption{\footnotesize{A schematic illustration of spin pumping in a FM/NM bilayer.  The magnetization, ${\bf M}$, of the FM layer precesses around the static field direction ${\bf H}$.  This precession "pumps" a spin current into the normal metal layer due to a non-equilibrium spin accumulation generated at the FM/NM interface.}}
\label{spSchem}
\end{figure}
\begin{table}[!b]
\def\arraystretch{1}
\caption{\footnotesize{A summary of key experimental results using bilayer (FM/NM) devices.  In Refs. \cite{Saitoh2006, Azevedo2011, Chen2013c} a microwave cavity is used to provide the rf magnetic field, while in Ref. \cite{Bai2013} a CPW is used.}}
\centering
\begin{tabular}{>{\centering\arraybackslash}m{4cm}>{\centering\arraybackslash}m{4cm}>{\centering\arraybackslash}m{6cm}}
\toprule   
\multicolumn{2}{l}{{\bf Spin Pumping and Spin Rectification}} \\ \toprule                 
Reference & Device Structure & Result \\ \midrule \\
Costache et al. \cite{Costache2006} & \includegraphics[width=2.2cm]{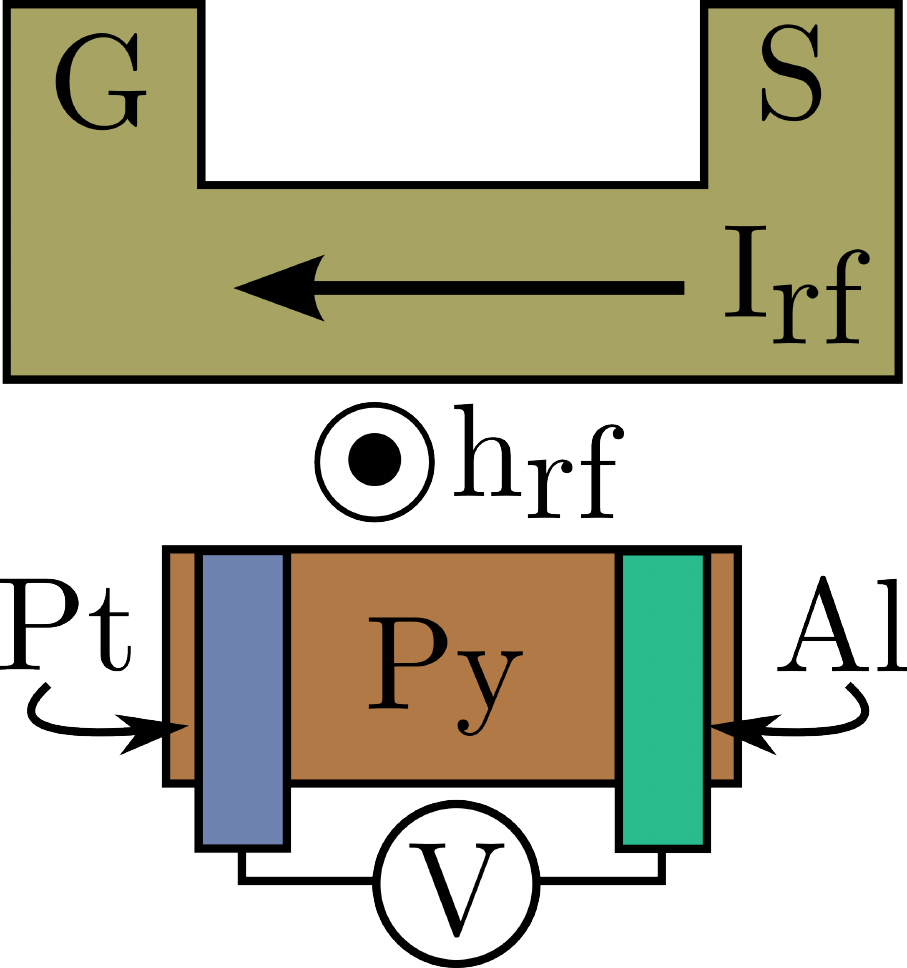} & Electrical detection of SP in "vertical" configuration (without ISHE) using Pt/Py/Al device\\ \\
Saitoh et al. \cite{Saitoh2006} & \includegraphics[width=2.2cm]{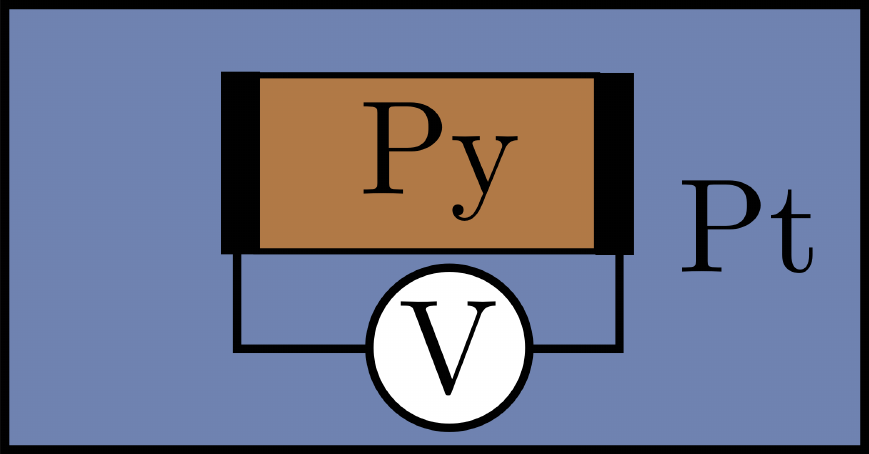} & Electrical detection of SP in "transverse" configuration through the ISHE using Py/Pt bilayer \\ \\
Mosendz et al. \cite{Mosendz2010} & \includegraphics[width=2.8cm]{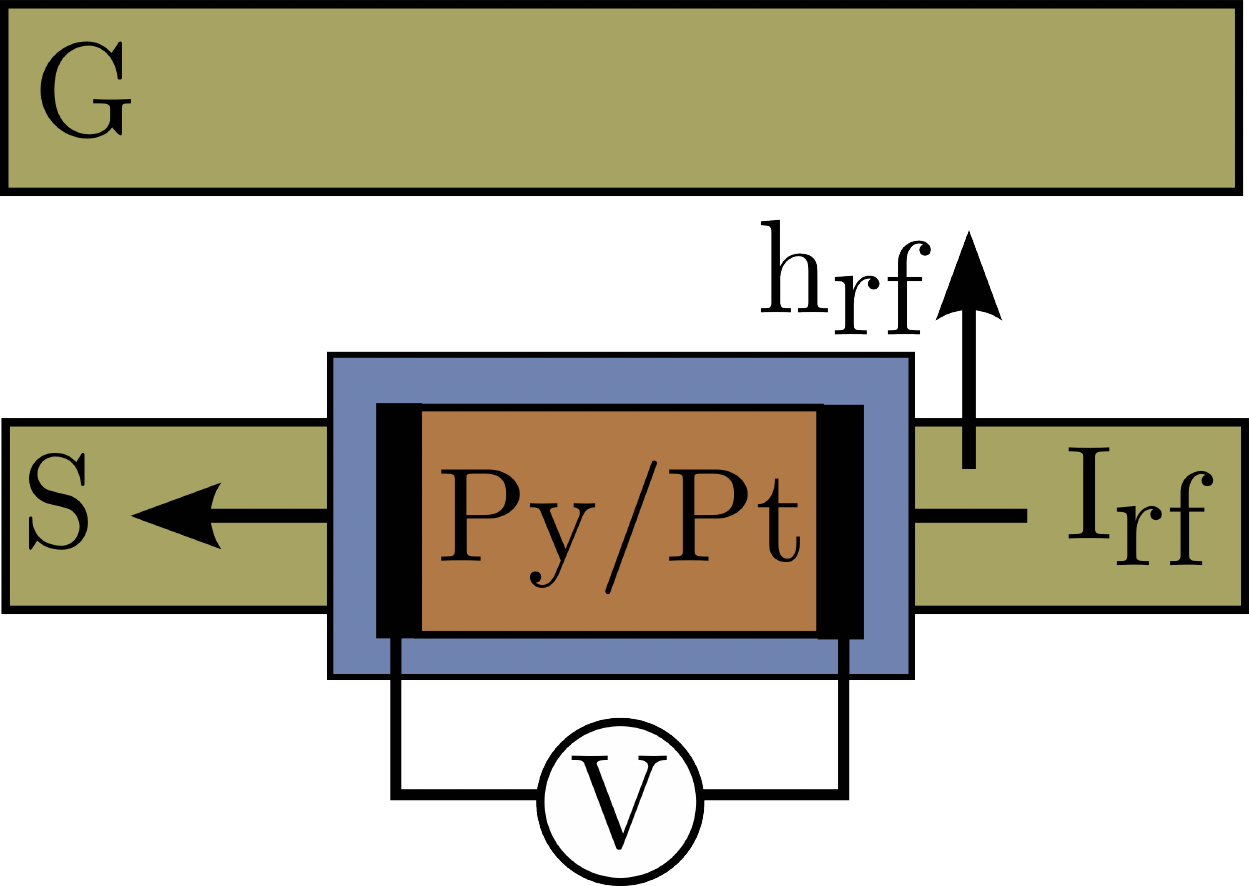} & Separation of SP and SR voltage signals using line shape symmetry (L vs D respectively) using Py/Pt bilayer \\ \\
Azevedo et al. \cite{Azevedo2011} & \includegraphics[width=2.2cm]{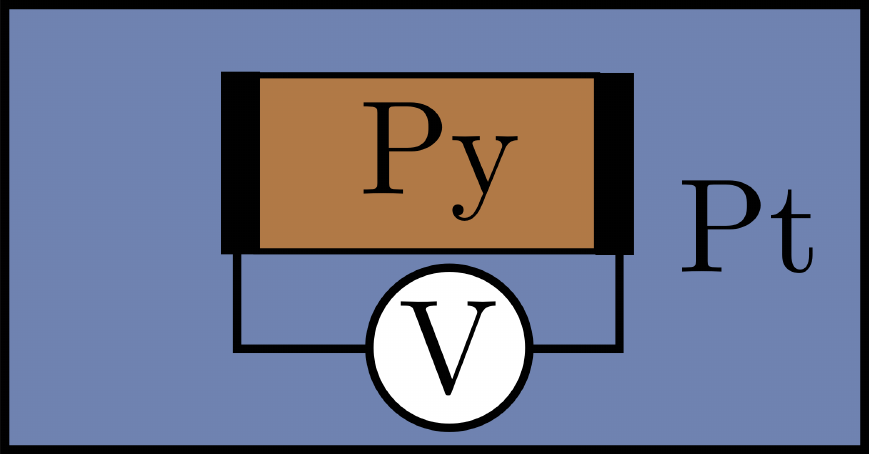} & Separation of SP and SR voltage signals using line shape symmetry and angular dependence using Py/Pt bilayer\\ \\
Chen et al. \cite{Chen2013c} & \includegraphics[width=2cm]{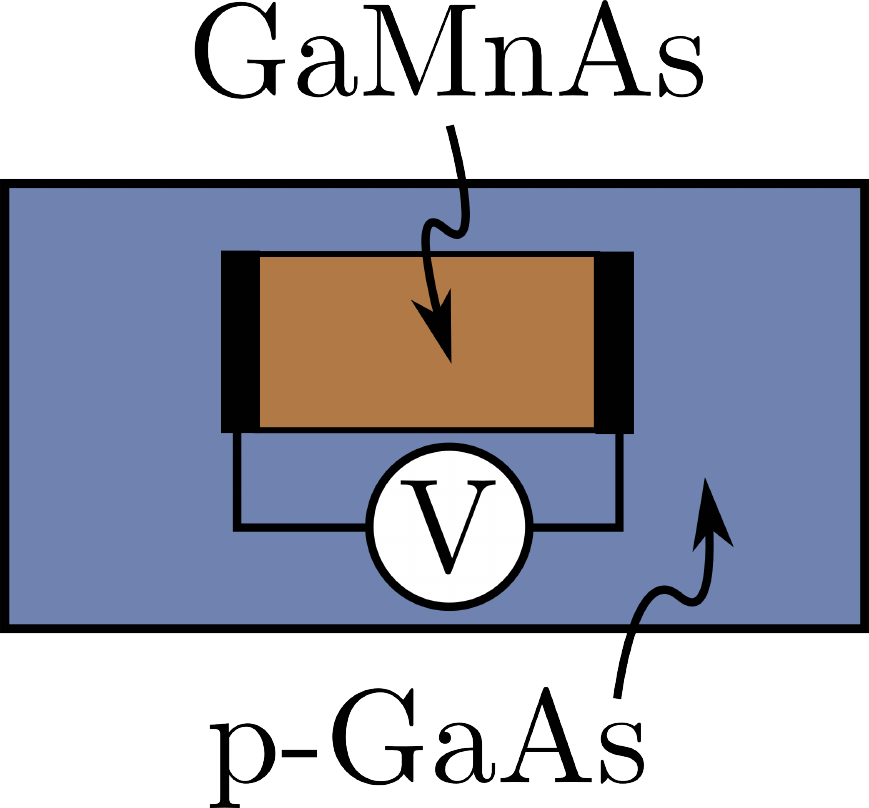} & Separation of SP and SR voltage signals using line shape symmetry and angular dependence in GaMnAs/p-GaAs bilayer \\ \\
Bai et al. \cite{Bai2013} & \includegraphics[width=3.2cm]{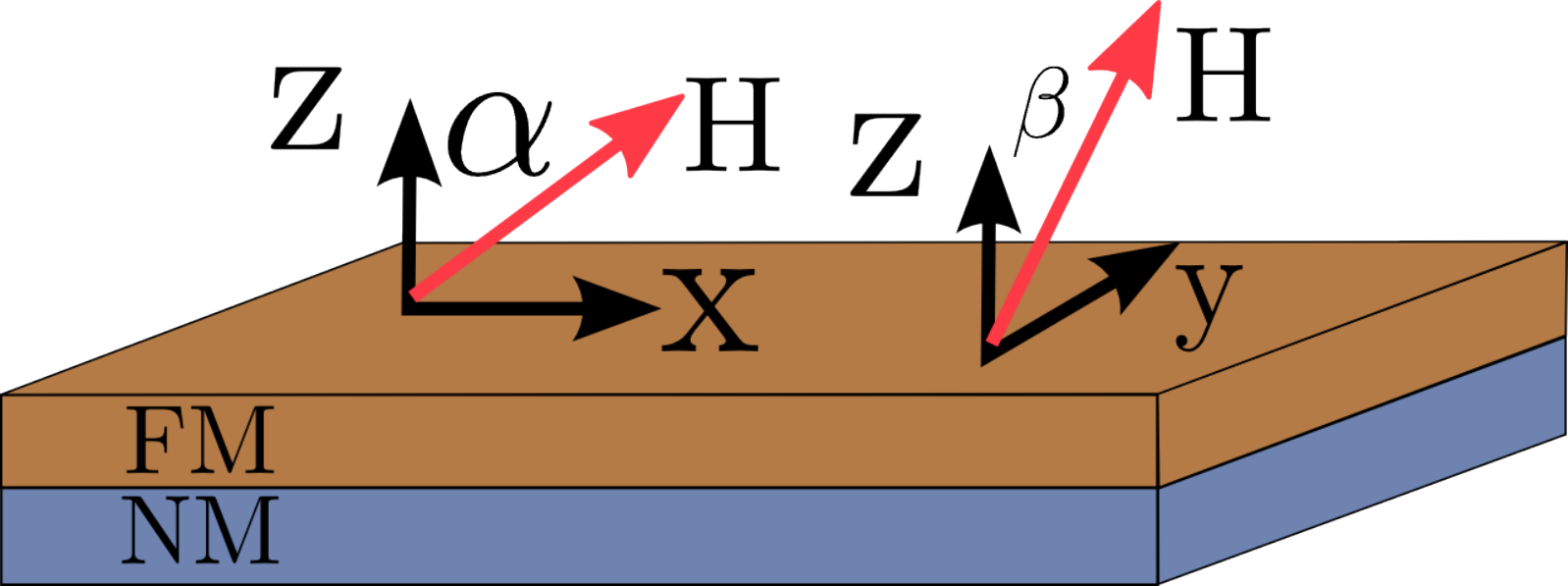} & Universal method of SP and SR separation based on symmetry considerations \\ \\ \toprule
\multicolumn{1}{l}{{\bf spin-transfer Torque}} \\ \toprule                  
Reference & Device Structure & Result \\ \midrule \\
Liu et al. \cite{Liu2011} & \includegraphics[width=2.2cm]{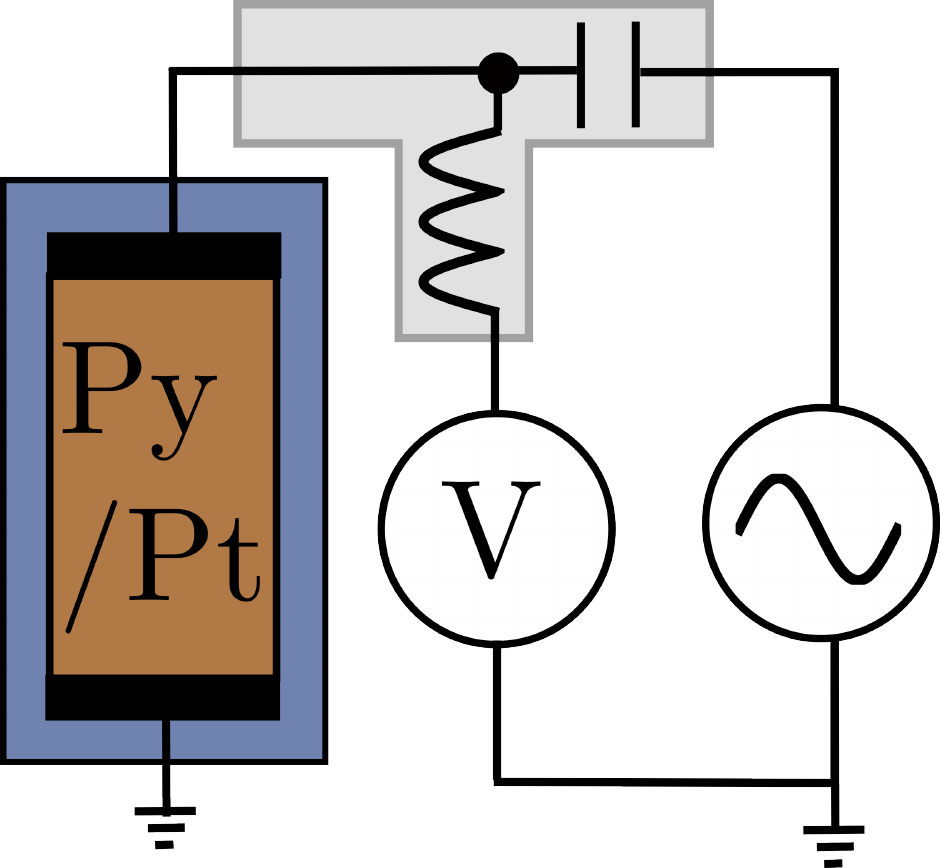} & Electrical detection of SR due to STT in Py/Pt bilayer \\ \\
\bottomrule
\end{tabular}
\label{spexperiments}
\end{table}
In a FM/NM bilayer, spin pumping can be understood schematically using Fig. \ref{spSchem}: At FMR, the precessional magnetization ${\bf M}(t)$ in the ferromagnet differs from the saturation magnetization ${\bf M}_0$ by an amount $\Delta {\bf M}(t)$, which corresponds to the non-equilibrium magnetization generated by FMR. Inside the FM layer the precessing magnetization relaxes towards its equilibrium position via magnetization damping, leading to the partial dissipation of the non-equilibrium magnetization $\Delta {\bf M}(t)$. At the interface of the FM/NM bilayer, the non-equilibrium magnetization diffuses from the FM into the NM as a flow of spin current, which contributes to additional dissipation of $\Delta {\bf M}(t)$. Magnetization precession, magnetization damping, and magnetization diffusion, these are the three key ingredients of spin pumping. As shown in  Fig. \ref{spSchem}, since $\Delta {\bf M}(t)$ involves both a static and a precessional component, the spin pumping effect simultaneously generates both dc and ac spin current in the FM/NM bilayer.

Although the concept and full theory of spin pumping \cite{Tserkovnyak2002, Brataas2002} were not developed until 2002, historically, the fact that the three key ingredients of magnetization precession, damping, and diffusion would lead to the generation of a spin current in FM/NM bilayers was first revealed in the transmission-electron-spin resonance (TESR) experiments performed in the late 1970's \cite{Janossy1976, Silsbee1979}. In such microwave absorption experiments, Silsbee et al. found that the TESR of copper foils was greatly enhanced in the presence of the ferromagnetic films (such as permalloy, iron, and nickel) deposited on one surface of the copper foil \cite{Silsbee1979}. By developing a phenomenological theory using the Bloch equations to describe magnetization precession and damping, coupled by the transport of non-equilibrium magnetization across the FM/NM interface, Silsbee et al. revealed the key physics of spin pumping, i.e, the generation of a spin current via the interplay of magnetization precession, damping, and diffusion in FM/NM bilayers \cite{Silsbee1979}.          

Microscopically, spin pumping is a consequence of spin dependent reflectivity and transmission parameters of NM electrons at the FM/NM interface. The microscopic theory of spin pumping was  derived by Tserkovnyak et al.  \cite{Tserkovnyak2002, Brataas2002}, using the spin mixing conductance as the main parameter to rigorously describe the spin current. According to this theory the spin current injection into the normal metal relaxes over a characteristic length scale $\lambda_\text{SD}$, the spin diffusion length, and the spin current density decays away from the interface as \cite{Tserkovnyak2002, Brataas2002}
\begin{equation}
\textbf{j}_s (y) = \textbf{j}_s\left(0\right) \frac{\sinh\left[\left(t_\text{NM} - n\right)/\lambda_\text{SD}\right]}{\sinh\left[t_\text{NM}/\lambda_\text{SD}\right]}, ~~\textbf{j}_s(0) = \frac{\hbar G_r}{ 4\pi M_0^2} \textbf{M} \times \frac{d\textbf{M}}{dt}. \label{spinpumpingcurrent}
\end{equation}
Here $G_r$ is the real part of the spin mixing conductance, $G^{\uparrow\downarrow} = G_r + i G_i$, $\hbar$ is the reduced Planck constant, $t_\text{NM}$ is the thickness of the NM layer, $n$ is the distance normal to the FM/NM interface (see Fig. \ref{spSchem}) and $\textbf{M}$ and $M_0$ are the full and dc magnetization of the FM layer respectively, just as in Sec. \ref{spinrectification}.  Only the spin polarization direction $\widehat{\sigma}$ is determined by the direction of $\textbf{j}_s$ in Eq. \ref{spinpumpingcurrent}, with the spatial direction normal to the interface (since the current diffuses away from the interface).  The process of spin pumping is also closely related to the earlier idea of a voltage generated by spin flip scattering at the FM/NM interface \cite{Berger1999} and depends on the spin mixing conductance in the same way as the spin Hall magnetoresistance discussed in Sec. \ref{sec:shmr} \cite{Weiler2013b}.  

In addition to TESR experiments that can measure the spin current, other methods have been demonstrated to detect spin pumping.  Comparing the spin current of Eq. \ref{spinpumpingcurrent} to the LLG equation in Eq. \ref{llg}, the spin current has the same form as the Gilbert damping term and therefore from the perspective of the FM the flow of spin current will result in an increased damping due to the transfer of angular momentum into the NM \cite{Tserkovnyak2002}.  Indeed such a signature of spin pumping enhanced magnetization damping was observed in Refs. \cite{Mizukami2001, Mizukami2002, Urban2001, Heinrich2003, Lenz2004}.  However, owing to the simplicity of charge transport measurements, the most common way to detect spin currents is through their conversion into a charge current via the inverse spin Hall effect (ISHE).  

\begin{figure}[!ht]
\centering
\includegraphics[width=6cm]{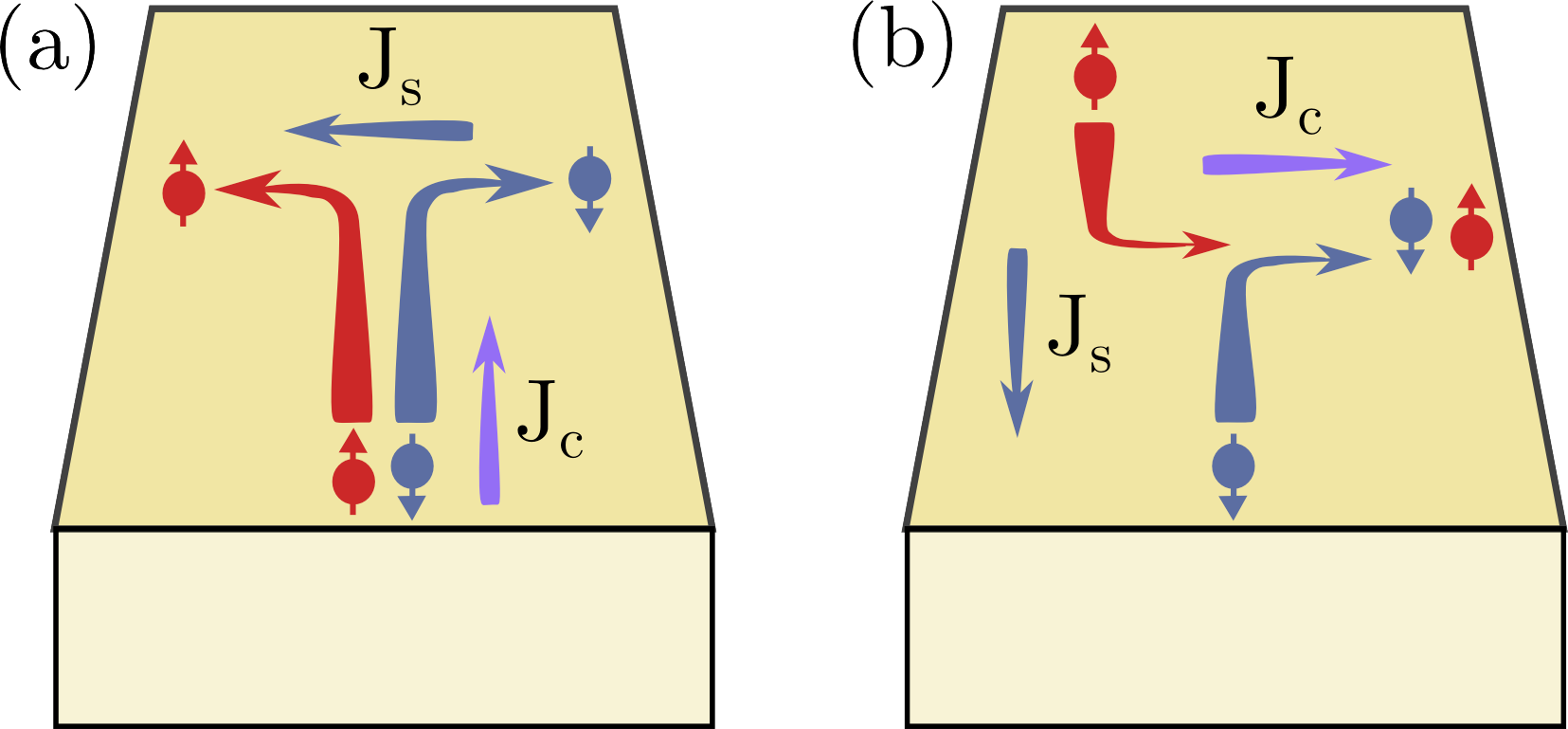}
\caption{\footnotesize{Schematic illustrations of (a) the spin Hall effect and (b) inverse spin Hall effect.  J$_c$ and J$_s$ denote the charge current and the spatial direction of the spin current respectively.  The spin polarization direction of the spin current is in the up-spin direction.}}
\label{sheSchem}
\end{figure}

Similar to spin rectification and spin pumping, in which the basic physics concepts were developed decades before modern nanotechnologies made them practically useful, spin Hall effects (SHEs) were initially studied in the 1970's \cite{Dyakonov1971} but they have only become the subject of intense interest recently after their theoretical rediscovery \cite{Hirsch1999}.  For an excellent review on spin Hall effects see Ref. \cite{Sinova2014}.  Fig. \ref{sheSchem} schematically illustrates both the SHE and ISHE which physically result from spin dependent scattering of charge carriers due to the spin-orbit interaction.  
In the spin Hall effect, a change current is converted to a spin current, and the resulting spin accumulation at the sample boundaries can be detected e.g. via Kerr rotation microscopy \cite{Kato2004} or polarized electroluminescence \cite{Wunderlich2005}.  The inverse spin Hall effect is the inverse process, where a spin current is converted into a charge current according to
\begin{equation}
\textbf{J}_c =\frac{e \theta_\text{SH} }{\hbar}  \textbf{J}_s \times \widehat{\sigma}. \label{eq:ISHE}
\end{equation}
Here $\theta_\text{SH}$ is the spin Hall angle which characterizes the efficiency of spin/charge current conversion and is dependent on the strength of the spin-orbit interaction in a material.  Due to its ability to convert spin to charge currents, for spin pumping experiments the inverse spin Hall effect can be exploited to convert the spin current generated through spin pumping into a charge current which can be electrically measured.

%%%%%%%%%%%%%%%%%%%%%%%%%%%%%%%%%%%%%%%%%%%%%%%%%%%%%%%%%%%%%%%%%%%%%%%%%%%%%%%%%%%%%%%%%%%%%%%%%%%%%%%%%%%%%%%%%%%%%%%%%%%%%%%%%%%%%%%%%%%%%%%%%%%%%%%%%%%%%%%%%%%%%%%%%%%%%%%%%%%%%%%%%%%%%%%%%%%%%%%%%%%%%%%%%%%%%%%%%%%

\subsubsection{Spin Battery and "Longitudinal" Spin Pumping Voltage}
Before the dc spin pumping voltage was measured via the inverse spin Hall effect, the first measurement of such a voltage \cite{Costache2006, Wang2006} used a device called the spin battery \cite{Brataas2002}, which was the spintronics analog of solar cells. As was known from the spin injection experiment \cite{Johnson1985} without microwave irradiation,  spin diffusion at the FM/NM interface leads to spin accumulation and depletion near the interface. From the viewpoint of the two-channel model for spin transport (see section 2.1), the physics of spin accumulation and depletion near the interface of a FM/NM is analogous to the charge accumulation and depletion in a semiconductor P/N junction.  Comparing the two, spin pumping at FMR in a FM/NM bilayer is analogous to a solar cell made of a P/N junction where, instead of the interband electrical dipole transition that converts energy of sunlight to the electrical energy of a charge current, the magnetic dipole transition of FMR converts energy from microwaves to produce spin current. 
\begin{figure}[!ht]
\centering
\includegraphics[width=14cm]{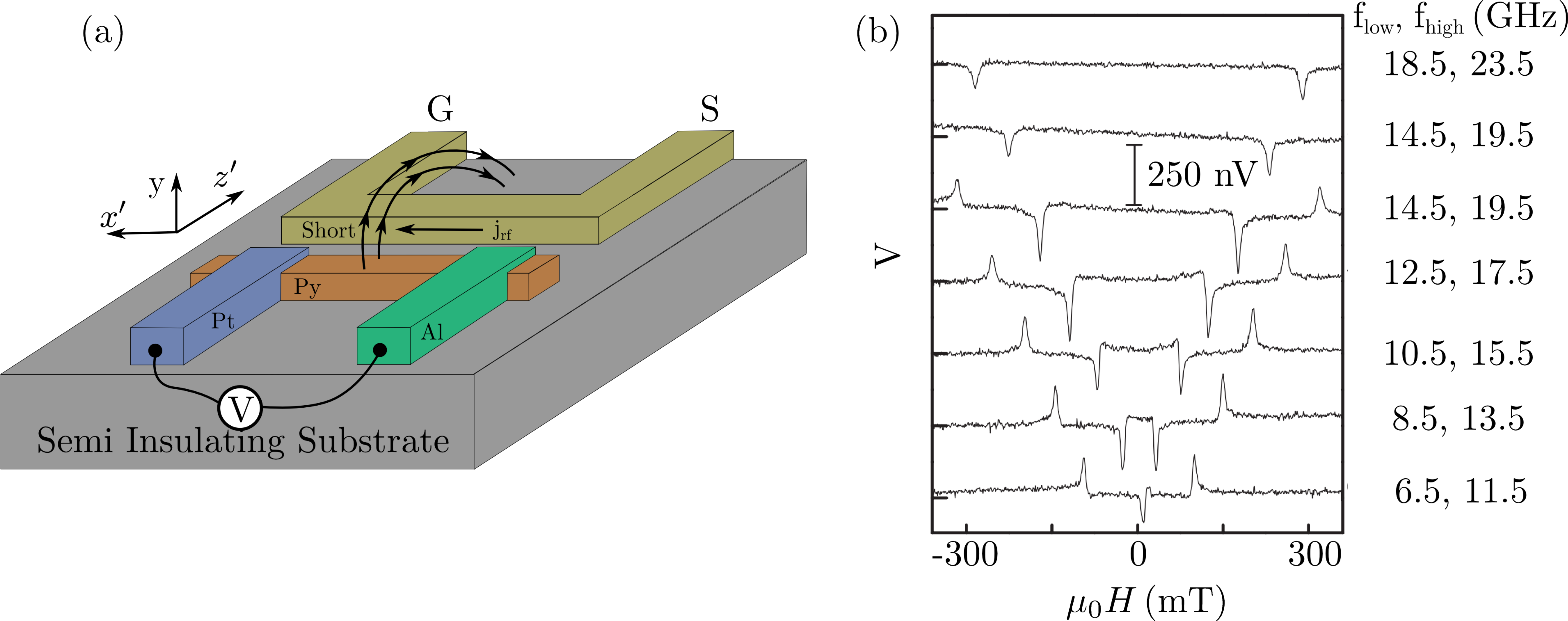}
\caption{\footnotesize{(a) A schematic diagram of the device used to observe longitudinal spin pumping.  A 25 nm thick Py strip with lateral dimensions 0.3 $\mu$m $\times$ 3 $\mu$m is placed at the short end of a CPW with 30 nm thick Pt and Al contacts.  Py FMR is driven by an $h_y$ field and a voltage is measured between the Pt and Al contacts using lock-in amplification.  (b) The dc voltage measured due to longitudinal spin pumping.  Here $f_\text{low}$ and $f_\text{high}$ are the two frequencies of the rf field used during the lock-in measurement.  The peaks and dips correspond to the resonance at $f_\text{high}$ and $f_\text{low}$ respectively.  $Source:$ Panel (b) from Ref. \cite{Costache2006}.}}
\label{vspdata}
\end{figure}

In the spin battery, when the spin injection rate (via spin pumping) is lower than the spin relaxation rate, the NM acts as a pure spin sink.  However if the spin injection rate exceeds the spin relaxation rate the spin accumulation at the interface may allow a back flow of spin current into the FM.   Due to spin dependent conductivities of the FM this back flow results in the production of a dc voltage.  To observe this effect Costache et al. \cite{Costache2006} used the device structure shown in Fig. \ref{vspdata} (a).  FMR was generated in a Py strip by the Oersted field of a CPW.  The Py strip was connected at both ends to normal metals.  To measure the dc voltage that was longitudinally generated across the interface by spin pumping, two different contact metals were used.  Pt, which is an excellent spin sink, will not generate a voltage, whereas Al, with its small spin flip relaxation rate, will produce a dc voltage.  Such an asymmetry between the two contacts is expected to produce a net voltage across the Py strip. Typical dc voltage curves measured on such a Pt/Py/Al device are plotted in Fig. \ref{vspdata} (b), showing the measured voltage of about $\sim$ 200 nV. In contrast, measurements on reference samples made of Pt/Py/Pt structure exhibit only weak signals up to 20 nV. Such an experimental contrast, combined with a detailed theoretical investigation \cite{Wang2006}, supported the case that the dc voltage longitudinally generated across the Py/Al interface was observed in the Pt/Py/Al device \cite{Costache2006}. However, it should be mentioned that in 2006 when such a "longitudinal" spin pumping experiment was performed \cite{Costache2006}, the method for line shape analysis of the rectification voltage was not yet established \cite{Mecking2007, Mosendz2010, Wirthmann2010, Harder2011a, Azevedo2011}. Hence, the asymmetric line shape shown in Fig. \ref{vspdata} (b), which indicates a contribution to the voltage from spin rectification, was not analyzed. In 2007, the issue of how to distinguish the dc voltage signal caused by spin pumping \cite{Costache2006} and spin rectification \cite{Gui2007} was raised and discussed via private communications among the groups of Refs. \cite{Costache2006, Wang2006, Gui2007}. In section \ref{sec:spsrseparation}  we will address this previously controversial subject and review the methods and solutions developed in the past decade to solve this problem.  

%%%%%%%%%%%%%%%%%%%%%%%%%%%%%%%%%%%%%%%%%%%%%%%%%%%%%%%%%%%%%%%%%%%%%%%%%%%%%%%%%%%%%%%%%%%%%%%%%%%%%%%%%%%%%%%%%%%%%%%%%%%%%%%%%%%%%%%%%%%%%%%%%%%%%%%%%%%%%%%%%%%%%%%%%%%%%%%%%%%%%%%%%%%%%%%%%%%%%%%%%%%%%%%%%%%%%%%%%%%

\subsubsection{Inverse Spin Hall Effect and "Transverse" Spin Pumping Voltage} \label{lateralsp}
\begin{figure}[!ht]
\centering
\includegraphics[width=12cm]{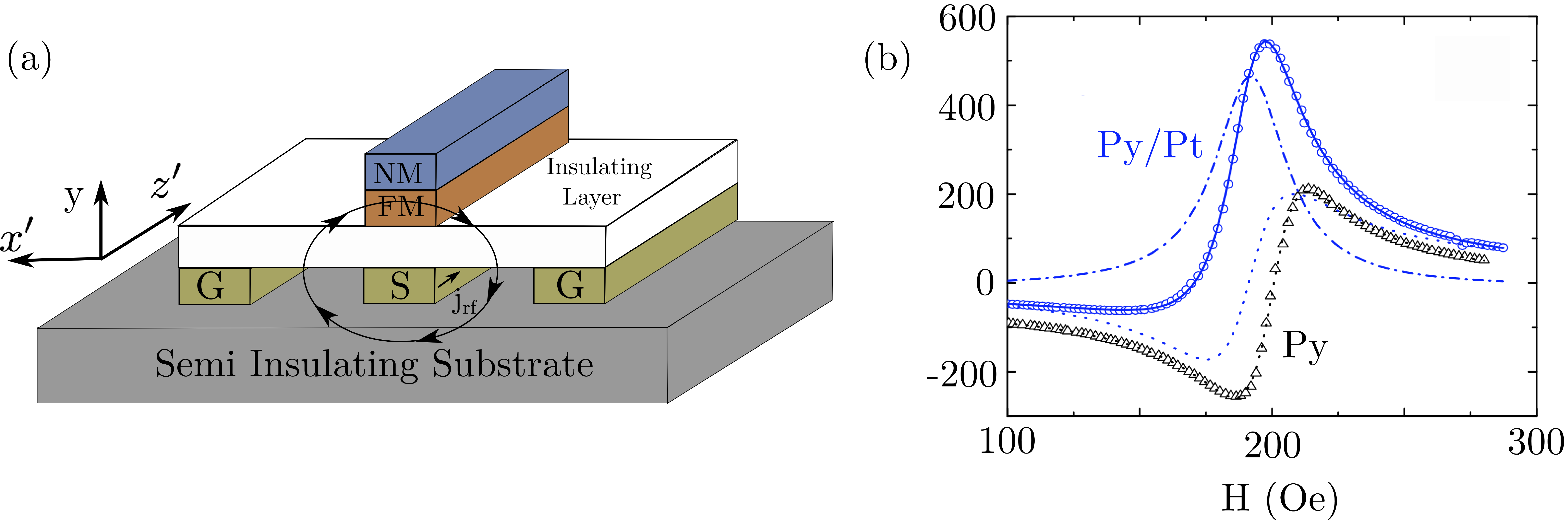}
\caption{\footnotesize{(a) A schematic diagram of a typical bilayer device.  The device is similar to the spin dynamo, with the microstrip replaced by the FM/NM bilayer.  (b) Experimental data for a Py/Pt bilayer \cite{Mosendz2010}.  The lateral dimensions of the bilayers are 2.92 mm $\times ~20~\mu$m and they are each 15 nm thick.  The insulating layer separating the 30 $\mu$m wide, 200 nm thick Au CPW from the bilayer was 100 nm thick MgO.  The FMR is driven by an $h_{x^\prime}$ field and due to the capacitative coupling the relative phase is zero, which means that the AMR in Py will be purely dispersive, as shown with the black triangles in the data.  The blue circles show the signal from the Py/Pt which contains both the Lorentz (dotted-dashed line) and dispersive (dotted line) components.}}
\label{lspdata}
\end{figure}

Compared to the "longitudinal" spin pumping configuration, the use of a "transverse" configuration using the inverse spin Hall effect, as introduced by Saitoh et al. \cite{Saitoh2006}, has several practical advantages, such as the ease of charge current detection and larger dc voltages, typically several $\mu$V even at low microwave powers.  Hence, in contrast to the "longitudinal" spin pumping voltage, the "transverse" spin pumping voltage induced by the inverse spin Hall effect has been studied by many different groups.  Fig. \ref{lspdata} shows a typical experimental setup which may be used to detect spin pumping via the inverse spin Hall effect.  Normally a FM/NM bilayer is placed on top of a CPW, allowing the $h_{x^\prime}$ field generated by the current in the CPW to drive magnetization precession in the FM.  This magnetization precession produces a spin current via the spin pumping mechanism which is then converted into a charge current in the NM via the inverse spin Hall effect and is detected as a voltage across the bilayer in the $z^\prime$ direction.
Due to its large spin-orbit coupling Pt has a large spin Hall angle, $\theta_\text{SH} \sim 0.05$ \cite{Bai2013} (although the exact value has been the subject of some controversy \cite{Liu2011a}, and care must be taken to either separate spin pumping and spin rectification effects as discussed in Sec. \ref{sec:spsrseparation}, or to eliminate AMR induced SR altogether \cite{Obstbaum2014}.) and therefore Pt is a common NM used in spin pumping experiments.  

With the charge current determined by the ISHE in Eq. \ref{eq:ISHE}, the spin pumping voltage can be found by integrating the time averaged charge current density, 
\begin{equation*}
V_\text{SP} = R_{NM} \int \langle \textbf{j}_c \left(n, t\right) \cdot \widehat{q}\rangle  dA = R_{NM} \frac{e \theta_\text{SH}}{\hbar} \left[\int \langle j_s \left(n, t\right)\rangle dA\right] \left(\widehat{n} \times \widehat{\sigma}\right)\cdot \widehat{q}
\end{equation*}
where $\widehat{q}$ is the measurement direction (either along the length or width of the sample -- in the configuration shown in Fig. \ref{lspdata} $\widehat{q} = \widehat{z}^\prime$) and $dA = dn dw$ with $dw$ along the width of the sample.  To determine $V_\text{SP}$ we will use the coordinate systems in Fig. \ref{srgeo}.  In either the in-plane or out-of-plane configurations, taking $\textbf{M} = \left(m_x^t, m_y^t, M_0\right)$ will yield the same expression for the spin current density
\begin{equation*}
j_s \left(n, t\right) = \frac{\hbar G_r}{4 \pi M_0^2} \frac{\sinh \left[\left(t_\text{NM} - n\right)/\lambda_\text{SD}\right]}{\sinh \left[t_\text{NM}/\lambda_\text{SD}\right]} \omega \text{Im}\left(m_x^* m_y\right)
\end{equation*}
and therefore 
\begin{equation}
V_\text{SP} = \kappa \text{Im} \left(\frac{m_x^* m_y}{M_0^2}\right) \left(\widehat{n} \times \widehat{\sigma}\right) \cdot \widehat{q} \label{spvoltage}
\end{equation}
where the proportionality constant $\kappa = \frac{\theta_\text{SH} \lambda_\text{SD}}{\sigma_\text{NM}} \left(\frac{e\omega}{4\pi}\right) \frac{G_r}{t_\text{NM} w} \tanh \left(\frac{t_\text{NM}}{2\lambda_\text{SD}}\right)$ depends on material and device specific parameters, such as the normal metal conductivity $\sigma_\text{NM}$, but is not important for a line shape analysis.  We can immediately see that because the voltage depends on $m_x^* m_y$ there will be no dependence on the spin resonance phase and the line shape will be completely Lorentzian.  Using the solution to the LLG equation found in Eq. \ref{polder}, $V_\text{SP}$ for in-plane magnetic fields is 
\begin{align*}
\left(V_\text{SP}\right)_\theta^l &= \kappa \left(A_{xx} A_{xy} h_{x^\prime}^2 \sin\theta_H \cos^2\theta_H  + A_{yy} A_{xy} h_y^2 \sin\theta_H  +A_{xx} A_{xy} h_{z^\prime}^2 \sin^3 \theta_H\right) L,\\
\left(V_\text{SP}\right)_\theta^t &= - \kappa \left(A_{xx} A_{xy} h_{x^\prime}^2 \cos^3\theta_H  + A_{yy} A_{xy} h_y^2 \cos\theta_H  +A_{xx} A_{xy} h_{z^\prime}^2 \cos\theta_H \sin^2 \theta_H\right) L.
\end{align*}
For the two out-of-plane rotations the spin pumping voltage is
\begin{align*}
\left(V_\text{SP}\right)_\phi^l &= 0, \\
\left(V_\text{SP}\right)_\phi^t &= \kappa \left(A_{xx} A_{xy} \sin\phi_H \cos^2\phi_H h_{x^\prime}^2 + A_{xy} A_{yy} \sin\phi_H h_y^2 + A_{xx} A_{xy} \sin^3\phi_H h_{z^\prime}^2\right)L, \\
\left(V_\text{SP}\right)_\psi^l &= \kappa \left(A_{xx} A_{xy} \sin\psi_H \cos^2 \psi_H h_{x^\prime}^2 + A_{xy} A_{yy} \sin\psi_H h_y^2 + A_{xx} A_{xy} \sin^3\psi_H h_{z^\prime}^2\right)L,\\
\left(V_\text{SP}\right)_\psi^t &= 0.
\end{align*}
The angular dependence of spin pumping is summarized in Table \ref{spangular} for each measurement/rotation configuration.  

In the analysis presented here we have directly used the rf magnetization determined from the LLG equation.  However it should be noted that is also possible to replace the $\text{Im}\left(m_x^* m_y\right)$ term in Eq. \ref{spvoltage} with $\sin^2\theta_c$ where $\theta_c$ is the cone angle.  The resulting expression assumes that the ellipticity of the precession is small, but provides a convenient way to characterize the behaviour at low microwave powers \cite{Mosendz2010}.

\begin{table}[!ht]
\def\arraystretch{1}
\caption{\footnotesize{The angular dependence of spin pumping.}}
\centering
\begin{tabular}{>{\centering\arraybackslash}m{1.94cm}>{\centering\arraybackslash}m{1.94cm}>{\centering\arraybackslash}m{1.94cm}>{\centering\arraybackslash}m{1.94cm}>{\centering\arraybackslash}m{1.94cm}>{\centering\arraybackslash}m{1.94cm}>{\centering\arraybackslash}m{1.94cm}>{\centering\arraybackslash}m{0pt}@{}}
\cmidrule[0.75pt](l{0.3em}r{0.3em}){1-7}                       
& \includegraphics[width=1.94cm]{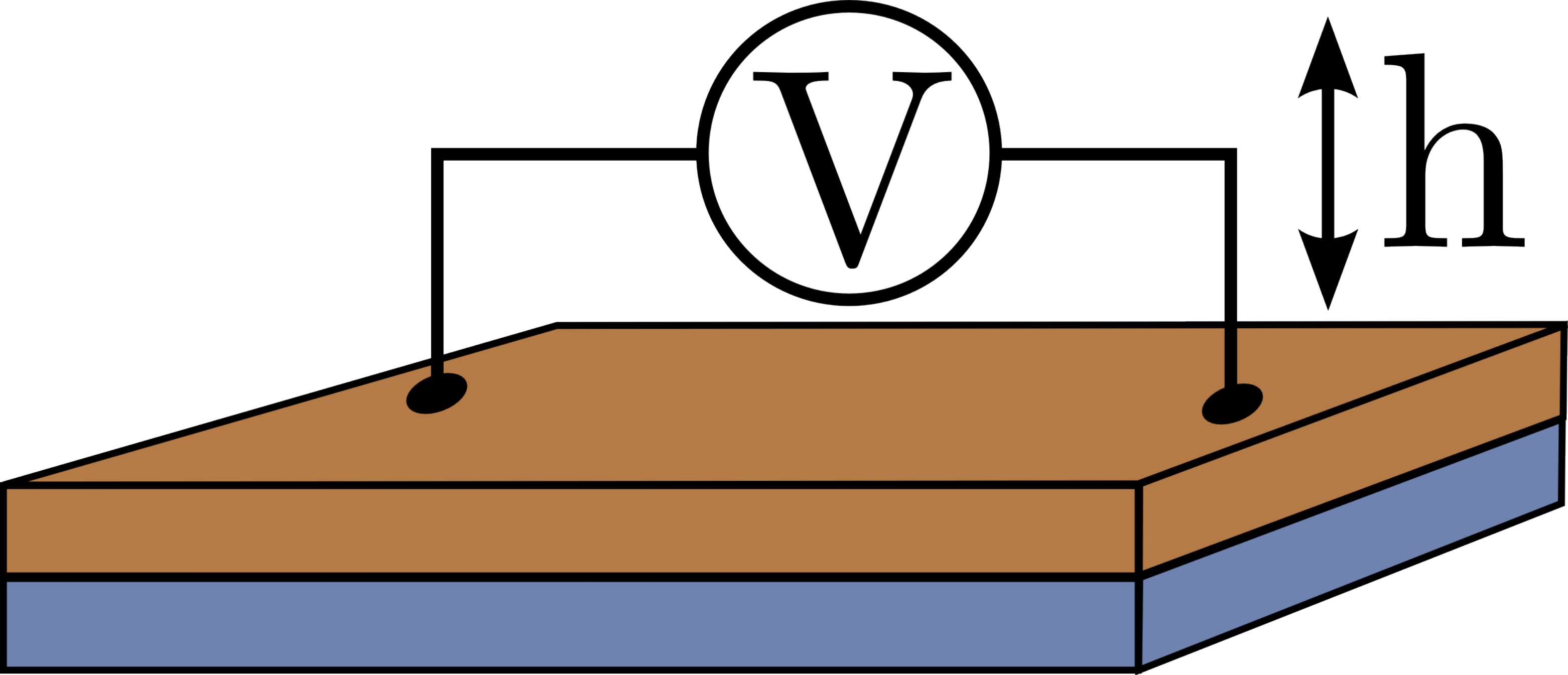} & \includegraphics[width=1.94cm]{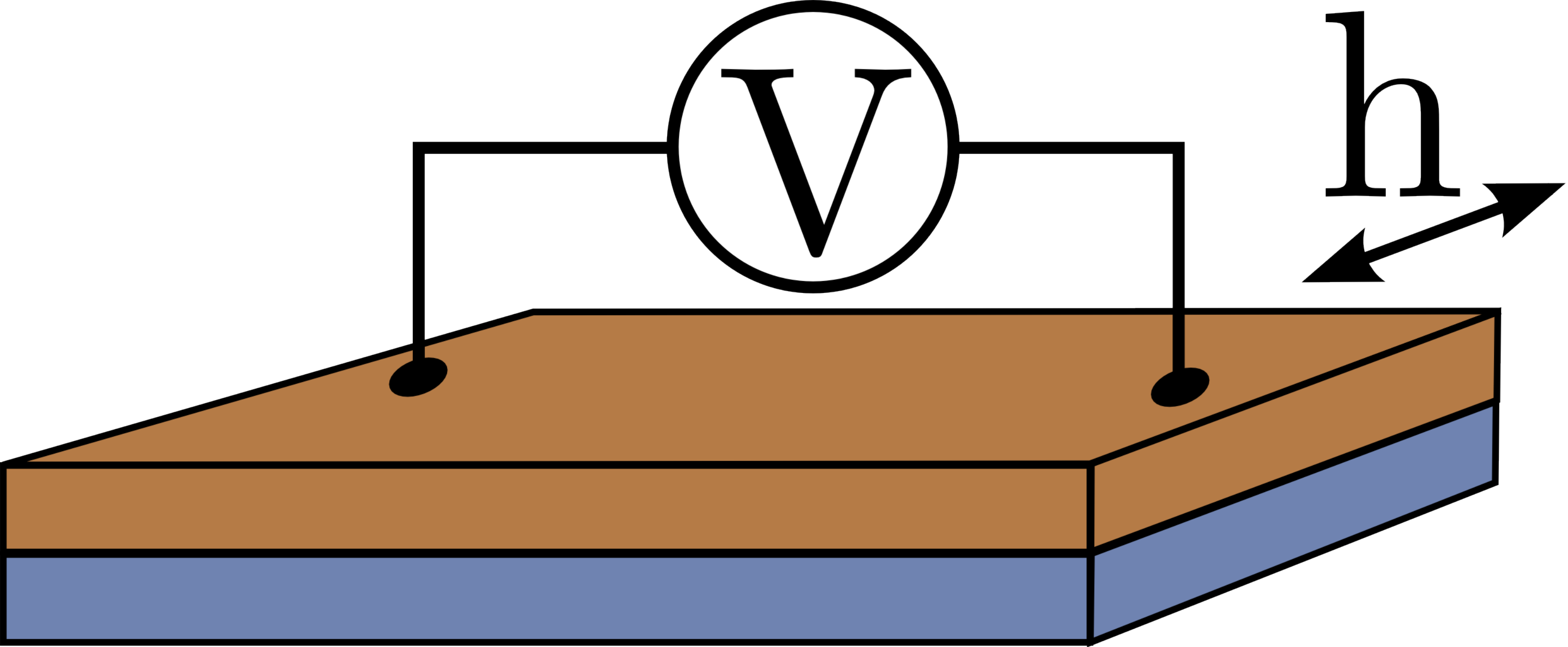} &\includegraphics[width=1.94cm]{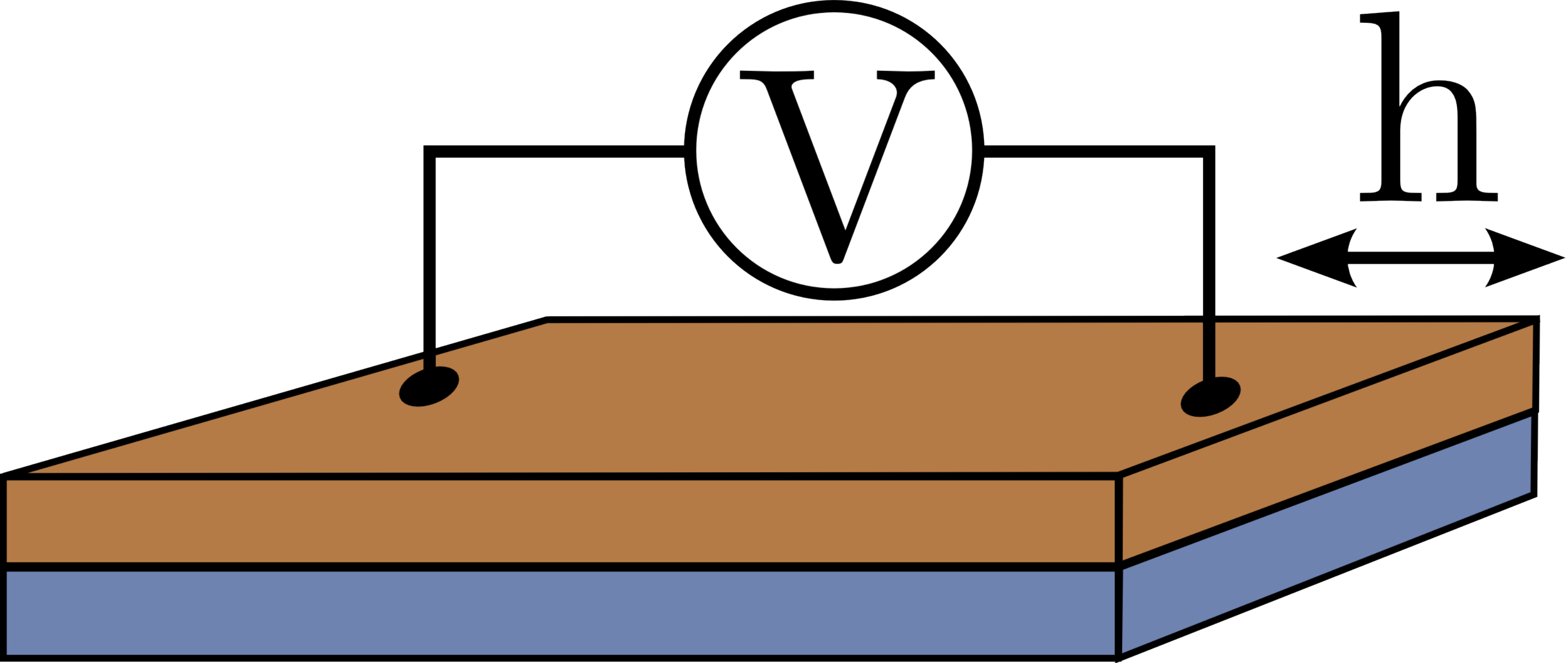}&\includegraphics[width=1.94cm]{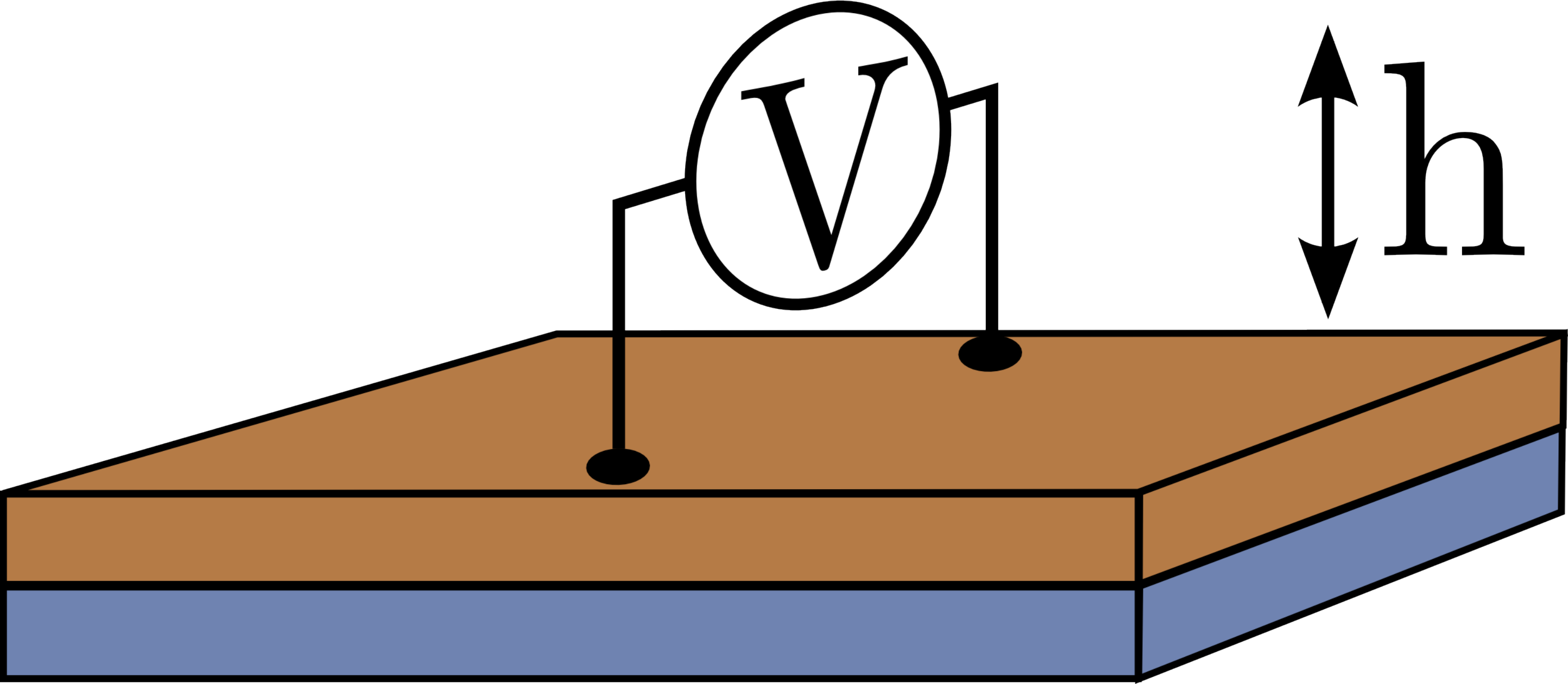}&\includegraphics[width=1.94cm]{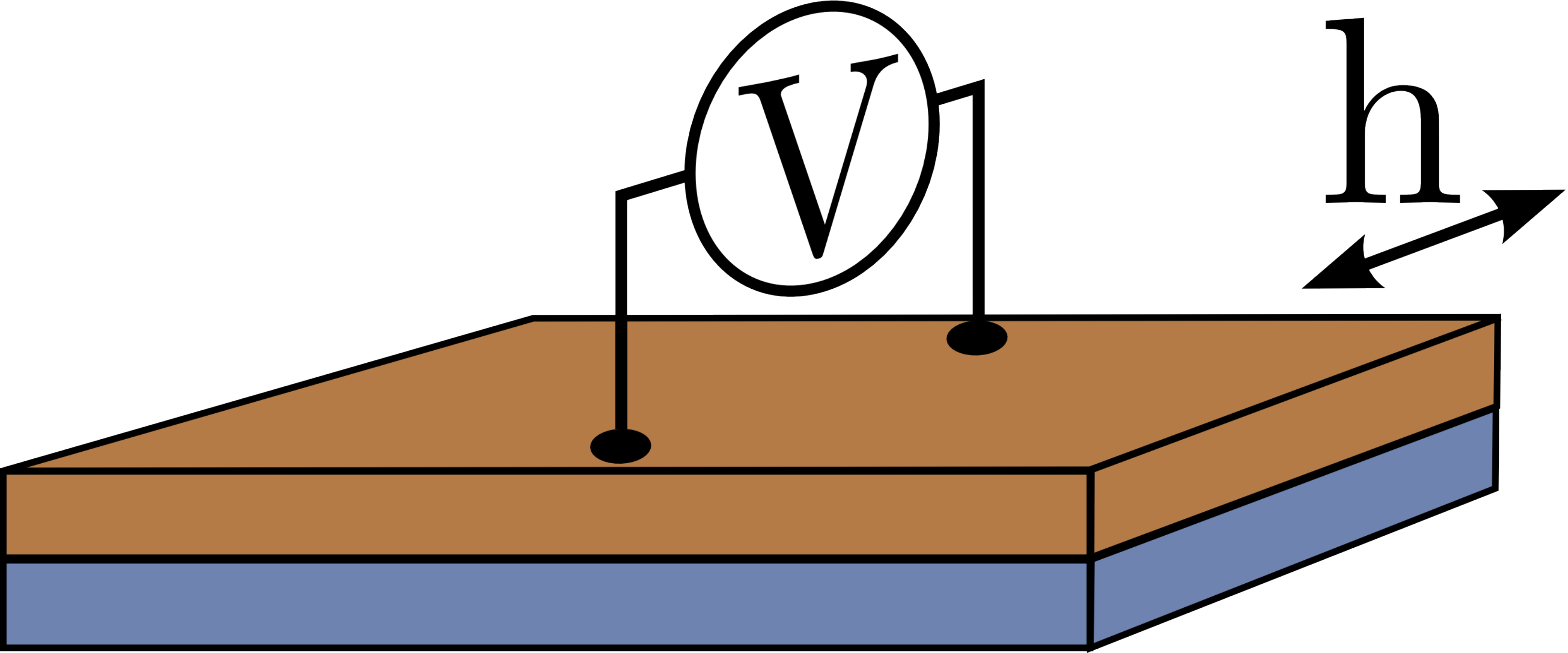}&\includegraphics[width=1.94cm]{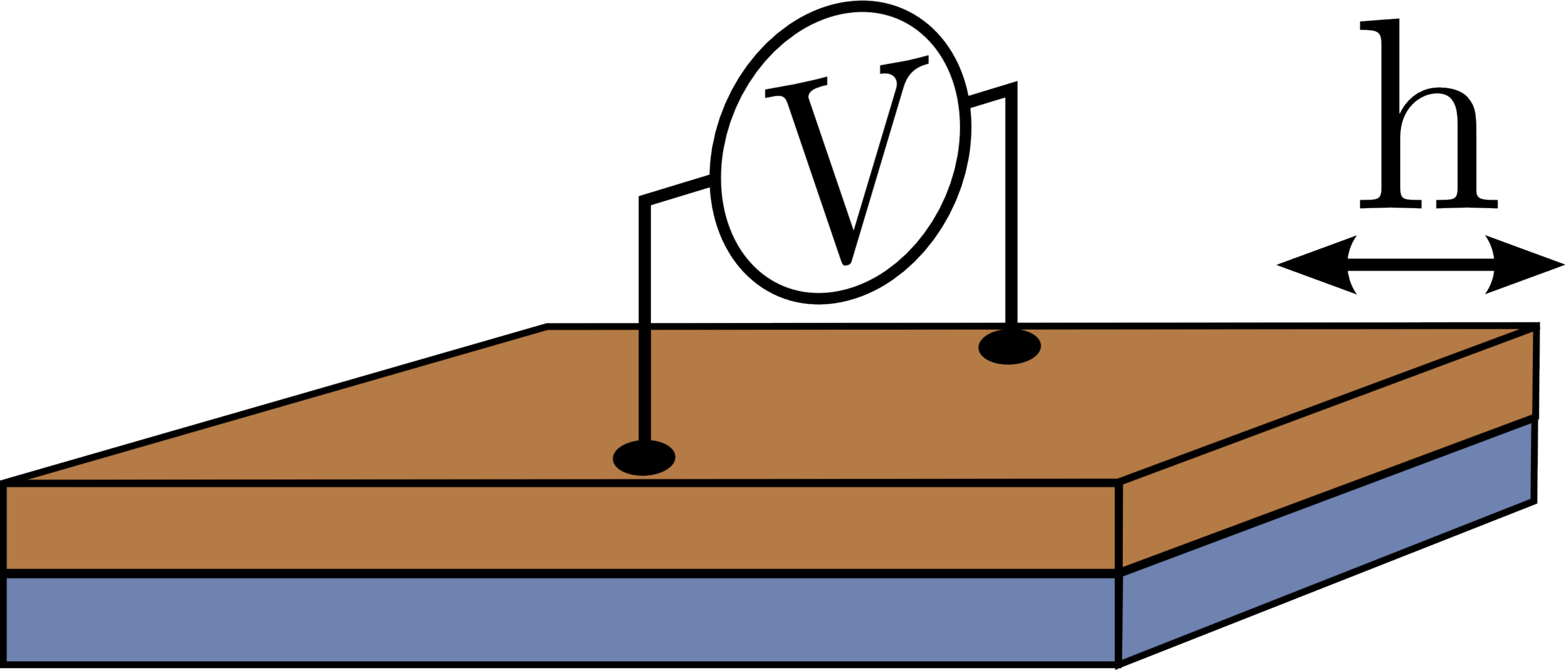}\\ \cmidrule(l{0.3em}r{0.3em}){1-7} 
\multirow{2}{*}[-1em]{\includegraphics[width=2cm]{Table45C1.pdf}} & $\sin\theta_H$ & $\sin\theta_H\cos^2\theta_H$ & $\sin^3\theta_H$ & $\cos\theta_H$ & $\cos^3\theta_H$ & $\cos\theta_H\sin^2\theta_H$ &\\[0.5cm]
&\includegraphics[width=1.94cm]{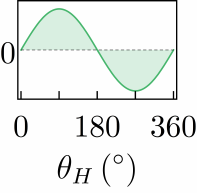}&\includegraphics[width=1.94cm]{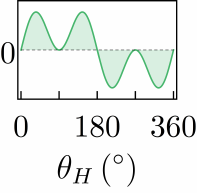}&\includegraphics[width=1.94cm]{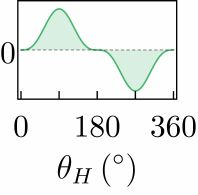} &\includegraphics[width=1.94cm]{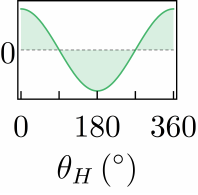}&\includegraphics[width=1.94cm]{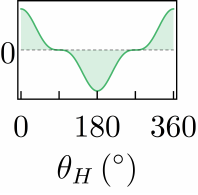}&\includegraphics[width=1.94cm]{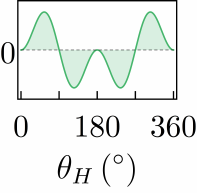}\\
\cmidrule(l{0.3em}r{0.3em}){1-7}
\multirow{2}{*}[-1em]{\includegraphics[width=1.9cm]{Table45C2.pdf}} & $0$ & $0$ & $0$ & $\sin^3\theta_H$ & $\sin\theta_H$ & $\sin\theta_H\cos^2\theta_H$ &\\[0.5cm]
&\includegraphics[width=1.94cm]{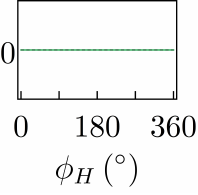}&\includegraphics[width=1.94cm]{Table45Phi1.pdf}&\includegraphics[width=1.94cm]{Table45Phi1.pdf} &\includegraphics[width=1.94cm]{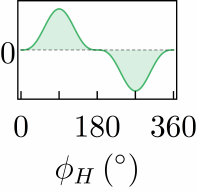}&\includegraphics[width=1.94cm]{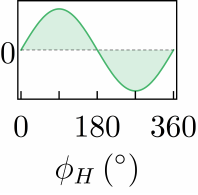}&\includegraphics[width=1.94cm]{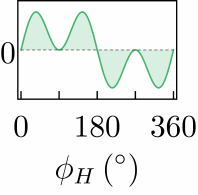}\\
\cmidrule(l{0.3em}r{0.3em}){1-7}
\multirow{2}{*}[-1em]{\includegraphics[width=1.94cm]{Table45C3.pdf}} & $\sin^3\psi_H$ & $\sin\psi_H\cos^2\psi_H$ & $\sin\psi_H$ & $0$ & $0$ & $0$ &\\[0.5cm]
&\includegraphics[width=1.94cm]{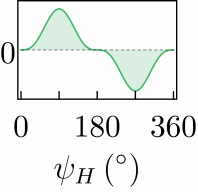}&\includegraphics[width=1.94cm]{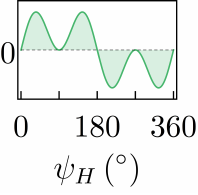}&\includegraphics[width=1.94cm]{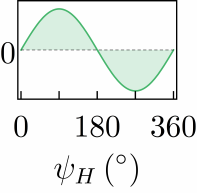} &\includegraphics[width=1.94cm]{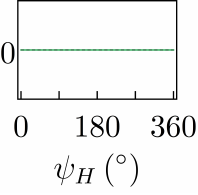}&\includegraphics[width=1.94cm]{Table45Psi4.pdf}&\includegraphics[width=1.94cm]{Table45Psi4.pdf}\\
\cmidrule[0.75pt](l{0.3em}r{0.3em}){1-7} 
\end{tabular}
\label{spangular}
\end{table}

Returning to the typical spin pumping experiment shown in Fig. \ref{lspdata}, since the microwave magnetic field is in the $h_{x^\prime}$ direction, when the relative phase is zero the line shape due the spin rectification will be purely dispersive.  This is what is observed in the experimental data shown in Fig. \ref{srdevice} (b).  The voltage in a Py monolayer is seen to be dispersive, while the voltage in the Py/Pt bilayer has both a dispersive (calculated with the dotted line) and a Lorentz (calculated with the dashed-dotted line) component.  Therefore the line shape difference may be used to separate spin rectification from spin pumping in this case.  However as we saw in Sec. \ref{spinrectification}, when the relative phase is nonzero the SR voltage will have both dispersive and Lorentz contributions, and therefore distinguishing SP from SR becomes a more complex.  Since the first experimental observations of spin pumping \cite{Azevedo2005, Costache2006, Saitoh2006}  were made almost concurrently with the first observations of spin rectification in microstructured devices, methods to separate the two effects were not yet fully developed.  Instead these initial studies attempted to reduce the effect of spin rectification by placing the sample at a position of minimal rf electric field so that there would not be any directly applied rf current.  However more robust methods that can actually separate SP from SR are desirable so that SP in various device structures can be reliably studied.  This is a necessary step in order to exploit spin pumping as a direct source of spin polarized current.  Accurately measuring spin pumping is also needed to reliably  determine the spin Hall angle, which is of practical interest in the development of spintronic devices.  Therefore shortly after the first spin pumping measurements it was realized that a robust method to separate spin pumping and spin rectification signals was necessary.  

%%%%%%%%%%%%%%%%%%%%%%%%%%%%%%%%%%%%%%%%%%%%%%%%%%%%%%%%%%%%%%%%%%%%%%%%%%%%%%%%%%%%%%%%%%%%%%%%%%%%%%%%%%%%%%%%%%%%%%%%%%%%%%%%%%%%%%%%%%%%%%%%%%%%%%%%%%%%%%%%%%%%%%%%%%%%%%%%%%%%%%%%%%%%%%%%%%%%%%%%%%%%%%%%%%%%%%%%%%%

\subsubsection{Spin Rectification vs Spin Pumping} \label{sec:spsrseparation}

Several methods have been developed to separate spin pumping and spin rectification in FM/NM bilayers (for a concise review see the supplementary material of Ref. \cite{Bai2013}).  The first method developed relies on a line shape analysis \cite{Mosendz2010}.  This method is effective when the spin rectification signal is purely dispersive so that it can clearly be distinguished from the Lorentzian spin pumping.  However this method becomes problematic when there is a relative phase shift between rf electric and magnetic fields.  Another method using the angular dependence has also been used \cite{Azevedo2011}.  As we see from Tables \ref{srangularip} and \ref{spangular}, for the case of an in-plane static field, when FMR is driven by a transverse $h$ field both spin rectification and spin pumping have a $\sin\theta_H \cos^2\theta_H$ angular dependence.  However when a normal $h$ field produces the voltage, spin pumping follows a $\sin\theta_H$ while spin rectification follows a $\sin\theta_H\cos\theta_H$ angular dependence, and therefore it is possible to distinguish the two effects. Using such a special measurement configuration, a pure spin pumping signal has been detected in the Py/Pt bilayer \cite{Feng2012}.  

\begin{figure}[!ht]
\centering
\includegraphics[width=14cm]{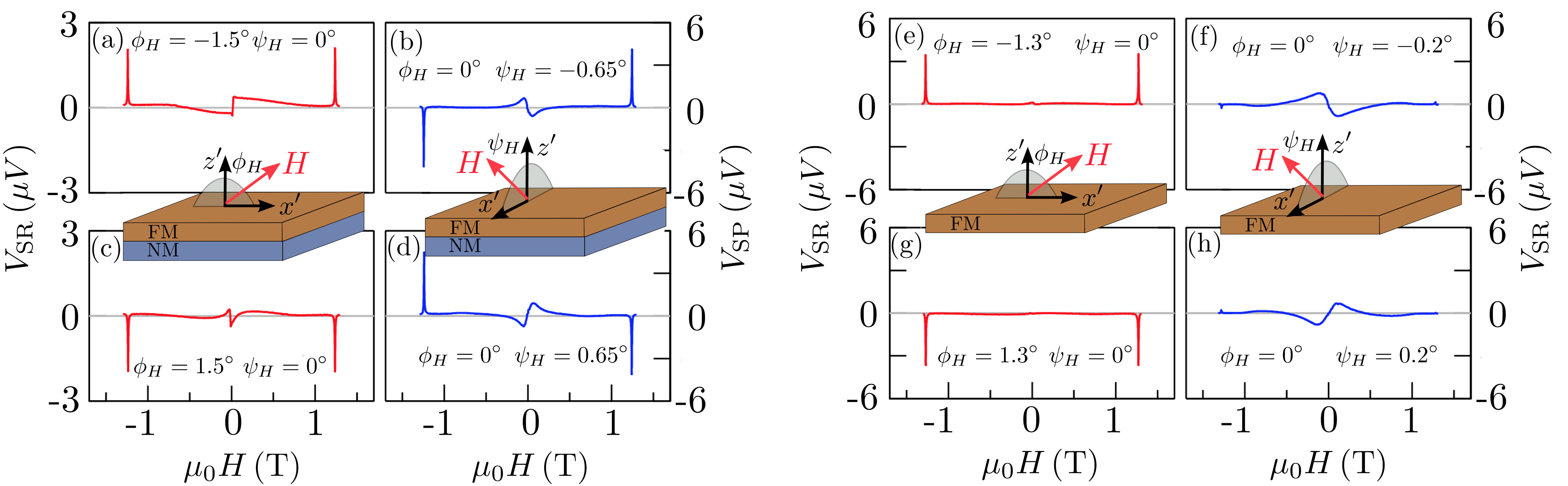}
\caption{\footnotesize{(a) - (d) dc voltage measurements on a Py/Pt bilayer as a function of the magnetic field \textbf{H} applied at angles (a) $\phi_H = -1.5^\circ, \psi_H = 0^\circ$ (b) $\phi_H = 0^\circ, \psi_H = -0.65^\circ$ (c) $\phi_H = 1.5^\circ, \psi_H = 0^\circ$ (d) $\phi_H = 0^\circ, \psi_H = 0.65^\circ$.  When $\psi_H = 0^\circ$ the signal is purely spin rectification, while at $\phi_H = 0^\circ$ the signal is purely spin pumping (e) - (h) dc voltage measurements on a Py monolayer at angles (e) $\phi_H = -1.3^\circ, \psi_H = 0^\circ$ (f) $\phi_H = 0^\circ, \psi_H = -0.2^\circ$ (g) $\phi_H = 1.3^\circ, \psi_H = 0^\circ$ (h) $\phi_H = 0^\circ, \psi_H = 0.2^\circ$.for comparison.  At $\psi_H = 0^\circ$ the spin rectification is still present.  However at $\phi_H = 0^\circ$ there is no voltage signal since there is no spin pumping in a monolayer.  $Source:$ Adapted from Ref. \cite{Bai2013}.}}
\label{spseparation}
\end{figure}

A method combining line shape analysis with the even and odd contributions of the voltage signal with respect to the static magnetic field has also been developed \cite{Rousseau2012}.  However this method also cannot be used when the microwave field is parallel to the sample, which is the same limitation encountered when using the angular dependence to separate the effects.  More recently a universal method based on the magnetic field symmetries has been implemented \cite{Bai2013}.  This method exploits the symmetries of $V_\text{SP}$ and $V_\text{SR}$ when a magnetic field is applied nearly perpendicular to the sample plane, enabling a reliable measurement configuration in which \textit{only} spin pumping contributes to the voltage signal.  As a result the separation step can effectively be performed \textit{during} the experiment, reducing the need for complex line shape and angular analyses.  From Tables \ref{srangularip} and \ref{spangular} we can see that for an out-of-plane magnetic field and longitudinal voltage measurements  

\begin{equation}
\text{at $\psi_H=0^\circ$ there is \textit{no spin pumping} and}~V_\text{SR}\left(\phi_H, H\right) = V_\text{SR} \left(\phi_H, - H\right) = -V_\text{SR} \left(-\phi_H, H\right) \label{vsrsymmetry}
\end{equation}
and 
\begin{equation}
\text{at $\phi_H = 0^\circ$ there is \textit{no spin rectification} and}~V_\text{SP} \left(\psi_H, H\right) = - V_\text{SP} \left(\psi_H, -H\right) = - V_\text{SP} \left(-\psi_H, H\right) \label{vspsymmetry}.
\end{equation}
Therefore as shown in Fig. \ref{spseparation}, $V_\text{SP}$ and $V_\text{SR}$ can be distinguished experimentally based on their $\textbf{H}$ field symmetries.  This method enables an accurate determination of the spin Hall angle and the characterization of a materials spin/current conversion efficiency.  Since the work of Bai et al. in 2013 \cite{Bai2013} a method for the spin pumping and spin rectification separation based on the difference in symmetry dependence of $V_\text{SP}$ and $V_\text{SR}$  on the spin diffusion direction was used by Zhang et al. \cite{Zhang2016}, which allows separation of the effects without line shape fitting.  It is also worth noting that in addition to the methods discussed here, the rectification signal in a Schottky diode (due to tunnelling anisotropic magnetoresistance) \cite{Gould2004, MatosAbiague2009} may be distinguished from spin pumping based on its bias dependence \cite{Liu2014}.

%%%%%%%%%%%%%%%%%%%%%%%%%%%%%%%%%%%%%%%%%%%%%%%%%%%%%%%%%%%%%%%%%%%%%%%%%%%%%%%%%%%%%%%%%%%%%%%%%%%%%%%%%%%%%%%%%%%%%%%%%%%%%%%%%%%%%%%%%%%%%%%%%%%%%%%%%%%%%%%%%%%%%%%%%%%%%%%%%%%%%%%%%%%%%%%%%%%%%%%%%%%%%%%%%%%%%%%%%%%

\subsubsection{Spin-Transfer Torque Rectification in Bilayers}\label{standshe} 

The final voltage producing effect observed in FM/NM bilayers is spin-transfer torque induced rectification.  As shown in Fig. \ref{vinbilayers} (c) STT induced rectification requires the conversion of an rf charge current in the NM into an ac spin current via the spin Hall effect.  This ac spin current then produces a torque on the magnetization of the FM via the exchange interaction, subsequently generating a dynamic resistance and coupling to the rf charge current to create a dc voltage.  The key ingredient in all spin torque induced effects is the generation of a spin polarized current, and while the focus of this section will be on spin Hall induced STT, it is important to note that there are other physical mechanism which may generate spin polarized currents.  First, a spin polarized current may result from the flow of an electric current between non collinear magnetic structures.  This form of spin-transfer torque may occur within domain wall structures \cite{Yamaguchi2007, Bedau2007} and is the basis of spin rectification in MTJs which will be discussed further in Sec. \ref{srinmtjs}.  Second, in crystalline structures lacking inversion symmetry, spin-orbit coupling may induce a polarization of the conduction electrons in the presence of an electric current \cite{Gambardella2011}.  This so-called spin-orbit torque has recently attracted much attention from the spintronics community \cite{Kim2013, Fan2013, Garello2013, Jamali2013, Fan2014, Skinner2014, Kurebayashi2014, Mellnik2014, Ciccarelli2014, Skinner2015, Wang2015, Pai2015, Ciccarelli2015, Nan2015} and results in an effective wave vector dependent magnetic field.  The spin-orbit torque may arise due to either Rashba (interface) or Dresselhaus (bulk) spin-orbit interactions in certain FM/NM devices and may produce a field-like term which can even oppose the field-like torque \cite{Skinner2014}.  In such situations the spin-torque line shape may contain a dispersive contribution, unlike the case of pure spin Hall induced spin torque, and care must be taken to separate the effects \cite{Mellnik2014, Pai2015}.  Here we will not discuss the recent and developing subject of spin-orbit torques in great detail, choosing to focus on the spin Hall induced STT and its distinction from SR and SP.  

\begin{figure}[!ht]
\centering
\includegraphics[width=12cm]{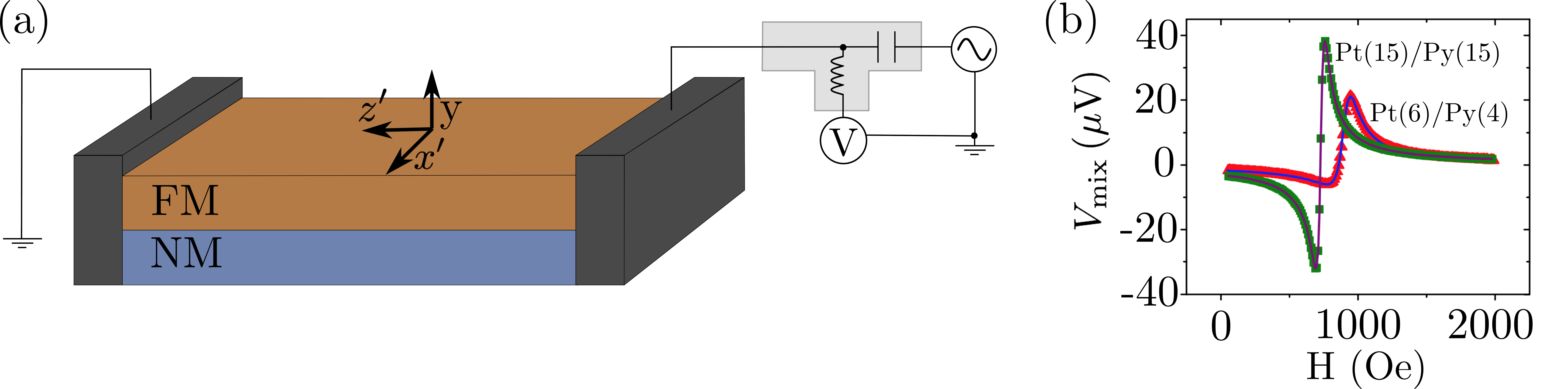}
\caption{\footnotesize{(a) A schematic diagram of a typical spin-transfer torque rectification measurement setup.  A bias tee is used to provide both an rf and dc current as well as measure the dc voltage produced in the bilayer. (b) Typical experimental data for Py/Pt bilayers of different thicknesses (shown in nm) \cite{Liu2011}.  The voltage is made of a mixture of dispersive and symmetric line shapes due to the field and spin-transfer torques respectively.  $Source$: Adapted from Ref. \cite{Liu2011}.}}
\label{stdevice}
\end{figure}

A typical measurement setup used to detect the spin-transfer torque rectification via the spin Hall effect is shown in Fig. \ref{stdevice} (a).  Unlike in the detection of field torque or spin pumping, a typical STT rectification measurement setup will apply the rf current directly to the bilayer sample, avoiding the use of a CPW or cavity to apply an rf field and instead uses a bias tee to allow the direct application of both rf and dc currents as well as the measurement of the dc voltage across the sample.  As we discuss below this means that the contribution due to spin-transfer torque rectification will be purely symmetric.  Typical experimental data, measured on Pt(15 nm)/Py(15 nm) and Pt(6 nm)/Py(4 nm) bilayers at 8 GHz, is shown in Fig. \ref{stdevice} (b).  As would be expected the line shape contains both symmetric and dispersive contributions due to both field and spin-transfer torque rectification effects.  

The idea of current induced magnetization dynamics was proposed in the works of Berger \cite{Berger1996} and Slonczewski \cite{Slonczewski1996} and was first seen in Co/Cu multilayers by observing the magnetoresistance changes when large current densities ($\sim 10^8$ A/cm$^2$) were injected through point contacts \cite{Tsoi1998, Myers1999}.  For a detailed review of spin-transfer torque see e.g. Ref. \cite{Brataas2011, Ralph2007}.  The effect of a spin polarized current on the magnetization dynamics can be described by modifying the LLG equation to include an extra torque term \cite{Slonczewski1996, Zhang2002} which may be written as \cite{Liu2011}, 
\begin{equation}
\frac{d\textbf{M}}{dt} = - \gamma \textbf{M} \times \textbf{H} + \frac{\alpha}{M_0} \textbf{M} \times \frac{d\textbf{M}}{dt} - \Gamma j_s \textbf{M} \times \left(\widehat{\sigma} \times \textbf{M} \right) \label{llgstbilayer}
\end{equation}
where $\Gamma = -\left(\gamma \hbar\right)/\left(2e\mu_0 M_0^2 t_\text{FM}\right)$ with $t_\text{FM}$ being the thickness of the FM layer.  The first two terms are just the unmodified LLG equation describing precession about $\textbf{H}$ and Gilbert damping respectively, and the last term is the modification due to a the spin-transfer torque resulting from a current with spin polarization along $\widehat{\sigma}$.  Here we have neglected the field-like term arising from the spin-orbit torque.  For a discussion of the LLG modification due to spin-orbit torques, see e.g. Ref. \cite{Pai2015}.  

The STT term provides an additional torque in the $\textbf{M}-\widehat{\sigma}$ plane.  Since the magnitude of $\textbf{M}$ is taken to be constant, there is no spin-transfer torque component along $\textbf{M}$, although in principle there may be a component perpendicular to the $\textbf{M}-\widehat{\sigma}$ plane which would act as a field-like torque.  However in FM/NM bilayers the out-of-plane spin-transfer torque may be neglected \cite{Liu2011, Wang2014c} (see also the discussion in \cite{Brataas2011} and \cite{Zhang2004, Li2004, Li2004a, He2006}).  

The voltage generated along the length of the FM can now be determined by solving the LLG equation with the STT term in analogy with the approach taken in Sec. \ref{spinrectification}.  When the field torque is due to $h_{x^\prime}$ the voltage is given by  
\begin{equation}
V_{\text{mix}} = \frac{1}{2} \frac{\Delta R}{M_0} I_{z^\prime} \sin2\theta_H\cos\theta_H \left(A_{xx} h_{x^\prime} D + S j_s L \right) \label{vmix}
\end{equation}
where $S = \left(\Gamma M_0^2\right)/\left(2 \Delta H \left(d\omega_r/dH\right)|_{H=H_r}\right)$ and $V_\text{mix}$ indicates that the voltage is due to both the field torque and STT rectifications.  In a FM/NM bilayer AMR provides the magnetoresistance which produces the dc signal, however a more general expression may be obtained by replacing $-\Delta R \sin 2\theta_H$ with $dR/d\theta$ \cite{Liu2011}, which is obtained by expanding the voltage as a function of $\theta$ ($\theta$ may be $\theta_H$ as it is for AMR, however it may also be another relevant angle, depending on the system in question, see Table \ref{mrsummary} and Refs. \cite{Sankey2007, Kubota2008}).  

The first term in Eq. \ref{vmix} is just the field torque rectification and has a dispersive line shape.  However the presence of a spin-transfer torque also has a residual effect on this term by increasing the line width (for the same reasons as the spin pumping induced line width enhancement).  The lack of a symmetric component to the field torque line shape is due to the fact that there is no phase shift between the rf current, $j_{\text{rf}}$ and the Oersted field.  This is expected, since the Oersted field is produced by the current flowing in the NM layer, which is exactly the same current injected into the FM.  This situation differs from the measurement of the field torque rectification using a CPW, where the Oersted field is generated by the current flowing in the CPW which is electrically isolated from the FM microstrip and may therefore have a phase shift compared to $j_{\text{rf}}$.  On the other hand the second term, which is the voltage due to spin-transfer torque rectification, is purely symmetric as expected, since the spin current, $j_s$, generated through the spin Hall effect will be in phase with the rf current with which it couples to produce the dc signal.  This is the same situation as that observed for spin pumping.

%%%%%%%%%%%%%%%%%%%%%%%%%%%%%%%%%%%%%%%%%%%%%%%%%%%%%%%%%%%%%%%%%%%%%%%%%%%%%%%%%%%%%%%%%%%%%%%%%%%%%%%%%%%%%%%%%%%%%%%%%%%%%%%%%%%%%%%%%%%%%%%%%%%%%%%%%%%%%%%%%%%%%%%%%%%%%%%%%%%%%%%%%%%%%%%%%%%%%%%%%%%%%%%%%%%%%%%%%%%

\subsubsection{Spin Hall Magnetoresistance Rectification}\label{smrrect} 

Thus far we have examined in detail the phenomena of spin rectification in FM/NM bilayers, taking care to highlight the distinction between spin pumping and field or spin-transfer torque driven rectification.  In such systems the magnetoresistance required for rectification results from the current flow in the FM layer, however, as discussed in Sec. \ref{sec:shmr}, small magnetoresistance effects can also be observed in FMI/NM bilayers due to spin dependent scattering at the interface, which means that rectification effects can also play a role in FMI/NM bilayers.  Both theoretical \cite{Chiba2014a, Chiba2015} and experimental \cite{Iguchi2014, Rao2015, Schreier2015, Sklenar2015, Jungfleisch2016} investigations of SMR rectification have been performed, revealing new information about the magnetization dynamics of magnetic insulators.  The first studies focussed on the direct rectification of the ac spin Hall effect by means of SMR \cite{Iguchi2014, Rao2015}, with theoretical descriptions including the effect of a relative phase shift between rf microwave field and current as well as multiple driving fields \cite{Iguchi2014}.  In these studies the angular dependence was used to distinguish SMR rectification from spin pumping, however due to extremely small SMR magnetoresistance ratios, the electrical signal due to the ac spin Hall effect was overwhelmed by the dc spin Hall effect signal induced by spin pumping.  

In order to more easily observe SMR rectification, another approach is to use the spin-torque FMR in a FMI/NM bilayer.  The theoretical description of such systems was provided by Chiba et al. using a drift-diffusion model with quantum mechanical boundary conditions at the FMI/NM interface.  They found a rectified voltage consisting of both symmetric and asymmetric contributions,   
\begin{equation}
V_\text{SMR} = - \frac{\Delta R}{2} I_{z^\prime} \left[C \left(H_\text{STT} + \alpha h\right) L + C_+ h D\right] \sin\theta_H\cos^2\theta_H. \label{eq:smr}
\end{equation}
Here
\begin{equation*}
C = \frac{\gamma}{\alpha\sqrt{\left(2 \pi M_0 \gamma\right)^2 + \omega^2}},~C_+ = \frac{\gamma}{\alpha\omega}\left[1 + 2 \pi M_0 \alpha C\right],
\end{equation*}
$H_\text{STT}$ is an effective field accounting for the spin-transfer torque,
\begin{equation*}
H_\text{STT} = \frac{\hbar \theta_\text{SH} J_c}{2eM_0 t_\text{FMI}} \text{Re}\left(\eta\right), ~\eta = \frac{2\lambda_\text{SD} \rho G^{\uparrow\downarrow} \tanh \left(\frac{t_\text{NM}}{2\lambda_\text{SD}}\right)}{1+ 2\lambda_\text{SD} \rho G^{\uparrow\downarrow} \coth \left(\frac{t_\text{NM}}{\lambda_\text{SD}}\right)},
\end{equation*}
and the magnetoresistance ratio, $\Delta R = \Delta \rho l/A = \rho \theta_\text{SH}^2 \left(2\lambda_\text{SD}/t_\text{NM}\right) \tanh\left(t_\text{NM}/2\lambda_\text{SD}\right) \text{Re}\left(\eta/2\right)$, depends on the normal metal thickness, $t_\text{NM}$.  Note that due to the interface nature of SMR, the ratio of the Lorentz and dispersive contributions is strongly dependent on sample thickness, with large dispersive contributions present for small thickness, and also that, unlike other rectification effects we have considered which are proportional to the microwave power $P$, SMR rectification is proportional to $\sqrt{P}$.

\begin{figure}[!ht]
\centering
\includegraphics[width=10cm]{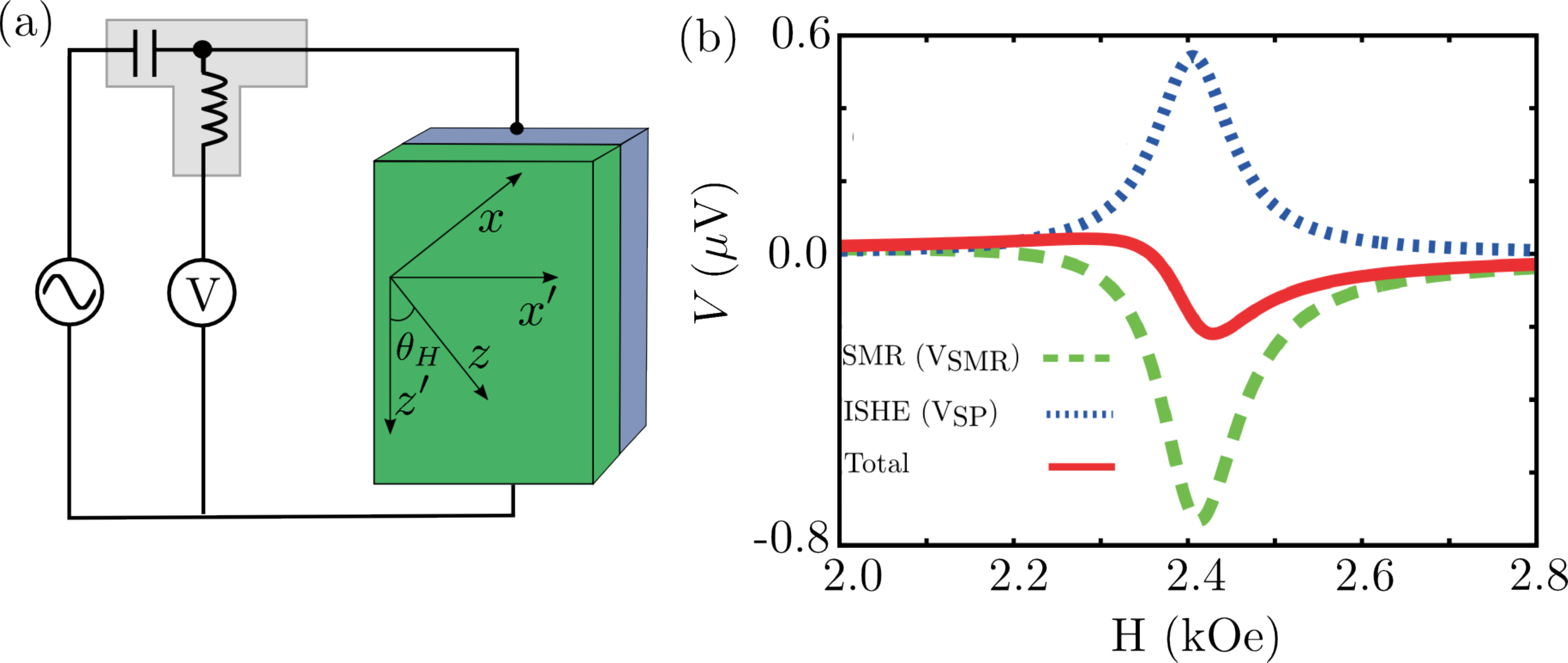}
\caption{\footnotesize{(a) The experimental setup used to observe SMR rectification via spin-transfer torque driven magnetization dynamics. (b) Calculated SMR and SP contributions to the voltage signal.  While the $V_\text{SP}$ is symmetric, $V_\text{SMR}$ has a large dispersive component.  $Source$: Adapted from Ref. \cite{Chiba2014a}.}}
\label{smrrectification}
\end{figure}

The experimental setup for STT SMR experiments is shown in Fig. \ref{smrrectification} (a), with a calculation according to Eq. \ref{eq:smr} including the effect of spin pumping in panel (b) for $t_\text{FMI} = 4$ nm.  For simplicity, the treatment leading to Eq. \ref{eq:smr} includes only an $x^\prime$ component of the rf driving field (we have dropped the subscript so that comparing to our previous descriptions of spin pumping and spin rectification $h = h_{x^\prime}$) and also has assumed that the relative phase $\Phi = 0$.  From Table \ref{spangular} we see that the spin pumping voltage in this configuration will also have a $\sin\theta_H\cos^2\theta_H$ angular dependence, however the contributions can be distinguished based on a line shape analysis, assuming the relative phase shift is 0.  Based on such line shape differences recent experiments have provided evidence of SMR rectification \cite{Schreier2015, Sklenar2015, Jungfleisch2016}, taking into account the effect of non-zero relative phase shifts and finding good agreement with the predicted thickness dependence of the dispersive contribution. 

%%%%%%%%%%%%%%%%%%%%%%%%%%%%%%%%%%%%%%%%%%%%%%%%%%%%%%%%%%%%%%%%%%%%%%%%%%%%%%%%%%%%%%%%%%%%%%%%%%%%%%%%%%%%%%%%%%%%%%%%%%%%%%%%%%%%%%%%%%%%%%%%%%%%%%%%%%%%%%%%%%%%%%%%%%%%%%%%%%%%%%%%%%%%%%%%%%%%%%%%%%%%%%%%%%%%%%%%%%%

\subsection{Magnetic Tunnel Junctions: The Spin Diode} \label{srinmtjs}

\subsubsection{Spin-Transfer Torque Rectification in Magnetic Tunnel Junctions} \label{sec:strmtj}

spin-transfer torque is not only relevant in bilayer devices but also plays an important role in the voltage produced in spin valves and magnetic tunnel junctions \cite{Tulapurkar2005, Sankey2006, Sankey2007, Ralph2007, Kubota2008, Moriyama2008}.  When a current flows through the pinned layer of an MTJ it develops a spin polarization due to the exchange interaction.  The transfer of spin angular momentum carried by this spin polarized current will then exert a torque on the free layer magnetization, as we have just discussed for bilayers, leading to magnetization dynamics and rectification.  This rectification effect can be interpreted as a form of spin diode, where high and low resistance states are controlled by the direction of current flow (which changes the spin polarization of the spin current and therefore modifies the spin-transfer torque) and high switching rates are standard \cite{Tulapurkar2005}.  Although the initial power sensitivity of such diodes was only 1.4 mV/mW$^{-1}$ (still well below the 3 800 mV/mW$^{-1}$ of semiconductor diodes) recent breakthroughs in MTJ design have enabled power sensitivities
\begin{figure}[!ht]
\centering
\includegraphics[width=13cm]{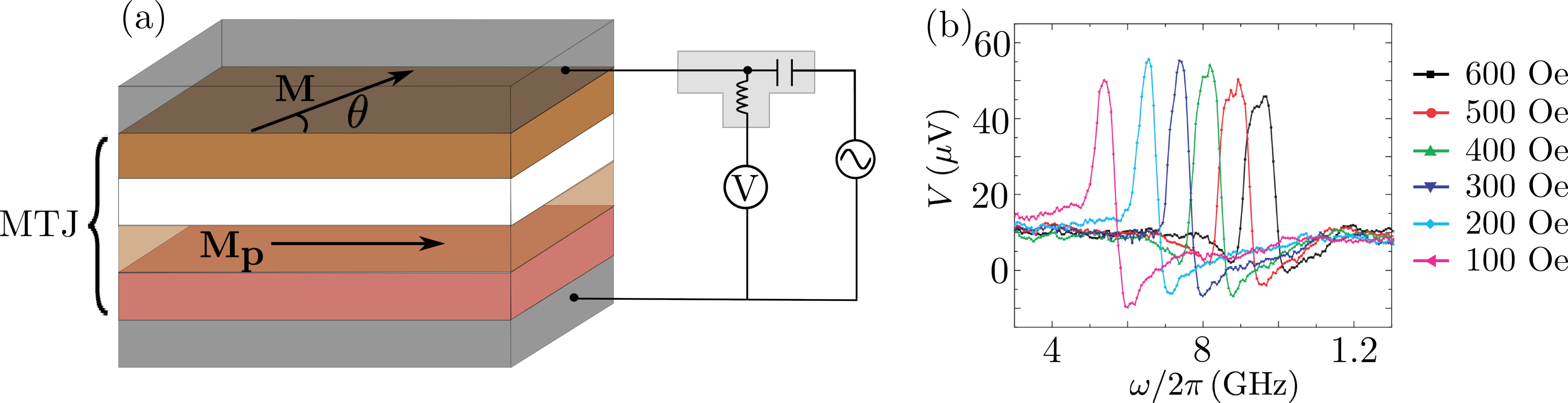}
\caption{\footnotesize{(a) A typical MTJ spin-transfer torque measurement configuration.  A bias tee allows the application of both an rf and dc current as well as the measurement of the voltage across the MTJ.  The applied current becomes spin polarized by the pinned layer and the spin polarized current can then apply a torque to the free layer magnetization.  (b) Typical experimental voltage signal as a function of frequency for several magnetic field strengths \cite{Tulapurkar2005}.  The line shape has both dispersive and symmetric contributions (from the $\beta_{\text{FT}}$ and $\beta_{\text{STT}}$ terms in Eq. \ref{mtjlineshape} respectively).}}
\label{mtjrectification}
\end{figure}
\begin{table}[h!]
\def\arraystretch{1}
\caption{\footnotesize{A summary of key experimental results using magnetic tunnel junctions (spin diodes).}}
\centering
\begin{tabular}{>{\centering\arraybackslash}m{4cm}>{\centering\arraybackslash}m{4cm}>{\centering\arraybackslash}m{6cm}}
\toprule               
Reference & Device Structure & Result \\ \midrule \\
Tulapurkar et al. \cite{Tulapurkar2005} & \includegraphics[width=2.3cm]{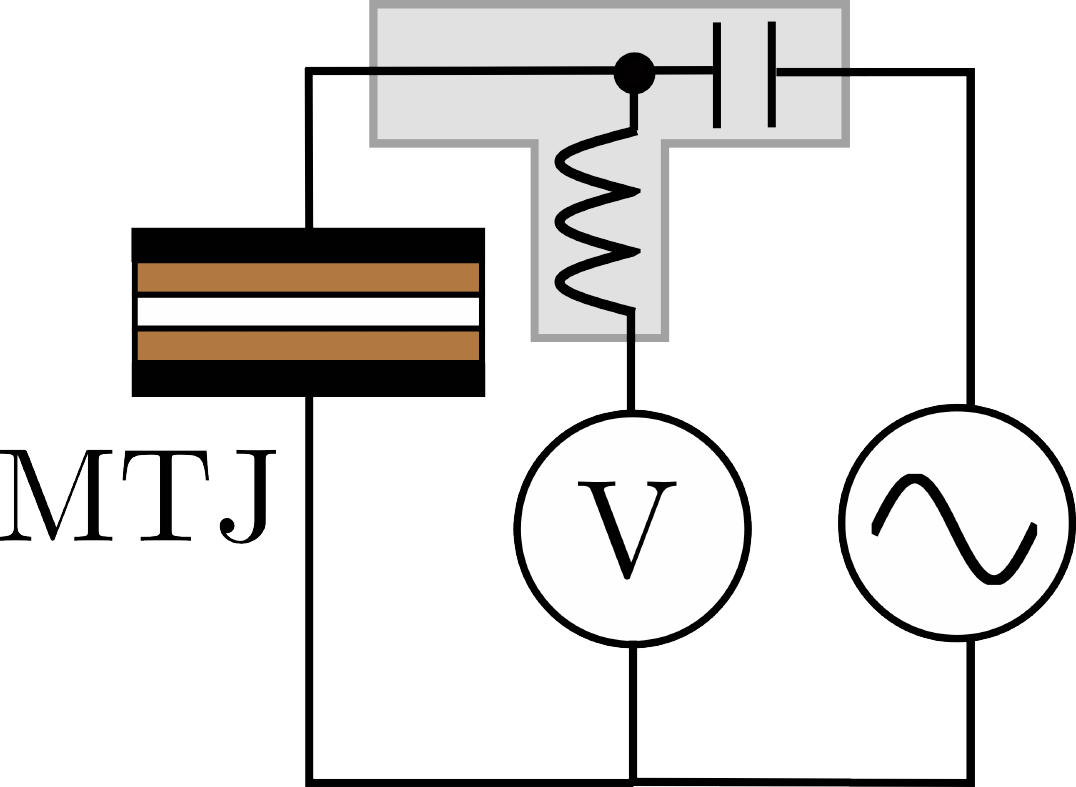} & Electrical detection of SR due to STT in a CoFeB/MgO/CoFeB MTJ \\ \\
Sankey et al. \cite{Sankey2007} & \includegraphics[width=2.3cm]{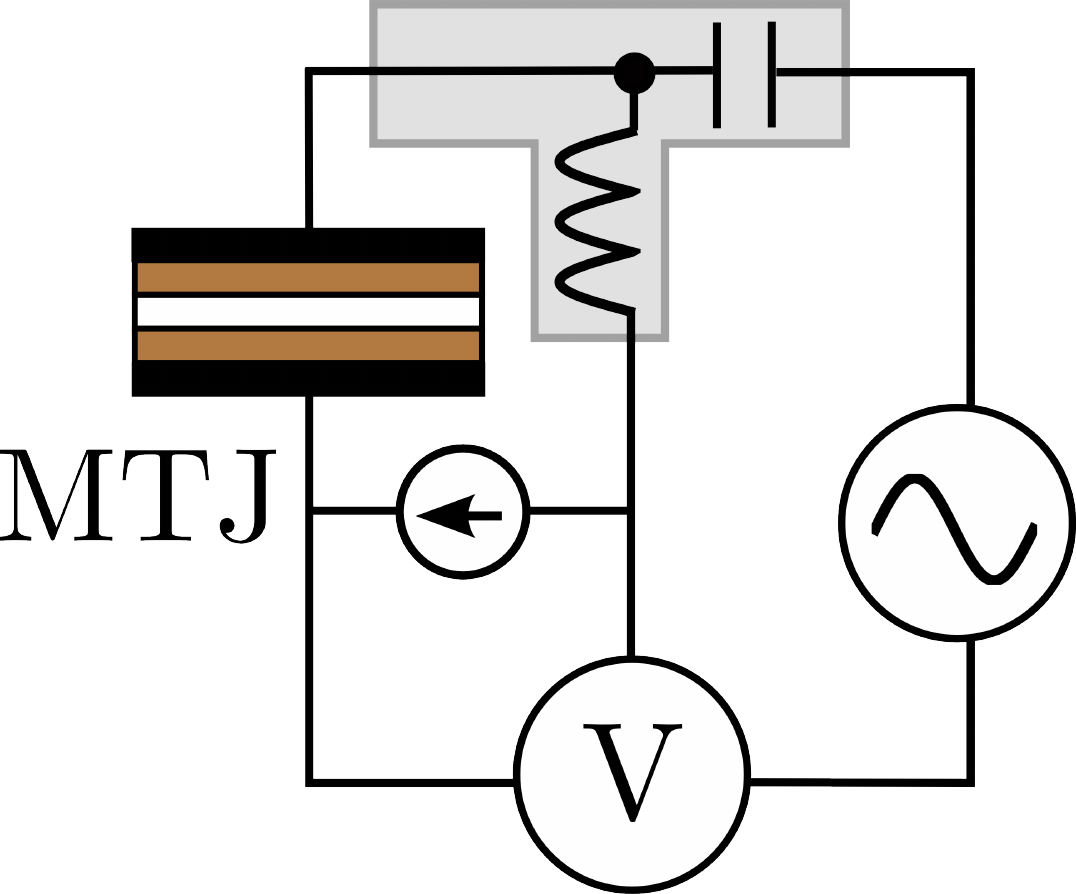} & Use of SR to measure bias and angular dependence of STT vector in CoFeB/MgO/CoFeB MTJ \\ \\
Miwa et al. \cite{Miwa2014}\linebreak Fang et al. \cite{Fang2016}& \includegraphics[width=2.3cm]{Table7Fig3.pdf} & Room temperature spin diode with large power sensitivity due to optimization of MTJ structure (CoFeB/MgO/FeB and CoFe/Ru/CoFeB MTJs). 
\\ \bottomrule
\end{tabular}
\label{mtjsummary}
\end{table}
of 25 000 mV/mW$^{-1}$ at low temperatures by Cheng et al. \cite{Cheng2013}, and at room temperature 12 000 mV/mW$^{-1}$ by Miwa et al. \cite{Miwa2014}, and 75 400mV/mW$^{-1}$ by Fang et al. \cite{Fang2016}.  While Fang's method does not require a bias magnetic field,  Miwa's method has a better signal-to noise ratio.  Such high power sensitivities opens the door to enhanced microwave imaging and sensing techniques which will be discussed in Sec. \ref{imagingsec}.  It should be noted that an analogous diode effect due to GMR has also been observed \cite{Kleinlein2014, Zietek2015, Zietek2015b}.

Table \ref{mtjsummary} summarizes some of the key studies in the development of the spin transfer based spin diode and MTJ rectification and Fig. \ref{mtjrectification} shows a typical spin diode experiment in detail.  Similar to the spin-transfer torque measurement in bilayer systems, a bias tee is used to allow the application of both an rf and dc current/voltage as well as to measure the voltage across the device.  

The current becomes spin polarized after passing through the fixed layer and can then act as a torque on the free layer.  Typical experimental voltage curves as a function of the current frequency are shown in Fig. \ref{mtjrectification} (b) for several magnetic field strengths.  This data was collected from an MTJ with a pinned layer made of CoFeB and CoFe layers antiferromagnetically coupled across a Ru spacing layer.  The free CoFeB layer was separated from the pinned layer by an MgO tunnel barrier and a 0.55 mA rf current was applied with no bias current/voltage.  The line shape is seen to have both Lorentz and dispersive components as would be expected from Eq. \ref{mtjlineshape}.

Since the spin current polarization will be along the direction of the pinned layer, it is convenient to describe the magnetization dynamics in an MTJ by writing the STT modified LLG equation as
\begin{equation}
\frac{d \textbf{M}}{dt} = - \gamma \textbf{M} \times \textbf{H}_i + \frac{\alpha}{M_0} \left(\textbf{M} \times \frac{d \textbf{M}}{dt}\right) + \frac{\gamma \beta_{\text{STT}} I}{M_0} \textbf{M} \times \left(\textbf{M} \times \textbf{M}_p\right) + \gamma \beta_\text{FT} I \textbf{M} \times \textbf{M}_p \label{llgstmtj}.
\end{equation}
Here $\textbf{M}$ is the magnetization of the free layer, $\textbf{M}_p$ is the magnetization direction of the pinned layer, which polarizes the current and $I$ is the amplitude of the rf current applied to the MTJ.  Finally $\beta_{\text{STT}}$ and $\beta_{\text{FT}}$ are the spin-transfer torque and field-like torque amplitudes per unit current respectively.  The $\beta_{\text{STT}}$ term is the same as the in-plane modification made to the LLG equation in Eq. \ref{llgstbilayer} while the $\beta_{\text{FT}}$ term is an explicit out-of-plane torque which is necessary for MTJs \cite{Diao2007, Suzuki2008}.  

In the bilayer system, where the rectification was due to AMR, the voltage could be determined using the generalized Ohm's law.  In an MTJ where the magnetoresistance is due to TMR, the voltage must be found by solving the modified LLG equation to determine the angle $\theta$ between the pinned and free layer magnetizations and then expanding the magnetoresistance as a function of this angle.  Following the initial study by Tulapurkar et al. \cite{Tulapurkar2005}, here we will analyze in detail the case of zero bias voltage/current.  We take the $y$ axis perpendicular to the easy axis plane and the magnetizations in the $x-z$ plane with the $z$ axis along the easy axis of the MTJ.  The demagnetization fields will be along the $y$ axis since the thickness of the MTJ is small compared to its lateral dimensions.  We take $\textbf{M}$ along $\widehat{z}$ which is at an angle $\theta$ to the pinned layer magnetization along $\widehat{z}^\prime$.  If we assume the magnetization of $\textbf{M}$ is constant, then when we expand $\textbf{M}$ around its equilibrium value $\textbf{M}_0$ it can have two components perpendicular to $\textbf{M}_0$.  It is convenient to write these two directions as $\textbf{M} \times \textbf{M}_p$ and $\textbf{M} \times \left(\textbf{M} \times \textbf{M}_p\right)$ and expand $\textbf{M}$ as
\begin{equation*}
\textbf{M} = \textbf{M}_0 + a e^{-i\omega t} \frac{\textbf{M}_0 \times \textbf{M}_p}{|\textbf{M}_0 \times \textbf{M}_p|} + b e^{-i\omega t} \frac{ \textbf{M}_0 \times \left(\textbf{M}_0 \times \textbf{M}_p\right)}{|\textbf{M}_0 \times \left(\textbf{M}_0 \times \textbf{M}_p\right)|}.
\end{equation*}
Taking $I = I_0 e^{-i\omega t}$ the LLG equation can be solved for the coefficients $a$ and $b$ in analogy with how $m_x$ and $m_y$ were found for the field torque rectification.  In the case of the spin dynamo, since we were measuring a voltage along the current direction, only $m_x$ contributed to the rectified voltage, but $m_x$ had a contribution from both $h_x$ and $h_y$ due to the matrix form of the susceptibility.  By analogy, since the resistance depends on $\widehat{\textbf{M}} \cdot \textbf{M}_p$, only $b$ will contribute to the spin diode voltage, but will depend on both the amplitudes $\beta_{\text{STT}}$ and $\beta_{\text{FT}}$ by the solution to Eq. \ref{llgstmtj}.  We can write the resistance given by Eq. \ref{gmrtheta} as
\begin{equation*}
R = R_P + \frac{\Delta R}{2} \left(1-\widehat{\textbf{M}} \cdot \textbf{M}_P\right) = R_P +\frac{\Delta R}{2} \left[1-\cos\theta +|\sin\theta| \left(\text{Re}\left(b\right) \cos \left(\omega t\right) + \text{Im}\left(b\right) \sin \left(\omega t\right) \right)\right].
\end{equation*}
Therefore the voltage is
\begin{equation}
V = \frac{\Delta R}{2} I_0 |\sin\theta| \text{Re}\left(b\right) \langle \sin^2 \omega t\rangle = \frac{\Delta R}{4} \gamma I_0^2 \sin^2 \theta \text{Re}\left[\frac{\tilde{\omega} \beta_{\text{FT}} + i \omega \beta_{\text{STT}}/M_0}{\omega^2 - \omega_r^2 - i \omega \Delta}\right] \label{mtjlineshape}
\end{equation}
where $\omega_r = \gamma \sqrt{H\left(H+M_0\right)}$ is the resonance frequency which follows the Kittel formula, $\Delta = \alpha \gamma \left(2H + M_0\right)$ and $\tilde{\omega} = \gamma \left(H+ M_0\right)$.  For more discussion of the zero bias line shape see \cite{Kupferschmidt2006, Kovalev2007, Sankey2006}.

Here we have assumed that the spin-transfer torque coefficients $\beta$ are independent of the angle between the magnetization directions of the pinned and free layers.  In general this may not be true \cite{Boulle2007} with angular dependence appearing explicitly in the conductance and also in non-collinear systems which require a rigorous generalization of the two-current model \cite{Kovalev2007} (which is consistent with a scattering approach \cite{Kovalev2002}).  However it has been suggested \cite{Slonczewski2005} that the angular dependence of the spin transfer efficiency and the electric conductance cancel, leaving the coefficients independent of the angle $\theta$.  However the same cancellation does not occur for the bias voltage dependence since the conductivities of the spin channels have different bias voltage dependencies \cite{Theodonis2006, Suzuki2008} and therefore the $\beta$ coefficients will generally have a $V_\text{bias}$ dependence.
Kubota et al. \cite{Kubota2008} have performed a systematic study of the bias voltage dependence of the spin-transfer torque rectification in MTJs which have also been studied in Ref. \cite{Sankey2007}.  In the analysis of the biased samples, the torque dependence on the bias current dominates and the angular dependence is ignored in comparison.  Spin-transfer torque in spin valves is analogous \cite{Manchon2012}.  We should note that most STT experiments, including those discussed here have employed amplitude modulation.  However in 2013 Gon\c{c}alves et al. \cite{Goncalves2013} demonstrated that magnetic field modulation, often used in conventional FMR, could improve the sensitivity of STT-FMR.

In addition to the STT signal in MTJs in principle a field torque may also exist if an rf magnetic field is applied.  The voltage produced by field torque rectification in the free layer would then be the same as we have already discussed in Sec. \ref{spinrectification}.  Finally, apart from the spin-torque and field-like torque, a voltage-induced torque can also play an important role in driving FMR in the ultrathin ferromagnetic structures which may be present in MTJs.
%%%%%%%%%%%%%%%%%%%%%%%%%%%%%%%%%%%%%%%%%%%%%%%%%%%%%%%%%%%%%%%%%%%%%%%%%%%%%%%%%%%%%%%%%%%%%%%%%%%%%%%%%%%%%%%%%%%%%%%%%%%%%%%%%%%%%%%%%%%%%%%%%%%%%%%%%%%%%%%%%%%%%%%%%%%%%%%%%%%%%%%%%%%%%%%%%%%%%%%%%%%%%%%%%%%%%%%%%%%

\subsubsection{Voltage Torque Rectification} \label{sec:vtrec}

The origin of voltage torque rectification in MTJs is voltage-controlled magnetic anisotropy, and as this effect does not rely on the microwave current, it can dominant in MTJs with thick tunnel barriers.   In 2008 and 2009 theoretical work indicated that an electric field can substantially alter the interfacial magnetic anisotropy energy and even induce magnetization reversal in 3d transition ferromagnets \cite{Duan2008, Nakamura2009, Tsujikawa2009}. Later, in 2011, both Wang et. al \cite{Wang2011} and Shiota et. al \cite{Shiota2011} experimentally realized the direct resistance switching induced by a dc voltage in MgO-based MTJs. Since the estimated power consumption for single switching is well below that required for the spin-current-injection switching process, these results open a new avenue for exploring voltage-controlled spintronic devices analogous to those which are already ubiquitous in semiconductor technologies.

\begin{figure}[!ht]
\centering
\includegraphics[width=8cm]{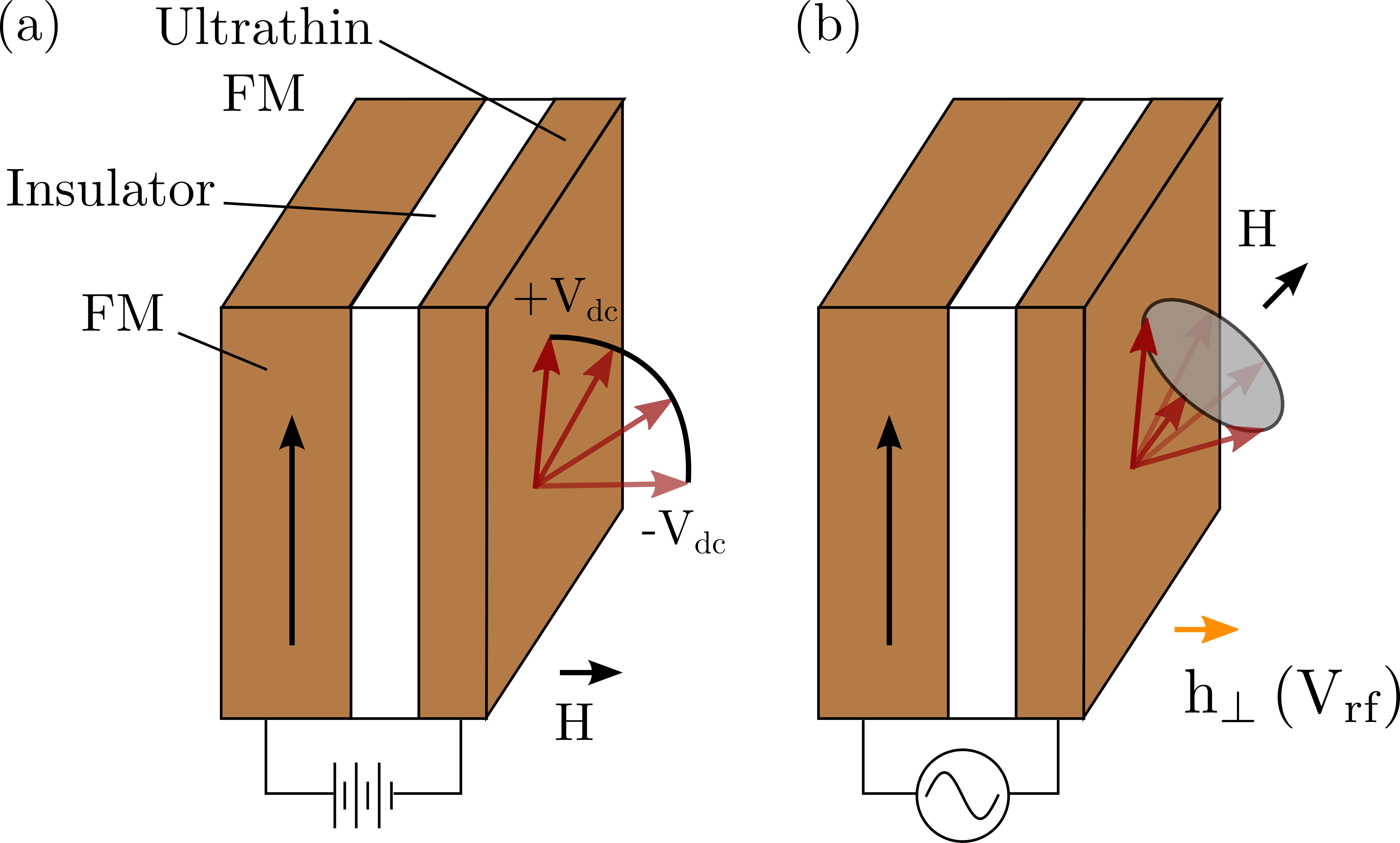}
\caption{\footnotesize{Magnetization dynamics in MTJs can also be driven by a voltage-torque due to electric field induced interfacial magnetic anisotropy changes.  (a) Application of a dc voltage can switch the easy axis from the in-plane to out-of-plane directions. (b) Application of an rf voltage can excite magnetization precession about an external magnetic field direction.  The yellow arrow indicates the effective rf magnetic field induced by the anisotropy control.  $Source$: Adapted from Ref. \citenum{Nozaki2012}.}}
\label{voltageTorqueSchem}
\end{figure}

\begin{figure}[!ht]
\centering
\includegraphics[width=12cm]{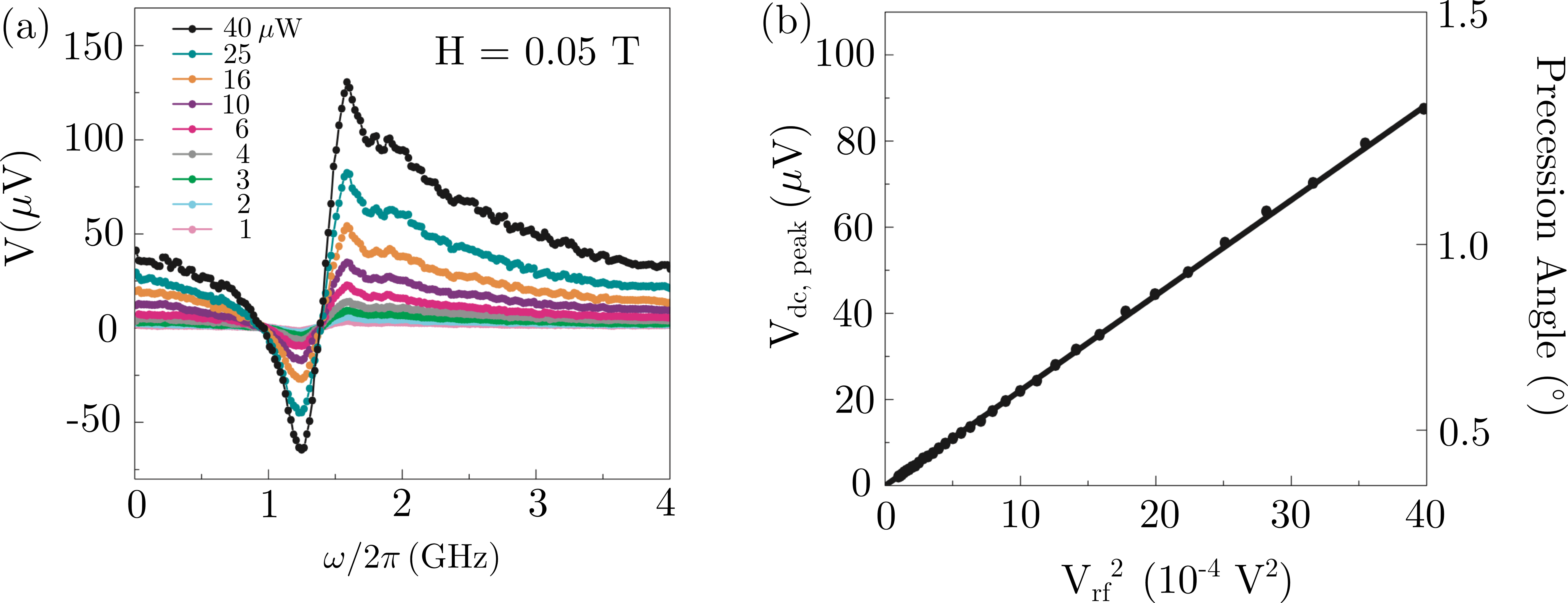}
\caption{\footnotesize{(a) Radiofrequency power dependence of the voltage spectra measured under a constant external magnetic field of 0.05 T at $\phi_H = 45^\circ$. (b) Intensity of the peak voltage signal, which is proportional to the square of the input rf voltage $V_\text{rf}$, as expected.  The FMR precession angle is shown on the right hand axis.  $Source$: Adapted from Ref. \citenum{Nozaki2012}.}}
\label{voltageTorqueData}
\end{figure}

A consequence of the voltage-controlled magnetic anisotropy is that magnetization oscillations in an MTJ device can also be excited by an applied microwave voltage, resulting in resistance oscillations. Fig. \ref{voltageTorqueSchem} shows the basic concept of voltage-induced FMR in an MTJ with an ultrathin ferromagnetic layer, where switching of the magnetic easy axis between the in-plane and the out-of-plane directions is demonstrated by controlling the sign of the dc voltage bias.  As the tunnelling resistance depends on the relative magnetization configuration, the excited FMR dynamics (Fig. \ref{voltageTorqueSchem} (b)) alter the configuration and generate an oscillating resistance. Mixing with the microwave current, a dc voltage is produced \cite{Nozaki2012, Zhu2012}.  As expected, the observed rectification voltage is linearly proportional to the square of the input microwave voltage and shown in Fig. \ref{voltageTorqueData} (b).  In MgO-based MTJs the magnitude of high-frequency spin-torque and voltage-torque can be similar, and thus quantitative descriptions of voltage-driven magnetization dynamics in MTJs should generally include both torque terms \cite{Zhu2012}.  It has also been shown that voltage-controlled magnetic anisotropy can be used in spin diode rf detectors, such as to improve sensitivity \cite{Zhu2012, Shiota2014, Skowronski2015}.

In this section we have discussed the dc voltage produced in monolayer devices with integrated CPWs, known as spin dynamos, in FM/NM bilayers, where field torque, spin pumping and spin-transfer torque all contribute, and in MTJs where the voltage signal is due to spin-transfer torque caused by the pinned magnetization.  These effects are summarized in Table \ref{voltageeffectssummary}.  

\begin{table}[h]
\def\arraystretch{1.2}
\caption{\footnotesize{Summary of the dc voltage observed in ferromagnetic thin films, FM/NM bilayers and MTJs.}}
\centering
\begin{tabular}{>{\centering\arraybackslash}m{2cm}>{\centering\arraybackslash}m{2.5cm}>{\centering\arraybackslash}m{4cm}>{\arraybackslash}m{5cm}}
\toprule                       
Device & Effect & Voltage & \multicolumn{1}{c}{Lineshape} \\ \midrule
Thin Film & Field Torque & $V= V_{\text{SR}}^h \left(\langle j_\text{rf} h_{\text{rf}}\rangle\right)$ & Lorentz and dispersive, controlled by relative phase and driving field component \\ \midrule
Bilayer & Field Torque, Spin-transfer torque, Spin Pumping & $\begin{aligned} V = ~&V_{\text{SR}}^h \left(\langle j_\text{rf} h_{\text{rf}}\rangle\right) \\ &+ V_{\text{SR}}^{\text{STT}} \left(\langle j_\text{rf} j_{\text{rf}}^s\rangle\right)\\ &+ V_{\text{SP}}\left(\langle h_{\text{rf}}^2\rangle\right)\end{aligned}$& Lorentz and dispersive from field torque, Lorentz from spin pumping, Lorentz from spin-transfer torque (if field-like spin-transfer torque is negligible)\\ \midrule
MTJ & Spin-Transfer Torque, Field Torque, Voltage Torque &$\begin{aligned} V = ~&V_{\text{SR}}^h \left(\langle j_\text{rf} h_{\text{rf}}\rangle\right)\\ & + V_{\text{SR}}^{\text{STT}} \left(\langle j_\text{rf} j_{\text{rf}}^s\rangle\right)\end{aligned}$ & Lorentz and dispersive from field torque, Lorentz from spin-transfer torque and dispersive from field-like spin torque (Voltage torque behaviour is analogous to spin-transfer torque)
\\ \bottomrule
\end{tabular}
\label{voltageeffectssummary}
\end{table}

%%%%%%%%%%%%%%%%%%%%%%%%%%%%%%%%%%%%%%%%%%%%%%%%%%%%%%%%%%%%%%%%%%%%%%%%%%%%%%%%%%%%%%%%%%%%%%%%%%%%%%%%%%%%%%%%%%%%%%%%%%%%%%%%%%%%%%%%%%%%%%%%%%%%%%%%%%%%%%%%%%%%%%%%%%%%%%%%%%%%%%%%%%%%%%%%%%%%%%%%%%%%%%%%%%%%%%%%%%%

\newpage
\section{Applications of Spin Rectification} \label{sec:applications}

Having described the physical mechanisms of spin rectification and its characteristics in various device structures, in this section we turn to the applications of spin rectification, both in the study and understanding of basic physics and in the development of imaging and material characterization techniques.  As we will see, spin rectification has proven to be a versatile technique which is used in a wide variety of studies due to its high sensitivity, phase sensitivity and ease of use.

%%%%%%%%%%%%%%%%%%%%%%%%%%%%%%%%%%%%%%%%%%%%%%%%%%%%%%%%%%%%%%%%%%%%%%%%%%%%%%%%%%%%%%%%%%%%%%%%%%%%%%%%%%%%%%%%%%%%%%%%%%%%%%%%%%%%%%%%%%%%%%%%%%%%%%%%%%%%%%%%%%%%%%%%%%%%%%%%%%%%%%%%%%%%%%%%%%%%%%%%%%%%%%%%%%%%%%%%%%%

\subsection{Electrical Detection of Spin Waves} \label{spinwaves}

Spin waves are collective excitations which occur in magnetic lattices due to the short-range exchange integration and/or the long-range dipole-dipole interaction.  Unlike the $\textit{uniform}$ precession of FMR, spin wave excitations are characterized by a \textit{spatially varying} spin precession phase and therefore have a propagation direction and a finite wavelength. From a fundamental physics perspective, since a set of discrete spin waves can easily be excited under microwave radiation (with FMR being the lowest order mode), spin wave spectroscopy provides a powerful tool for understanding both the exchange and dipole-diple interactions.  On the application side, the generation of spin waves is an important energy loss mechanism in modern spintronic devices operated at high frequencies and the inverse of the lowest spin wave frequency determines the switching timescale of devices.   

\subsubsection{Review of Spin Waves}

\begin{figure}[!ht]
\centering
\includegraphics[width=12cm]{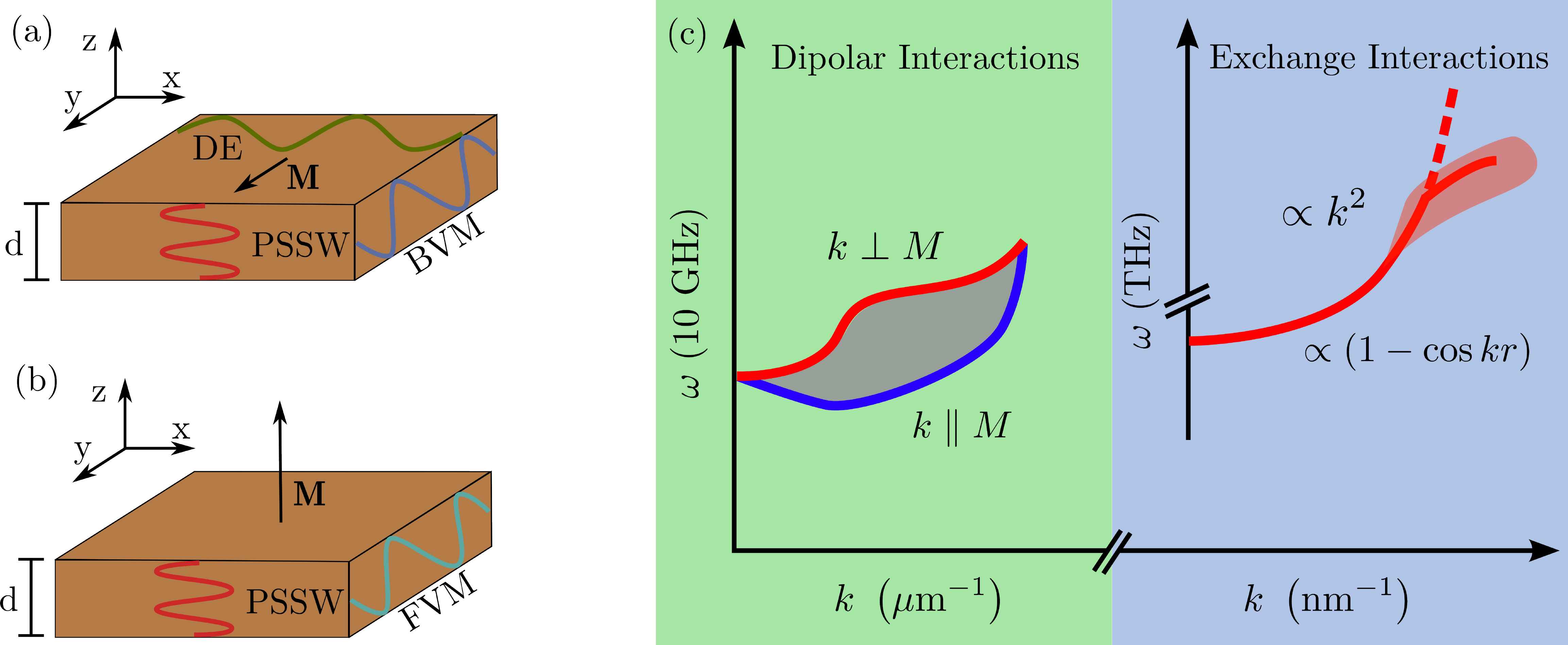}
\caption{\footnotesize{Schematic illustration of spin waves in a thin film in the case that exchange and dipole-dipole interactions can be treated separately.  For both in-plane and out-of-plane magnetizations the short-range exchange interaction will produce perpendicular standing spin waves normal to the thin film.  On the other hand the short-range dipole-dipole interaction is responsible for spin waves with an in-plane wave vector.  (a) For the case of in-plane magnetization the spin waves due to the dipole-dipole interaction will either be Damon-Eshbach modes, with in-plane wave vector perpendicular to {\bf M}, or backward volume modes, with in-plane wave vector parallel to {\bf M}.  (b) For the perpendicularly magnetized thin film, due to the in-plane symmetry, there are only one type of in-plane modes, the forward volume spin waves.  (c) Schematic spin-wave dispersion.  In the long wavelength regime ($k \sim \mu m^{-1}$) the Damon-Eshbach (red) and backward volume modes (blue) are shown, with the grey shaded area indicating arbitrary magnetization orientation.  In the nanometer wavelength regime the dispersion is cosine-like.  The red shaded area indicates a region of heavy spin wave damping.  $Source$: Panel (c) adapted from Ref. \cite{Lenk2011}.}}
\label{swSchem}
\end{figure}

When determining the nature of spin waves, the boundary conditions at the sample interface are of utmost importance.  In a thin film these pinning conditions will, roughly speaking, cause the spin waves to form standing waves.  Therefore the wavelength of spin waves with wave vector perpendicular to the film will be small and these so called perpendicular standing spin waves (PSSWs) will be dominated by the exchange interaction.  Conversely the comparatively large wavelength in the film plane results in dipole-dipole interaction dominated spin waves, so called magneto static modes.  The in-plane magneto static modes can be conveniently classified based on the orientation of the wave vector with respect to the magnetization.  When the wave vector is perpendicular to an in-plane magnetization, Damon-Eshbach (DE) spin waves (also known as surface spin waves) will form, while if the wave vector is parallel to the in-plane magnetization back ward volume modes (BVM) will be produced.  On the other hand when the magnetization is normal to the film plane, due to the in-plane symmetry there is only one type of in-plane spin wave, the forward volume modes (FVM).  The different types of spin waves in thin magnetic films are sketched in Fig. \ref{swSchem}, with panels (a) and (b) illustrating the various types of spin waves classified by the magnetization and wave vector orientation, and panel (c) schematically showing the spin wave dispersions.
  
Ignoring the effects of damping, the dispersion of the resonance frequency of PSSWs can be determined from the LLG equation by including an additional exchange interaction term \cite{Herring1951}, leading to the PSSW dispersion relations 

\begin{align*}
\text{PSSW} - {\bf M}_0 \text{ in-plane:}~~\omega_r^2 &= \gamma^2 \left(H + M_0 + 2 A k_z^2/\mu_0 M_0 \right) \left(H + 2A k_z^2/\mu_0 M_0\right), \\
\text{PSSW} - {\bf M}_0 \text{ out-of-plane:}~~\omega_r &= \gamma \left(H- M_0 + \frac{2 A k_z^2}{\mu_0} M_0\right).
\end{align*} 

where $A$ is the exchange stiffness constant characterizing the strength of the exchange interaction \cite{Schmool2015}.

The spin wave dispersion relations for the DE, BVM and FVM modes can be determined using a magneto static approximation \cite{Walker1957}, leading to
\begin{align*}
\text{BVM:}~~\omega_r^2 &= \gamma^2 H \left[H + M_0 \left(\frac{1-e^{-k_z d}}{k_z d}\right)\right], \\
\text{DE:} ~~ \omega_r^2 &= \gamma^2 \left[H\left(H+M_0\right) + \left(\frac{M_0}{2}\right)^2 \left(1-e^{-2 k_x d}\right)\right], \\
\text{FVM:} ~~ \omega_r^2 &= \gamma^2 \left[ \left(H - M_0\right)^2 + M_0 \left(H-M_0\right)\left(1-\frac{e^{-k_T d}}{k_T d}\right)\right].
\end{align*}
where $k_T = \sqrt{k_x^2 + k_y^2}$ is the in-plane wave vector.  Notice that all of the spin wave modes, except for BVM modes, are shifted to higher frequencies than the FMR and therefore require more energy to excite.  Interestingly, due to the nature of the dipole-dipole interaction, the BVM modes actually require less energy than FMR due to their negative group velocity.  

The spin wave discussed above consider the limits where either exchange or dipolar interactions are dominant.  Of course in general both interactions will contribute to the formation of spin waves, leading to so called dipole exchange spin waves (DESW) \cite{Sparks1970, Sparks1970a, Kalinikos1986} which follow the dispersion \cite{Kalinikos1986},
\begin{subequations}
\begin{align}
\omega_r^2 &= \gamma^2 \left(H_i + \frac{2 A k^2}{\mu_0 M_0}\right) \left(H_i + \frac{2Ak^2}{\mu_0 M_0} + M_0 F_p\right) \label{spinwavedispersion} \\
\intertext{where}
F_p &= P_p  + \sin^2 \phi_H \left(1-P_p + \frac{M_0 P_p \left(1-P_p\right)}{H_i + \frac{2 A k^2}{\mu_0M_0}}\right), \\
\intertext{and}
P_p &= \frac{k_y}{2} \int_0^d \int_0^d \psi_p(z) \psi_p\left(z^\prime\right) \exp \left(-k_y |z-z^\prime|\right) dz dz^\prime. \label{spinwavep}
\end{align}
\label{desw}%
\end{subequations}
The geometry used here is that of Fig. \ref{swSchem} (b) with the magnetization tilted an angle $\phi_H$ from the $z$ axis and we denote the total internal field, including demagnetization factors, by $H_i$.  Here $k^2 = k_z^2 + k_y^2$ with $k_y = n\pi/w$ ($w$ being the width of the thin film) and $k_z = \left(p-\Delta p \right) \pi/d$.  The discrete number $p$ is the integer number of half wavelengths along the $z$ direction, normal to the sample with $\Delta p$ a correction factor determined by the pinning condition at the interfaces $z=0$ and $z=d$,
\begin{equation}
\left(2A \frac{\partial \psi_p}{\partial z} - K_s \psi_p\right)\bigg|_{z=0,d} = 0 \label{pincondition}.
\end{equation}
Finally $\psi_p$ is an eigenfunction of the PSSW with the form $\psi_p = C_p \left(\alpha \sin k_z z+\beta \cos k_z z\right)$ where $C_p$ is a normalization constant and $\alpha$ and $\beta$ are constraints determined by the surface anisotropy $K_s$ and the exchange stiffness.  In the limit $\alpha/\beta \to \infty$ the spins at the surface of the material are completely pinned and $\Delta p = 0$, while if $\alpha/\beta \to 0$, the surface spins are completely free and $\Delta p = 1$, therefore $p$ is constrained by $0\le p \le 1$.

Eq. \ref{spinwavedispersion} describes DESW that depend on $p$ ($\Delta p$) and $n$.  However this expression may be used to describe magneto static modes and PSSW as well by taking appropriate limits.  Setting the exchange stiffness, $A = 0$ the perpendicular profile of the spin waves becomes trivial and the DESW reduce to magneto static modes with quantization number $n$ and the exact form (BVM, FVM, DE) depending on the measurement geometry.  Alternatively, setting $n=0$ the dispersion of Eq. \ref{spinwavedispersion} reduces to the case of PSSW characterized by $p-\Delta p$.  Therefore the FMR ($p-\Delta p =0, n=0$), DESW ($p-\Delta p\ne0, n\ne0$), PSSW ($p-\Delta p\ne 0, n=0$) and magneto static modes ($p-\Delta p=0, n\ne 0$) can all be described by Eq. \ref{spinwavedispersion}, which makes Eq. \ref{spinwavedispersion} useful in the experimental identification of the spin wave modes across the whole $H$ range.  For more in depth discussion of spin waves see e.g. Refs. \cite{HillebrandsBookVolI, DemokritovBook, StancilBook}.

%%%%%%%%%%%%%%%%%%%%%%%%%%%%%%%%%%%%%%%%%%%%%%%%%%%%%%%%%%%%%%%%%%%%%%%%%%%%%%%%%%%%%%%%%%%%%%%%%%%%%%%%%%%%%%%%%%%%%%%%%%%%%%%%%%%%%%%%%%%%%%%%%%%%%%%%%%%%%%%%%%%%%%%%%%%%%%%%%%%%%%%%%%%%%%%%%%%%%%%%%%%%%%%%%%%%%%%%%%%

\subsubsection{Detection of Spin Waves}

\begin{table}[!t]
\def\arraystretch{1.2}
\caption{\footnotesize{A summary of experimental techniques used for the investigation of linear spin waves and nonlinear magnetization dynamics.  The schematic setups and non exhaustive list of references given here are meant only as a brief overview and starting point for further reading.}}
\centering
\begin{tabular}{>{\centering\arraybackslash}m{4cm}>{\centering\arraybackslash}m{3cm}>{\centering\arraybackslash}m{3cm}>{\centering\arraybackslash}m{3cm}}
\toprule                       
Technique & Experimental Setup & Linear Spin Wave Studies & Nonlinear Studies\\ \cmidrule[0.08em]{1-4}
Time-Domain Techniques & & \\ \cmidrule[0.08em]{1-4}
Pulse-Inductive Microwave Magnetometer & \includegraphics[width=3cm]{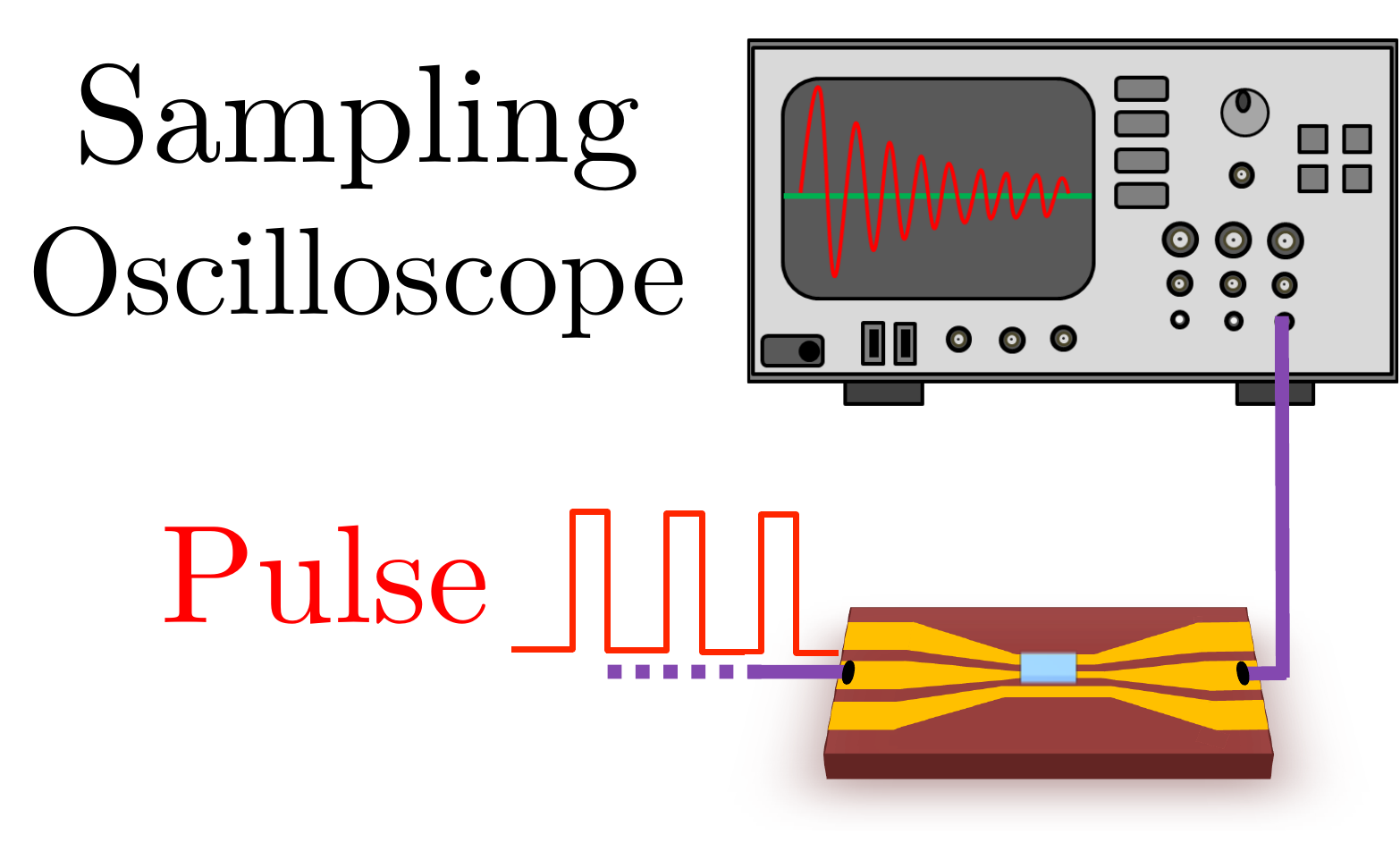} &  \cite{Silva1999, Covington2002, Counil2004, Bonin2005} & \cite{Nibarger2003, Gerrits2006}\\ \midrule
Time Resolved XMCD & \includegraphics[width=3cm]{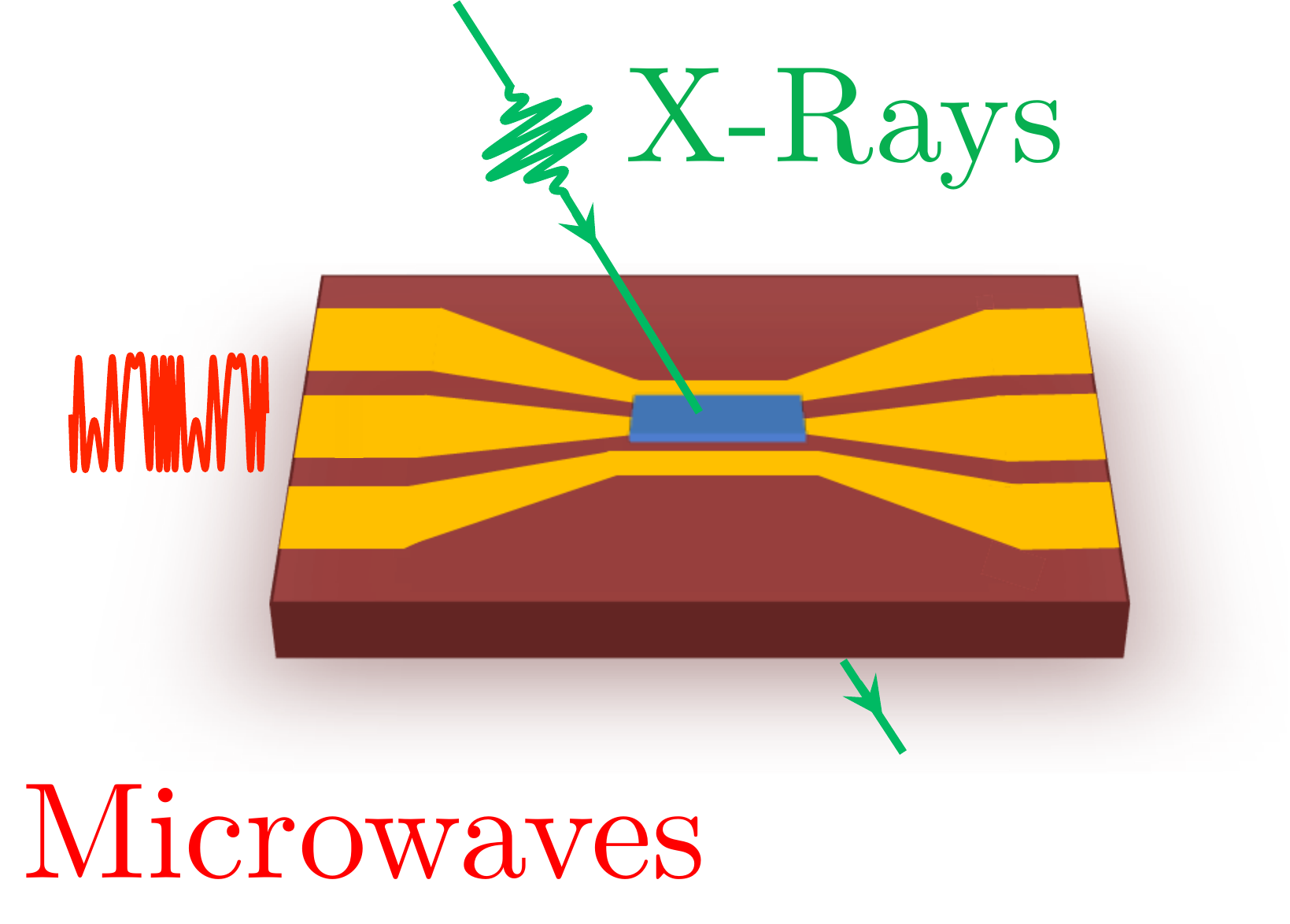} & \cite{Goulon2006, Boero2009, Salikhov2011} & \cite{Bauer2015a} \\ \midrule
Time and Spatially Resolved MOKE Microscopy & \includegraphics[width=3cm]{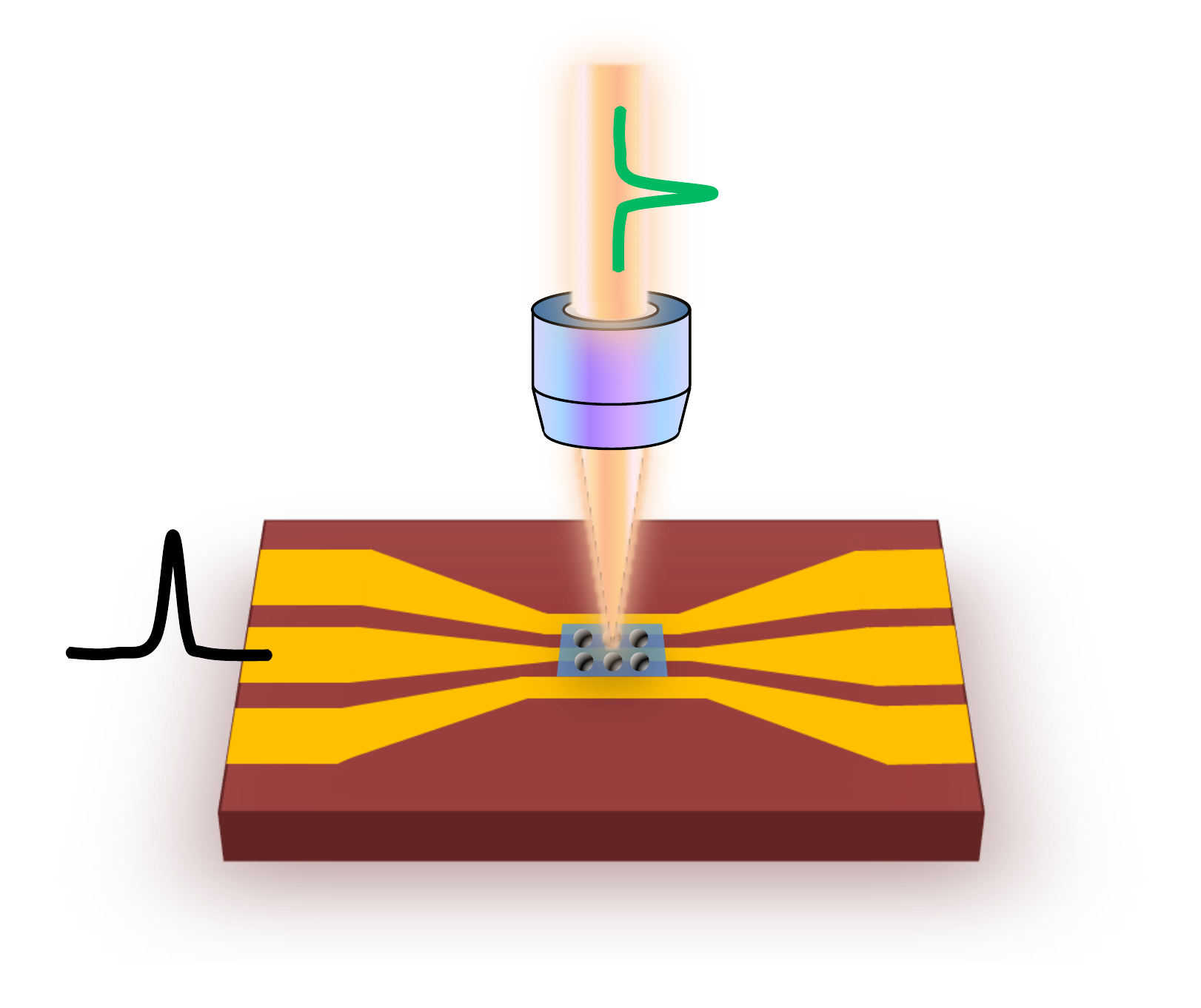}  & \cite{Park2002, Tamaru2002, Pechan2005, Neusser2010} & \cite{Gerrits2007}\\ \cmidrule[0.08em]{1-4}
 \multicolumn{2}{l}{Frequency and Field Domain} \\ \cmidrule[0.08em]{1-4}
Broadband Spin Wave Spectrometer & \includegraphics[width=3cm]{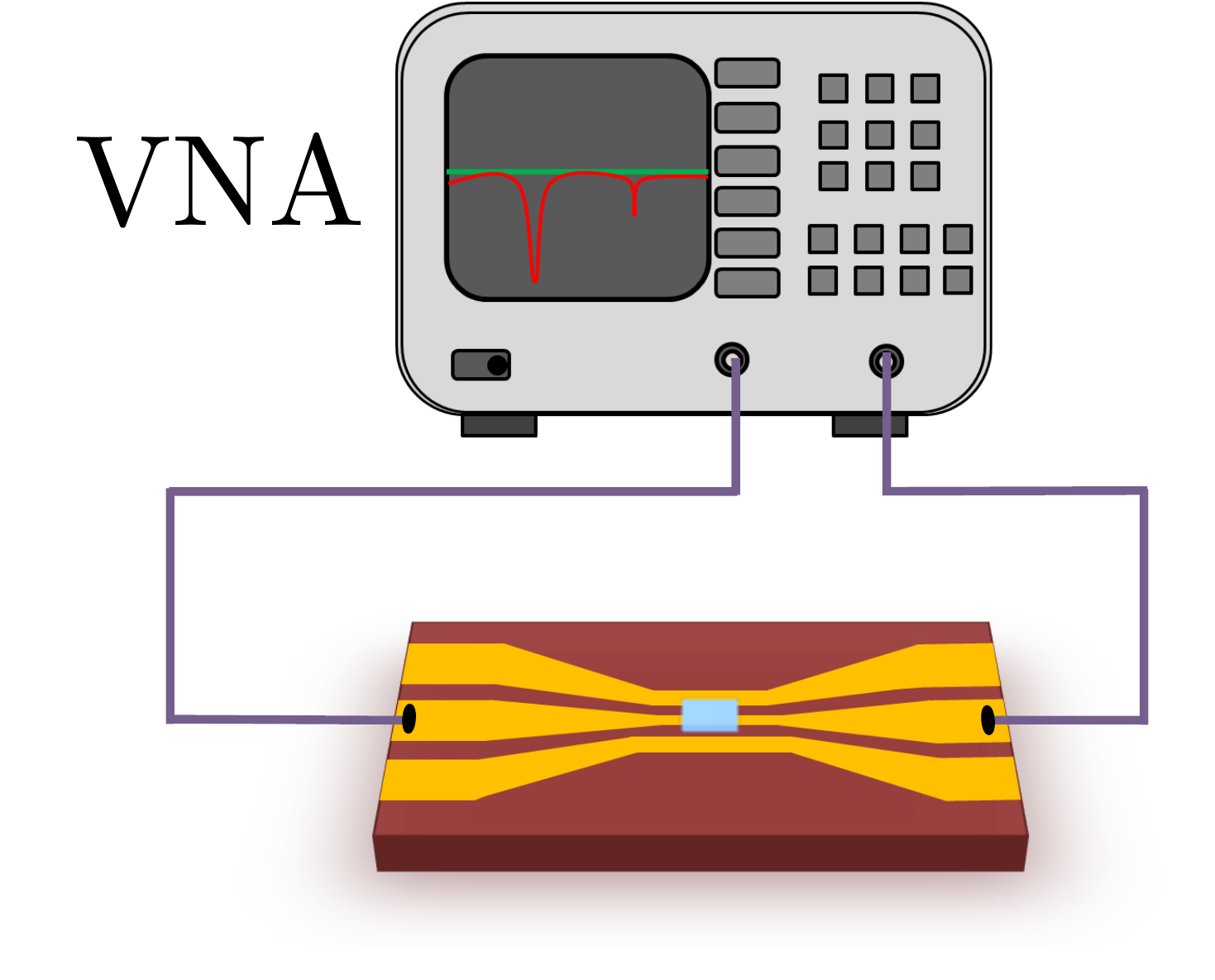}  &  \cite{Bailleul2001, Neudecker2006a, Podbielski2006} & \cite{Zhang1987, Cox2001, An2004}\\ \midrule
Brillouin Light Scattering & \includegraphics[width=3cm]{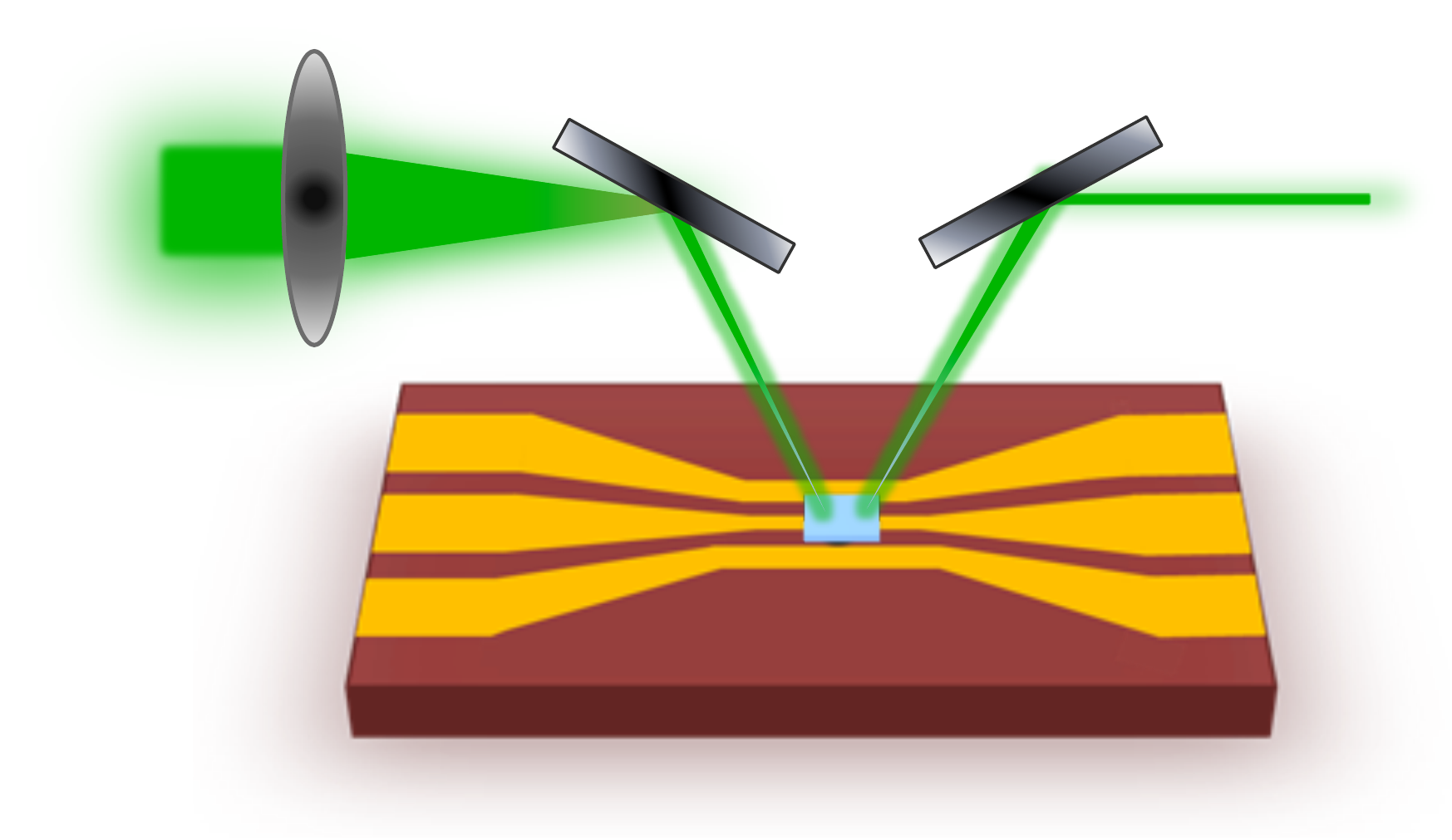} &  \cite{Demokritov2001, Jorzick2002, Sebastian2015} & \cite{Wettling1983, Kabos1996, Edwards2012}\\ \midrule
Ferromagnetic Resonance Force Microscopy & \includegraphics[width=3cm]{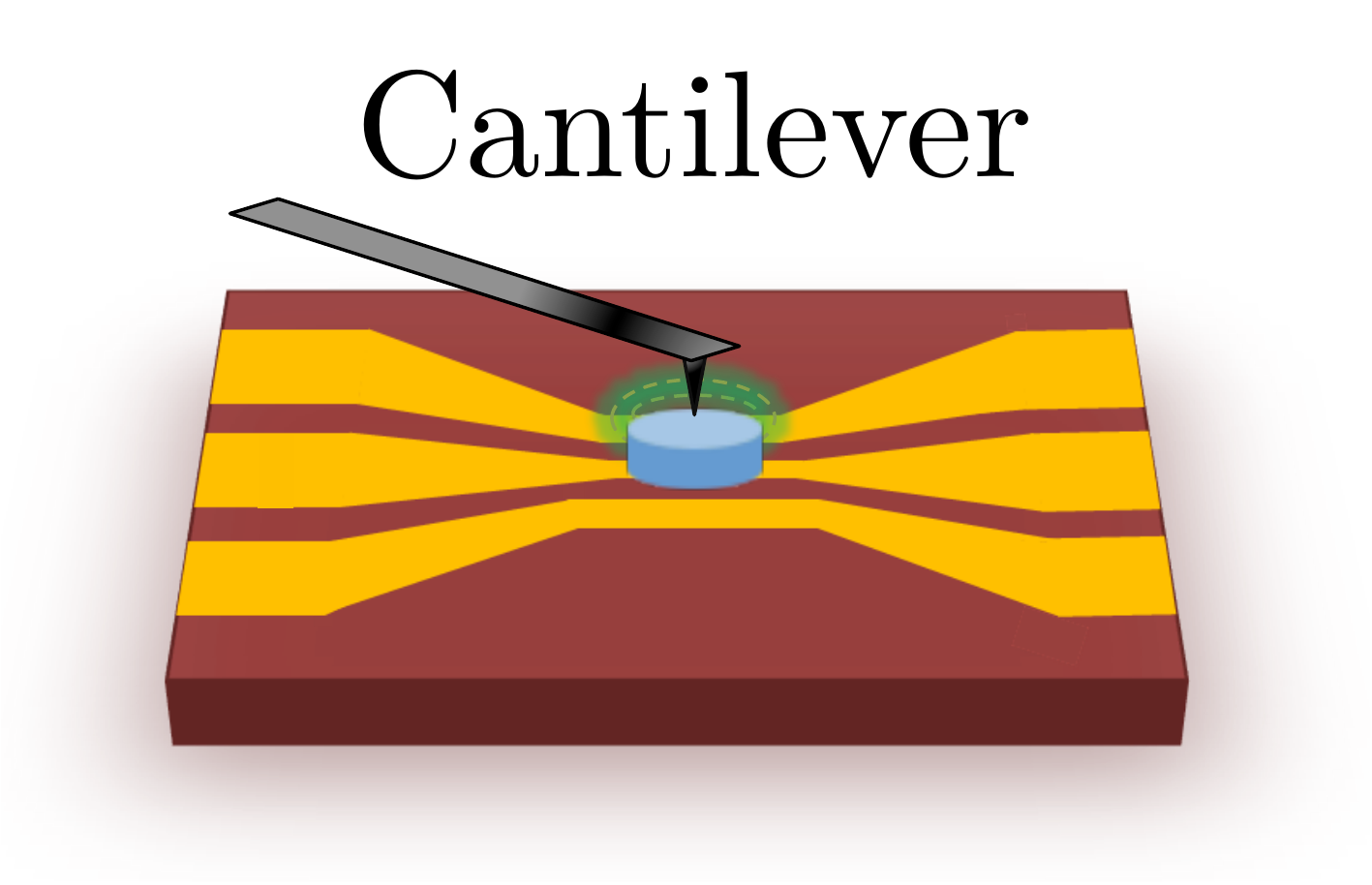} &  \cite{Guo2013} & \cite{Guo2015} \\ \midrule
Electrical Detection via Spin Rectification & \includegraphics[width=3cm]{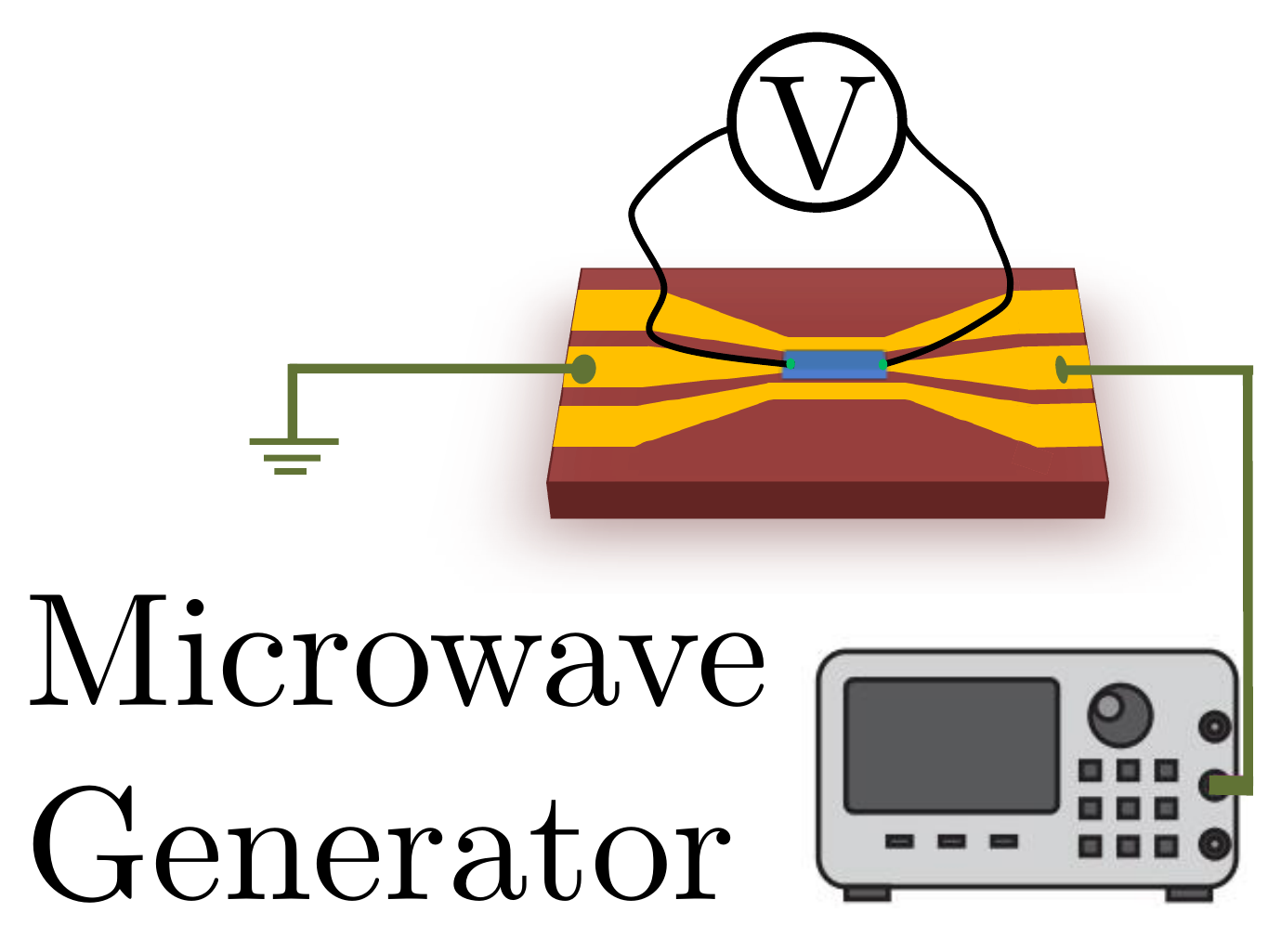} &  \cite{Gui2007a, Bailleul2007, Yamaguchi2009, Yamaguchi2011, Duan2015} & \cite{Gui2009}
\\ \bottomrule
\end{tabular}
\label{swdetectiontable}
\end{table}

Experimentally there are many techniques which have been adapted to study spin waves. Some of these techniques are summarized in Table \ref{swdetectiontable}. Generally these techniques fall into the categories of time-domain measurements, or field and frequency domain measurements.  Time domain techniques measure the temporal evolution of the magnetization typically using pulse-inductive microwave magnetometry (PIMM) or time and spatially resolved magneto-optic Kerr effect (MOKE) microscopy.  PIMM uses a pulsed excitation to generate spin waves, and the inductance induced voltage is measured in a stripline \cite{Silva1999}.  On the other hand MOKE techniques employ a femtosecond laser to carry out pump probe measurements.  As both the polarization and intensity of light reflected from a magnetized surface will change based on its magnetic properties, the location of a spin wave resonance can be directly measured using time-resolved scanning Kerr microscopy as a local spectroscopic probe.  More recently x-ray detected magnetic resonance (XDMR) which uses x-ray magnetic circular dichroism (XMCD) to probe the local magnetization in a microwave pump field has been used to investigate resonance processes with element specificity \cite{Goulon2006, Boero2009}.

\begin{figure}[!b]
\centering
\includegraphics[width=7cm]{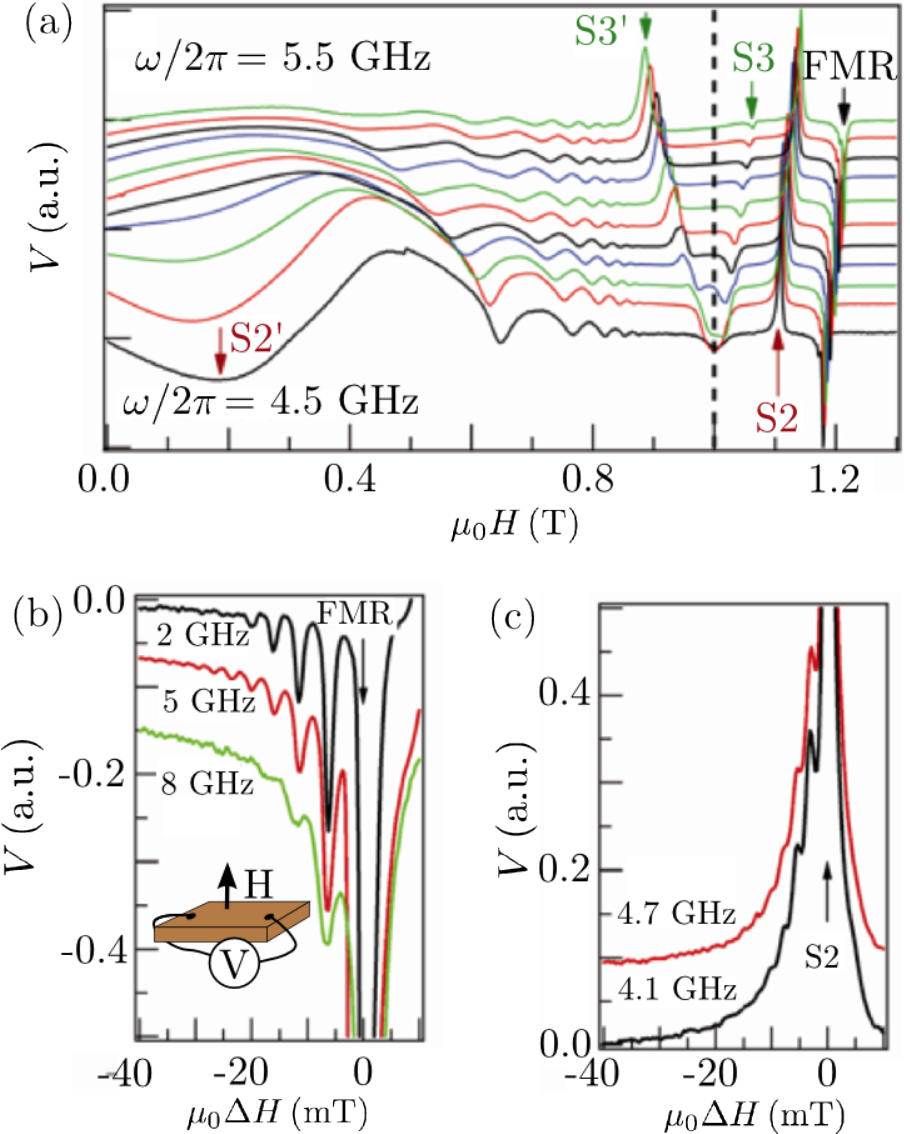}
\caption{\footnotesize{Typical voltage spectra focusing on different $H$ field ranges, measured under a nearly perpendicular applied magnetic field. All voltage spectra are normalized and vertically offset.  (a) Typical spectra over the whole $H$ field range at various frequencies (4.5 to 5.5 GHz in 0.1 GHz steps).  Arrows indicate the positions of FMR and perpendicular standing spin waves for $p=2$ (S2, S2$^\prime$) and $p=3$ (S3, S3$^\prime$).  (b) Spectra of forward volume modes found near FMR. The inset shows the detection and magnetization field static field configuration.  The rf field was provided by a CPW (not shown).  (c) Spectra of the dipole-exchange spin waves found near the $S2$ perpendicular standing spin wave.  Note that in (b) and (c) the 0 of the field axis is centred at the FMR and S2 spin wave position respectively, with $\Delta H$ indicating the deviation from this position.  $Source:$ Adapted from Ref. \cite{Gui2007a}.}}
\label{spinwave}
\end{figure}  

Alternatively, field and frequency domain techniques have also proven successful in the study of spin waves.  Such techniques include both direct measurements of the $\omega\left(k\right)$ spin wave dispersion using Brillouin light scattering (BLS) \cite{Ercole1998, Demokritov2001, Jorzick2002, Bayer2006, Sebastian2015} as well as broadband measurements of the microwave reflection/transmission/absorption as a function of frequency and external magnetic fields \cite{Kittel1958, Zhai2002, Bailleul2001, Neudecker2006a, Podbielski2006}.  In BLS the Stokes and anti-Stokes shifts produced by the creation and absorption of magnons are measured using a sensitive tandem Fabry-P\'erot interferometer, which can access both the spatial and temporal properties of spin wave packets, and simultaneously various frequencies in a wide frequency range \cite{Sebastian2015}.  More recently, ferromagnetic resonance force microscopy (FMRFM) of confined spin-wave modes with improved, 100 nm resolution, has been developed \cite{Guo2013}.  Ferromagnetic resonance spectra in Py disks (diameters ranging from 100 to 750 nm) has shown multiple distinguishable edge modes.  

Also falling into the field and frequency domain category is the use of spin rectification \cite{Gui2007a, Bailleul2007, Yamaguchi2009, Yamaguchi2011, Duan2015}.  A key difference in SR based spin wave detection is that the device itself acts as the detector.  This enables simple implementation, simplified data analysis and, increased sensitivity due to the lock-in techniques used.  One of the key results arising from SR detection of spin waves is the unambiguous observation of DESW \cite{Gui2007a, Duan2015}.  Although spin waves were experimentally studied as early as the 1950's, for many years only magneto static modes or PSSWs could be observed \cite{White1956, Dillon1958, Jorzick1999, Jorzick2002, Wang2002, Park2002} with the detection of DESW and the related question of the boundary condition dependence of spin dynamics \cite{Rado1959, Pincus1960, Soohoo1961, Soohoo1963} remaining elusive.  This initial inability to observe the DESW was mainly due to the weak signals compared to FMR and PSSW and the fact that the spacing between DESW resonances is only a few mT, which means they are easily washed out by Gilbert damping broadening of the resonance peak.  The high sensitivity of electrical detection based on SR allowed the DESW to be observed for the first time \cite{Gui2007a} and subsequent studies have further elucidated the nature of DESW, e.g. the pinning conditions in nanowires \cite{Duan2015}.

Fig. \ref{spinwave} nicely illustrates the versatility of spin rectification in the study of spin waves, showing PSSW in Fig. \ref{spinwave} (a), FVM modes in Fig. \ref{spinwave} (b) and DESW in Fig. \ref{spinwave} (c).  This sensitive data can be ready used to assign the order of each spin wave mode \cite{Gui2007a}.  For a more detailed review of spin-wave detection techniques see e.g. \cite{Lenk2011}.    

%%%%%%%%%%%%%%%%%%%%%%%%%%%%%%%%%%%%%%%%%%%%%%%%%%%%%%%%%%%%%%%%%%%%%%%%%%%%%%%%%%%%%%%%%%%%%%%%%%%%%%%%%%%%%%%%%%%%%%%%%%%%%%%%%%%%%%%%%%%%%%%%%%%%%%%%%%%%%%%%%%%%%%%%%%%%%%%%%%%%%%%%%%%%%%%%%%%%%%%%%%%%%%%%%%%%%%%%%%%

\subsection{Electrical Detection of Nonlinear Magnetization Dynamics} \label{nonlinear}

The onset of nonlinear dynamics produces a rich array of new physics compared to the linear regime, such as amplitude dependent resonance frequency, foldover effects and bistability.  These nonlinear fingerprints are found throughout nature, from mechanical \cite{LandauBookMechanics} to magnetic systems \cite{Anderson1955, Bertotti2001}.  In magnetic systems the key effects of nonlinear dynamics are 1) power dependent resonance position shifts of both the FMR and spin wave modes and 2) line width broadening.  The former is due to a nonlinearity of the driving force while the later is due to additional nonlinear damping.  The onset of the foldover and bistability effects also occur in magnetic systems when the resonance shift reaches a critical value which was initially investigated by Anderson and Suhl \cite{Anderson1955}.  However in the experimental search for foldover effects in magnetic systems, large discrepancies from the Anderson-Suhl model have been found \cite{Suhl1960, Gottlieb1960, Masters1960, Goldberg1979, Silber1983, Seagle1985, McKinstry1985, Zhang1986, Gnatzig1987, Zhang1988, Chen1989, Fetisov1999, Fetisov2004}.  These discrepancies were partially due to the presence of spin wave instabilities and thermal effects, which made the interpretation of early experiments difficult, but the primary reason that the critical frequency shift could not be reached is due to the effect of nonlinear damping, which effectively pushes the power and frequency required to observe foldover effects beyond the range initially investigated.  This later complication was identified by spin rectification studies which were subsequently used to study the foldover FMR \cite{Gui2009, Gui2009a}.   

\subsubsection{Review of Nonlinear Magnetization Dynamics}

The effects of resonance shift and line width broadening can be described by determining the cone angle of magnetization precession in the nonlinear regime.  Physically the shift in resonance frequency occurs due to the decrease in $M_z$ as the magnetization precession is tilted to larger cone angles by a high driving power.  However as the resonance frequency is shifted, nonlinear damping becomes more important.  Based on the analysis in Sec. \ref{magnetizationdynamics} the cone angle $\theta_c$ is determined by $\theta_c^2 \sim \left(m_x^2 + m_y^2\right)/M_0^2$, where we have assumed that $M_0^2 \gg \left(m_x^2 + m_y^2\right)$.  The solution for the LLG equation can then be used to relate the cone angle to the driving rf field.  For example when $\textbf{H}$ is applied nearly perpendicular to the sample plane $\theta_c^2 = h^2/\left[\left(H-H_r\right)^2 + \Delta H^2\right]$, where $h = h_x = h_y $.  However to be consistent with this higher order expansion, the nonlinear effects on the resonance field and the line width must also be taken into account, which can be done by making the replacements \cite{Gui2009, Gui2009a}

\begin{equation*}
\Delta H \to \Delta H_0 + \beta M_0 \theta_c^2 = \Delta H_\text{in} + \alpha \omega/\gamma + \beta M_0 \theta_c^2,
\end{equation*}
\begin{equation*}
H_r \to H_r^0 -\frac{1}{2} M_0 \theta_c^2.
\end{equation*}
Here $\Delta H_0 = \Delta H_\text{in} + \alpha \omega/\gamma$ and $H_r^0$ are the line width and resonance position without any nonlinear modifications, respectively, and $\beta$ is the nonlinear damping coefficient. 

\begin{figure}[!ht]
\centering
\includegraphics[width=14cm]{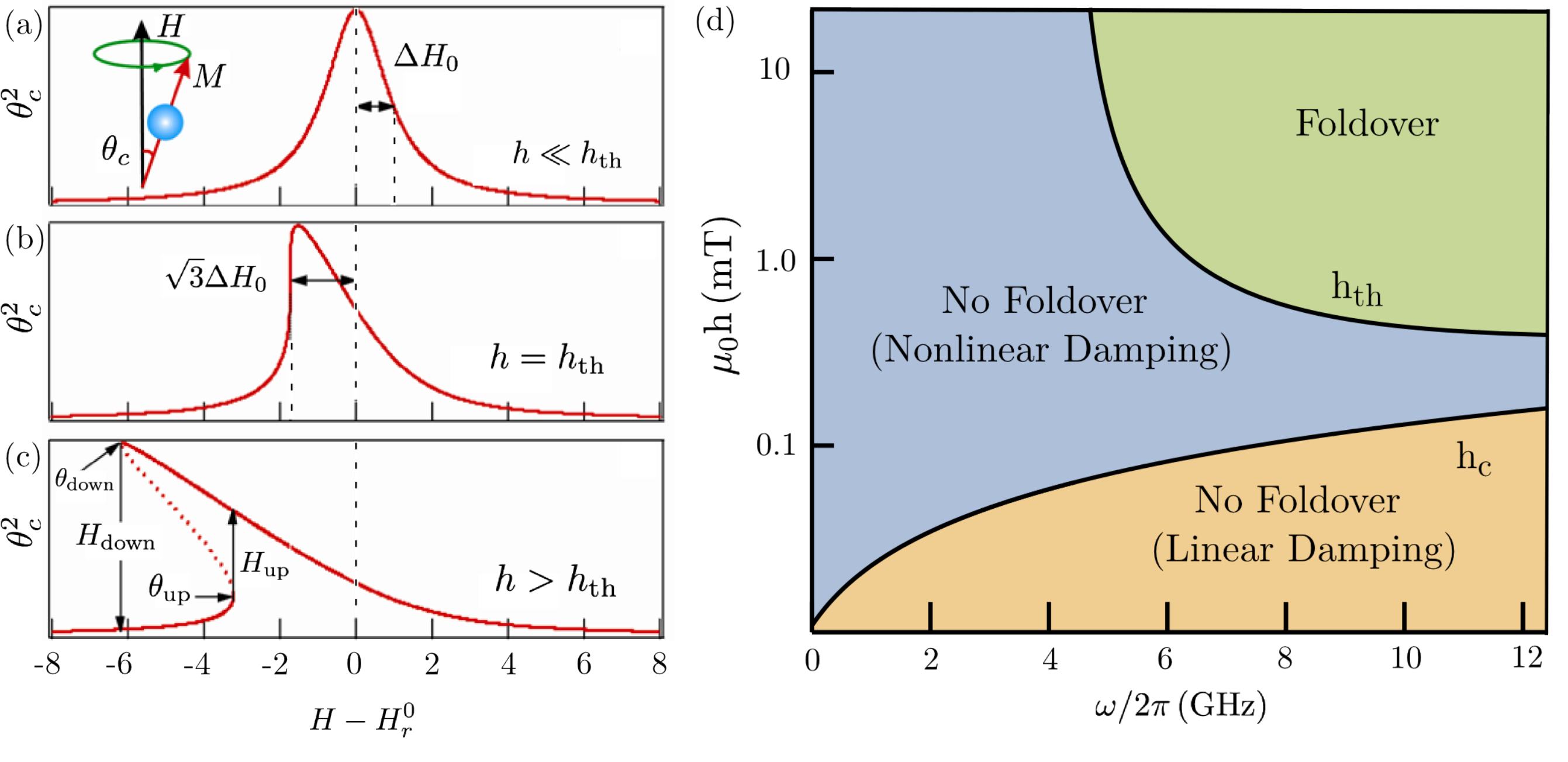}
\caption{\footnotesize{(a) - (c) A discontinuity in $\theta_c^2$ occurs when the rf driving field exceeds a threshold $h_\text{th}$.  (d) The $h - \omega$ phase diagram.  In the region below $h_c$, which is the critical field in the absence of nonlinear damping, no foldover is observed.  Even with nonlinear damping and high powers, below the threshold $h_\text{th}$ no foldover is observed.  $Source:$ Adapted from Ref. \cite{Gui2009a}.}}
\label{linphase}
\end{figure}

For sufficiently large microwave power $P$, the microwave field $h \propto \sqrt{P}$ exceeds a critical field strength $h_\text{th}$ where the FMR response shows a discontinuity in $\theta_c$, as shown in Fig. \ref{linphase} (a) - (c).  This discontinuity is the foldover effect.  In Fig. \ref{linphase} (a) - (c) the nonlinear damping term is ignored (i.e. $\beta \to 0$) in which case the above treatment reduces to that of Anderson and Suhl \cite{Anderson1955}.  However the introduction of nonlinear damping will result in a high power region where the foldover effect is suppressed as shown in Fig. \ref{linphase} (d).  Below the critical field in the absence of nonlinear damping, $h_c$ no foldover will be observed.  For $h_{th} > h > h_c$ nonlinear damping will suppress the onset of foldover effects, while for $h > h_{th}$ foldover can now be observed.  

%%%%%%%%%%%%%%%%%%%%%%%%%%%%%%%%%%%%%%%%%%%%%%%%%%%%%%%%%%%%%%%%%%%%%%%%%%%%%%%%%%%%%%%%%%%%%%%%%%%%%%%%%%%%%%%%%%%%%%%%%%%%%%%%%%%%%%%%%%%%%%%%%%%%%%%%%%%%%%%%%%%%%%%%%%%%%%%%%%%%%%%%%%%%%%%%%%%%%%%%%%%%%%%%%%%%%%%%%%%

\subsubsection{Detection of Nonlinear Magnetization Dynamics}

Table \ref{swdetectiontable} provides a brief summary of experimental techniques that have been used to study nonlinear magnetization dynamics.  The schematics shown here indicate the use of CPWs to deliver a driving rf field, which enables the application of high powers due the enhanced field confinement and close proximity to samples under investigation.  However early studies of nonlinear effects in magnetic materials often used microwave cavities, such as those in a pulsed microwave reflection cavity spectrometer system \cite{Zhang1987, Cox2001, An2004}.  In such systems the peak power levels with a pulse width of about 50 $\mu$s can be as high as 1 kW, allowing the measurements of spin wave instabilities in both ferrimagnetic (YIG) \cite{Zhang1987} and ferromagnetic (Py) films \cite{An2004}.  

To investigate the wave vector dependence of nonlinear effects, which is necessary to fully understand FMR at high microwave powers \cite{Suhl1957}, combined BLS and microwave pumping techniques have also been used \cite{Wettling1983, Kabos1996, Edwards2012}.  More recently micro focussed BLS has enabled sensitive local probing of micro- and/or nano-structured devices.  For example, Edwards et al. reported an experimental observation of parametric excitation of magnetization oscillations in a Py micro disk adjacent to a Pt layer \cite{Edwards2012}, demonstrating control of nonlinear dynamic magnetic phenomena in microscopic structures via pure spin currents.  Additionally the longitudinal magneto-optical Kerr effect has been used to investigate large precession angles of up to $\theta_c = 20^\circ$ \cite{Gerrits2007}, demonstrating the Suhl threshold effect for parametric spin wave generation \cite{Suhl1957}. Meanwhile, ferromagnetic resonance force microscopy \cite{Guo2015} has also proven to be a useful tool for nonlinear studies, allowing simplified sample fabrication of single magnetic nano-structures by removing the requirement of electrical contacts.  Finally the phase resolved ability of XMCD measurements has allowed new nonlinear phenomena to be observed at low magnetic bias fields resulting in the generalization of spin-wave turbulence theories \cite{Bauer2015a}.

\begin{figure}[!ht]
\centering
\includegraphics[width=13cm]{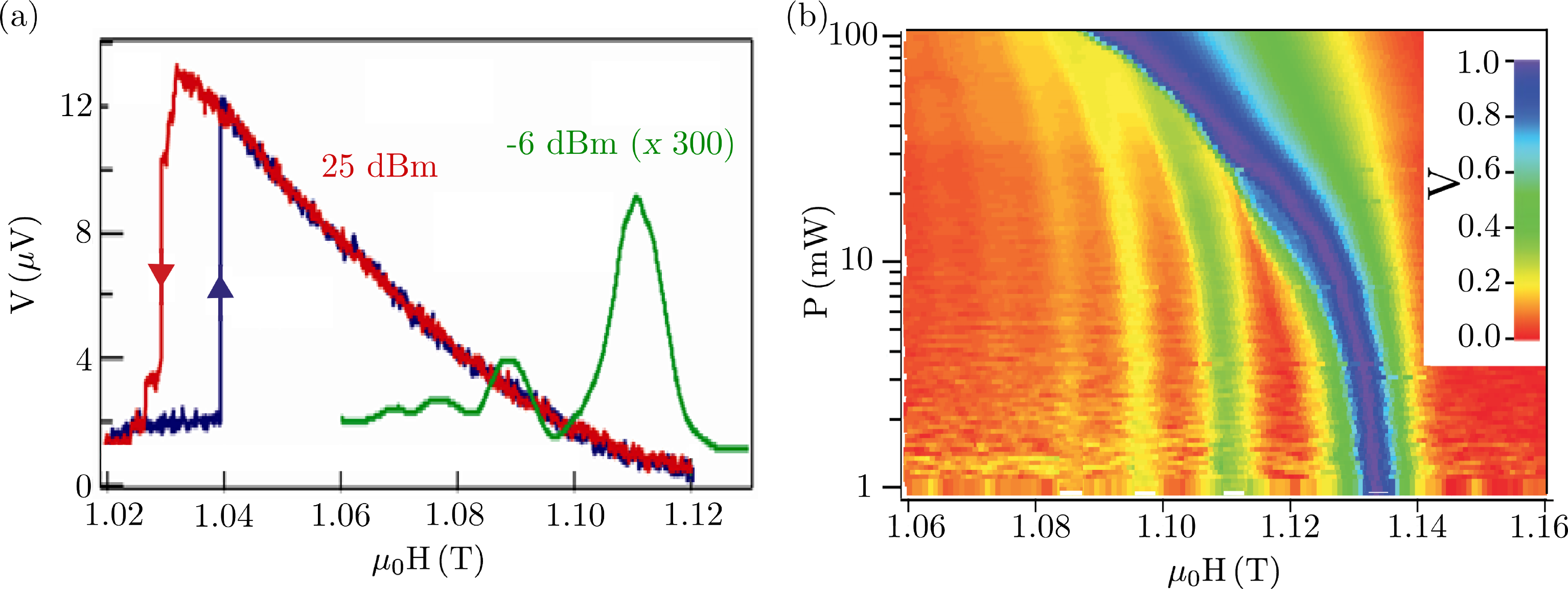}
\caption{\footnotesize{Nonlinear magnetization dynamics detected using spin rectification. (a) Observation of the foldover effect at high power (25 dBm).  The green curve shows the rectification signal at low power for comparison. (b) Linear to nonlinear transition observed in the normalized spin rectification voltage over a power range, 1 mW $< P < 100$ mW.  FMR (dark blue) and several perpendicular standing spin wave modes (at lower field) can be observed.    $Source:$ Adapted from Ref. \cite{Gui2009}.}}
\label{nlData}
\end{figure}

Electrical detection techniques based on spin rectification have also proven useful in the study of nonlinear magnetization dynamics in a variety of magnetic structures \cite{Gui2009, Gui2009a, Bi2011a, Hirayama2015, Zhou2016}.  For example, nonlinear FMR in a micro-structured MTJ  has been reported \cite{Bi2011a} and the large precession angle magnetization dynamics in a nanoscale MTJ have been used to measure the electric-field modulation ratio of magnetic anisotropy energy density \cite{Hirayama2015}.  Zhou et al. have also studied nonlinear effects in YIG/Pt bilayers where foldover has been observed and the power dependent cone angle due to spin pumping can be explained by accounting for nonlinear damping \cite{Zhou2016}.  Finally nonlinear phenomena has also been investigated in a Py strip using a second generation spin dynamo, as shown in Fig. \ref{nlData}.  Electrical detection enables several key nonlinear behaviours to be easily observed, such as the foldover effect, shown in Fig. \ref{nlData} (a) at high powers (25 dBm) with the blue (red) curves measured while sweeping the field up (down), and the onset of nonlinear behaviour as a function of microwave power, seen in panel (b), where, in addition to FMR (dark blue), several perpendicular standing spin wave modes can be observed at lower field.    
 
%%%%%%%%%%%%%%%%%%%%%%%%%%%%%%%%%%%%%%%%%%%%%%%%%%%%%%%%%%%%%%%%%%%%%%%%%%%%%%%%%%%%%%%%%%%%%%%%%%%%%%%%%%%%%%%%%%%%%%%%%%%%%%%%%%%%%%%%%%%%%%%%%%%%%%%%%%%%%%%%%%%%%%%%%%%%%%%%%%%%%%%%%%%%%%%%%%%%%%%%%%%%%%%%%%%%%%%%%%%

\subsection{Electrical Detection of ac Spin Current}

\begin{figure}[!ht]
\centering
\includegraphics[width=14.7cm]{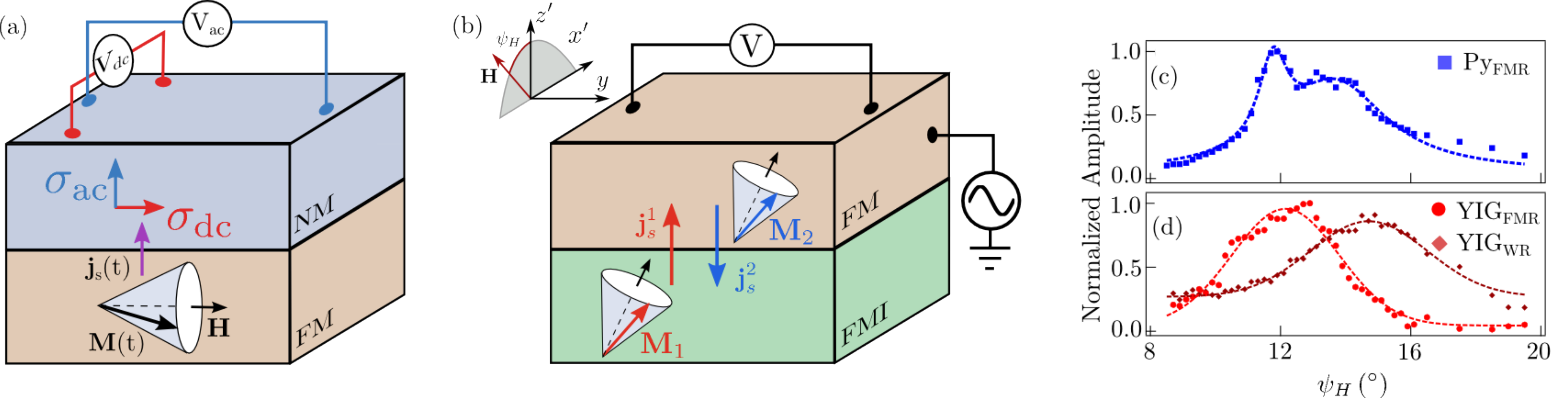}
\caption{\footnotesize{(a) The spin polarization of the pumped spin current has a small dc component (red) along the static field direction and a larger transverse ac component (blue).  The dc voltage can be measured in-plane transverse to the static field while the ac voltage will be along the field direction.  The dc spin pumping has been studied more intensively since it can easily be measured through the inverse spin Hall effect. (b) The experimental setup used to detect the ac spin current through the simultaneous enhancement of spin rectification and spin pumping. (c) By tuning the angle of the external field, the Py and YIG resonance can be made to occur simultaneously, showing enhanced Py rectification and (d) enhanced YIG spin pumping at the mutual resonance condition.  $Source:$ Panels (c) and (d) adapted from Ref. \cite{Hyde2014}.}}
\label{accurrent}
\end{figure}

Spin pumping is perhaps the most common technique used to generate pure spin currents and has been employed in NM/FM bilayers \cite{Saitoh2006}, NM/ferromagnetic semiconductor bilayers \cite{Chen2013c} and even NM/ferromagnetic insulator (FMI) bilayers \cite{Kajiwara2010}.  The use of FMI is especially intriguing since it explicitly demonstrates the flow of a pure spin current in the absence of charge current (see Ref. \cite{SpinCurrentBook} for an in depth discussion of spin currents in FMI).  In the treatment of spin pumping presented in Sec \ref{lateralsp} we tacitly assumed that the spin polarization $\boldsymbol{\sigma}$ was time independent and directed along $\textbf{M}_0$.  This assumption of a dc spin polarization results in a dc spin current which, after being converted to a charge current via the inverse spin Hall effect, can be directly detected as a dc voltage.  To date, most work on spin pumping and spin currents has focused on this dc effect.  However the assumption that $\boldsymbol{\sigma}$ is constant neglects the fact that the direction of spin polarization is actually determined by the full time dependent magnetization, and therefore $\boldsymbol{\sigma}$ is actually time dependent.  This means that in addition to a dc component directed along the external field direction, there is also a transverse ac spin current.  Recently spin pumping theory has been extended to include such an ac spin current \cite{Jiao2013} and proposals for an ac spin current source have been put forth \cite{Hofer2014}.  A sketch of the dc and ac spin currents is shown in Fig. \ref{accurrent} (a).  As we already know, the precessing magnetization will pump a spin current into an adjacent material layer due to a non-equilibrium spin distribution at the interface.  Here the dc (yellow) and ac (red) component of this spin current are explicitly shown.  In addition to the obvious dynamical
difference between the dc voltage, $V_\text{dc}$, generated through the ISHE and the ac voltage, $V_\text{ac}$, generated by the ac spin current, the other key differences between these two signals are: 1) $V_\text{ac}$ is an order of magnitude larger than $V_\text{dc}$ \cite{Jiao2013, Wei2014}.  This is due to the fact that, as we saw in Sec. \ref{lateralsp}, the dc spin pumping voltage is proportional to $\sin^2 \theta_c$, while the ac spin pumping voltage is proportional to $\sin\theta_c \cos\theta_c$.  Therefore the dc signal is maximized at $\theta_c = 90^\circ$ while the ac signal is largest at $\theta_c = 45^\circ$.  2) The ac voltage scales as the root of the microwave power, $\sqrt{P}$, while the dc spin pumping is linearly proportional to $P$ \cite{Wei2014}.  3) As shown in Fig. \ref{accurrent} (a), $V_\text{dc}$ is measured transverse to the applied static field direction, whereas the $V_\text{ac}$ will be generated parallel to the static magnetic field.  

In general detection of ac spin pumping is more difficult than measuring the dc effect, since the ac signal will have the same high frequency as the microwave field and parasitic effects such as $h$ field induced eddy currents and the FMR excitation at the same frequency produce large backgrounds which must be separated.  Nevertheless a variety of approaches have succeeded in measuring the ac spin current, using for example rectifying diodes \cite{Hahn2013, Wei2014}, a vector network analysis \cite{Weiler2014} or more recently direct detection via x-ray techniques \cite{Li2016}.  However it turns out that the ac spin current can also be measured through electrical detection by exploiting the spin torque exerted by an ac spin current \cite{Hyde2014}.  The schematic setup for such an approach is shown in Fig. \ref{accurrent} (b).  In a conventional dc spin pumping experiment the spin current flowing from the FMI (or FM metal) to the NM will increase the FMI (FM) damping \cite{Heinrich2003}.  However if the NM is replaced by a FM layer the ac spin current generated in the FMI can act as a spin torque on the FM, causing the precession amplitude to increase and pump an ac spin current back into the FMI, thus compensating the initially increased damping of the FMI (the dc spin current cannot produce this effect since it will be parallel to the static field direction, and therefore will not produce a torque on the FM).  Therefore when both the FMI and FM are undergoing resonance simultaneously, the damping will be reduced and both the spin rectification signal from the FM and the spin pumping from the FMI will be enhanced.  As shown in Fig. \ref{accurrent} (c) the simultaneous resonance condition can be achieved by tuning the angle of the external magnetic field as shown.  An enhancement of both spin rectification and spin pumping is seen at $\psi_H \sim 12^\circ$.  (The second, smaller peak, near $\psi_H \sim 15^\circ$ results from an additional weak resonance signal in the YIG due to a spin wave mode).  In addition to the ac spin current detection via SP and SR enhancement, this technique also shows that the ac spin current can be used as a spin torque and could therefore be applied to magnetization switching techniques.    

%%%%%%%%%%%%%%%%%%%%%%%%%%%%%%%%%%%%%%%%%%%%%%%%%%%%%%%%%%%%%%%%%%%%%%%%%%%%%%%%%%%%%%%%%%%%%%%%%%%%%%%%%%%%%%%%%%%%%%%%%%%%%%%%%%%%%%%%%%%%%%%%%%%%%%%%%%%%%%%%%%%%%%%%%%%%%%%%%%%%%%%%%%%%%%%%%%%%%%%%%%%%%%%%%%%%%%%%%%%

\subsection{Electrical Detection of Magnetization Dynamics in Ferromagnetic Semiconductors} \label{fmSemiConductors}

In a 1931 letter to Rudolf Perierls (who pioneered the ``hole" concept in semiconductors), Wolfgang Pauli offered the following advice regarding the sensitivity of semiconductor properties to impurities: ``One shouldn't work on semiconductors, that is a filthy mess; who knows whether any semiconductors exist"  (English translation of the original German: ``\"{U}ber Halbleiter soll man nicht arbeiten, das ist eine Schweinerei, wer wei\ss, ob es \"{u}berhaupt Halbleiter gibt)?  \cite{PauliLetters}.  Today the deliberate introduction of impurities into the crystal structure of a semiconductor (``doping"), which alters the conducting properties and permits the creation of semiconductor junctions between differently-doped regions of the crystal, plays a fundamental role in modern semiconductor technology and related IT industries.  Interestingly, when the paramagnetic element Mn is substituted for Ga in a GaAs lattice, it not only acts as an acceptor, making it a p-type material, but also the produced holes mediate a ferromagnetic interaction between the local moments of the open $d$ shells in the Mn atoms, resulting in ferromagnetic properties at low temperatures \cite{Ohno1992} (for a review of ferromagnetic semiconductors see e.g. Refs. \cite{Ohno1998, MacDonald2005, Jungwirth2006, Dietl2014, Jungwirth2014}). Therefore ferromagnetic semiconductors, such as GaMnAs, offer a unique combination of material properties, which are amenable to both data processing (due to p-type conducting behaviours) and data storage (due to hysteretic magnetization) in a single material system. 

\begin{figure}[!ht]
\centering
\includegraphics[width=14cm]{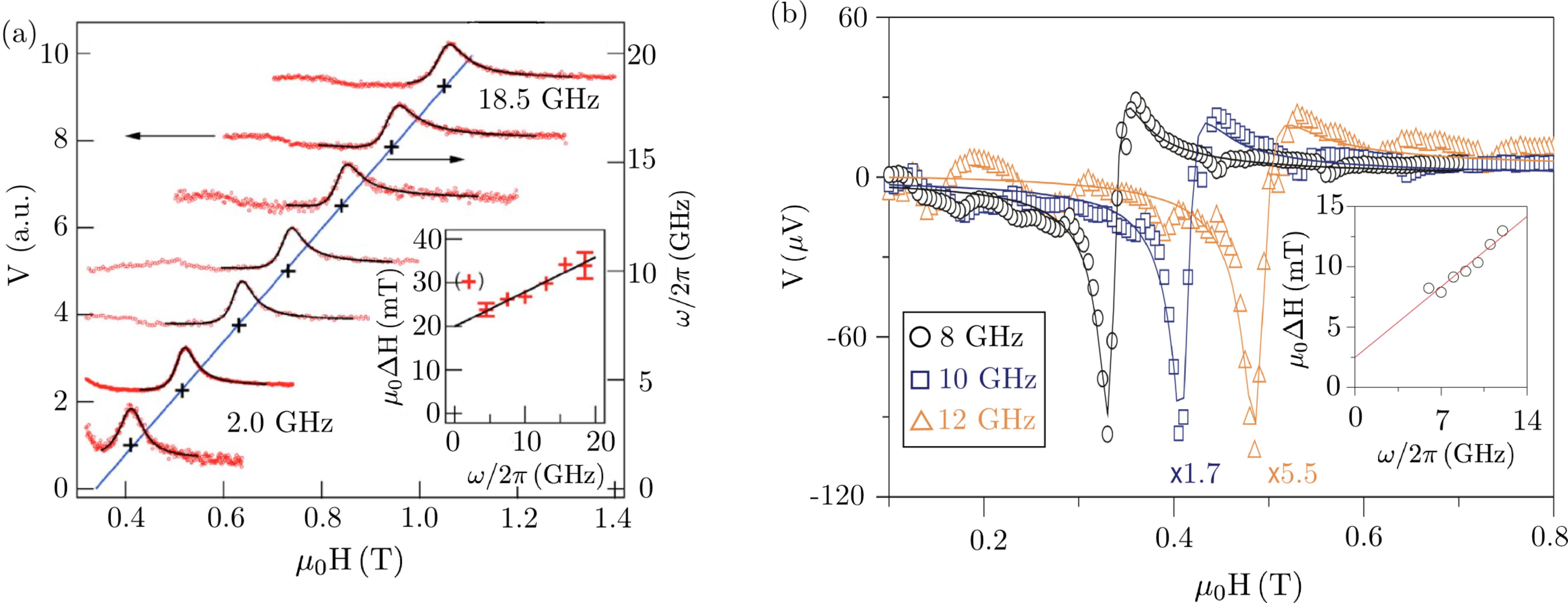}
\caption{\footnotesize{(a) SR voltage signal at different frequencies measured on a GaMnAs Hall bar placed between the ground and signal lines of a CPW.  The inset shows the linear frequency dependence of the FMR line width.  (b) .  $Source:$ Panel (a) adapted from Ref. \cite{Wirthmann2008}.  Panel (b) adapted from Ref. \cite{Fang2011}.}}
\label{scRecWirthmann}
\end{figure}

Similar to ferromagnetic metals, SR also exists in ferromagnetic semiconductors and just as in metals may have several origins, e.g. AMR \cite{Wirthmann2008} and spin-orbit torques \cite{Fang2011}.  As in other material systems, SR allows simple, broadband FMR measurements enabling, for example, determination of damping characteristics.  Fig. \ref{scRecWirthmann} (a) shows a typical voltage signal to due to AMR induced SR in a GaMnAs Hall bar as shown in Table \ref{spindynamosummary}.  By fitting the line shape to Eq. \ref{eq:outofplaneV} both the resonance position (shown by crosses) and the line width (shown in the inset) can be determined just as they are for SR in a ferromagnetic metal.  In general, a better understanding of damping in ferromagnetic semiconductors is necessary for applications to data storage technologies.  An interesting step in this direction was made in 2011 by Fang et al. \cite{Fang2011} who fabricated a nanostructured GaMnAs device and introduced a form of spin-orbit-driven FMR originating from the effective magnetic field in the magnetic material induced by an rf electric current.  Due to the improved homogeneity of the nano-structured sample they found an exceptionally low extrinsic damping, implying that the GaMnAs could be used for data storage if further homogeneity refinements could be made.  Typical voltage measurements from such a device are shown in Fig. \ref{scRecWirthmann} (b) where the inset displays the frequency dependent line width from which the Gilbert damping can be determined.

\begin{figure}[!ht]
\centering
\includegraphics[width=14cm]{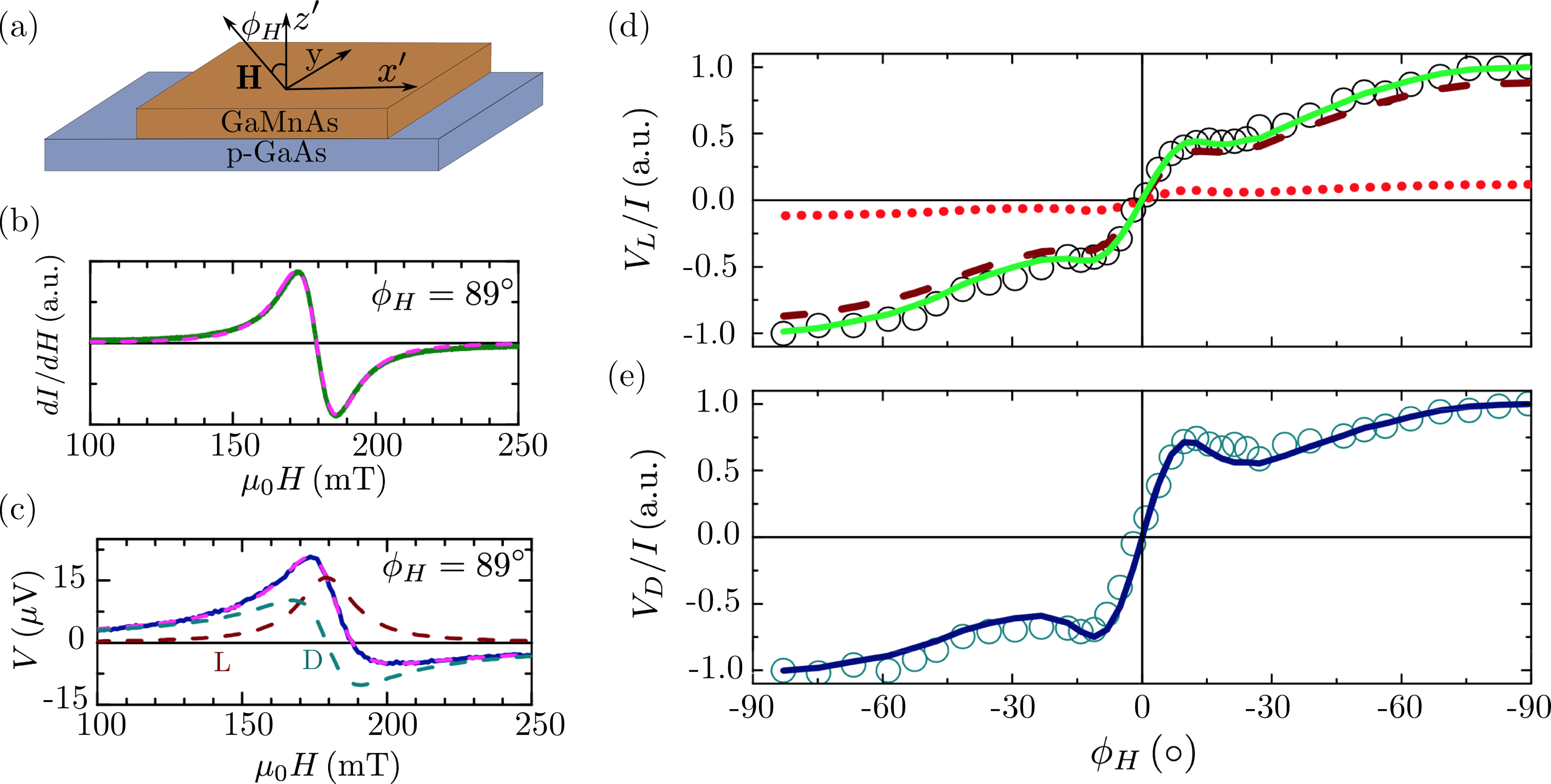}
\caption{\footnotesize{Measurements of the microwave absorption and spin rectification/spin pumping voltage in a GaMnAs/p-GaAs bilayer.  (a) Device structure.  Voltage measurements are made in the $x^\prime$ direction. (b) Derivative of the microwave absorption and (c) dc voltage for a magnetic field nearly perpendicular to the bilayer.  (d) The Lorentz and (e) dispersive amplitude of the voltage signal as a function o $\phi_H$ normalized to the microwave absorption $I$.  In (d) the dotted and dashed lines show the contribution of spin pumping and spin rectification respectively, with the solid fitted curve the total voltage due to both effects.  In (e) the solid curve is the calculated voltage due only to spin rectification (as spin pumping does not contribute to the dispersive component).  The theoretical angular dependence follows from our discussion in Sec. \ref{spinrectification}, with deviations from the expected sinusoidal dependence the result of normalization by the microwave absorption, which also has an angular dependence.  $Source:$ Adapted from Ref. \cite{Chen2013c}.}}
\label{chenFig}
\end{figure}

Not only are AMR and spin-orbit torque induced spin rectification observed in ferromagnetic semiconductors, additionally in 2013, Chen et al. \cite{Chen2013c} reported the observation of spin pumping in a GaMnAs/p-GaAs bilayer structure, again using electrical detection of FMR.  Just like the case for the FM/NM bilayers discussed in Sec. \ref{srinbilayers} the precession of the magnetization in the ferromagnetic layer, GaMnAs, produces a pure spin current which is pumped into the adjacent non-magnetic GaAs layer.  This spin current is subsequently converted to a change current by the ISHE and hence generates a voltage drop across the device.  Of course spin rectification is also present and therefore the techniques of line shape and angular dependence analysis must be used to separate the two effects.  Fig. \ref{chenFig} shows the angular dependent measurements made in this ferromagnetic semiconductor system.  The angular dependence shown in panels (d) and (e) follow the expectations based on our discussion in Sec. \ref{spinrectification} (with deviations from the expected sinusoidal behaviour due to normalization by the microwave absorption, $I$, which also has an angular dependence).  More recently Nakayama et al. \cite{Nakayama2015} investigated the rectified voltage in a Pt/GaMnAs bilayer structure using a similar analysis method.  Interestingly they found an exceptionally large spin mixing conductance at the Pt/GaMnAs interface, 6.2 $\times$ 10$^{19}$ cm$^{-2}$ which is ten times greater than that of the GaMnAs/p-GaAs interface. This result indicates that hybrid structures with ferromagnetic semiconductors and metals may be of interest in the investigation of spin-current related phenomena due to their large efficiency of spin-current generation at the interface. 

%%%%%%%%%%%%%%%%%%%%%%%%%%%%%%%%%%%%%%%%%%%%%%%%%%%%%%%%%%%%%%%%%%%%%%%%%%%%%%%%%%%%%%%%%%%%%%%%%%%%%%%%%%%%%%%%%%%%%%%%%%%%%%%%%%%%%%%%%%%%%%%%%%%%%%%%%%%%%%%%%%%%%%%%%%%%%%%%%%%%%%%%%%%%%%%%%%%%%%%%%%%%%%%%%%%%%%%%%%%

\subsection{Electrical Detection of Domain Wall Dynamics}

A magnetic domain is a spatial transition region between different magnetic moments with an angular displacement of $90^\circ$ or $180^\circ$ \cite{HubertBook}.  The spatial dimensions of such a region depends strongly on the material, and can be a few to thousands of atoms in soft materials.  Micro magnetic calculations based on the LLG equation have revealed two types of domain walls which can be classified by the canting direction of magnetic moments: in a N\'{e}el wall the moments rotate in a direction parallel to i.e. within the plane of, the domain wall while for a Bloch wall they rotate in a direction perpendicular to i.e. through the plane of, the domain wall.  In thin films without perpendicular magnetic anisotropy, so that the easy axis lies in the film plane, this means that a N\'{e}el wall has moments which rotate in-plane, while a Bloch wall has moments which rotate out-of-plane.  In thin films with low anisotropies N\'{e}el walls are energetically favourable \cite{McMichael1997} and vortex structures may form when two N\'{e}el walls intersect.  From a fundamental physics perspective studying magnetic domains can teach us about the link between the microscopic and macroscopic properties of ferromagnetic materials \cite{HubertBook}.  On the other hand the ability to directly introduce domain walls through the design of pinning sites and to control domain wall motion lays the foundation for magnetic domain wall logic \cite{Allwood2005, Parkin2008, Malinowski2011, Boulle2011, Boone2010} which promises a non-volatile memory device with the high performance and reliability of conventional solid-state memory, but at the low cost of conventional magnetic disk drive storage.

\begin{figure}[!ht]
\centering
\includegraphics[width=10cm]{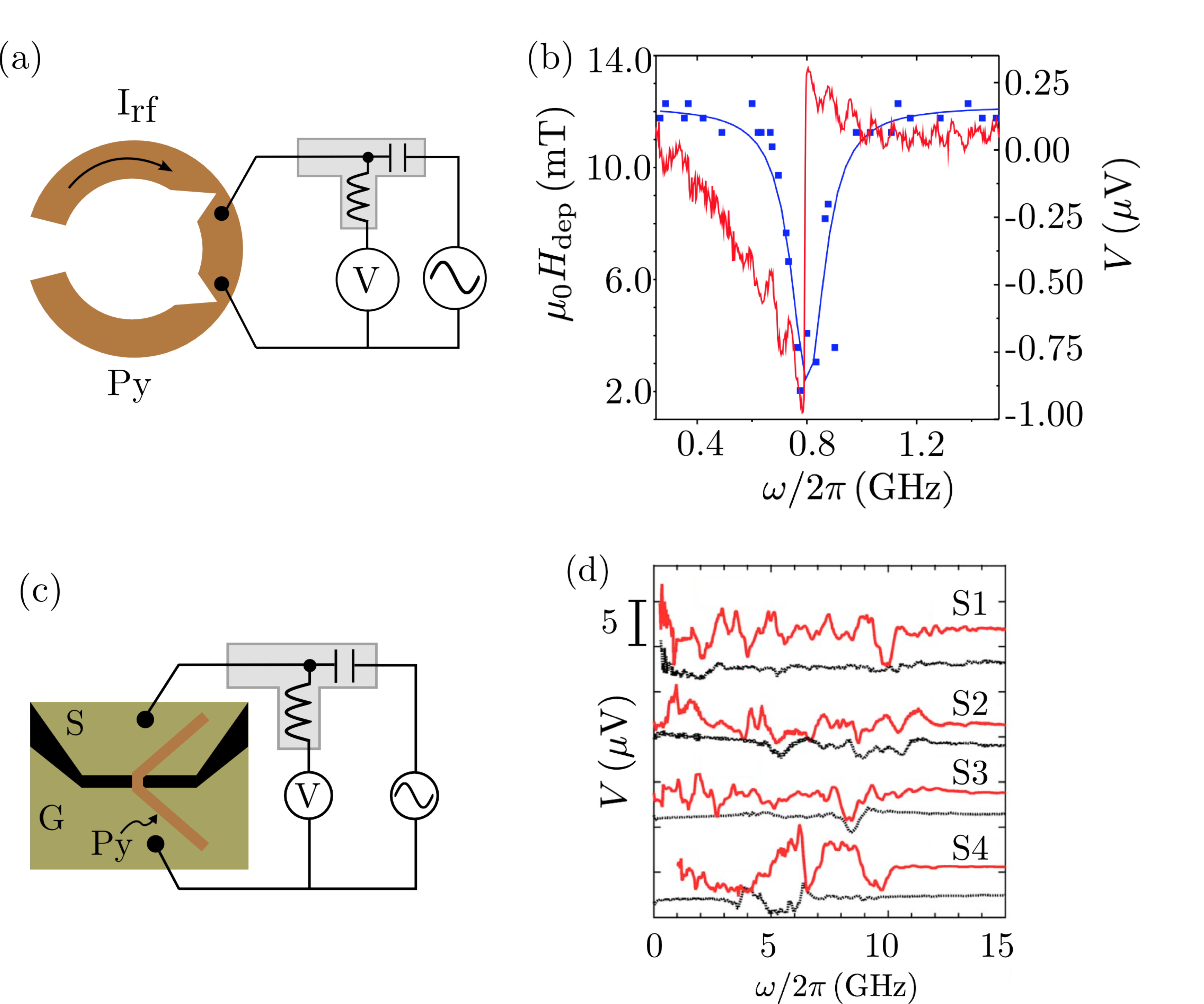}
\caption{\footnotesize{(a) Device used to measure the domain wall resonance frequency and depinning field via spin rectification. A domain wall is pinned near the constriction sight formed by notches in a Py ring.  (b)  Measurements of the pinning field (blue) of a vortex domain wall and the voltage (red) due to AMR rectification.  (c) The experimental setup to perform a magnetic fingerprinting experiment of a ferromagnetic nanowire.  (d) The voltage spectrum as a function of the rf current frequency for Py semicircular (S2, S3, S4) and ``boomerang" shaped (S1) nanowires.  The black curves are measured in the absence of a domain wall, while the red curves are measured after a domain wall has been introduced by saturating the magnetization perpendicular to the easy axis along the strip.  $Source:$ Adapted from Refs. \cite{Bedau2007} and \cite{Yamaguchi2007a}.}}
\label{dwfig}
\end{figure}

In order to study domain wall properties, e.g. inertia and resonance frequency, and to manipulate domain walls for magnetic logic purposes, it is important to study domain wall motion.  For this purpose rf excitations have been found to be exceptionally efficient, moving domain walls at far lower current densities than pulsed voltages \cite{Saitoh2004, Yamaguchi2004, Buchanan2006, Neudecker2006, VanWaeyenberge2006}.  Since magnetoresistance occurs during the domain wall motion, which is efficiently driven by rf currents, spin rectification effects will be present during domain wall motion and have become a useful tool to study the domain dynamics in nano-structured magnets, such as the determination of the nonlinear pining potential of a domain wall \cite{Bedau2008}. 
 
In general rectification effects and the study of domain walls intersect in three key ways.  First, more indirectly related to rectification but directly related to our discussion of magnetoresistance in Sec. \ref{magnetotransport}, is the direct use of magnetoresistance effects, such as GMR or AMR, to study domain wall properties.  For example, GMR has been used to study the time dependence of the domain wall position \cite{Ono1999, Ono2001} and to determine the domain wall mass \cite{Saitoh2004}, while domain wall pinning conditions have been investigated through AMR \cite{Klaui2003}, and time resolved AMR measurements have provided insight into the effect of current and spin-transfer torque on domain wall motion \cite{Hayashi2006}.  

Second, domain walls display their own magnetoresistance, which can yield rectification effects and can be manipulated by controlling the spin structure at the pinning site \cite{VonBieren2013a}.  In analogy with GMR, the switching of domains will produce a magnetoresistance due to a mistracking effect, where the orientation of the electron spins comprising the current lag behind the orientation of the local magnetization inside the domain wall \cite{Viret1996, Gregg1996, Levy1997, Ebels2000} resulting in a spin accumulation effect \cite{Ebels2000}.  Direct measurements of domain magnetoresistance have been used to study e.g. spin current asymmetry \cite{Marrows2004}.

Finally, and most relevant to our discussion of spin rectification, due to their general effect on magnetization dynamics, domain walls will influence AMR rectification in ferromagnetic nano structures.  Fig. \ref{dwfig} (a) shows a Py ring structure where constriction sites, introduced by notches in the ring, will lead to domain wall formation.  A bias T allows the injection of an rf current and detection of the dc voltage produced through AMR rectification.  Fig. \ref{dwfig} (b) shows the measurement results with the depinning field (the field required to move the domain wall past the pinning site) shown in blue and the rectification voltage shown in red.  Thus SR measurements can be used to determine the domain wall resonance of both transverse and vortex domain walls and have even been used to observe multiple resonances within a magnetic domain \cite{Bedau2007, Sangiao2014}.  

Another interesting use of rectification in nanostructures with domain walls is in ``magnetic fingerprinting" of a sample.  When a current is passed through a ferromagnetic nanowire a voltage will be generated due to STT  as discussed in Sec. \ref{standshe}.  In the presence of a domain wall the voltage spectrum will strongly depend on the internal spin structure of the wall, and therefore can be used as a fingerprint of the sample \cite{Yamaguchi2007}.  An example of such ``fingerprints" is shown in Fig. \ref{dwfig} (d).  The voltages are measured using either a semicircular nanowire of varying thicknesses (S2, S3, S4) or a ``boomerang" structured nanowire (S1), as shown in Fig. \ref{dwfig} (a).  The red (black) curves show the rectification voltage in the presence (absence) of a domain wall. 

%%%%%%%%%%%%%%%%%%%%%%%%%%%%%%%%%%%%%%%%%%%%%%%%%%%%%%%%%%%%%%%%%%%%%%%%%%%%%%%%%%%%%%%%%%%%%%%%%%%%%%%%%%%%%%%%%%%%%%%%%%%%%%%%%%%%%%%%%%%%%%%%%%%%%%%%%%%%%%%%%%%%%%%%%%%%%%%%%%%%%%%%%%%%%%%%%%%%%%%%%%%%%%%%%%%%%%%%%%%

\subsection{Spintronic Microwave Sensing and Imaging} \label{imagingsec}

%%%%%%%%%%%%%%%%%%%%%%%%%%%%%%%%%%%%%%%%%%%%%%%%%%%%%%%%%%%%%%%%%%%%%%%%%%%%%%%%%%%%%%%%%%%%%%%%%%%%%%%%%%%%%%%%%%%%%%%%%%%%%%%%%%%%%%%%%%%%%%%%%%%%%%%%%%%%%%%%%%%%%%%%%%%%%%%%%%%%%%%%%%%%%%%%%%%%%%%%%%%%%%%%%%%%%%%%%%%

\begin{figure}[!b]
\centering
\includegraphics[width=11cm]{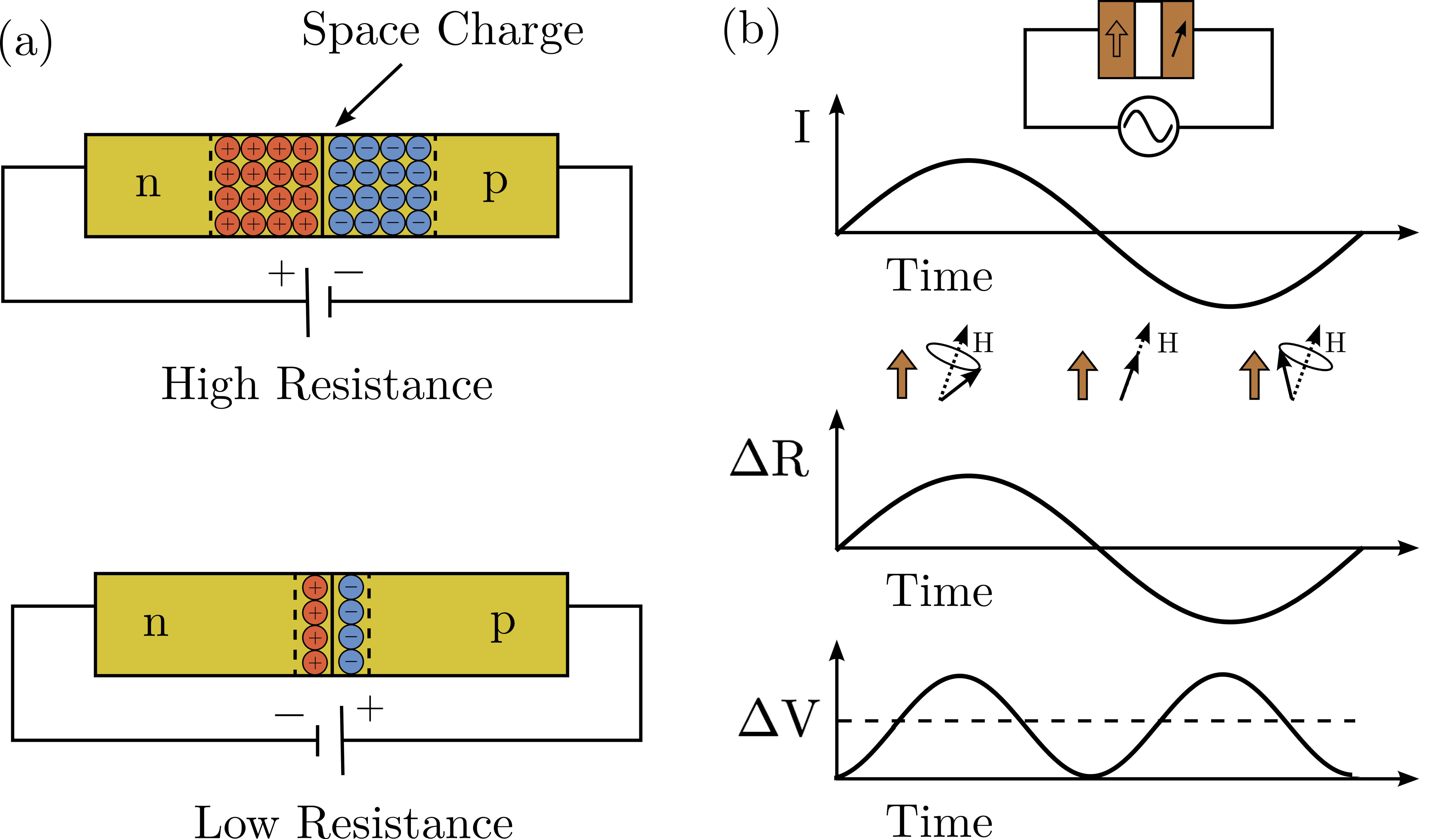}
\caption{\footnotesize{(a) The behaviour of a semiconductor p-n junction diode compared to (b) an MTJ based spin-transfer torque diode.  $Source:$ Adapted from Ref. \cite{Tulapurkar2005}.}}
\label{diodeSchem}
\end{figure}

In addition to applications for magnetic storage and logic devices which we have already discussed, spin rectification is playing an increasingly important role in the development of novel microwave sensing and imaging techniques.  High speed telecommunications require highly sensitive microwave devices that are operational at room temperatures.  In particular the rectifying behaviour of semiconductor diodes is of key importance, making conventional diodes a foundation of modern electronics.  However, although semiconductor diodes satisfy the high sensitivity and room temperature requirements, they have poor signal-to-noise ratios which means they cannot operate at low input power.  Spin rectification on the other hand behaves remarkably different from semiconductor rectification and has been used to develop the spin-transfer torque diode, which can form the basis of nano meter scale radio frequency detectors for telecommunications.  Fig. \ref{diodeSchem} (a) illustrates a semiconductor p-n junction diode.  When current flows from the n to p side the space charge region is enlarged resulting in a high resistance state.  On the other hand when the current flows from the p to n side the space charge region is reduced, producing a low resistance state.  For the MTJ based spin-transfer torque diode shown in Fig. \ref{diodeSchem} (b) the varying resistance states are produced by the TMR of an MTJ.  For a positive current the angle between the fixed and free layers is maximized, resulting in a high resistance state, while for a negative current the angle is minimized, producing a low resistance state.  Due to this oscillating resistance a rectified voltage is produced.

Since the dc voltage generated, and hence the efficiency of microwave rectification, is proportional to the magnetoresistance ratio of the device, an MTJ is logically the best choice for microwave detection and in 2005, using a conventional CoFeB/MGO/CoFeB MTJ, Tulapurkar et al. \cite{Tulapurkar2005} first demonstrated the concept of a spin-transfer torque diode.  However the sensitivity achieved at the time, up to 1.4 mV/mW, was still roughly three orders of magnitude lower than commercially available semiconductor Schottky diode detectors (3 800 mV/mW).  More recently, using an improved understanding of spin dynamics in such devices to improve device fabrication, highly sensitive nanoscale spin-transfer torque diodes with sensitivity up to 75 400 mV/mW have been developed \cite{Cheng2013, Miwa2014, Fan2014a, Fang2016}.  This order of magnitude sensitivity enhancement over semiconductor devices, while still meeting the required functionalities, indicates the exciting potential of spintronic and spin rectification based devices for microwave sensing.  One downside to MTJ based sensing is the complex device fabrication required.  In this regard the recently realized GMR based diode effect may prove advantageous \cite{Kleinlein2014, Zietek2015}.

\begin{figure}[!b]
\centering
\includegraphics[width=8cm]{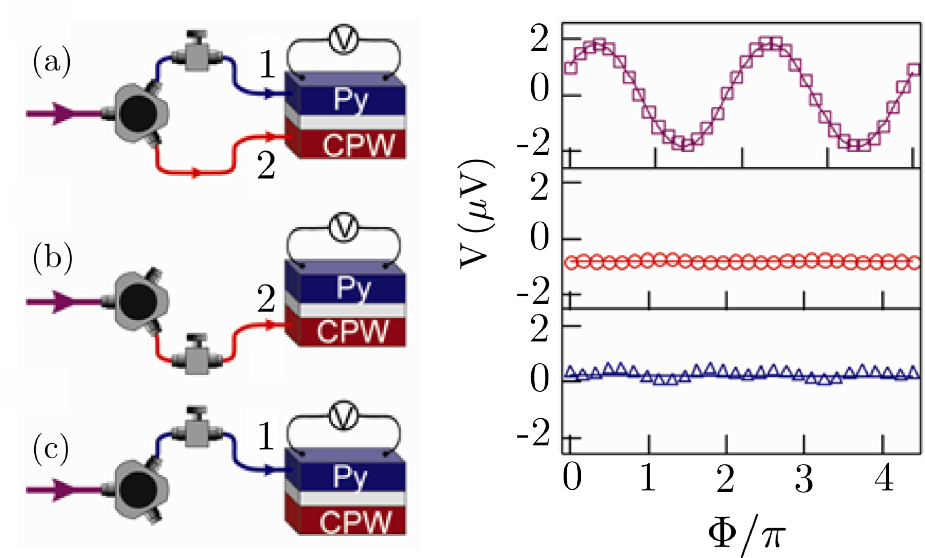}
\caption{\footnotesize{The rectification voltage measured by adjusting an rf phase shifter inserted into (a) one of the two microwave paths, and (b) only one microwave path which is connected to either the CPW or (c) the Py strip of a spin dynamo.  The strong sinusoidal dependence of $V\left(\Phi\right)$ is observed in (a) when both paths are coupled in the spintronic device, demonstrating the controllable influence of the relative electromagnetic phase on the rectified voltage.  $Source:$ Adapted from Ref. \cite{Wirthmann2010}.}}
\label{smi}
\end{figure}

Microwave detection also has important applications for novel imaging techniques.  A key advantage of microwave imaging compared to conventional optical imaging is the increased penetration depth of microwaves, enabling imaging behind barriers/obstacles and the potential to ``see the invisible".  However unlike optical imaging there are no microwave lenses, which necessitates the use of both \textit{amplitude} and \textit{phase} information in order to produce accurate images.  The simplest way to obtain the required phase information is through an interferometry based technique, known as spintronic Michelson interferometry \cite{Wirthmann2010}.  The basis of this technique is illustrated in Fig. \ref{smi}.  By using two coherently split microwave paths to drive the current and field, respectively, in a spintronic device, any change in path length, due to either a purposely introduced phase shift or changes in dielectric path length introduced by obstacles, can be observed.  The spintronic device used as a sensor can be either a spin dynamo \cite{Wirthmann2010} or an MTJ \cite{Fan2014a, Yao2014}.  Other techniques, such as XMCD, have also been successful in phase resolved measurements \cite{Guan2007, Martin2008, Arena2009, Bailey2013}, however an important advantage of spintronic Michelson interferometry is its simplicity.  When compared to other microwave imaging techniques spintronic Michelson interferometry is inexpensive and avoids complex reconstruction algorithms.

As illustrated in Fig. \ref{smi}, the rectified voltage in the spintronic Michelson interferometer depends directly on the phase difference between the CPW and Py signals.  Since microwaves have the ability to penetrate dielectric objects, and interact with their internal structures and defects, this means that an object inserted into path-2 will change the microwave phase and thus influence the measured voltage.  Since microwaves are non ionizing and, at the power levels required for imaging, have not been shown to cause any long term damage to human tissue, microwave imaging techniques have great potential potential for use in medical imaging technologies, particularly for breast cancer imaging due to the good dielectric contrast of breast tissue \cite{Fear2002, Fear2003, Amineh2011, Nikolova2011}.  Additionally the fact that working in the near field allows the Abbe diffraction limit to be overcome \cite{Ash1972}, means that microwaves have the capability to detect structures even smaller than their $\sim$ cm wavelength.

Following the proof of concept for such near-field spin rectification based microwave imaging systems \cite{Wirthmann2010, Zhu2010a, Cao2012a} these techniques have been used for a variety of imaging purposes, including breast cancer \cite{Fu2014} and ground penetrating radar detection \cite{Yao2013}.  
In addition to the actual imaging of objects, rectification based sensors have been used in a variety of other contexts.  For example, it has been demonstrated that spin rectification can be used as an effective method to detect the full $h$ field vector polarization \cite{Bai2008, Wang2013}.  Phase resolved techniques have also enabled the identification of non resonant rectification \cite{Zhu2011c}, and have been used to develop non destructive dielectric characterization techniques \cite{Zhu2011b}.  Also, imaging has been performed using Seebeck induced rectification \cite{Zhang2012}  in MTJs \cite{Fu2012, Gui2014}.

%%%%%%%%%%%%%%%%%%%%%%%%%%%%%%%%%%%%%%%%%%%%%%%%%%%%%%%%%%%%%%%%%%%%%%%%%%%%%%%%%%%%%%%%%%%%%%%%%%%%%%%%%%%%%%%%%%%%%%%%%%%%%%%%%%%%%%%%%%%%%%%%%%%%%%%%%%%%%%%%%%%%%%%%%%%%%%%%%%%%%%%%%%%%%%%%%%%%%%%%%%%%%%%%%%%%%%%%%%%

\subsection{Spintronic Rectifier for Electromagnetic Energy Harvesting}

In the move towards renewable energy a multifaceted approach is necessary to provide both sufficient energy levels and sustainability.  Well known green energy techniques such as solar, wind, geothermal and ocean energy are of course important contributors to this vision, however with the ubiquitous presence of WiFi and cell phone signals, the use of energy harvesting from ambient electromagnetic fields is also poised to be a key contributor in future green energy technologies.  The feasibility of energy harvesting relies on the efficient operation of rectifiers, which use a nonlinear voltage-current relationship to convert microwave frequency fields into a dc power source.  While there are a variety of potential devices, such as the Schottky diode, semiconductor tunnel junction, metal-insulator-metal (MIM) diode and spin diode based on a MTJ, the spin diode is believe to have the most long term potential \cite{Hemour2014}.  

\begin{figure}[!ht]
\centering
\includegraphics[width=8cm]{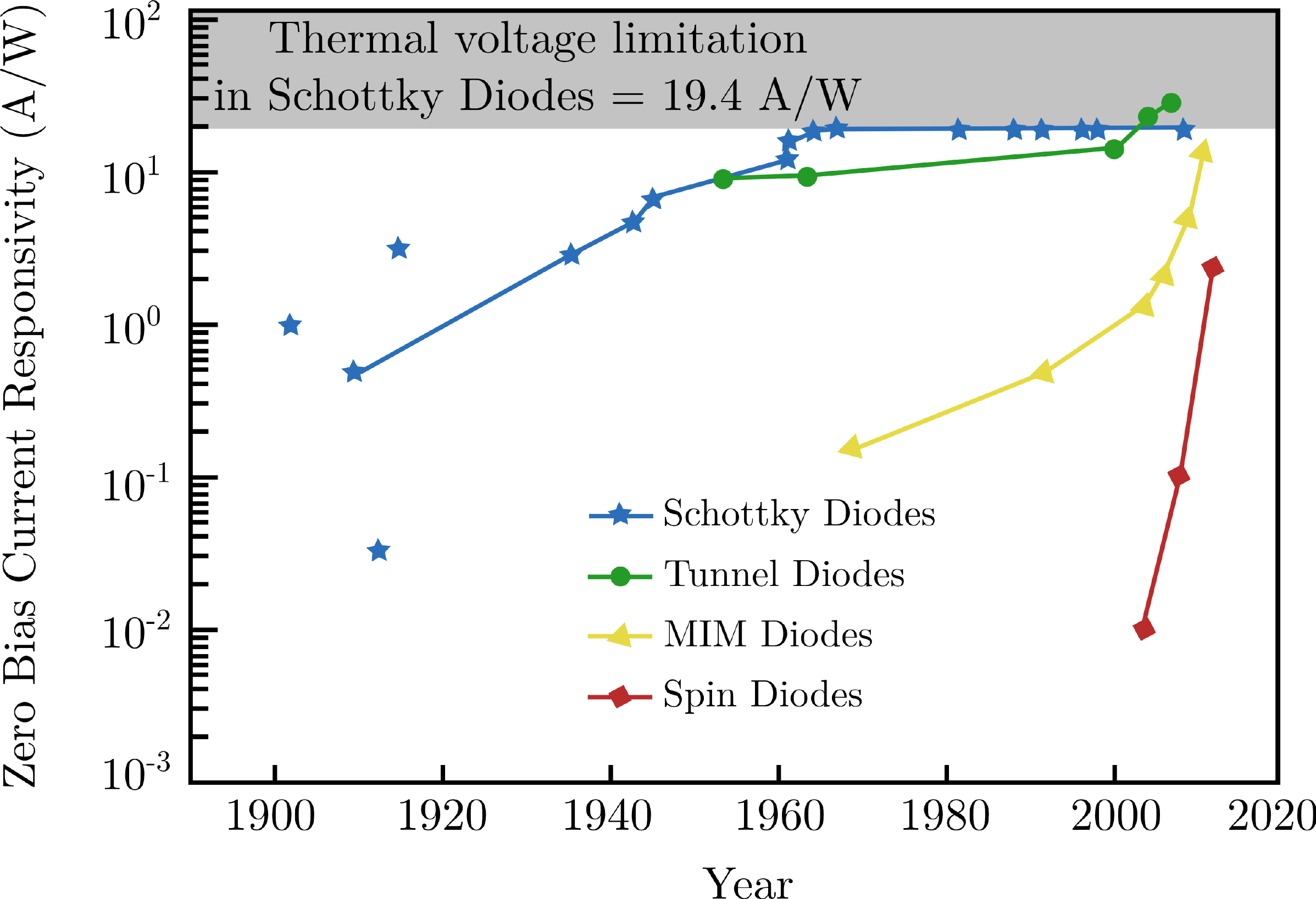}
\caption{\footnotesize{Improvement of the rectification capabilities of various diode technologies over the years.  $Source:$ Adapted from Ref. \cite{Hemour2014}.}}
\label{fig:enharv}
\end{figure}

Undoubtedly the Schottky diode will play an important role in the next decade, as it is currently the most commonly used device in rectifiers, with the most matured technology.  Unfortunately, as summarized in Fig. \ref{fig:enharv}, the Schottky diode already reached its maximum responsivity of 19.4 A/W (at 300 K) in the 1970s, with fundamental limitations set by the laws of thermionic emission which govern the electric current flow.  In contrast, the current flow due to quantum tunnelling in MTJs, created by separating two conductors with a very thin insulator, are mainly limited by the barrier thickness, and still have much room for growth.  Although the MTJ fabrication process is very complex, it has benefited from the push for magnetoresistive random-access memory (MRAM) motivated by the huge consumer electronics market.  In the coming years this push will likely result in a maturation of fabrication techniques analogous to that experienced by Schottky diodes in the mid 20$^\text{th}$ century.  As a consequence, spin diode technology represents the most promising device platform in the development of effective ambient energy harvesting technology \cite{Hemour2014}.  

Fig. \ref{fig:enharv} shows the evolution of the responsivity in different rectifier devices.  Despite the fact that the first realization of spin diodes in 2005 achieved only 0.01 A/W responsivity at zero current bias \cite{Tulapurkar2005}, by 2014 this value had been improved by two orders of magnitude to 2.1 A/W \cite{Miwa2014}.  Just as semiconductor tunnel junctions have already surpassed the Schottky diode limitations, the trend in Fig. \ref{fig:enharv} suggests that in the near future spin diodes may do the same.  Although it is still too early to conclusively predict which kind of tunnelling device will dominate future recitifier technology, certainly the performance and feasibility of microwave energy harvesting devices will greatly benefit from this technological competition.

%%%%%%%%%%%%%%%%%%%%%%%%%%%%%%%%%%%%%%%%%%%%%%%%%%%%%%%%%%%%%%%%%%%%%%%%%%%%%%%%%%%%%%%%%%%%%%%%%%%%%%%%%%%%%%%%%%%%%%%%%%%%%%%%%%%%%%%%%%%%%%%%%%%%%%%%%%%%%%%%%%%%%%%%%%%%%%%%%%%%%%%%%%%%%%%%%%%%%%%%%%%%%%%%%%%%%%%%%%%

\section{Summary and Outlook: Three Generations of FMR Spectroscopy} \label{sec:conclusions}

In this review we have examined the role of electrical detection in the study of magnetization dynamics.  We saw that, due to spin rectification, which produces a dc voltage via the nonlinear coupling of a dynamic magnetoresistance with a dynamic microwave current, the magnetization dynamics of magnetic materials, including metals, semiconductors and even insulators, can be probed through electrical detection.  Although the examination of rectification spectra may be complicated by extraneous effects, such as spin pumping, well established techniques such as angular 
\begin{figure}[!ht]
\centering
\includegraphics[width=8cm]{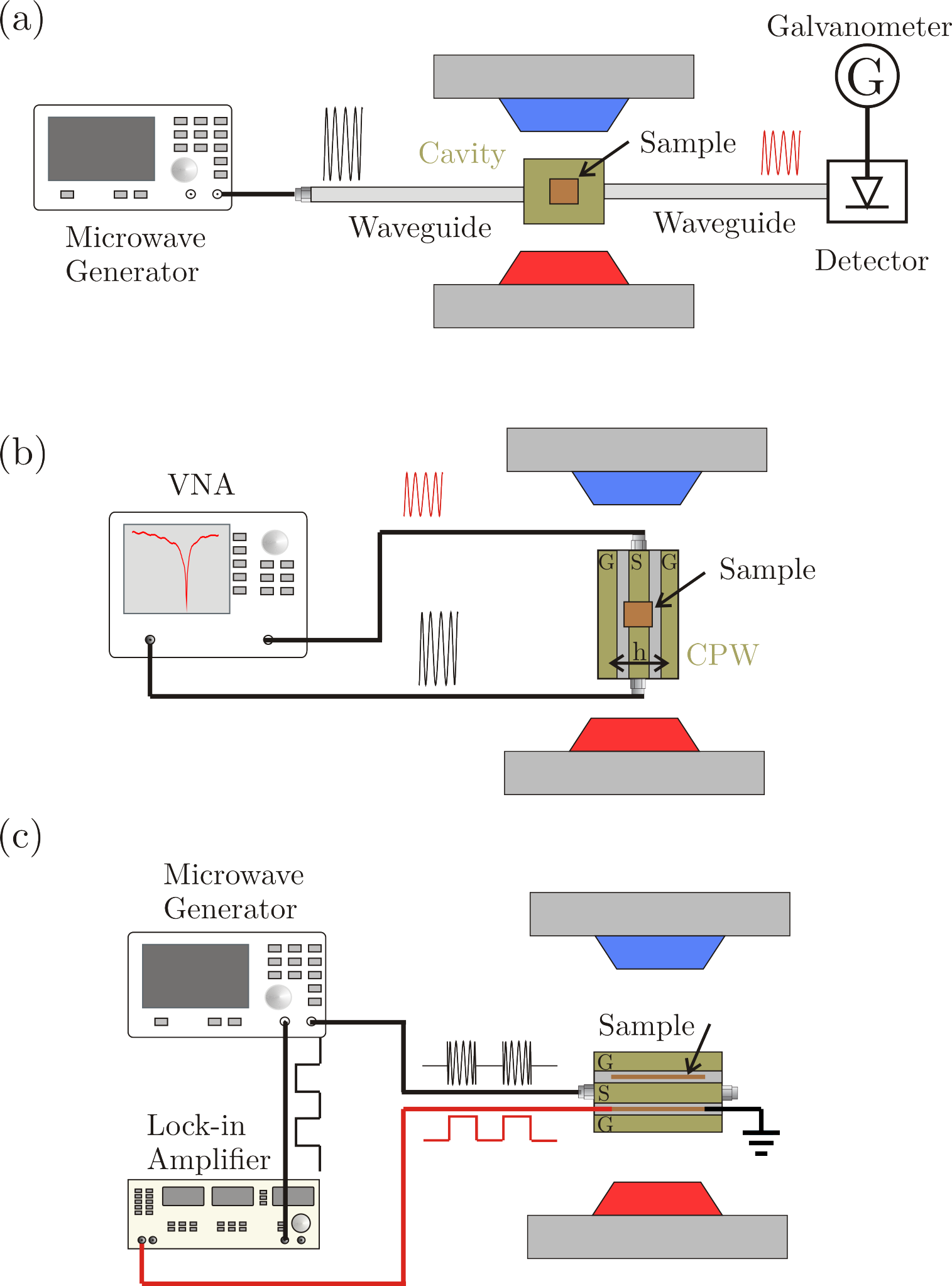}
\caption{\footnotesize{The three generations of FMR spectroscopy techniques.  (a) Early investigations of spin dynamics were based on ferromagnetic resonance cavities.  The microwave signal through the cavity was detected electrically using a rectifying diode.  (b) The second generation of FMR spectroscopy arose with the invention of the CPW and VNA.  The CPW provides an easy way to generate localized, on chip microwave fields, while a VNA can directly measure microwave transmission/reflection properties. (c) Finally, the third and most recent generation is based on electrical detection of spin rectification using lock-in techniques.}}
\label{conclusionFig}
\end{figure}
and line shape analyses enable a wealth of information to be extracted from the rectified voltage, making such techniques powerful for both fundamental physics studies and novel spintronic applications.  

From a historical perspective, electrical detection represent the third generation in a long history of FMR spectroscopy, dating back to the early 20$^{th}$ century \cite{Arkadyev1912, Griffiths1946}.  The first generation of ferromagnetic resonance spectroscopy made use of microwave cavities, typically waveguides, as shown in Fig. \ref{conclusionFig} (a).  With the ferromagnetic sample placed at the position of strongest microwave magnetic field (usually the centre of the cavity) the microwave signal through the cavity can be detected using a rectifying semiconductor diode.  This technology was commercially developed by the German company Bruker and is actually most widely used today in electron paramagnetic resonance studies, although it is also a powerful tool in studies of magnetization dynamics in bulk ferromagnetic materials and films \cite{Heinrich2003}.  However, since microwave cavities work in a very narrow frequency range, measurements of the dispersion relation are difficult and time consuming.  Additionally as the absorption of microwaves is proportional to the volume of ferromagnetic materials, the technology is not suitable for the study of micro- and nano-structured samples, due to extremely weak signals.

As a result of advances in microwave transmission and detection technology, the second generation of FMR spectroscopy was born, combining the broadband capabilities of vector network analyzers and coplanar waveguides.  In 1969, Cheng Wen invented the CPW at RCA's Sarnoff Laboratories in New Jersey \cite{Wen1969} (interestingly CPW is also the acronym of his English name, Cheng P. Wen).  This invention compactified traditional three-dimensional microwave devices into two-dimensions while still maintaining high performance.  This not only greatly reduced the size of microwave devices but also allowed the on chip integration of microwave devices and other electronics.  The invention of the CPW, which allowed improved microwave transmission, was complemented by the development of the scalar network analyzer by Hewlett-Packard in 1973, which enabled the direct measurement of microwave amplitudes over a frequency range of 15 MHz to 18 GHz.  The rapid development of  the telecommunications industry at the end of the 20$^{th}$ century further demanded improved high-frequency measurement systems, leading to the commercialization of the modern vector network analyzer, which can not only measure the amplitude but also the phase of broadband microwaves.   As shown in Fig. \ref{conclusionFig} (b), the second generation of broadband ferromagnetic resonance measurement technology combines the CPW and VNA \cite{Giesen2007, Kuhlmann2012}.   The magnetic sample is placed on the signal (S) strip of a GSG CPW and the microwave field carried by the CPW drives FMR precession.  By fixing the magnetic field the S-parameter can be measured by the VNA as a function of microwave frequency producing an FMR spectra.  This technique is widely used to study thin magnetic films, especially ultra-thin films \cite{Grenda2013}.  In addition, a related technique combining a modulated magnetic field and broadband stripline with pulsed inductive microwave magnetometry (PIMM) has been developed in some laboratories to measure FMR \cite{Silva1999, Kalarickal2006}.

The first two generations of FMR spectroscopy indirectly measure the energy loss due to FMR, and continue to be used as powerful tools to study magnetization dynamics.  However due to the inevitable energy loss in a transmission line, which is very sensitive to the microwave frequency as well as the measurement environment, these techniques require careful calibration.  Additionally for many applications the expense of a VNA may be prohibitive.   This lead to the development of the third generation of FMR spectroscopy in the beginning of the $21^{st}$ century, electrical detection, as shown in Fig. \ref{conclusionFig} (c). Electrical detection has proven to be a powerful tool in the study of spin excitations in micro- and nano-structured magnetic samples \cite{Tsoi2000, Kiselev2003, Gui2005a, Tulapurkar2005, Costache2006a, Costache2006, Saitoh2006, Sankey2006, Kubota2008, Gui2007, Mecking2007, Yamaguchi2007, Gui2007a, Bedau2007, Sankey2007}.  In this third generation of FMR spectroscopy, microwave fields are still produced by a CPW, however lock-in amplification is used to measure the spin rectification voltage resulting from the spin dynamics of the sample.  Since this technique 1) has a high microwave field density due to the CPW, 2) has a voltage signal which is independent of sample volume, and is instead dependent on sample resistance which can be tailored by careful device fabrication, and 3) uses the highly sensitive measurements of lock-in amplification which can detect voltages down to the several nV scale, electrical detection through spin rectification has been used to study a variety of nanostructured samples from FM nanowires \cite{Yamaguchi2007} to MTJs \cite{Tulapurkar2005}.  In addition, due to its high sensitivity and ease of use, spin rectification has been used to study a wide range of materials -- from FM metals, to FM semiconductors, to FM insulators -- a wide range of physics -- from magnetization dynamics, to spin pumping and the spin Hall effect -- and to develop new approaches to microwave sensing and imaging.  

\begin{figure}[!ht]
\centering
\includegraphics[width=11.5cm]{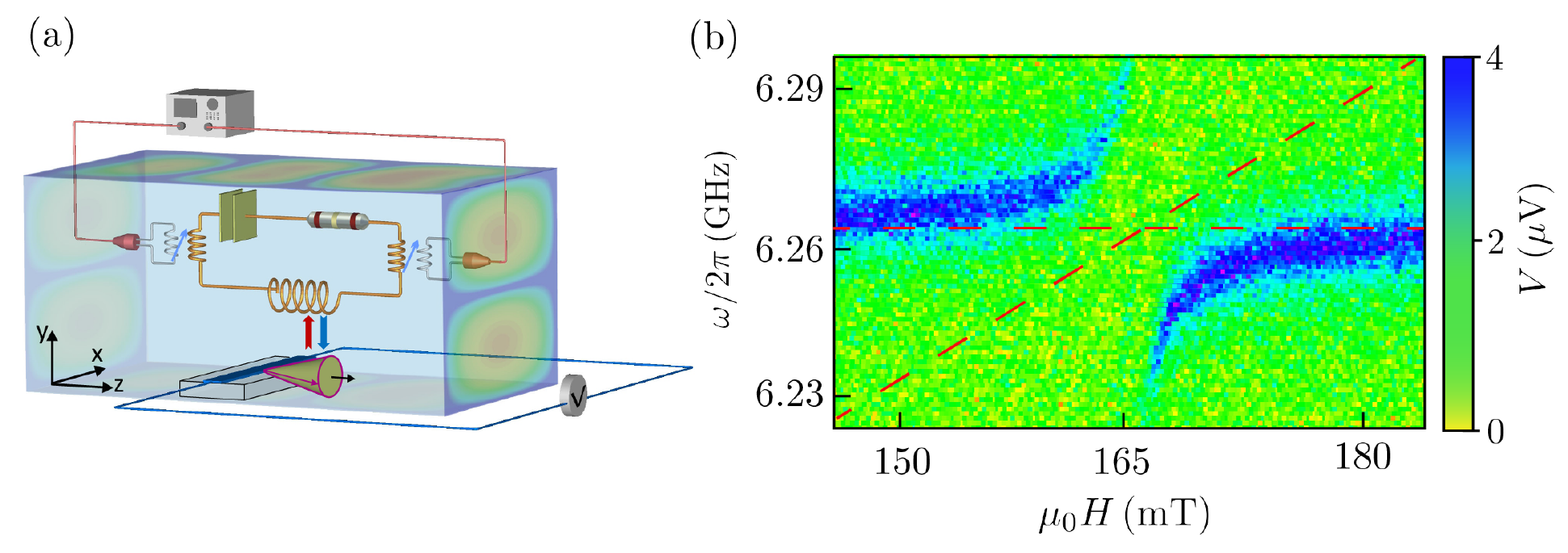}
\caption{\footnotesize{(a) A schematic picture of the cavity spintronic system, consisting of a YIG/Pt bilayer inside of a microwave cavity.  The strong electrodynamic coupling between the magnetization dynamics, governed by the LLG equation, and the electrodynamics, governed by Maxwell's equations, lead to the creation of a new quasiparticle, the cavity-magnon-polariton. (b) The CMP will modify the spin current generated in the YIG/Pt bilayer, allowing characteristic coupling signatures, such as a large avoided crossing, to be electrically detected.  The horizontal and diagonal dashed lines indicate the uncoupled cavity and FMR dispersions respectively.  $Source:$ Panel (a) adapted from Ref. \cite{Bai2015}.}}
\label{cavitySpintronicsFig}
\end{figure}

While at the time of writing, in 2015, the third generation of FMR spectroscopy has become a well established technique for the study of microstructured magnetic devices, a new generation of FMR spectroscopy is just forming.  Known as cavity spintronics \cite{Hu2015}, this emerging field incorporates elements of first and third generation FMR spectroscopy techniques to perform both microwave transmission and electrical detection of high quality ferromagnetic samples in low loss microwave cavities/resonators, and promises to reveal important insights into the physics of strongly coupled spin-photon systems.  As predicted by Soykal and Flatte\'e \cite{Soykal2010, Soykal2010a} high quality YIG samples, combined with high quality microwave cavities can exhibit coherent spin-photon interactions with exceptionally long decoherence times.  Such systems were first studied experimentally by Huebl et al. \cite{Huebl2013a} using microwave transmission and various foundational studies have confirmed key predictions about these strongly coupled systems \cite{Tabuchi2014, Zhang2014}, which can be accurately described by coupling the Maxwell and LLG equations to form a cavity-magnon-polariton (CMP) \cite{Bai2015, Cao2014, Rameshti2015, Yao2015, Harder2016}.  An exciting possibility for future spintronic technologies is the electrical detection of the CMP \cite{Bai2015, MaierFlaig2016}, which indicates the influence of strong coupling on the spin current in bilayer devices and could enable the coherent control of spin current.  With the theoretical and experimental foundation in place, a flurry of recent applications have been proposed e.g. magnon dark mode memories \cite{Zhang2015g}, magnon qubit coupling \cite{Tabuchi2015, Tabuchi2015b}, and the coupling of optical and microwave frequency regimes through whispering gallery modes \cite{Osada2015, Zhang2015b, Haigh2015b, Bourhill2015}.  On this new frontier of strong magnon-photon coupling the electrical detection of magnetization dynamics is poised to play a key role due to its versatile capabilities as we have reviewed in this article. In this perspective, the twilight of the fourth generation of FMR spectroscopy, with its focus on cavity spintronics and quantum magnetism, is on the horizon.

\section*{Acknowledgements}
This review article has been based on a decade of collaboration and communication with many colleagues as identified in the reference section. We thank all members and alumni of the Dynamic Spintronics Group at the University of Manitoba, especially N. Mecking, A. Wirthmann, X.L. Fan, L.H. Bai, Y. Hou, B.M. Yao, L. Fu, Z.H. Zhang, and P. Hyde for their contributions. We are also thankful for the encouragement and/or feedback provided during the public screening of our preprint by W.E. Bailey, G.E.W. Bauer, J.W. Cai, C.L. Chien, M. Costache, V. Flovik, B. Heinrich, T. Jungwirth, M. Kl\"aui, O. Klein, A. Kovalev, I. Krivorotov, G.J. Schmidt, D. Stancil, T. Stobiecki, Y. Tserkovnyak, Y.Z. Wu, J.Q. Xiao, H. Yang, S. Yuasa, and W.X. Zhang. Specially, the authors would like to thank the pioneer of this field, Hellmut Juretschke at the age of 92, for surprising us by sending an inspiring note. This work was supported by NSERC, CFI, CBCF, UofM, and NSFC (No. 11429401) Grants. 

\newpage
\section*{References}
\bibliography{mainText.bbl}
%\bibliography{/users/michaelharder/desktop/bibliography/library.bib}

\end{document}